\documentclass[%
preprint,
superscriptaddress,
amsmath,amssymb,
longbibliography,
]{revtex4-1}
\usepackage{amsmath}
\usepackage{amssymb}
\usepackage{makecell}
\usepackage{graphicx}
\usepackage{braket}
\usepackage{color}
\usepackage{xcolor}
\usepackage{siunitx}
\usepackage{upgreek}
\usepackage[utf8]{inputenc}
\usepackage{physics}
\usepackage{multirow}
\usepackage{array}
\usepackage{url}
\usepackage{dcolumn}
\usepackage{bm}
\usepackage{hyperref}
\usepackage{natbib}
\usepackage{gensymb}
\usepackage{placeins}
\usepackage{xcolor}
\usepackage{adjustbox}
\usepackage{soul}
\usepackage{float}
\usepackage[T1]{fontenc}
\begin{document}



\title{Defects in hexagonal boron nitride for quantum technologies}


\author{Tobias Vogl}
\affiliation{Department of Computer Engineering, TUM School of Computation,
Information and Technology, Technical University of Munich, 80333 Munich, Germany}

\author{Viktor Ivády}
\affiliation{Department of Physics of Complex Systems, Eötvös Loránd University, Egyetem tér 1-3, H-1053 Budapest, Hungary}
\affiliation{MTA–ELTE Lend\"{u}let "Momentum" NewQubit Research Group, Pázmány Péter, Sétány 1/A, 1117 Budapest, Hungary}

\author{Isaac J. Luxmoore}
\affiliation{Department of Engineering, University of Exeter, Exeter, EX4 4QF, UK}

\author{Hannah L. Stern}
\affiliation{Department of Materials, University of Oxford, OX1 3PH, UK}




\begin{abstract}
Atomic defects in solid-state materials are building blocks for future quantum technologies, such as quantum communication networks, computers, and sensors. Until recently, a handful of defects in a small selection of host materials have been possible candidates. Recent developments have revealed that hexagonal boron nitride, a wide-bandgap two-dimensional material, hosts single-photon-emitting atomic defects with access to optically addressable electronic and nuclear spins at room temperature. Now, atomically thin quantum devices that operate at ambient conditions are a possibility. In this perspective, we discuss the recent progress, and challenges, in understanding the fundamental photophysics of defects in hBN, as well as specific opportunities they present for the development of quantum technologies.

\end{abstract}

\maketitle


\section{Introduction}

A major challenge in the realisation of widespread quantum technologies is the development of material platforms where quantum states can be reliably accessed and controlled. A range of systems are being explored, including trapped atoms/ions, quantum dots, molecules and atomic scale impurities in solids, each with their own benefits and drawbacks \cite{Becher2023}. Of these, it is impurities, or point defects, in solid-state materials that offer access to quantum coherent states at room temperature and are highly compatible with solid-state device engineering \cite{Awschalom2018,Atature,Wolfowicz2021}. While point defects are ubiquitous to extended solids, it is isolated point defects in wide-bandgap host materials with quantised electronic transitions, at optical and microwave frequencies, that can be used to implement qubits. Single-photon emitting defects provide photonic qubits for quantum computing \cite{O'Brien2007} and quantum communication protocols, such as quantum cryptography \cite{Hanson2008,Couteau2023}. Whereas defects with optically addressable electronic and nuclear spins fulfill many of the DiVicenzo criteria for quantum computing \cite{DiVincenzo2000}, and are building blocks for quantum sensors \cite{Schirhagl2014,Vaidya2023}, and repeaters and memories in quantum optical networks \cite{Atature,Kimble2008}. 

Until recently, remarkably few defects in a small range of host materials have been intensively pursued for quantum information technologies. Most focus has been on diamond, where the large bandgap (5.5 eV) and spin-free carbon lattice provides a near-ideal host for defects. Research over decades on the negatively charged nitrogen vacancy centre (NV$^{-}$) \cite{Doherty2013} has resulted in numerous influential reports, including entanglement between an optical photon and the electronic spin of the defect \cite{Togan2010}, entanglement of the multiple electronic spins \cite{Bernien2013,Humphreys2018,Pompili2021}, quantum control over individual nuclear spins, extending to demonstration of multi-qubit quantum registers \cite{Dutt,Taminiau2014,Bradley2019}, and quantum microscopy using the spin of the NV$^{-}$ centre as a nanoscale sensor \cite{Chernobrod2005,Taylor2008,Degen2008,Degen2017}. These demonstrations have propelled defects to the forefront of solid-state platforms for quantum applications. However, scalability is critical if lab-based demonstrations are to be translated to a new technologies. As such, focus is increasingly turning to the identification and development of defects in other materials, such as silicon carbide (SiC) \cite{Koehl2011,Bourassa2020}, silicon \cite{Redjem2020,Higginbottom2022, Simmons2024,Durand2024_silicon}, and galium nitride (GaN) \cite{Luo2024}, as these materials may offer easier defect integration to solid-state devices than diamond. 

Over the past ten years, it has been discovered that hexagonal boron nitride (hBN), a wide-bandgap two-dimensional (2D) material, hosts optically- and spin-active defects \cite{Tran2016a,Aharonovich2016, Gottscholl2020,Chejanovsky2021,Stern2022,Stern2024}. Unlike 3D counterparts, hBN offers the potential for easier and more scalable integration to solid-state devices. Over the past 15 years the huge interest in 2D material electronics has meant the growth, transfer and device fabrication capabilities with hBN have become well established. hBN can be grown epitaxially at wafer scale, on a range of substrates, with well-controlled thickness (from monolayer to 100's nm) \cite{Kim2015,Chugh2018,Lee2018, Wang2019,Wang2024}. hBN can be transferred to a range of substrates and other (bulk and 2D materials) \cite{Novoselov2016, Fukamachi2023}, meaning that hBN layers can be integrated with on-chip optical components such as waveguides \cite{Li2021}, cavities \cite{Kim2018,Vogl2019, Frch2021} and nanoantennas \cite{PalomboBlascetta2020}. Moreover, electrical devices can be built around defect-active hBN layers to enable Stark tuning \cite{Noh2018}, charge control \cite{Akbari2022} and electrical excitation of defects \cite{Zhigulin2025}. 

These attractive capabilities are only relevant if hBN defects possess competitive magneto-optical properties. It is our view that this is now the case. To date, defects in hBN show a mix of promising optical properties, even at room temperature, including high zero-phonon line (ZPL) fraction (as high as 0.8) \cite{Tran2016a, Jungwirth2017, Ari2025}, high brightness (MHz count rates) \cite{Grosso2017,Liu2020,Guo2023} and high single-photon purity \cite{Kumar2024}. Subclasses of optically active hBN defects possess optically addressable electronic and nuclear spins (on the ensemble and single-defect level) that can be coherently controlled, with microsecond coherence times at room temperature \cite{Gottscholl2020, Gottscholl2021a,Exarhos2019,Stern2022,Chejanovsky2021,Ivady2020, Stern2024, Gong2024, Scholten2024, Gao2025}. These recent discoveries provide a platform to explore new defect physics and device engineering. However, research into hBN defects is still in its infancy and faces significant fundamental and practical challenges before it can be regarded as a leading solid-state platform for quantum technologies. This perspective discusses the current progress and future challenges, experimental and theoretical, facing the next phase of development of hBN defects for optical and spin-based quantum technologies.


\section{Quantum optical technologies}

\begin{figure}[t]
  \includegraphics[width=\textwidth]{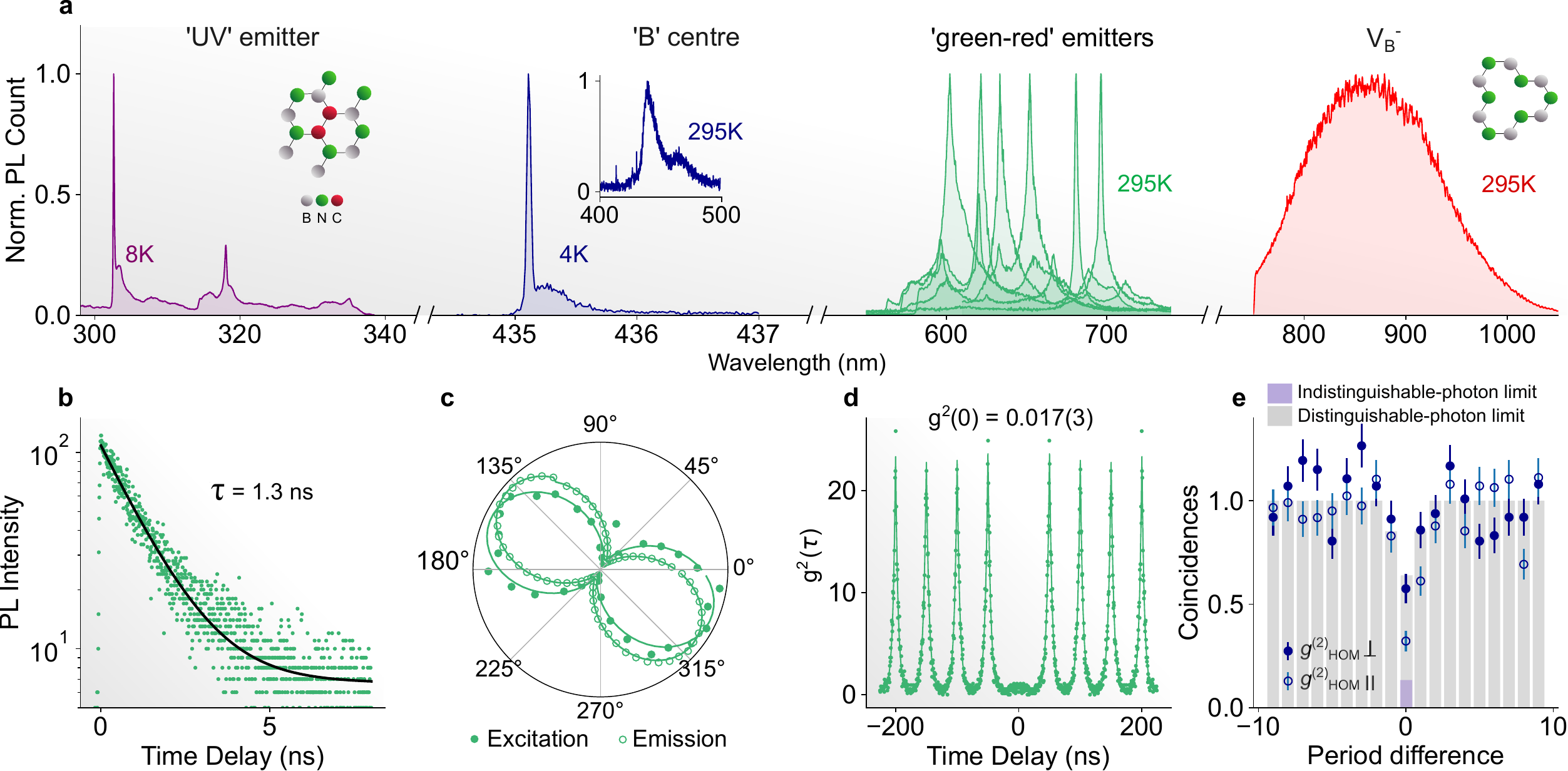}
  \caption{
  \label{Fig::fig1}}
\end{figure}

Single-photon-emitting defects are quantum light sources that provide a source of photonic qubits. An ideal on-demand single-photon-emitting defect emits strictly one photon at a time (single-photon purity) into a given spatiotemporal mode, with every emitted photon being identical (indistinguishable). Such systems are either required, or beneficial for, the development of quantum computing schemes, quantum simulation and quantum communication \cite{Aharonovich2016}. In this section we address the progress on applying hBN defects to two specific technologies of different levels of maturity: quantum key distribution (QKD) and quantum computing. QKD is a form of quantum communication where information is encoded in single photon states of light, which allows one to detect and quantify any eavesdropping attempt \cite{Gisin2002}. It is already a commercially available technology however currently typically uses attenuated classical light sources. The use of quantum light sources for QKD would enhance the secure data rate, but sources with high single-photon purity are strictly required. On the other hand, photonic quantum computing schemes utilise entanglement of single photons  to perform logic gates and demands defects that can emit high rates of indistinguishable photons \cite{O'Brien2007}.

hBN presents a new material system to explore for both technologies. Defect emission has been observed in hBN ranging from the UV to near-IR, and covering the entire spectral range in between \cite{Bourrellier2016, Tran2016a, Tran2016b, Camphausen2020, Gottscholl2020, Gale2022, Kumar2023} (Figure~\ref{Fig::fig1}a). With the exception of the negatively charged boron vacancy (V$_{B}^{-}$), the consensus is building that the majority of these defects are related to carbon substitutions in the hBN lattice, but  differentiation of their atomic structures is still under intense theoretical and experimental investigation (see Section 4). In this article, we broadly classify hBN defects by the energy of their ZPL (Figure \ref{Fig::fig1}a) (in Section 3 by their spin signatures). Important groups include the UV emitter (ZPL = ~302 nm) \cite{Bourrellier2016, Vuong2016,Li2022}, the B-centre (ZPL = 437 nm) \cite{zhigulin_photophysics_2023,Yamamura2024} and the emitters that span the green-red spectral region (ZPL = 550-700 nm) \cite{Tran2016a, Jungwirth2017, Mendelson2021,Kumar2023}, all of which are bright, single-photon emitters. For the green-red group, magnetic resonance studies indicate that multiple different atomic defects emit in this region (see Section 3), while the UV and B-centres are associated with a single defect structure (proposed to be a carbon dimer \cite{Plo2025} and carbon tetramer respectively \cite{Maciaszek2024}). In contrast, the V$_{B}^{-}$ is a weakly emissive defect without an observable ZPL which is measured on the ensemble level, making it unsuitable for quantum information processing (Figure~\ref{Fig::fig1}a).

\subsection{Quantum key distribution}

\subsubsection{Single-Photon Purity}

The ideal photon source for QKD is one that can produce high rates ($>$100 MHz) of high-purity single photons ideally at ambient conditions, a combination of properties that have so far have only been demonstrated by self-assembled quantum dots at cryogenic temperatures. A high single-photon purity means a high quantum efficiency (QE) combined with a low multi-photon probability (ie., a high overlap with the single photon Fock state). The overall QE of a defect is determined both by the internal QE and the ease of photon out-coupling, a challenge for defects in high-refractive index materials. For the green-red hBN defects, the sub-ns to ns PL lifetimes translate to $>$ GHz photon emission rates (Figure~\ref{Fig::fig1}b). In addition, the internal QE can reach 40-80\% \cite{Nikolay2019,Vogl2019,Yamamura2024}  and photon out-coupling and collection can approach 100\% due to the relatively low refractive index combined with the in-plane dipole of defects (Figure~\ref{Fig::fig1}c) \cite{Kumar2024}.  

The multi-photon emission probability of a defect is determined via second-order intensity correlation measurement, or $g^{(2)}(t)$. The reported $g^{(2)}(0)$ values for hBN defects vary significantly \cite{Tran2016a, Bommer2019, Grosso2017}. This may be affected by the wide range of setups and background correction methods used to conduct this experiment, as well as how clean the sample is in terms of background emission (noise). Notably, $g^{(2)}(0)$ of $<$ 0.015 have been reported across for the green-red emitters \cite{Grosso2017, Chatterjee2025} and polarisation-resolved measurements have improved $g^{(2)}(0)$ down to \textless 0.02 into free space at room temperature and without any background correction \cite{Kumar2024} (Figure~\ref{Fig::fig1}d). 

\subsubsection{System Efficiency}

 Quantum key distribution is currently almost exclusively performed using weak laser pulses, as they can be generated with a low technical complexity and with low cost. The most efficient protocols use weak laser pulses with a varying intensity in so-called decoy protocols \cite{Dong2023}. To be competitive with decoy protocols, the system efficiency of a quantum emitter (the probability that a photon is generated into the mode of the collection optics when a photon emission is triggered) should be similar to the mean photon number in the decoy protocol, typically around 0.4-0.5 \cite{Li2022b}. To achieve this with defect systems, high quantum efficiency, low internal optical losses and low $g^{(2)}(0)$ values ($g^{(2)}(0)<10^{-2}$) are required. Initial demonstrations of QKD with green-red emitting hBN defects show lower secret key rate (the ultimate performance metric in QKD) of the hBN system compared to state-of-the-art laser-based systems \cite{Samaner2022,Al-Juboori2023}. However, the quantum efficiency, emission mode directionality and $g^{(2)}(0)$ for hBN defects can be improved via resonant cavity systems  such that a system efficiency of the single-photon source exceeds 50\% \cite{Vogl2019,Häußler2021,Vogl2021}. Problematic, even for a cavity-enhanced system is if the non-radiative decay happens via a pathway with a long-lived dark state. In this case, the singe photon source (and therefore the QKD system) would be out-of-operation for that time and not generate any signal. Further QKD demonstrations should implement improved optical cavities \cite{Vogl2019} and cleaner crystals with less scattering \cite{Häußler2021}, both of which are feasible to enhance the QKD performance. Numerical simulations have shown that the optical properties of cavity-coupled hBN emitters is theoretically sufficient to outperform laser-based protocols on short and medium distances (low-medium loss channels), eg., up to 1000 km in a space-to-ground quantum channel, while the latter become more efficient for very long distances (high loss) if one assumes similar link parameters in terms of repetition rate (typically 100's MHz- GHz) and channel loss which are compatible for most of the hBN defects \cite{Vogl2019,Abasifard2024}. 

Despite relatively few reports of microcavity-enhanced demonstrations with hBN single-photon emitting defects (small Purcell factors 1-4) \cite{Vogl2019,Häußler2021}, the 2D host provides advantages for cavity coupling over defects in 3D crystals. Specifically, the in-plane dipole simplifies the alignment with rotation symmetric cavities and 2D material transfer should be beneficial for cavity fabrication.

\subsubsection{Emission Wavelength}
 
Besides these fundamental requirements on the single-photon emitter for QKD, the mode of photon propagation, and thus photon wavelength, are also  important considerations. Photons can be transported in QKD protocols via existing optical fibre infrastructure or via free-space. For fiber-based channels, the photon wavelength should be within the telecom C-band (1550 nm) or in the O-band (1330 nm) to minimise transmission losses. This presents a potential restriction for hBN as a platform, as so far most known hBN optical defects emit in the visible region. While telecom emitters in hBN have been theoretically predicted (eg., O$_\text{N}$S$_\text{N}$ is predicted to emit in the telecom O-band \cite{Cholsuk2022}, or by Stark tuning \cite{Dhu-al2024}), there is no experimental demonstration of this yet. The same challenge applies to other host materials which is why there are few reports of defect emitters in the O- \cite{Redjem2020,Wang2020,Higginbottom2023} and C-band \cite{Zhao2021}. Future work with setups that are designed to detect NIR photons, combined with ion implantation to create defects, should be performed to identify new defects compatible with the telecom C-band. Once these emitters have been fabricated, the direct fiber integration has potential to be straightforward due to the 2D geometry \cite{Vogl_2017}. 

QKD can also be carried out via free space \cite{Schmitt-Manderbach2007,Liao2017}. In this case, any transmission window of the atmosphere can be used. Due to the broad emission range of hBN defects, there are many emission wavelengths of hBN coinciding with an atmospheric transmission window. Free-space QKD typically uses photon polarisation as the quantum encoding mechanism, which makes it directly compatible with the linear polarisation of hBN defects without any loss (which would occur if one has to polarise unpolarised light). Unfortunately this requires fast polarisation modulators which are not commercially available. This has provided the key limitation to the speed of QKD experiments with green-red emitting hBN defects (with modulator speeds of 1 MHz \cite{Samaner2022} and 500 kHz \cite{Al-Juboori2023}), where the short excited lifetime on the order of 300 ps \cite{Vogl2018} to 3 ns \cite{Tran2016a} in theory allows for lifetime-limited repetition rates well exceeding 100 MHz. Routes forward should either include faster modulation schemes or the spatial overlap of differently polarised hBN defects that are identical in all other degrees of freedom (spectral shape, lifetime, etc.) \cite{Vest2022}. 

Despite the apparent challenges to deliver free-space QKD, hBN quantum emitters (green-red) have been proposed for daylight QKD schemes \cite{Abasifard2024} and have been qualified for use in space applications \cite{Vogl2019_radiation}, which allows one to use them as quantum light sources for satellite-based QKD \cite{Ahmadi2024}. Success of this daylight QKD scheme will rely on future demonstrations of strain or Stark tunability of the emission wavelength, as specific wavelengths have to be matched within the fraction of a nanometre. Such precise tuning would also allow one to have multiple sources operating at slightly offset emission wavelengths which enables multiplexing and therefore enhances the QKD data rates. To date, several promising demonstrations of such ZPL tuning have been reported. For the green-red emitters, room temperature tuning via Stark (a fraction of an meV) \cite{Noh2018} and strain tuning (several meV) \cite{Grosso2017}. At low temperature, Stark tuning has demonstrated tunability of several GHz, for the B-centre \cite{Zhigulin2023}. In addition, for satellite-based QKD, fast on-demand excitation (that can out-compete the excited state lifetime to prevent re-excitation during one pulse) will be needed to maintain a very low $g^{(2)}(0)$, which is not easy on a satellite. To address this, one can imagine fast and compact, integrated electrical excitation schemes with hBN defects embedded in heterostructure devices, which are currently under development \cite{Grzeszczyk2023,Yu2024}, but with far inferior optical properties compared to optically excited emitters. Due to the insulating nature, this only works with very thin hBN, so the challenge is to maintain a high photostability of the emitters despite the thin host crystal.

\subsection{Optical quantum computing}

\subsubsection{Indistinguishable Photons}

For optical quantum computing, in addition to photon purity, indistinguishable single photons are required to generate entangled cluster states \cite{Raussendorf2001,O'Brien2007}. Indistinguishability here means a high two-photon interference contrast. The number of qubits that can be entangled efficiently scales with this visibility and values exceeding 99\% are required to achieve fault-tolerant quantum computing \cite{Raussendorf2001}. Two-photon interference has been demonstrated for B-centre defects in hBN, even though the experiment was done at cryogenic temperatures (4 K) and a partial visibility of 0.44(11) (ie.\ lower than the ideal value of 1) was achieved \cite{Fournier2023} (Figure~\ref{Fig::fig1}e).  Generally, there have been few reports of optical linewidths approaching the lifetime limit for hBN defects. At room temperature it is hardly imaginable to achieve visibility \textgreater 99\% with the visible hBN single-photon emitters without addressing the mechanisms that give rise to linewidth broadening \cite{White2021}. To date, reports have indicated charge-related spectral wandering and phonon coupling play a dominant roles in optical broadening for hBN defects \cite{Jungwirth2017,White2021,  Akbari2022,Koch2024}. It is reported that mechanical decoupling from phonons can result in Fourier transform-limited linewidths below 100 MHz at room temperature \cite{Hoese2020,Dietrich2020}. It should be mentioned that this has only been reported for a specific defect type (speculated to be an interlayer defect) and as of yet has not been observed for other hBN defects. Moreover, this results in an in-plane emission directionality, which is then difficult to collect with a high efficiency.

Cavity funneling (ie.\ forcing the emitter to emit into the resonnant mode by blocking any off-resonant excitation) can be used to enhance the indistinguishability and directionality simultaneously \cite{Grange2015}, but this requires a cavity linewidth on the order of 100 MHz to achieve a visibility of 90\% \cite{Vogl2019}. A large free spectral range of the cavity is required to block any phonon modes (which means one cannot simply make the cavity longer to reduce the linewidth). Otherwise, one would sample the free space emission spectrum at the free spectral range and only photons from the same peak are indistinguishable and one has to filter out other peaks. This in turn results in a low efficiency (or photon rate) of the single-photon source. This is the reason why there has been no demonstration of quantum computing with hBN, though there have been collective measurements on single qubits performed on single photons from hBN \cite{Conlon2023}.



\section{Spin-based quantum technologies}

\subsection{Spin-active hBN defects}

\begin{figure}[t]
  \includegraphics[width=\textwidth]{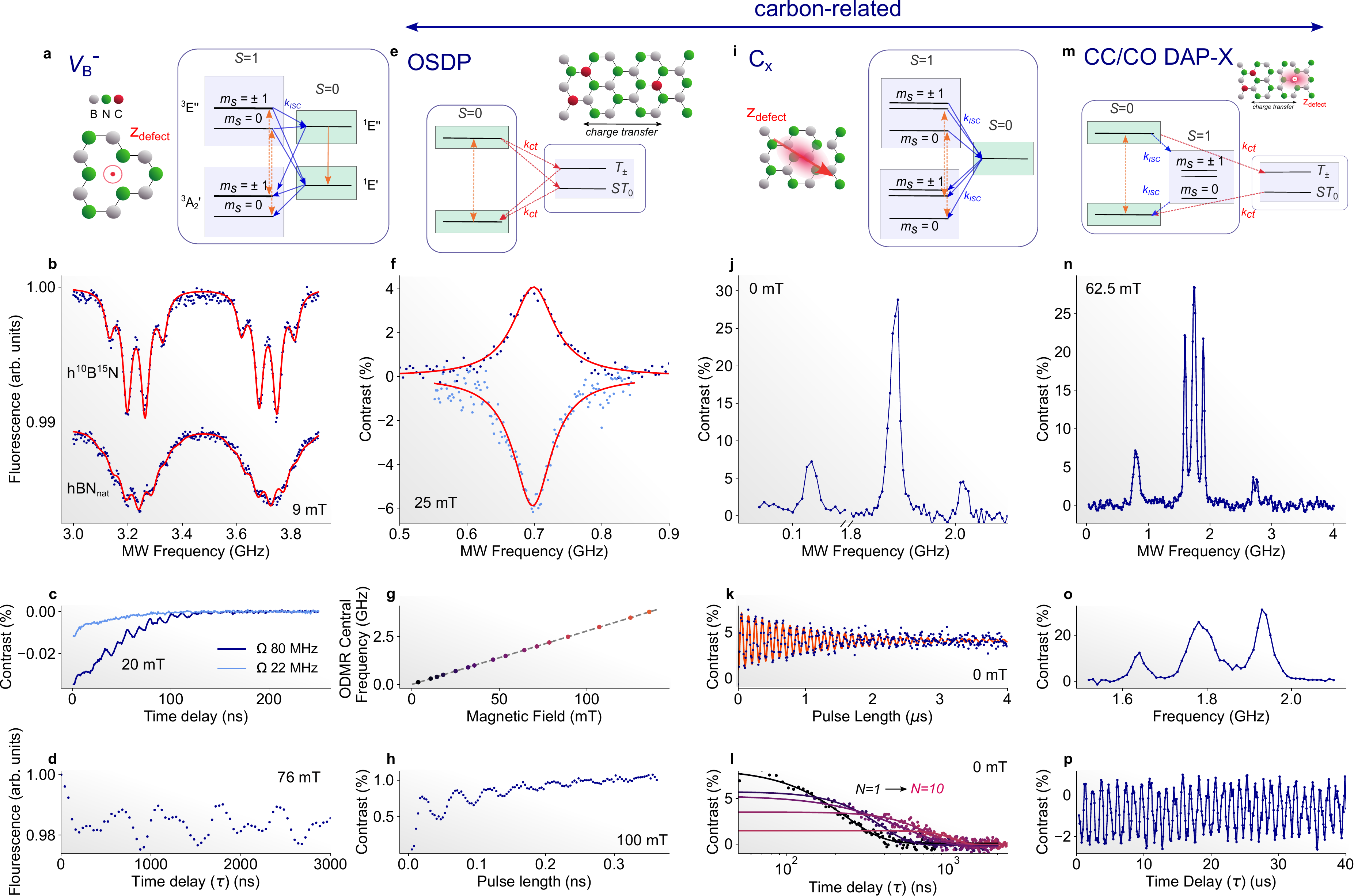}
  \caption{
    \label{Fig::fig2}}
\end{figure}

Defects with optically addressable electronic and nuclear spins are attractive systems for spin-based quantum technologies, such as repeaters for optical networking and quantum sensors \cite{Atature}. Hexagonal boron nitride represents the first 2D material platform where optically addressable spin defects have been identified at room temperature. Since 2020, multiple classes of hBN spin defects have emerged (Figure \ref{Fig::fig2} and Table \ref{tab:spin}). To date, the most studied is the V$_B^-$ \cite{Gottscholl2020,Ivady2020,Gottscholl2021a}, which is created via laser \cite{Gao2020_LaserWritingVB}, ion  \cite{Kianinia2020,Guo2022} or neutron \cite{Gottscholl2020} irradiation. The V$_B^-$ is a radiative ground state spin triplet (\textit{S} = 1) system. In the absence of hyperfine coupling, \textit{S} = 1 spin systems are described by a Hamiltonian of the type:

\begin{align}
    H &= H\textsubscript{ZF} + H\textsubscript{ZE},
    \label{Eq::HamiltonianI}\\
    H\textsubscript{ZF} &= D S_z^2 + E (S_x^2 - S_y^2), \label{Eq::HamiltonianII} \\
    H\textsubscript{Ze} &= \frac{\gamma_e}{2 \pi} \mathbf{B} \cdot \mathbf{S},
    \label{Eq::HamiltonianIII}
\end{align}

\noindent where $H_{\text{ZF}}$ is the zero-field splitting term, $H_{\text{ZE}}$ is the Zeeman term, $D$ and $E$ are the longitudinal and transverse zero-field splitting (ZFS) parameters respectively that define the defect's $x,y,z$ principal axes in units of Hz, $\mathbf{S}$ is the $S=1$ operator with cartesian components $S_{x,y,z}$, $\gamma_e$ is the electron gyromagnetic ratio and $\mathbf{B}$ is the applied magnetic field. For the V$_B^-$ the ground- and excited-state $D$ are $\sim3.5~\mathrm{GHz}$ \cite{Gottscholl2020} and $\sim2.1~\mathrm{GHz}$ \cite{baber2021}, respectively, and the spin quantisation (z) axis is out-of-plane (Figure~\ref{Fig::fig2}a). Similar to the NV$^{-}$ centre, spin-initialisation is achieved through optical pumping. The optically detected magnetic resonance (ODMR) spectra of the V$_B^-$ reveal hyperfine coupling to the nearest three nitrogen nuclei (Figure~\ref{Fig::fig2}b). The room temperature spin relaxation ($T_1$) and Hahn Echo spin coherence ($T_2^{echo}$) timescales are of the order of 10 $\mu$s \cite{Gottscholl2021a} and 100 ns, respectively (Fig. 2c) \cite{Gottscholl2020,Haykal2022,Ramsay2023}. 

The V$_B^-$ has a low quantum efficiency due to nanosecond forward intersystem crossing \cite{Ivady2020}, combined with a slow radiative decay, predicted to be $\sim 10~\mathrm{\mu s}$ \cite{Reimers2020}, but not yet measured experimentally. This has prevented optically addressing single defects, despite attempts using nanocavities \cite{Mendelson2022_Vbcavity,Xu2023_VBcavity}. The reverse intersystem crossing has recently been reported to be around 10 ns \cite{Clua-provost2024}, much shorter than theoretical predictions \cite{Reimers2020}, but which explains why photoluminescence can be measured despite the low quantum efficiency of the triplet state. 

Control of single electronic spins in hBN has been demonstrated with defects that emit in the green-red spectral range, suspected carbon-related defects \cite{Mendelson2021, Chejanovsky2021,Stern2022,Stern2024,Gao2025}. Curiously, a range of different ODMR signatures have been observed for defects that emit in this region, meaning the ODMR itself provides currently the best tool to distinguish different carbon-related defects (Figure~\ref{Fig::fig2}a and Table \ref{tab:spin}). In general, three different species have been observed. The first show no appreciable zero-field splitting \cite{Chejanovsky2021,Stern2022} (Figure~\ref{Fig::fig2}e-h). For these defects, the ODMR is characterised by a central ODMR frequency that follows a g-factor of $\sim$2, a broad linewidth without clear hyperfine structure, and ODMR contrast that can be positive or negative. Coherent control of these spins is difficult \cite{Scholten2024} and $T_1$ is $\sim$ 10 - 15 $\mu$s  (Figure~\ref{Fig::fig2}f)  \cite{Chejanovsky2021,Stern2022,Guo2023,Scholten2024}. ODMR spectra fitting this description have now been observed for wide range of carbon-doped hBN samples \cite{Scholten2025} on both the ensemble and single defect level, and, in some cases, co-existing with \textit{S} = 1 signatures on a single defect \cite{Mendelson2021,Chejanovsky2021,Stern2022,Guo2023,Scholten2024,Robertson2024, Gao2025}. Recently, a model invoking charge transfer between a pair of separate spin defects (dubbed the optical spin defect pair (OSDP) model) in the 2D lattice has been proposed to explain these findings \cite{Robertson2024,Scholten2024}. In this model, charge transfer from an optically excited \textit{S} = 0 defect to a nearby dark \textit{S} = 0 defect produces two weakly coupled \textit{S} = 1/2 defects that can give rise to an ODMR-active metastable state, in a similar way to radical pair molecular systems \cite{Woodward2002} (Figure~\ref{Fig::fig2}e). This proposed mechanism would explain the wide spread of emission wavelengths associated with defects that show these `S -1/2'-like ODMR signatures. If correct, the prevalence of spin pairs in hBN that can give rise to ODMR reveals a unique system where spatial proximity between defects may be engineered to tune ODMR properties, such a zero field splitting parameters or ODMR contrast. 

The other two groups are strongly coupled \textit{S} = 1 systems with GHz zero-field splitting \cite{Stern2024,Gao2025}. The first reported, denoted C$_{x}$ in this article, possess a \textit{S} = 1 ground state with $D$ $\sim$ 1.97 GHz and $E$ $\sim$ 60 MHz, an in-plane quantisation axis and high (up to 90\% reported) ODMR contrast \cite{Stern2024, Gilardoni2025} (Figure~\ref{Fig::fig2}i-l). This defect is formed in metal organic vapour phase epitaxy (MOVPE)-grown hBN multilayers in the presence of a carbon-based precursor (triethyl boron) \cite{Chugh2018}. The narrow and consistent distribution of zero-field splitting parameters measured for this defect indicate a specific atomic structure. Efficient coherent spin control is possible at zero magnetic field, with $T_1$ $\sim$ 100's of $\mu$s and $T_2^{echo}$ $\sim$ 200 ns that can be extended beyond a microsecond via decoupling pulse protocols \cite{Stern2024} (Figure~\ref{Fig::fig2}k,l).  

Distinct, single carbon-related defects with strongly coupled \textit{S} = 1 signatures and an out-of-plane quantisation axis have recently been reported \cite{Gao2025, Whitefield2025} (Figure~\ref{Fig::fig2}m-p). These signatures are observed in hBN crystals that have been irradiated with $^{13}$CO$_2$ and annealed. The ODMR reveals a spin with \textit{D} $\sim$ 1 GHz and \textit{E} $\sim$ 100-400 MHz with similar electronic spin coherence properties as C$_{x}$. The \textit{S} = 1 ODMR resonances are typically accompanied by an \textit{S} = 1/2 ODMR resonance, which reveals clear hyperfine coupling of 100-300 MHz, associated with coupling to a neighbouring $^{13}$C nuclear spin (Figure~\ref{Fig::fig2}n). The coexistence of both \textit{S} = 1 and \textit{S} = 1/2 signatures on the same defect suggests the presence of ODMR is due to both intersystem crossing and charge transfer processes (proposed to be the OSDP mechanism) that take place on the same defect (Figure~\ref{Fig::fig2}m) \cite{Gao2025}. The \textit{S} = 1/2 ODMR lines are assigned to a carbon donor accepter pair (DAP) (C$^{+}_{B}$C$^{0}_{N}$-DAP-2) or carbon-oxygen complex (C$_{B}$O$_{N}$), while the identity of the structure giving rise to the \textit{S} = 1 signatures is unknown at this stage, hence we denote this defect type CC/CO DAP-X \cite{Gao2025}. 

Many of these spin hBN defects have been observed for the first time in the last couple of years. There is much work left to do to establish their full photophysical properties, atomic structure and reproducible growth processes. In the following sections we highlight opportunities presented by these spin defects for various spin-based quantum technologies. 

\subsection{Optical networks}

Future global optical quantum networks will consist of a distribution of quantum states and entanglement to deliver secure global quantum communication, connectivity of quantum devices, and potentially a quantum internet \cite{Kimble2008,Azuma2023}. In such a network, quantum information and entanglement will be distributed across a web of nodes via photons \cite{Kimble2008,Atature}. Each node will be composed of a physical system (the quantum repeater) that can send and retrieve quantum photonic information and store it in a long-lived quantum memory (typically spin states). Quantum repeaters based on solid-state defects with optically addressable spins have been explored theoretically \cite{Childress2005} and experimentally \cite{Pompili2021,Fang2024}. In the majority of these schemes, entanglement of distance spins is achieved via optical quantum erasure measurements which removes the which-path information on the origin of the photon. 

To act as a repeater, a defect must have excellent optical properties, namely efficient and coherent optical transitions that are spin-dependent and spectrally resolvable \cite{Childress2005, Atature,Cholsuk2024}. On the spin side, the electronic spin must couple coherently to emitted photons, as well as to neighbouring nuclear spins so the defect can store quantum information while simultaneously setting up a new photonic link. This multi-qubit quantum register of electronic and nuclear spins can facilitate distributed entanglement over \textgreater than 2 nodes, as demonstrated for the NV$^{-}$ centre \cite{Pompili2021}. In terms of the spin coherence timescales, the electronic spin $T_{2}$ must be long enough long enough for propagation of the photon to the next node to establish entanglement \cite{Wehner2018, McMahon2015}. Practically, this mean the single- and two- gate timescales, entanglement efficiency and photon propagation time all set the minimum feasible timescale for $T_{2}$ \cite{McMahon2015}. Recently, it has been shown that electronic spin coherence times of 100's $\mu$s are not prohibitive for long-distance (10's km) entanglement, provided coupling to longer-lived nuclear spins is present \cite{Knaut2024}. 

The strongly coupled \textit{S} = 1 hBN defects (C$_{x}$ and CC/CO DAP-X) present the first feasible 2D spin systems to explore for quantum repeaters. Both show single qubit gates times of 10's ns and the potential for reasonable optical efficiency, however the spin coherence timescales are far shorter than for the NV$^{-}$ centre. Major spin-challenges include overcoming the dense nuclear spin bath and access to individual nuclear spins that can provide quantum memories.


\subsubsection{Mitigating spin decoherence}

A challenge for all hBN spin defects is mitigating the magnetic noise from the spin-rich BN lattice, which is the main contributor to spin decoherence \cite{Ye2019,Tarkanyi2025}. Standard approaches for reducing nuclear spin noise in other defect systems include isotopic purification \cite{Itoh2014,Bourassa2020} and dynamical decoupling protocols \cite{Viola1998}. Removal of nuclear spins altogether is not possible in hBN (all B and N isotopes are spin active), but decoupling the electronic and nuclear spins has been demonstrated for both the V$_{B}^{-}$ \cite{Ramsay2023,Rizzato2023_coherence}, single C$_x$ defects \cite{Stern2024} and, very recently, the CC/CO DAP-X defects \cite{Gao2025arxiv}. To date, these efforts have extended the electronic spin coherence to $\sim$ 50 $\mu$s, achieved at room temperature. These timescales are shorter (2-3 orders of magnitude) than can be achieved in diamond \cite{Bar-Gill2013} and silicon carbide \cite{Seo2016}. 
Experimental and theoretical studies of V$_B^-$ have, to date, provided the most insight to spin decoherence mechanisms in hBN \cite{Gottscholl2020,Gottscholl2021a,Tarkanyi2025}. At low magnetic field, cluster correlation expansion (CCE) modeling methods have indicated that short-range hyperfine interactions with the neighbouring $^{14}$N, as well as nuclear spin-nuclear spin flip-flops drive decoherence \cite{Haykal2022,Tarkanyi2025}. Enrichment of the host hBN with $^{10}$B has been theoretically and experimentally confirmed to extend the coherence time from $\sim$ 90 ns to $\sim$ 190 ns \cite{Gong2024}. This is because the reduced gyromagnetic ratio of $^{10}$B over the naturally more abundant $^{11}$B results in a weaker hyperfine interaction and boron nuclear spin flip-flop rates \cite{Haykal2022}. 

At fields as strong as a few Tesla, the coherence time $T_2^{echo}$ extends to 36 $\mu$s \cite{Murzakhanov2022} in agreement with theoretical predictions \cite{Lee2022,Tarkanyi2025}. In this high-field regime decoherence is dominated by dipole-dipole interaction-induced nuclear spin flip-flop interactions of homonuclear spin pairs \cite{Tarkanyi2025}. Theoretical predictions show that magnetic dipolar coupling between the nuclear spins is partially suppressed in hBN via the nuclear quadrupole interaction \cite{Lee2022}. 

A similar theoretical investigation of decoherence of the carbon-related single spin defects in hBN is not currently possible because their atomic structures are not yet confirmed. The decoherence mechanisms are likely to be qualitatively similar to those for V$_B^-$, however the different hyperfine interactions and defect symmetry will give rise to important quantitative differences, such as the field strengths required for changes in $T_{2}$. For example, for C$_x$ the significant transverse zero field splitting ($\textit{E}$) (related to the defects' symmetry), relative to the strongest hyperfine interaction, produces a clock transition at zero magnetic field \cite{Wolfowicz2013} which decouples the electronic spin from the nuclear spin bath in a small region around zero magnetic field \cite{Stern2024}. 

Detrimental spin noise can also be due to impurity electronic spins in the material. Cross relaxation between V$_B^-$ and \textit{S} = 1/2 impurities in the hBN lattice has been reported in several ODMR studies \cite{baber2021,Scholten2024}. Similar cross relaxation between bright and dark spin defects is likely to be prevalent in a range of carbon-doped hBN materials where the defect density is high, although there have been no detailed studies of this to date.  

\subsubsection{Coherent control of nuclear spins}
Despite being a major of source of decoherence, strongly coupled nuclear spins can present an important resource for quantum networking applications. Individual nuclear spins that are strongly coupled (when the hyperfine coupling is greater than the dephasing rate (1/$T_{2}^{*}$)) to the central electronic spin can be individually initialised and controlled, enabling access to longer-lived quantum memory and multi-qubit entanglement \cite{Dutt,Neumann2008}. 

While there is an abundance of spin-active nuclei in hBN, the challenge is to identify individually strongly coupled candidates that can be controlled in isolation from the bath. This was first demonstrated in hBN with the V$_B^-$ \cite{Gao2022}. Here, polarisation of the three $^{14}$N nuclei surrounding the V$_B^-$ can be achieved at the excited/ground state level anti crossings (ESLAC/GSLAC) of the defect \cite{Gao2022}. Using tailored pulse sequences to control both electronic and nuclear spins, spectroscopy of the $^{14}$N nuclei revealed an effective nuclear-nuclear coupling constant of 3.4 MHz at the GSLAC. Coherent control of the $^{14}$N nuclei resulted in an inhomogeneous dephasing time ($T_{2}^{*}$) of 3.5 $\mu$s \cite{Gao2022}. Dynamic nuclear polarisation of isotopically engineered h$^{10}$B$^{15}$N was more recently demonstrated, revealing improved resolution of the hyperfine coupling and coherent control of the $^{15}$N nuclei \cite{Gong2024} (Figure~\ref{Fig::fig2}d). While these demonstrations were the first examples of nuclear spin polarisation in a 2D material, the low quantum efficiency of the V$_B^-$ precludes the system from being applied to networking. 

In a breakthrough result, single nuclear spin control has recently been demonstrated via single hBN spin defects \cite{Gao2025}. This was demonstrated using the newly-identified carbon-related defect that is assigned to a DAP in the lattice which is coupled via charge transfer to a second defect (Figure~\ref{Fig::fig2}m-p). In this study, hyperfine coupling to a neighbouring $^{13}$C nucleus is observed as a 300 MHz-splitting in the ODMR (Figure~\ref{Fig::fig2}n). Via optical polarisation and a SWAP gate to intialise the $^{13}$C nuclear spin, coherent control revealed $T_{1}$ = 144 $\mu$s, $T_{2}^{*}$ = 16.6 $\mu$s and $T_{2}^{echo}$ = 162 $\mu$s, setting a new record for nuclear spin coherence in the hBN system \cite{Gao2025}. 

In general, the microsecond spin coherence timescales that can be accessed for the electronic spin of the hBN defects is not prohibitive for use in optical networking, as long as entanglement generation rates are faster \cite{Wei2022}. For the hBN spins, microwave control of the spin is relatively slow (single qubit gate times typically 10 - 50 ns) which may present challenges for systems where this is long compared to $T_{2}$. More work is needed to determine the effect of magnetic field strength, isotopic composition, layer number, strain and defect density on the electronic spin coherence that can be achieved via decoupling approaches. For nuclear spin coherence, the timescales demonstrated so far (100's $\mu$s) are promising but still shorter than what can be accessed in diamond-based systems (typically seconds) \cite{Knaut2024}. Future work is required to determine the true limit to nuclear spin coherence, when this is not limited by the electronic $T_{1}$ \cite{Gao2025}.



\subsection{Quantum sensing}

Spin defects are rapidly becoming established as a leading quantum sensing platform for biomedical applications \cite{Aslam2023,Rizzato2023_sensors} and materials characterisation \cite{Rovny2024}. Their spin states can be sensitive to small changes in magnetic field, as well as temperature, pressure, strain, stress and electric field. Combined with their atomic length scale, this enables mapping of these physical quantities with nanoscale spatial resolution. Meanwhile, their coherence properties and optical readout can enable unprecedented sensitivity. 

\begin{figure}[t]
  \includegraphics[width=\textwidth]{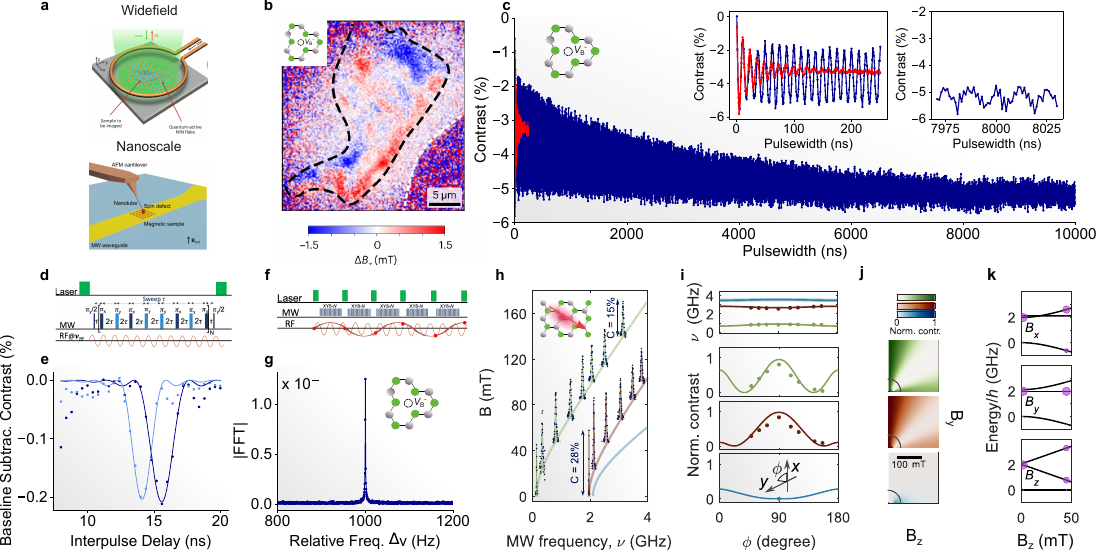}
  \caption{
    \label{Fig::fig3}}
\end{figure}

Spin defect-based sensors utilise either a single defect or a large ensemble of defects as the active sensor. Ensemble-based sensing is typically performed in a widefield optical setup where the spatial resolution is diffraction limited, but where the use of many defects gives higher sensitivity (Figure~\ref{Fig::fig3}a). In contrast, the spatial resolution for single-defect sensors is fundamentally limited by the nanoscale spatial extent of the electronic wavefunction, but in practice by the defect-sample distance. Sensors employed for nanoscale sensing are typically scanned over a sample using an atomic force microscope (AFM) (Figure~\ref{Fig::fig3}b). This sensing modality is most developed with NV$^{-}$ centres \cite{Budker2007,Taylor2008,Degen2017} where scanning probe magnetometers using single defects are now well developed and even commercially available \cite{Balasubramanian2008,Degen2008,Maze2008,Rondin2012,Rondin2014, Barry2020,Rovny2024}. 

While hBN defects are unlikely to compete with NV-based sensors with regards to spin coherence time, they display other advantages that may make them competitive sensors for particular applications. The atomically-thin nature of the hBN host material presents opportunities for nanoscale sensing and imaging with nm-spatial resolution \cite{Vaidya2023}. Meanwhile, the ability to easily integrate hBN layers with other materials opens doors to \textit{in-situ} imaging of 2D devices. Like diamond, hBN is also bio-compatible \cite{Kavi2024}, and early results show the defects have high susceptibility to a range of physical factors which could see the defects being useful probes in biological context. Many of these areas are under rapid investigation and it is likely that quantum sensing will be the first area where hBN defects are industrially applied. 

\subsubsection{Wide-field Quantum Sensing}

Progress towards applications with V$_B^-$ ensembles has been rapid, with proof of concept experiments exploring their use in temperature \cite{Gottscholl2021b,liu2025}, pressure \cite{He2025} and strain \cite{Yang2022,Lyu2022} sensing, and particularly in wide-field magnetometry (Figure~\ref{Fig::fig3}b) \cite{Kumar2022,Healey2023,Zhou2024}. V$_B^-$ ensembles have been integrated with ferromagnetic materials for in-situ magnetometry, for example in sensing the DC stray field from 2D ferromagnets \cite{Kumar2022,Huang2022,Healey2023}, spin noise relaxometry \cite{Huang2022,Zang_AFM2024_VBmagnon} and sensing spin waves \cite{Zhou2024,Manas-Valero2025,Zang_AFM2024_VBmagnon}. To date, the best reported DC-magnetic field sensitivity is on the order of $\mathrm{\mu T/\sqrt{Hz}}$ \cite{Zhou_NL2023_VBsensitivity}, inferior to the state of the art for ensemble NV$^{-}$ centre sensors (nT/$\mathrm{\sqrt{Hz}}$) \cite{Tang2023}. However, this is partly explained by the fact that the sensitivity scales with the square root of the number of defects in the ensemble \cite{Barry2020}, and much larger sample volumes are typically used in the NV$^{-}$ centre experiments. Whilst this can be addressed in future, the spin coherence time and quantum efficiency are also superior for NV$^{-}$ centres, and it may be difficult for V$_B^-$ ensembles to directly compete purely in terms of sensitivity in DC magnetometry. 

In the case of AC magnetometry, dynamic decoupling pulse sequences can be implemented to improve the spin coherence and provide a tuneable sensor bandwidth for identification of a signal frequency. Pulsed \cite{Rizzato2023_coherence, Gong2023} and continuous \cite{Ramsay2023} dynamic decoupling sequences have been shown to be effective for V$_B^-$ ensembles and applied to sensing of MHz \cite{Rizzato2023_coherence} and GHz \cite{Patrickson2024} range magnetic signals, where the sensitivity ($\sim$ 1 $\mathrm{\mu T/\sqrt{Hz}}$) can be comparable to that achieved with NV$^{-}$ centres, when scaled by the volume of the sensor \cite{Patrickson2024} (Figure~\ref{Fig::fig3}c-g). These sensing schemes have been extended to include advanced quantum heterodyne protocols that enable sensor frequency resolution far beyond the limit imposed by the spin coherence \cite{Rizzato2023_coherence, Patrickson2025}.

While the V$_B^-$ does not exhibit higher magnetic field susceptibility than the NV$^{-}$ centre, recent results suggest it performs competitively, if not better, for sensing of temperature, pressure and electric field \cite{Gottscholl2021b,Udvarhelyi2023,He2025}. For temperature sensing, the longitudinal zero field splitting term (\textit{D}) undergoes a 25-fold greater variation between room and cryogenic temperature than it does for the NV$^{-}$ \cite{Gottscholl2021b,Liu2025_sensing}. This large effect is assigned to both temperature-dependent lattice distortion and second-order spin-phonon coupling \cite{Liu2025_sensing}. \textit{D} for the V$_B^-$ is sensitive to in-plane pressure, but remarkably insensitive to out-of-plane pressure \cite{Gottscholl2021b}. This has led to use of the  V$_B^-$ in diamond anvil cells for investigation of pressure-driven magnetism in other 2D materials  \cite{He2025, Mu2025}. In contrast, it is the transverse zero field splitting parameter (\textit{E}) for the V$_B^-$ which is associated with sensitivity to electric field \cite{Udvarhelyi2023}. The electric field susceptibility is $d_{\perp}$ $\sim$ 40 Hz/Vcm$^{-1}$, roughly double that of the NV$^{-}$ centre \cite{Dolde2011}.

An emerging property of the V$_B^-$ system for sensing, is that the defects retain their magneto-optical and sensing properties even in the few-layer ($\sim$ 3-5) limit \cite{Durand2023}. This is unlike diamond where $T_{2}^{echo}$ for NV$^{-}$ centres starts to decrease when a defect is within 100 nm of the diamond surface \cite{Sangtawesin2019}. This provides opportunities to develop atomically-thin sensing foils with hBN sensors. 



\subsubsection{Nanoscale Quantum Sensing}

Nanoscale quantum sensing has been dominated by the NV$^{-}$ centre with few other defect systems displaying adequate sensitivity on the single defect level, or the versatile operating conditions, to be serious candidates. The single carbon-related defects in hBN with \textit{S} = 1 ODMR signatures provide a feasible alternative to the NV$^{-}$ centre. In particular, the C$_x$ spin defect shows high brightness and ODMR on the single-defect level, which translates to $\mathrm{\mu T/\sqrt{Hz}}$ DC sensitivity \cite{Stern2024,Gilardoni2025}, on par with the routinely achieved sensitivity for NV$^{-}$ centres \cite{Rovny2024}. Unusually, the ODMR contrast for C$_x$ is not completely quenched with high ($\textgreater$ 150 mT) off-axis magnetic field,  which significantly increases the dynamic range for DC magnetometry over the NV$^{-}$ centre \cite{Tetienne2012} (Figure~\ref{Fig::fig3}h). As a result, C$_x$ may open possibilities for the imaging of magnetic materials that require large, arbitrarily oriented bias fields to access emerging magnetic phenomena. The photophysical origin of the retention of ODMR contrast is related to the low symmetry of the defect (a property shared with other defects in diamond \cite{Lee2013,Foglszinger2022}), which manifests as three measurable ODMR resonances (Figure~\ref{Fig::fig3}i-k) \cite{Gilardoni2025}. Each resonance displays a different dependence (in terms of quenching and sensitivity) to magnetic field orientation, providing the capability to perform vectorial magnetic field sensing \cite{Gilardoni2025}. 

Omnidirectional magnetic field sensing has been demonstrated with other single carbon-related defects that possess \textit{S} = 1/2 spin signatures (the OSDP class) \cite{Gao2024_sensing,Scholten2024}. Without a quantisation axis, these systems offer opportunities for facile detection of magnetic field strength, which might have useful applications in RF receivers for example \cite{Robertson2025_RF}, but no information on orientation, meaning extrapolation to magnetisation information is not possible, for example.

\section{Defect structure identification}\label{Sec:4}

An immediate major challenge towards the development of defects in hBN for all technologies discussed so far is the elucidation of their atomic structure. Apart from the V$_B^-$ and the C$_B$C$_N$ UV emitter, the atomic structure of most defects is still under investigation. The assignment is complicated due to the disparity in optical properties and/or the lack of clear hyperfine signatures in ODMR spectra. However, conclusive confirmation of defect structure is critical for optimising operating conditions, refining the determinisitic fabrication process of these defects (which is discussed in detail in other reviews \cite{Wolfowicz2021,Freysoldt2014,Ivady2018}), and advancing the fundamental understanding of their electronic properties. This demands joint efforts from theory and experiment.

\subsection{\textit{Ab initio} Studies}

First-principles calculations offer valuable microscopic insights into the electronic structure and the resulting spin and optical properties of solid state defects, including hBN \cite{Ivady2018,gali_recent_2023}. The workhorse of \emph{ab initio} point defect studies in semiconductors is density functional theory (DFT), which has already proven its predictive power for formation energies \cite{Freysoldt2014}, spin properties \cite{Ivady2018}, and vibrational properties \cite{alkauskas_first-principles_2014} in wide band gap semiconductors, including hBN \cite{Ivady2020,Plo2025}. However, one needs to be cautious when dealing with the excited state properties of defects. DFT is a mean-field ground state theory designed to accurately predict the ground state's total energy, density, spin density, and forces. Whereas, the measurable properties of spin-less single-photon emitters in hBN, ie., the UV \cite{Bourrellier2016}, blue \cite{zhigulin_photophysics_2023} and green-red  \cite{Mendelson2021} emitters, are linked to the excited state(s) of the defects.

Excited state properties can be approximated in DFT by forcing the occupation of the defect state to mimic an optical excitation of the centre. This procedure works reasonably well when the excited state can be described in the single-electron picture, ie., exciting an electron from an occupied defect orbital to an empty defect orbital. The single electron picture may not be applicable for all defects, as the true many-particle wavefunction of the excited state can be formed as a linear combination of multiple single-particle transitions. Disregarding such state mixing in the optically excited states leads to an enhanced error in predicting ZPL energies, lifetimes, and decay rates, which can alter the identification of defects' microscopic structures.

\begin{figure}[t]
  \includegraphics[width=\textwidth]{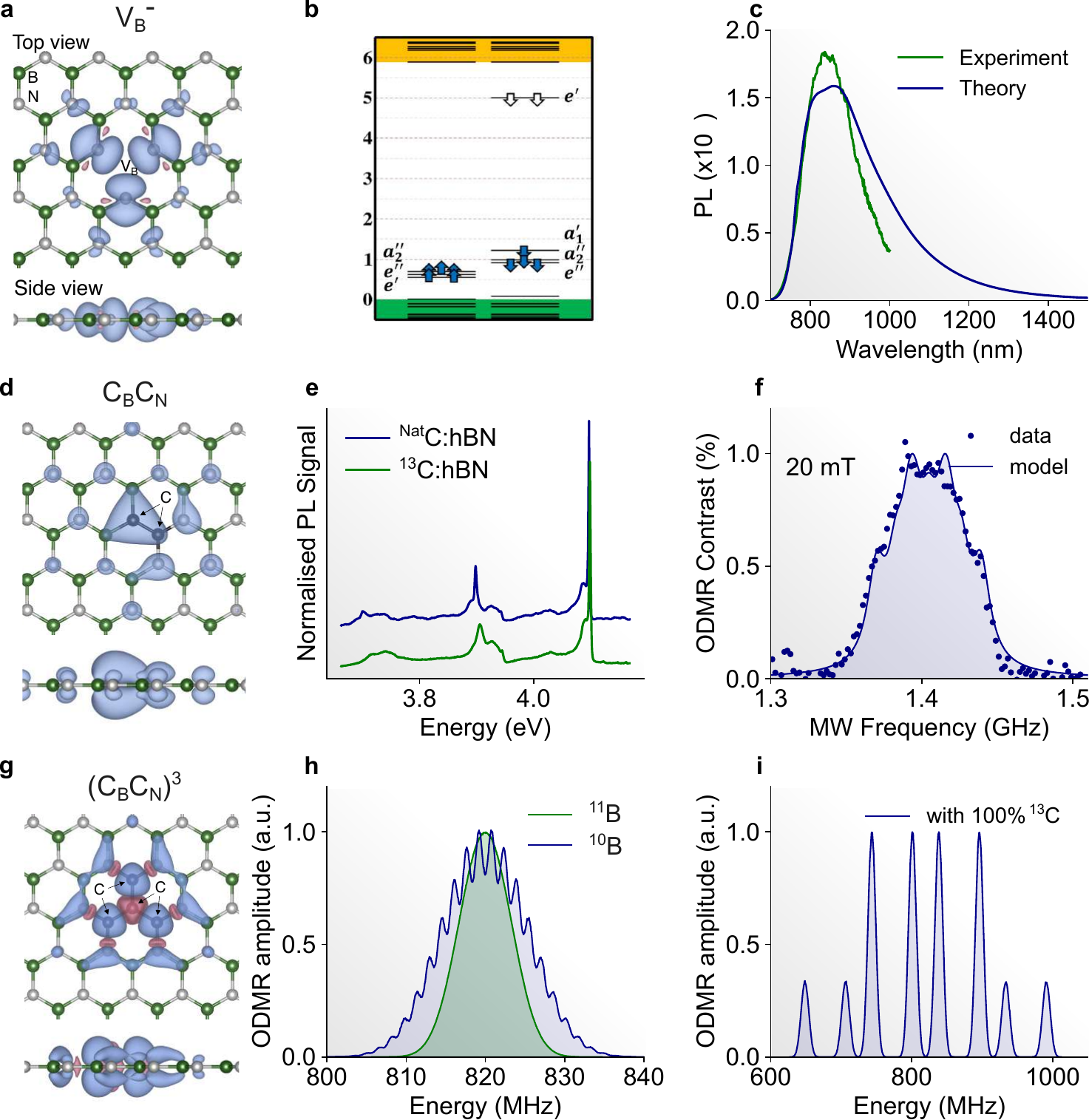}
  \caption{
    \label{Fig::fig4}}
\end{figure}

Hexagonal boron nitride presents particular challenges in this respect. hBN features two types of states: the in-plane sp$^{2}$ hybrid bonding states ($\sigma$) and the out-of-plane p$_z$ states ($\pi$) of the BN layers. A point defect can perturb both of these symmetrically distinguishable electronic states, often giving rise to a multitude of defect orbitals in the band-gap of hBN. For instance, the V$_B^-$ center in hBN has twice as many active defect orbitals as the NV$^{-}$ center in diamond, making the theoretical characterisation of the V$_B^-$ center's excited state challenging \cite{Ivady2020,Reimers2020,barcza_dmrg_2021} (Figure ~\ref{Fig::fig4}a,b). Generally, the number of single-particle defect states in the band gap indicates the complexity of the electronic structure and complex electronic structures may necessitate more advanced theoretical methods.
The development and testing of post-DFT methods for hBN are currently at the forefront of theoretical studies \cite{Reimers2020,barcza_dmrg_2021,Cholsuk2023}. Consolidating the necessary theoretical framework and solidifying theoretical predictions are expected to significantly contribute to successfully identifying the microscopic structure of the color centers in the not-so-distant future.

The rapid pace of experimental development, however, demands fast answers from theory. A viable approach is to focus on quantities that can be both reliably predicted by DFT methods and are experimentally accessible. While ZPL energies can be inaccurate in DFT, polarisation angles and the fine structure of the phonon sideband can be obtained with reasonable accuracy \cite{alkauskas_first-principles_2014, davidsson_theoretical_2020, Kumar2024} (Figure ~\ref{Fig::fig4}c). Characteristic features of the polarisation of emitted/absorbed light are determined by the symmetry and orientation of the defect, providing relevant information about the underlying structure. The angle of the transition dipole and the lattice vectors for low-symmetry defect structures are parameters that can be compared with theoretical predictions \cite{doi:10.1021/acs.jpcc.4c03404}. The main features of the phonon sideband can be correlated with atomic masses, bond strength, and symmetry, serving as an optical fingerprint of the defect. When compared with high-resolution, low-temperature, and low-strain PL spectra, this can aid in the identification of the defect's microscopic structure. Furthermore, since the features of the phonon sideband depend on atomic masses, isotope-engineered samples can provide further evidence of the atomic composition of the underlying structure \cite{alkauskas_first-principles_2014}. These arguments motivate conducting in-depth experimental analyses of polarisation data and phonon sidebands, and placing greater emphasis on related DFT calculations in hBN. A very recent example of such a joint experimental theoretical work has been published in Ref.\ \cite{Plo2025}, where the microscopic structure of the UV emitter was studied and assigned to the C$_{B}$C$_{N}$ complex (Figure ~\ref{Fig::fig4}d,e). Here, this assignment was possible due to shifts in the phonon side bands in hBN isotopically enriched with $^{13}$C, which were well modeled by simulation. 

For spin-active defects, the hyperfine interaction with nearby nuclear spins and potentially the zero field splitting interaction provide another, very reliable means of identification. Indeed, the direction of the preferential quantisation axis of the electron spin and value of $D$ and $E$ carry important information on the symmetry of the defect and the degree of spin density localisation. The hyperfine splitting of the spin sublevels due to strongly coupled nuclear spin(s) serves as a unique fingerprint of the defect that can be accurately calculated in reliable ground state DFT calculations \cite{szasz_hyperfine_2013,takacs_accurate_2024}. This enabled identification of the V$_B^-$ center \cite{Gottscholl2020, Ivady2020}. 

Assignment of the single carbon-related spin defects has been more difficult. The C$_x$ defects show a narrow ESR/ODMR linewidth at zero field \cite{Stern2024} compared to the V$_B^-$. This is consistent both with the low symmetry of C$_x$ combined with localisation of the electron spin density on spin-less $^{12}$C carbon atoms which appears in $\sim0.99$ probability in natural abundance. However, the hyperfine structure at higher field has been difficult to assign to a particular structure (Figure ~\ref{Fig::fig4}f). For carbon-based defects, the use of \textit{S} = 1/2 $^{13}$C enriched carbon sources during growth or implantation could give rise to a resolvable hyperfine structure. Such a hyperfine spectrum alone could declare the number of carbon atoms in the structure, while DFT calculations could nail down the microscopic configuration. This has been demonstrated by theoretical studies of carbon tetramers (not yet assigned to experimentally observed spin defects) \cite{Benedek2023} (Figure~\ref{Fig::fig4}h-i). This approach has also been successful experimentally; Gao \textit{et al.} has observed hyperfine splittings in $^{13}$C implanted and annealed hBN, which helped to narrow down the possible defect configurations to DAP complexes \cite{Gao2025}.

\subsection{Electron and X-Ray Beam Imaging}

Beyond DFT, electron and X-ray microscopies have been used - in some cases correlated with photoluminescence - to gain insight to atomic structure \cite{Hayee2020,Singla2024,Pelliciari2024}. In general, the challenge here is correlation of optical and X-ray/electron microscopy images, which have different spatial resolution limits. A combined study of cathodoluminescence (CL), photoluminescence (PL) and electron microscopy presented a complex picture of multiple defect types contributing to the green-red emitter family, but could not propose specific chemical structures \cite{Hayee2020}. Resonant inelastic X-Ray scattering (RIXS) has been recently used to identify a common elementary excitation at 285 meV for hBN green-red single-photon emitters \cite{Pelliciari2024}. These results point towards the importance of the N $\pi^{*}$ anti-bonding orbitals in defining the optical properties of hBN defects that emit in the visible region. High resolution scanning transmission electron microscopy (HAADF-STEM) in conjunction with confocal imaging and elemental mapping (EELS) has imaged carbon complexes formed in 60-nm thick exfoliated hBN flakes \cite{Singla2024}. It was observed that carbon readily replaces nitrogen and boron in the hBN lattice, and several potential carbon-related structures (carbon dimers and single carbon substitutions) were proposed. To date, all correlative studies of this nature have struggled with unambiguous identification of atomic structure of the emissive/spin active defects due to the density of non-emissive defects. Future attempts will need to utilise other ways to get around this problem, possibly with ultra high purity hBN, or using nanoscale probes/markers to aid the correlation.

\section{Conclusion}

We consider that hBN defects present an exciting novel platform to explore for future quantum technologies. While significant further technical development is required before hBN defects will be deployed for quantum computing or repeaters for optical networks, in our view there are several applications where hBN defects can outperform alternative candidates in the near term.

The first is quantum sensing, specifically in-situ quantum sensing in 2D heterostructures and nanoscale sensing. With regards to the former, hBN already forms a critical dielectric component of most 2D heterostructure devices. If high concentrations of defects (either V$_B^-$ or single defects) can be engineered atomically close to the surface of hBN layers, that are then incorporated within devices, \textit{in-situ} hBN quantum sensors could be game changing for exploration of magnetic phenomena, stress gradients, or charge flow in these devices as well as routine methods of device quality control, for example. 

For nanoscale sensing, single carbon-related hBN defects offer the potential for nm-spatial resolution. This is an extremely attractive proposition, as for NV$^{-}$ scanning magnetometry, typical spatial resolution is 20 - 30 nm \cite{Rovny2024}, due to the difficulty in fabricating single NVs close to the apex of a diamond tip and the degradation of magneto-optical properties for shallow NV$^{-}$ centres. In addition, single $C_{x}$ hBN defects operate under off-axis bias magnetic field of 100's mT. This may enable investigation of new material systems, or reveal new insight to nanoscale spin textures in other materials. Although the full picture is still emerging, the range of different carbon-related single spin defects points towards future opportunities to engineer specific sensors where the spin multiplicity, zero field splitting magnitude and orientation of the spin quantisation axis could be designed for the sensing application. For example, the presence of defects with quantisation axes aligned to or perpendicular to the 2D layers makes the alignment of bias fields more straightforward. 

A second key area is likely to be free-space and satellite-based QKD. The high repetition rates, linear polarisation and broad range of emission wavelengths make hBN defects key contenders in this area. However, electrical excitation as well as ZPL tuneability will need to become established as routine capabilities. 

Finally, future progress across all areas will be contingent on understanding how to control the creation of hBN defects in different hBN materials. Although out of the scope of this perspective, on-demand defect creation is critical for future device fabrication. There are already promising signs that defect creation can be controlled via external treatment (ion/electron beam/annealing) of hBN exfoliated crystals, or carbon-doping of CVD material. Early efforts to create defects in hBN with spatial control are promising \cite{Ziegler2019,Gale2022,Kumar2023}, but  precise site-control is still needed. Defect creation attempts would benefit from identification of unknown defect structures, namely the green-red emitting family that shows ODMR. It is our view that coordinating experimental spin resonance, with \textit{Ab initio} studies and materials characterisation, should lead to many of these defects being assigned structures in the near future.

\section{Acknowledgments}

We acknowledge Andrew Ramsey for helpful discussions. 

T.V.\ acknowledges acknowledges support from the Munich Quantum Valley, which is supported by the Bavarian state government with funds from the Hightech Agenda Bayern Plus.

V.I.\ acknowledges support from the National Research, 
Development and Innovation Office of Hungary (NKFIH) within the Quantum Information National Laboratory of Hungary (Grant No.\ 2022-2.1.1-NL-2022-00004) and the project FK 145395. This project is funded by the European Commission within Horizon Europe projects (Grant Nos.\ 101156088 and 101129663). 

H.L.S.\ acknowledges support from a Royal Society University Research Fellowship (URF\ RI \ 221326) and the UKRI EPSRC projects (Grant No.\ APP17698 and APP51824). 

\section{Figure Captions}

\textbf{Figure One}\\
\textbf{Optical properties of hBN defects} (a) Integrated photoluminescence (PL) spectra for different classes of hBN defects including the C$_B$C$_N$ `UV emitter' (purple trace, 8 K)  \cite{Plo2025}, the ‘B’ centre (4 K) (blue traces, inset shows 295 K) \cite{Fournier2023}, `green-red' emitting defects (green traces, 295 K) \cite{Tran2016a}, and the V$_{B}^{-}$ (red trace, 295 K) \cite{Ramsay2023}. (b) Time-resolved photoluminescence of a ‘green-red’ emitting hBN defect. Reproduced from \cite{Vogl2019}. (c) Polarisation of excitation (bold circles) and emission (open circles) of hBN green-red defects. Reproduced from \cite{Kumar2024}.  (d) Second-order intensity correlation measurement ($g^{(2)}$(t)) of a ‘green-red’ hBN defect, displaying a $g^{(2)}$(0) of 0.017(3). Reproduced from \cite{Kumar2024} (e) Two-photon coincidences from a Hong-Ou-Mandel measurement of a B-centre. The orange (blue) dots denote the normalised coincidence rate for the parallel-polarisation (orthogonal-polarisation) configuration of the input ports of the second beam splitter. The center peak provides the values of $g^{(2)}_{HOM}$(0) = 0.32 ± 0.05 (parallel case) and $g^{(2)}_{HOM}$(0) = 0.58±0.07 (orthogonal case). The light-gray bars mark the theoretical values for distinguishable photons of $g^{(2)}$(0) = 0.14. The purple bar indicates the theoretical value for fully indistinguishable photons of $g^{(2)}$(0) = 0.14. Reproduced from \cite{Fournier2023}.\\

\textbf{Figure Two}\\

\textbf{hBN spin defects} (a) Right: The simplified electronic structure of the V$_{B}^{-}$ defect, modified from \cite{Ivady2020}. The orange arrows represent optical excitation/emission between ground and excited states, and blue arrows represent intersystem crossing. Left: A schematic of the V$_{B}^{-}$ defect, where z$_{defect}$ indicates an out-of-plane quantisation axis. (b) Measured ODMR spectra of V$_{B}^{-}$ in h$^{10}$B$^{15}$N and naturally abundant hBN$_{nat}$ at magnetic field B$_{z}$ $\sim$ 9 mT. Solid lines represent multi-peak Lorentzian fits, reproduced from \cite{Gong2024}. (c) Comparison of spin echo for V$_{B}^{-}$ ensembles at low (22 MHz, black) and high (80 MHz, red) Rabi frequency, $\Omega$. At high Rabi frequency, the contrast is stronger since a higher proportion of spins are controlled, reproduced from \cite{Ramsay2023}. (d) Measurement of $^{15}$N nuclear spin Rabi oscillations for the V$_{B}^{-}$, reproduced from \cite{Gong2024}. (e) An `optical spin defect pair’ (OSDP) model used to explain the presence of \textit{S} = 1/2-like ODMR signatures that are measured across different hBN materials. Orange arrows represent optical excitation/emission between ground and excited states and red dashed arrows represent charge transfer. The electronic model is modified from \cite{Scholten2025}. The upper part of the figure shows a potential atomic structures for a OSDP, where charge transfer occurs between two separate defects. (f) ODMR spectra for two single defects in carbon-doped multilayer hBN, measured with an applied field of 25 mT, reproduced from \cite{Stern2022}. (g) The central resonance frequency for a single carbon-related defect in carbon-doped multilayer hBN, reproduced from \cite{Stern2022}. (h) Rabi oscillations measured for ensembles of carbon-related defects in an exfoliated single-crystal hBN flake, at B = 100 mT. The resonance frequency being driven is 2.9 GHz. Reproduced from \cite{Scholten2024}. (i) The proposed electronic structure for C$_{x}$, a carbon-related \textit{S} = 1 defect in hBN with \textit{D} = 1.97 GHz and \textit{E} = 60 MHz. The optical spin polarisation mechanism includes spin-dependent intersystem crossing (blue arrows) via a single metastable state. The orange arrows represent optical excitation/emission. The left hand image represents the in-plane quantisation axis for the defect, of unknown atomic structure. (j) cw-ODMR spectra for a single C$_{x}$ defect. (k) Rabi oscillations for C$_{x}$ defect, fit to a function of the form $\exp(-t/T_{Rabi})
\sin(2\pi t \Omega-\phi)$ (orange curve), where $\phi$ is the phase offset, $\Omega$ is the Rabi frequency and $T_{Rabi}$ is the decay lifetime of the Rabi oscillations. Measured at 0 mT with drive frequency of 1.9 GHz. (l) Dynamic decoupling measurements with N$\pi$ refocusing pulses, where each measurement is fit to $\exp[-(t/T_{DD})^{\alpha}]$. (j-l) are reproduced from \cite{Stern2024}. (m) The proposed electronic structure of a \textit{S} = 1 carbon related defect formed in exfoliated hBN via CO$_{2}$ implantation and thermal annealing. This defect type shows both \textit{S} = 1 and \textit{S} = 1/2 like signatures and has been assigned to a pair of spin defects (\textit{S} = 1/2 and \textit{S} = 1) where the optical spin polarisation mechanism includes both spin-dependent intersystem crossing and charge transfer. One of the defects of the pair is predicted to be a C-C or C-O donor acceptor pair (DAP). Schematic modified from \cite{Gao2025}. The schematic shows a possible atomic structure for the two defects of the pair, where one has an out-of-plane quantisation axis in its triplet state. (n) An ODMR spectrum of one of the defects of this category, measured with an out of plane magnetic field of 62.5 mT. (o) The ODMR spectrum after nuclear spin initialisation revealing polarisation of the $^{15}$N nuclear spin. (p) An example of nuclear spin Rabi oscillation, persisting for 40 $\mu$s without significant decay. (n-p) are reproduced from \cite{Gao2025}. All data were measured at 295 K. \\

\textbf{Figure Three}\\
\textbf{Quantum sensing with hBN defects} (a) Schematic of widefield and nanoscale quantum sensing, reproduced from \cite{Healey2023,Gao2024_sensing}. (b) Magnetic field map of a CrTe$_{2}$ flake, imaged using the V$_{B}^{-}$, reproduced from \cite{Healey2023}. (c) Comparison of unprotected Rabi oscillation (red trace) and amplitude modulated CCD scheme, for the V$_{B}^{-}$ (blue trace), showing stabilisation of Rabi oscillation on $\mu$s-timescale. Inset figures show zoomed regions of the Rabi oscillations. Reprodcued from \cite{Ramsay2023}. (d) Pulsed dynamical decoupling protocol used for sensing RF signals (V$_{B}^{-}$). (e) Response of the V$_{B}^{-}$ sensor as fluorescence contrast dips occurring at matching of the interpulse delay ($\tau$) with the RF signal period according to: $\tau$ = 1/(4$\nu_{RF}$). The data points (dots) are best fitted with a (sinc)$^{2}$ function. (f) Coherently Averaged Synchronized Readout (CASR) protocol. XY8-2 subsequences are synchronized to the sensing frequency for 2 seconds. (g) Transformation of the signal resulting from the measurement sequence (f) yields a sharp peak at the relative frequency $\Delta\nu$. (d-g) are reproduced from \cite{Rizzato2023_coherence}. (h) Persistence of saturated cwODMR contrast for C$_{x}$ with magnetic field applied along the defect y direction. The solid curves represent the transition frequencies of the three ODMR transitions as a function of B$_{y}$ amplitude. The measured cwODMR contrast as a function of MW frequency is represented by circles. (i) Angular magnetic-field dependence of cwODMR frequency (top panels) and normalised contrast of the three ODMR resonances of the C$_{x}$ defect (top to bottom), with 50-mT bias magnetic field applied in the xy plane. Data are presented as circles and curves indicate the simulated cwODMR contrast. The inset indicates the direction of rotation of the bias magnetic field. (j) Calculated contrast of each resonance as a function of B$_{y}$ and B$_{z}$. (k) Evolution of spin eigenstates with applied magnetic field along the x,y,z axes of the defect, from top to bottom. Calculated amplitudes of the optically initialised population of each spin sublevel are indicated by the size of the purple circles. (h-k) are reproduced from \cite{Gilardoni2025}. All of the data was measured at 295 K. \\

\textbf{Figure Four} \\

\textbf{Ab initio modeling of hBN defects} (a) Spin density of V$_{B}^{-}$ in hBN as obtained by spin-polarised HSE06 DFT. (b) Single particle electronic structure of the V$_{B}^{-}$, reproduced from \cite{Ivady2020}. (c) Theoretical photoluminescence spectrum of the V$_{B}^{-}$ compared to the experimental spectrum, reproduced from \cite{Ivady2020} and \cite{Gottscholl2020}. To produce the theoretical curve the calculated Huang-Rhys factor of 3.5 is used whereas the position of ZPL at 765 nm (1.62 eV) was aligned to match to the highest intensities of the experimental spectrum which corresponds to about 0.09 eV redshift compared to the KS DFT result (1.71 eV). (d) Spin density of C$_{B}$C$_{N}$. (e) Photoluminescence spectra measured at 8K of the `UV emitter' in two different isotopically enriched hBN samples, C$_{nat}$:hBN and $^{13}$C:hBN. Reproduced from \cite{Plo2025}. (f) ODMR spectrum of the C$_{x}$ defect, measured with 20-mT magnetic field parallel to the defect z axis. The dots are the data and the shaded curve is the computed spectrum. Reproduced from \cite{Stern2024}. (g) Spin density of (C$_{B}$C$_{N}$)$^{3}$. (h) ODMR signal of the (C$_{B}$C$_{N}$)$^{3}$ defect at B = 0 mT with no carbon nuclear spins in $^{10}$B (blue) and $^{11}$B (green) containing samples. (i) ODMR signal of the (C$_{B}$C$_{N}$)$^{3}$ defect when it includes four $^{13}$C nuclear spins. (i,h) are reproduced from \cite{Benedek2023}.\\

\clearpage

\clearpage

\clearpage

\clearpage

\begin{table}[h]
    \centering
    \begin{adjustbox}{width=1\textwidth}
    \small
    \begin{tabular}{|c|c|c|c|c|c|c|c|}
        \hline
        \textbf{Defect} & \textbf{ZPL (nm)} & \textbf{\textit{D, E} (GHz)} & \textbf{Electronic spin coherence} & \textbf{T$_{1}$ ($\mu$s)} & \textbf{E-N coupling} & \textbf{Nuclear spin coherence} & \textbf{Refs} \\ \hline
        V$_{B}^{-}$ & 750-1000 & 3.48, 0.06 & \makecell{$T_{2}$* = 50 ns, $T_{2}^{echo}$ = \textless 100 ns,  \\ $T_{2}^{DD}$ = 4 $\mu$s (N=1000)} & 10 & yes & $T_{2}^{*}$ = 3.5 $\mu$s & \cite{Gottscholl2020,Ivady2020,Gottscholl2021a,Ramsay2023,Gao2022} \\ \hline
        
        OSDP & 400-800 & \textless 50 MHz & - & 10-15 & no & - & \cite{Chejanovsky2021,Stern2022,Scholten2024, Guo2023, Robertson2024, Singh2025} \\ \hline
        
        C$_{x}$ & 590(15) & 1.971(25), 62(10) & \makecell{$T_{2}$* = 106(12) ns, $T_{2}^{echo}$ = 228(11) ns, \\ $T_{2}^{DD}$ = 1.08(4) $\mu$s (N=10)} & 30-200 & no & - & \cite{Stern2024, Gilardoni2025} \\ \hline
        
        C$_{B}^{+}$C$_{N}^{0}$-DAP - X & 570-700 & $\sim$ 1, 0.1-0.4 & \makecell{$T_{2}^{echo}$ = 100-200 ns, \\ $T_{2}^{DD}$ = 38 $\mu$s (N=1024)} & 17-216 & yes & \makecell{$T_{2}^{*}$ = 16.6 $\mu$s, \\ $T_{2}^{echo}$ = 162 $\mu$s}  & \cite{Gao2025, Gao2025arxiv} \\ \hline

    \end{tabular}
    \end{adjustbox} 
    \caption{Spin-active hBN defects}
    \label{tab:spin}
\end{table}

\clearpage
\bibliography{ref}

\begin{thebibliography}{198}%
\makeatletter
\providecommand \@ifxundefined [1]{%
 \@ifx{#1\undefined}
}%
\providecommand \@ifnum [1]{%
 \ifnum #1\expandafter \@firstoftwo
 \else \expandafter \@secondoftwo
 \fi
}%
\providecommand \@ifx [1]{%
 \ifx #1\expandafter \@firstoftwo
 \else \expandafter \@secondoftwo
 \fi
}%
\providecommand \natexlab [1]{#1}%
\providecommand \enquote  [1]{``#1''}%
\providecommand \bibnamefont  [1]{#1}%
\providecommand \bibfnamefont [1]{#1}%
\providecommand \citenamefont [1]{#1}%
\providecommand \href@noop [0]{\@secondoftwo}%
\providecommand \href [0]{\begingroup \@sanitize@url \@href}%
\providecommand \@href[1]{\@@startlink{#1}\@@href}%
\providecommand \@@href[1]{\endgroup#1\@@endlink}%
\providecommand \@sanitize@url [0]{\catcode `\\12\catcode `\$12\catcode `\&12\catcode `\#12\catcode `\^12\catcode `\_12\catcode `\%12\relax}%
\providecommand \@@startlink[1]{}%
\providecommand \@@endlink[0]{}%
\providecommand \url  [0]{\begingroup\@sanitize@url \@url }%
\providecommand \@url [1]{\endgroup\@href {#1}{\urlprefix }}%
\providecommand \urlprefix  [0]{URL }%
\providecommand \Eprint [0]{\href }%
\providecommand \doibase [0]{http://dx.doi.org/}%
\providecommand \selectlanguage [0]{\@gobble}%
\providecommand \bibinfo  [0]{\@secondoftwo}%
\providecommand \bibfield  [0]{\@secondoftwo}%
\providecommand \translation [1]{[#1]}%
\providecommand \BibitemOpen [0]{}%
\providecommand \bibitemStop [0]{}%
\providecommand \bibitemNoStop [0]{.\EOS\space}%
\providecommand \EOS [0]{\spacefactor3000\relax}%
\providecommand \BibitemShut  [1]{\csname bibitem#1\endcsname}%
\let\auto@bib@innerbib\@empty
\bibitem [{\citenamefont {Becher}\ \emph {et~al.}(2023)\citenamefont {Becher}, \citenamefont {Gao}, \citenamefont {Kar}, \citenamefont {Marciniak}, \citenamefont {Monz}, \citenamefont {Bartholomew}, \citenamefont {Goldner}, \citenamefont {Loh}, \citenamefont {Marcellina}, \citenamefont {Goh}, \citenamefont {Koh}, \citenamefont {Weber}, \citenamefont {Mu}, \citenamefont {Tsai}, \citenamefont {Yan}, \citenamefont {Huber-Loyola}, \citenamefont {Höfling}, \citenamefont {Gyger}, \citenamefont {Steinhauer},\ and\ \citenamefont {Zwiller}}]{Becher2023}%
  \BibitemOpen
  \bibfield  {author} {\bibinfo {author} {\bibfnamefont {C.}~\bibnamefont {Becher}}, \bibinfo {author} {\bibfnamefont {W.}~\bibnamefont {Gao}}, \bibinfo {author} {\bibfnamefont {S.}~\bibnamefont {Kar}}, \bibinfo {author} {\bibfnamefont {C.~D.}\ \bibnamefont {Marciniak}}, \bibinfo {author} {\bibfnamefont {T.}~\bibnamefont {Monz}}, \bibinfo {author} {\bibfnamefont {J.~G.}\ \bibnamefont {Bartholomew}}, \bibinfo {author} {\bibfnamefont {P.}~\bibnamefont {Goldner}}, \bibinfo {author} {\bibfnamefont {H.}~\bibnamefont {Loh}}, \bibinfo {author} {\bibfnamefont {E.}~\bibnamefont {Marcellina}}, \bibinfo {author} {\bibfnamefont {K.~Eng~Johnson}\ \bibnamefont {Goh}}, \bibinfo {author} {\bibfnamefont {T.~Seng}\ \bibnamefont {Koh}}, \bibinfo {author} {\bibfnamefont {B.}~\bibnamefont {Weber}}, \bibinfo {author} {\bibfnamefont {Z.}~\bibnamefont {Mu}}, \bibinfo {author} {\bibfnamefont {J-Y.}\ \bibnamefont {Tsai}}, \bibinfo {author} {\bibfnamefont {Q.}~\bibnamefont {Yan}}, \bibinfo {author} {\bibfnamefont {T.}~\bibnamefont
  {Huber-Loyola}}, \bibinfo {author} {\bibfnamefont {S.}~\bibnamefont {Höfling}}, \bibinfo {author} {\bibfnamefont {S.}~\bibnamefont {Gyger}}, \bibinfo {author} {\bibfnamefont {S.}~\bibnamefont {Steinhauer}}, \ and\ \bibinfo {author} {\bibfnamefont {V.}~\bibnamefont {Zwiller}},\ }\bibfield  {title} {\enquote {\bibinfo {title} {2023 roadmap for materials for quantum technologies},}\ }\href@noop {} {\bibfield  {journal} {\bibinfo  {journal} {Materials for Quantum Technology}\ }\textbf {\bibinfo {volume} {3}},\ \bibinfo {pages} {012501} (\bibinfo {year} {2023})}\BibitemShut {NoStop}%
\bibitem [{\citenamefont {Awschalom}\ \emph {et~al.}(2018)\citenamefont {Awschalom}, \citenamefont {Hanson}, \citenamefont {Wrachtrup},\ and\ \citenamefont {Zhou}}]{Awschalom2018}%
  \BibitemOpen
  \bibfield  {author} {\bibinfo {author} {\bibfnamefont {D.~D.}\ \bibnamefont {Awschalom}}, \bibinfo {author} {\bibfnamefont {R.}~\bibnamefont {Hanson}}, \bibinfo {author} {\bibfnamefont {J.}~\bibnamefont {Wrachtrup}}, \ and\ \bibinfo {author} {\bibfnamefont {B.~B.}\ \bibnamefont {Zhou}},\ }\bibfield  {title} {\enquote {\bibinfo {title} {Quantum technologies with optically interfaced solid-state spins},}\ }\href@noop {} {\bibfield  {journal} {\bibinfo  {journal} {Nature Photonics}\ }\textbf {\bibinfo {volume} {12}},\ \bibinfo {pages} {516–527} (\bibinfo {year} {2018})}\BibitemShut {NoStop}%
\bibitem [{\citenamefont {Atatüre}\ \emph {et~al.}(2018)\citenamefont {Atatüre}, \citenamefont {Englund}, \citenamefont {Vamivakas}, \citenamefont {Lee},\ and\ \citenamefont {Wrachtrup}}]{Atature}%
  \BibitemOpen
  \bibfield  {author} {\bibinfo {author} {\bibfnamefont {M.}~\bibnamefont {Atatüre}}, \bibinfo {author} {\bibfnamefont {D.}~\bibnamefont {Englund}}, \bibinfo {author} {\bibfnamefont {N.}~\bibnamefont {Vamivakas}}, \bibinfo {author} {\bibfnamefont {S.~Yun}\ \bibnamefont {Lee}}, \ and\ \bibinfo {author} {\bibfnamefont {J.}~\bibnamefont {Wrachtrup}},\ }\bibfield  {title} {\enquote {\bibinfo {title} {Material platforms for spin-based photonic quantum technologies},}\ }\href@noop {} {\bibfield  {journal} {\bibinfo  {journal} {Nature Reviews Materials}\ }\textbf {\bibinfo {volume} {3}},\ \bibinfo {pages} {38–51} (\bibinfo {year} {2018})}\BibitemShut {NoStop}%
\bibitem [{\citenamefont {Wolfowicz}\ \emph {et~al.}(2021)\citenamefont {Wolfowicz}, \citenamefont {Heremans}, \citenamefont {Anderson}, \citenamefont {Kanai}, \citenamefont {Seo}, \citenamefont {Gali}, \citenamefont {Galli},\ and\ \citenamefont {Awschalom}}]{Wolfowicz2021}%
  \BibitemOpen
  \bibfield  {author} {\bibinfo {author} {\bibfnamefont {G.}~\bibnamefont {Wolfowicz}}, \bibinfo {author} {\bibfnamefont {F.~J.}\ \bibnamefont {Heremans}}, \bibinfo {author} {\bibfnamefont {C.~P.}\ \bibnamefont {Anderson}}, \bibinfo {author} {\bibfnamefont {S.}~\bibnamefont {Kanai}}, \bibinfo {author} {\bibfnamefont {H.}~\bibnamefont {Seo}}, \bibinfo {author} {\bibfnamefont {A.}~\bibnamefont {Gali}}, \bibinfo {author} {\bibfnamefont {G.}~\bibnamefont {Galli}}, \ and\ \bibinfo {author} {\bibfnamefont {D.~D.}\ \bibnamefont {Awschalom}},\ }\bibfield  {title} {\enquote {\bibinfo {title} {Quantum guidelines for solid-state spin defects},}\ }\href@noop {} {\bibfield  {journal} {\bibinfo  {journal} {Nature Reviews Materials}\ }\textbf {\bibinfo {volume} {6}},\ \bibinfo {pages} {906–925} (\bibinfo {year} {2021})}\BibitemShut {NoStop}%
\bibitem [{\citenamefont {O'Brien}(2007)}]{O'Brien2007}%
  \BibitemOpen
  \bibfield  {author} {\bibinfo {author} {\bibfnamefont {J.~L.}\ \bibnamefont {O'Brien}},\ }\bibfield  {title} {\enquote {\bibinfo {title} {Optical quantum computing},}\ }\href@noop {} {\bibfield  {journal} {\bibinfo  {journal} {Science}\ }\textbf {\bibinfo {volume} {318}},\ \bibinfo {pages} {1567--1570} (\bibinfo {year} {2007})}\BibitemShut {NoStop}%
\bibitem [{\citenamefont {Hanson}\ and\ \citenamefont {Awschalom}(2008)}]{Hanson2008}%
  \BibitemOpen
  \bibfield  {author} {\bibinfo {author} {\bibfnamefont {R.}~\bibnamefont {Hanson}}\ and\ \bibinfo {author} {\bibfnamefont {D.~D.}\ \bibnamefont {Awschalom}},\ }\bibfield  {title} {\enquote {\bibinfo {title} {Coherent manipulation of single spins in semiconductors},}\ }\href@noop {} {\bibfield  {journal} {\bibinfo  {journal} {Nature}\ }\textbf {\bibinfo {volume} {453}},\ \bibinfo {pages} {1043–1049} (\bibinfo {year} {2008})}\BibitemShut {NoStop}%
\bibitem [{\citenamefont {Couteau}\ \emph {et~al.}(2023)\citenamefont {Couteau}, \citenamefont {Barz}, \citenamefont {Durt}, \citenamefont {Gerrits}, \citenamefont {Huwer}, \citenamefont {Prevedel}, \citenamefont {Rarity}, \citenamefont {Shields},\ and\ \citenamefont {Weihs}}]{Couteau2023}%
  \BibitemOpen
  \bibfield  {author} {\bibinfo {author} {\bibfnamefont {C.}~\bibnamefont {Couteau}}, \bibinfo {author} {\bibfnamefont {S.}~\bibnamefont {Barz}}, \bibinfo {author} {\bibfnamefont {T.}~\bibnamefont {Durt}}, \bibinfo {author} {\bibfnamefont {T.}~\bibnamefont {Gerrits}}, \bibinfo {author} {\bibfnamefont {J.}~\bibnamefont {Huwer}}, \bibinfo {author} {\bibfnamefont {R.}~\bibnamefont {Prevedel}}, \bibinfo {author} {\bibfnamefont {J.}~\bibnamefont {Rarity}}, \bibinfo {author} {\bibfnamefont {A.}~\bibnamefont {Shields}}, \ and\ \bibinfo {author} {\bibfnamefont {G.}~\bibnamefont {Weihs}},\ }\bibfield  {title} {\enquote {\bibinfo {title} {Applications of single photons to quantum communication and computing},}\ }\href@noop {} {\bibfield  {journal} {\bibinfo  {journal} {Nature Reviews Physics}\ }\textbf {\bibinfo {volume} {5}},\ \bibinfo {pages} {326–338} (\bibinfo {year} {2023})}\BibitemShut {NoStop}%
\bibitem [{\citenamefont {DiVincenzo}(2000)}]{DiVincenzo2000}%
  \BibitemOpen
  \bibfield  {author} {\bibinfo {author} {\bibfnamefont {David~P.}\ \bibnamefont {DiVincenzo}},\ }\bibfield  {title} {\enquote {\bibinfo {title} {The physical implementation of quantum computation},}\ }\href@noop {} {\bibfield  {journal} {\bibinfo  {journal} {Fortschritte der Physik}\ }\textbf {\bibinfo {volume} {48}},\ \bibinfo {pages} {771--783} (\bibinfo {year} {2000})}\BibitemShut {NoStop}%
\bibitem [{\citenamefont {Schirhagl}\ \emph {et~al.}(2014)\citenamefont {Schirhagl}, \citenamefont {Chang}, \citenamefont {Loretz},\ and\ \citenamefont {Degen}}]{Schirhagl2014}%
  \BibitemOpen
  \bibfield  {author} {\bibinfo {author} {\bibfnamefont {R.}~\bibnamefont {Schirhagl}}, \bibinfo {author} {\bibfnamefont {K.}~\bibnamefont {Chang}}, \bibinfo {author} {\bibfnamefont {M.}~\bibnamefont {Loretz}}, \ and\ \bibinfo {author} {\bibfnamefont {C.~L.}\ \bibnamefont {Degen}},\ }\bibfield  {title} {\enquote {\bibinfo {title} {Nitrogen-vacancy centers in diamond: Nanoscale sensors for physics and biology},}\ }\href@noop {} {\bibfield  {journal} {\bibinfo  {journal} {Annual Review of Physical Chemistry}\ }\textbf {\bibinfo {volume} {65}},\ \bibinfo {pages} {83--105} (\bibinfo {year} {2014})}\BibitemShut {NoStop}%
\bibitem [{\citenamefont {Vaidya}\ \emph {et~al.}(2023)\citenamefont {Vaidya}, \citenamefont {Gao}, \citenamefont {Dikshit}, \citenamefont {Aharonovich},\ and\ \citenamefont {Li}}]{Vaidya2023}%
  \BibitemOpen
  \bibfield  {author} {\bibinfo {author} {\bibfnamefont {S.}~\bibnamefont {Vaidya}}, \bibinfo {author} {\bibfnamefont {X.}~\bibnamefont {Gao}}, \bibinfo {author} {\bibfnamefont {S.}~\bibnamefont {Dikshit}}, \bibinfo {author} {\bibfnamefont {I.}~\bibnamefont {Aharonovich}}, \ and\ \bibinfo {author} {\bibfnamefont {T.}~\bibnamefont {Li}},\ }\bibfield  {title} {\enquote {\bibinfo {title} {Quantum sensing and imaging with spin defects in hexagonal boron nitride},}\ }\href@noop {} {\bibfield  {journal} {\bibinfo  {journal} {Advances in Physics: X}\ }\textbf {\bibinfo {volume} {8}},\ \bibinfo {pages} {2206049} (\bibinfo {year} {2023})}\BibitemShut {NoStop}%
\bibitem [{\citenamefont {Kimble}(2008)}]{Kimble2008}%
  \BibitemOpen
  \bibfield  {author} {\bibinfo {author} {\bibfnamefont {H.~J.}\ \bibnamefont {Kimble}},\ }\bibfield  {title} {\enquote {\bibinfo {title} {The quantum internet},}\ }\href@noop {} {\bibfield  {journal} {\bibinfo  {journal} {Nature}\ }\textbf {\bibinfo {volume} {453}},\ \bibinfo {pages} {1023–1030} (\bibinfo {year} {2008})}\BibitemShut {NoStop}%
\bibitem [{\citenamefont {Doherty}\ \emph {et~al.}(2013)\citenamefont {Doherty}, \citenamefont {Manson}, \citenamefont {Delaney}, \citenamefont {Jelezko}, \citenamefont {Wrachtrup},\ and\ \citenamefont {Hollenberg}}]{Doherty2013}%
  \BibitemOpen
  \bibfield  {author} {\bibinfo {author} {\bibfnamefont {M.~W.}\ \bibnamefont {Doherty}}, \bibinfo {author} {\bibfnamefont {N.~B.}\ \bibnamefont {Manson}}, \bibinfo {author} {\bibfnamefont {P.}~\bibnamefont {Delaney}}, \bibinfo {author} {\bibfnamefont {F.}~\bibnamefont {Jelezko}}, \bibinfo {author} {\bibfnamefont {J.}~\bibnamefont {Wrachtrup}}, \ and\ \bibinfo {author} {\bibfnamefont {L.~C.~L.}\ \bibnamefont {Hollenberg}},\ }\bibfield  {title} {\enquote {\bibinfo {title} {The nitrogen-vacancy colour centre in diamond},}\ }\href@noop {} {\bibfield  {journal} {\bibinfo  {journal} {Physics Reports}\ }\textbf {\bibinfo {volume} {528}},\ \bibinfo {pages} {1--45} (\bibinfo {year} {2013})}\BibitemShut {NoStop}%
\bibitem [{\citenamefont {Togan}\ \emph {et~al.}(2010)\citenamefont {Togan}, \citenamefont {Chu}, \citenamefont {Trifonov}, \citenamefont {Jiang}, \citenamefont {Maze}, \citenamefont {Childress}, \citenamefont {Dutt}, \citenamefont {S{\o}rensen}, \citenamefont {Hemmer}, \citenamefont {Zibrov},\ and\ \citenamefont {Lukin}}]{Togan2010}%
  \BibitemOpen
  \bibfield  {author} {\bibinfo {author} {\bibfnamefont {E.}~\bibnamefont {Togan}}, \bibinfo {author} {\bibfnamefont {Y.}~\bibnamefont {Chu}}, \bibinfo {author} {\bibfnamefont {A.~S.}\ \bibnamefont {Trifonov}}, \bibinfo {author} {\bibfnamefont {L.}~\bibnamefont {Jiang}}, \bibinfo {author} {\bibfnamefont {J.}~\bibnamefont {Maze}}, \bibinfo {author} {\bibfnamefont {L.}~\bibnamefont {Childress}}, \bibinfo {author} {\bibfnamefont {M.~V.G.}\ \bibnamefont {Dutt}}, \bibinfo {author} {\bibfnamefont {A.~S.}\ \bibnamefont {S{\o}rensen}}, \bibinfo {author} {\bibfnamefont {P.~R.}\ \bibnamefont {Hemmer}}, \bibinfo {author} {\bibfnamefont {A.~S.}\ \bibnamefont {Zibrov}}, \ and\ \bibinfo {author} {\bibfnamefont {M.~D.}\ \bibnamefont {Lukin}},\ }\bibfield  {title} {\enquote {\bibinfo {title} {Quantum entanglement between an optical photon and a solid-state spin qubit},}\ }\href@noop {} {\bibfield  {journal} {\bibinfo  {journal} {Nature}\ }\textbf {\bibinfo {volume} {466}},\ \bibinfo {pages} {730–734} (\bibinfo {year}
  {2010})}\BibitemShut {NoStop}%
\bibitem [{\citenamefont {Bernien}\ \emph {et~al.}(2013)\citenamefont {Bernien}, \citenamefont {Hensen}, \citenamefont {Pfaff}, \citenamefont {Koolstra}, \citenamefont {Blok}, \citenamefont {Robledo}, \citenamefont {Taminiau}, \citenamefont {Markham}, \citenamefont {Twitchen}, \citenamefont {Childress},\ and\ \citenamefont {Hanson}}]{Bernien2013}%
  \BibitemOpen
  \bibfield  {author} {\bibinfo {author} {\bibfnamefont {H.}~\bibnamefont {Bernien}}, \bibinfo {author} {\bibfnamefont {B.}~\bibnamefont {Hensen}}, \bibinfo {author} {\bibfnamefont {W.}~\bibnamefont {Pfaff}}, \bibinfo {author} {\bibfnamefont {G.}~\bibnamefont {Koolstra}}, \bibinfo {author} {\bibfnamefont {M.~S.}\ \bibnamefont {Blok}}, \bibinfo {author} {\bibfnamefont {L.}~\bibnamefont {Robledo}}, \bibinfo {author} {\bibfnamefont {T.~H.}\ \bibnamefont {Taminiau}}, \bibinfo {author} {\bibfnamefont {M.}~\bibnamefont {Markham}}, \bibinfo {author} {\bibfnamefont {D.~J.}\ \bibnamefont {Twitchen}}, \bibinfo {author} {\bibfnamefont {L.}~\bibnamefont {Childress}}, \ and\ \bibinfo {author} {\bibfnamefont {R.}~\bibnamefont {Hanson}},\ }\bibfield  {title} {\enquote {\bibinfo {title} {Heralded entanglement between solid-state qubits separated by three metres},}\ }\href@noop {} {\bibfield  {journal} {\bibinfo  {journal} {Nature}\ }\textbf {\bibinfo {volume} {497}},\ \bibinfo {pages} {86–90} (\bibinfo {year}
  {2013})}\BibitemShut {NoStop}%
\bibitem [{\citenamefont {Humphreys}\ \emph {et~al.}(2018)\citenamefont {Humphreys}, \citenamefont {Kalb}, \citenamefont {Morits}, \citenamefont {Schouten}, \citenamefont {Vermeulen}, \citenamefont {Twitchen}, \citenamefont {Markham},\ and\ \citenamefont {Hanson}}]{Humphreys2018}%
  \BibitemOpen
  \bibfield  {author} {\bibinfo {author} {\bibfnamefont {P.~C.}\ \bibnamefont {Humphreys}}, \bibinfo {author} {\bibfnamefont {N.}~\bibnamefont {Kalb}}, \bibinfo {author} {\bibfnamefont {J.~P.J.}\ \bibnamefont {Morits}}, \bibinfo {author} {\bibfnamefont {R.~N.}\ \bibnamefont {Schouten}}, \bibinfo {author} {\bibfnamefont {R.~F.L.}\ \bibnamefont {Vermeulen}}, \bibinfo {author} {\bibfnamefont {D.~J.}\ \bibnamefont {Twitchen}}, \bibinfo {author} {\bibfnamefont {M.}~\bibnamefont {Markham}}, \ and\ \bibinfo {author} {\bibfnamefont {R.}~\bibnamefont {Hanson}},\ }\bibfield  {title} {\enquote {\bibinfo {title} {Deterministic delivery of remote entanglement on a quantum network},}\ }\href@noop {} {\bibfield  {journal} {\bibinfo  {journal} {Nature}\ }\textbf {\bibinfo {volume} {558}},\ \bibinfo {pages} {268–273} (\bibinfo {year} {2018})}\BibitemShut {NoStop}%
\bibitem [{\citenamefont {Pompili}\ \emph {et~al.}(2021)\citenamefont {Pompili}, \citenamefont {Hermans}, \citenamefont {Baier}, \citenamefont {Beukers}, \citenamefont {Humphreys}, \citenamefont {Schouten}, \citenamefont {Vermeulen}, \citenamefont {Tiggelman}, \citenamefont {dos Santos~Martins}, \citenamefont {Dirkse}, \citenamefont {Wehner},\ and\ \citenamefont {Hanson}}]{Pompili2021}%
  \BibitemOpen
  \bibfield  {author} {\bibinfo {author} {\bibfnamefont {M.}~\bibnamefont {Pompili}}, \bibinfo {author} {\bibfnamefont {S.~L.N.}\ \bibnamefont {Hermans}}, \bibinfo {author} {\bibfnamefont {S.}~\bibnamefont {Baier}}, \bibinfo {author} {\bibfnamefont {H.~K.C.}\ \bibnamefont {Beukers}}, \bibinfo {author} {\bibfnamefont {P.~C.}\ \bibnamefont {Humphreys}}, \bibinfo {author} {\bibfnamefont {R.~N.}\ \bibnamefont {Schouten}}, \bibinfo {author} {\bibfnamefont {R.~F.L.}\ \bibnamefont {Vermeulen}}, \bibinfo {author} {\bibfnamefont {M.~J.}\ \bibnamefont {Tiggelman}}, \bibinfo {author} {\bibfnamefont {L.}~\bibnamefont {dos Santos~Martins}}, \bibinfo {author} {\bibfnamefont {B.}~\bibnamefont {Dirkse}}, \bibinfo {author} {\bibfnamefont {S.}~\bibnamefont {Wehner}}, \ and\ \bibinfo {author} {\bibfnamefont {R.}~\bibnamefont {Hanson}},\ }\bibfield  {title} {\enquote {\bibinfo {title} {Realization of a multinode quantum network of remote solid-state qubits},}\ }\href@noop {} {\bibfield  {journal} {\bibinfo  {journal} {Science}\
  }\textbf {\bibinfo {volume} {372}},\ \bibinfo {pages} {259--264} (\bibinfo {year} {2021})}\BibitemShut {NoStop}%
\bibitem [{\citenamefont {Dutt}\ \emph {et~al.}(2007)\citenamefont {Dutt}, \citenamefont {Childress}, \citenamefont {Jiang}, \citenamefont {Togan}, \citenamefont {Maze}, \citenamefont {Jelezko}, \citenamefont {Zibrov}, \citenamefont {Hemmer},\ and\ \citenamefont {Lukin}}]{Dutt}%
  \BibitemOpen
  \bibfield  {author} {\bibinfo {author} {\bibfnamefont {M.~V.~Gurudev}\ \bibnamefont {Dutt}}, \bibinfo {author} {\bibfnamefont {L.}~\bibnamefont {Childress}}, \bibinfo {author} {\bibfnamefont {L.}~\bibnamefont {Jiang}}, \bibinfo {author} {\bibfnamefont {E.}~\bibnamefont {Togan}}, \bibinfo {author} {\bibfnamefont {J.}~\bibnamefont {Maze}}, \bibinfo {author} {\bibfnamefont {F.}~\bibnamefont {Jelezko}}, \bibinfo {author} {\bibfnamefont {A.~S.}\ \bibnamefont {Zibrov}}, \bibinfo {author} {\bibfnamefont {P.~R.}\ \bibnamefont {Hemmer}}, \ and\ \bibinfo {author} {\bibfnamefont {M.~D.}\ \bibnamefont {Lukin}},\ }\bibfield  {title} {\enquote {\bibinfo {title} {Quantum register based on individual electronic and nuclear spin qubits in diamond},}\ }\href@noop {} {\bibfield  {journal} {\bibinfo  {journal} {Science}\ }\textbf {\bibinfo {volume} {316}},\ \bibinfo {pages} {1312--1316} (\bibinfo {year} {2007})}\BibitemShut {NoStop}%
\bibitem [{\citenamefont {Taminiau}\ \emph {et~al.}(2014)\citenamefont {Taminiau}, \citenamefont {Cramer}, \citenamefont {Sar}, \citenamefont {Dobrovitski},\ and\ \citenamefont {Hanson}}]{Taminiau2014}%
  \BibitemOpen
  \bibfield  {author} {\bibinfo {author} {\bibfnamefont {T.~H.}\ \bibnamefont {Taminiau}}, \bibinfo {author} {\bibfnamefont {J.}~\bibnamefont {Cramer}}, \bibinfo {author} {\bibfnamefont {T.~Van~Der}\ \bibnamefont {Sar}}, \bibinfo {author} {\bibfnamefont {V.~V.}\ \bibnamefont {Dobrovitski}}, \ and\ \bibinfo {author} {\bibfnamefont {R.}~\bibnamefont {Hanson}},\ }\bibfield  {title} {\enquote {\bibinfo {title} {Universal control and error correction in multi-qubit spin registers in diamond},}\ }\href@noop {} {\bibfield  {journal} {\bibinfo  {journal} {Nature Nanotechnology}\ }\textbf {\bibinfo {volume} {9}},\ \bibinfo {pages} {171–176} (\bibinfo {year} {2014})}\BibitemShut {NoStop}%
\bibitem [{\citenamefont {Bradley}\ \emph {et~al.}(2019)\citenamefont {Bradley}, \citenamefont {Randall}, \citenamefont {Abobeih}, \citenamefont {Berrevoets}, \citenamefont {Degen}, \citenamefont {Bakker}, \citenamefont {Markham}, \citenamefont {Twitchen},\ and\ \citenamefont {Taminiau}}]{Bradley2019}%
  \BibitemOpen
  \bibfield  {author} {\bibinfo {author} {\bibfnamefont {C.~E.}\ \bibnamefont {Bradley}}, \bibinfo {author} {\bibfnamefont {J.}~\bibnamefont {Randall}}, \bibinfo {author} {\bibfnamefont {M.~H.}\ \bibnamefont {Abobeih}}, \bibinfo {author} {\bibfnamefont {R.~C.}\ \bibnamefont {Berrevoets}}, \bibinfo {author} {\bibfnamefont {M.~J.}\ \bibnamefont {Degen}}, \bibinfo {author} {\bibfnamefont {M.~A.}\ \bibnamefont {Bakker}}, \bibinfo {author} {\bibfnamefont {M.}~\bibnamefont {Markham}}, \bibinfo {author} {\bibfnamefont {D.~J.}\ \bibnamefont {Twitchen}}, \ and\ \bibinfo {author} {\bibfnamefont {T.~H.}\ \bibnamefont {Taminiau}},\ }\bibfield  {title} {\enquote {\bibinfo {title} {A ten-qubit solid-state spin register with quantum memory up to one minute},}\ }\href@noop {} {\bibfield  {journal} {\bibinfo  {journal} {Physical Review X}\ }\textbf {\bibinfo {volume} {9}},\ \bibinfo {pages} {031045} (\bibinfo {year} {2019})}\BibitemShut {NoStop}%
\bibitem [{\citenamefont {Chernobrod}\ and\ \citenamefont {Berman}(2005)}]{Chernobrod2005}%
  \BibitemOpen
  \bibfield  {author} {\bibinfo {author} {\bibfnamefont {B.~M.}\ \bibnamefont {Chernobrod}}\ and\ \bibinfo {author} {\bibfnamefont {G.~P.}\ \bibnamefont {Berman}},\ }\bibfield  {title} {\enquote {\bibinfo {title} {Spin microscope based on optically detected magnetic resonance},}\ }\href@noop {} {\bibfield  {journal} {\bibinfo  {journal} {Journal of Applied Physics}\ }\textbf {\bibinfo {volume} {97}},\ \bibinfo {pages} {014903} (\bibinfo {year} {2005})}\BibitemShut {NoStop}%
\bibitem [{\citenamefont {Taylor}\ \emph {et~al.}(2008)\citenamefont {Taylor}, \citenamefont {Cappellaro}, \citenamefont {Childress}, \citenamefont {Jiang}, \citenamefont {Budker}, \citenamefont {Hemmer}, \citenamefont {Yacoby}, \citenamefont {Walsworth},\ and\ \citenamefont {Lukin}}]{Taylor2008}%
  \BibitemOpen
  \bibfield  {author} {\bibinfo {author} {\bibfnamefont {J.~M.}\ \bibnamefont {Taylor}}, \bibinfo {author} {\bibfnamefont {P.}~\bibnamefont {Cappellaro}}, \bibinfo {author} {\bibfnamefont {L.}~\bibnamefont {Childress}}, \bibinfo {author} {\bibfnamefont {L.}~\bibnamefont {Jiang}}, \bibinfo {author} {\bibfnamefont {D.}~\bibnamefont {Budker}}, \bibinfo {author} {\bibfnamefont {P.~R.}\ \bibnamefont {Hemmer}}, \bibinfo {author} {\bibfnamefont {A.}~\bibnamefont {Yacoby}}, \bibinfo {author} {\bibfnamefont {R.}~\bibnamefont {Walsworth}}, \ and\ \bibinfo {author} {\bibfnamefont {M.~D.}\ \bibnamefont {Lukin}},\ }\bibfield  {title} {\enquote {\bibinfo {title} {High-sensitivity diamond magnetometer with nanoscale resolution},}\ }\href@noop {} {\bibfield  {journal} {\bibinfo  {journal} {Nature Physics}\ }\textbf {\bibinfo {volume} {4}},\ \bibinfo {pages} {810–816} (\bibinfo {year} {2008})}\BibitemShut {NoStop}%
\bibitem [{\citenamefont {Degen}(2008)}]{Degen2008}%
  \BibitemOpen
  \bibfield  {author} {\bibinfo {author} {\bibfnamefont {C.~L.}\ \bibnamefont {Degen}},\ }\bibfield  {title} {\enquote {\bibinfo {title} {Scanning magnetic field microscope with a diamond single-spin sensor},}\ }\href@noop {} {\bibfield  {journal} {\bibinfo  {journal} {Applied Physics Letters}\ }\textbf {\bibinfo {volume} {92}},\ \bibinfo {pages} {243111} (\bibinfo {year} {2008})}\BibitemShut {NoStop}%
\bibitem [{\citenamefont {Degen}\ \emph {et~al.}(2017)\citenamefont {Degen}, \citenamefont {Reinhard},\ and\ \citenamefont {Cappellaro}}]{Degen2017}%
  \BibitemOpen
  \bibfield  {author} {\bibinfo {author} {\bibfnamefont {C.~L.}\ \bibnamefont {Degen}}, \bibinfo {author} {\bibfnamefont {F.}~\bibnamefont {Reinhard}}, \ and\ \bibinfo {author} {\bibfnamefont {P.}~\bibnamefont {Cappellaro}},\ }\bibfield  {title} {\enquote {\bibinfo {title} {Quantum sensing},}\ }\href@noop {} {\bibfield  {journal} {\bibinfo  {journal} {Reviews of Modern Physics}\ }\textbf {\bibinfo {volume} {89}},\ \bibinfo {pages} {035002} (\bibinfo {year} {2017})}\BibitemShut {NoStop}%
\bibitem [{\citenamefont {Koehl}\ \emph {et~al.}(2011)\citenamefont {Koehl}, \citenamefont {Buckley}, \citenamefont {Heremans}, \citenamefont {Calusine},\ and\ \citenamefont {Awschalom}}]{Koehl2011}%
  \BibitemOpen
  \bibfield  {author} {\bibinfo {author} {\bibfnamefont {W.~F.}\ \bibnamefont {Koehl}}, \bibinfo {author} {\bibfnamefont {B.~B.}\ \bibnamefont {Buckley}}, \bibinfo {author} {\bibfnamefont {F.~Joseph}\ \bibnamefont {Heremans}}, \bibinfo {author} {\bibfnamefont {G.}~\bibnamefont {Calusine}}, \ and\ \bibinfo {author} {\bibfnamefont {D.~D.}\ \bibnamefont {Awschalom}},\ }\bibfield  {title} {\enquote {\bibinfo {title} {Room temperature coherent control of defect spin qubits in silicon carbide},}\ }\href@noop {} {\bibfield  {journal} {\bibinfo  {journal} {Nature}\ }\textbf {\bibinfo {volume} {479}},\ \bibinfo {pages} {84–87} (\bibinfo {year} {2011})}\BibitemShut {NoStop}%
\bibitem [{\citenamefont {Bourassa}\ \emph {et~al.}(2020)\citenamefont {Bourassa}, \citenamefont {Anderson}, \citenamefont {Miao}, \citenamefont {Onizhuk}, \citenamefont {Ma}, \citenamefont {Crook}, \citenamefont {Abe}, \citenamefont {Ul-Hassan}, \citenamefont {Ohshima}, \citenamefont {Son}, \citenamefont {Galli},\ and\ \citenamefont {Awschalom}}]{Bourassa2020}%
  \BibitemOpen
  \bibfield  {author} {\bibinfo {author} {\bibfnamefont {A.}~\bibnamefont {Bourassa}}, \bibinfo {author} {\bibfnamefont {C.~P.}\ \bibnamefont {Anderson}}, \bibinfo {author} {\bibfnamefont {K.~C.}\ \bibnamefont {Miao}}, \bibinfo {author} {\bibfnamefont {M.}~\bibnamefont {Onizhuk}}, \bibinfo {author} {\bibfnamefont {H.}~\bibnamefont {Ma}}, \bibinfo {author} {\bibfnamefont {A.~L.}\ \bibnamefont {Crook}}, \bibinfo {author} {\bibfnamefont {H.}~\bibnamefont {Abe}}, \bibinfo {author} {\bibfnamefont {J.}~\bibnamefont {Ul-Hassan}}, \bibinfo {author} {\bibfnamefont {T.}~\bibnamefont {Ohshima}}, \bibinfo {author} {\bibfnamefont {N.~T.}\ \bibnamefont {Son}}, \bibinfo {author} {\bibfnamefont {G.}~\bibnamefont {Galli}}, \ and\ \bibinfo {author} {\bibfnamefont {D.~D.}\ \bibnamefont {Awschalom}},\ }\bibfield  {title} {\enquote {\bibinfo {title} {Entanglement and control of single nuclear spins in isotopically engineered silicon carbide},}\ }\href@noop {} {\bibfield  {journal} {\bibinfo  {journal} {Nature Materials}\ }\textbf
  {\bibinfo {volume} {19}},\ \bibinfo {pages} {1319–1325} (\bibinfo {year} {2020})}\BibitemShut {NoStop}%
\bibitem [{\citenamefont {Redjem}\ \emph {et~al.}(2020)\citenamefont {Redjem}, \citenamefont {Durand}, \citenamefont {Herzig}, \citenamefont {Benali}, \citenamefont {Pezzagna}, \citenamefont {Meijer}, \citenamefont {Kuznetsov}, \citenamefont {Nguyen}, \citenamefont {Cueff}, \citenamefont {Gérard}, \citenamefont {Robert-Philip}, \citenamefont {Gil}, \citenamefont {Caliste}, \citenamefont {Pochet}, \citenamefont {Abbarchi}, \citenamefont {Jacques}, \citenamefont {Dréau},\ and\ \citenamefont {Cassabois}}]{Redjem2020}%
  \BibitemOpen
  \bibfield  {author} {\bibinfo {author} {\bibfnamefont {W.}~\bibnamefont {Redjem}}, \bibinfo {author} {\bibfnamefont {A.}~\bibnamefont {Durand}}, \bibinfo {author} {\bibfnamefont {T.}~\bibnamefont {Herzig}}, \bibinfo {author} {\bibfnamefont {A.}~\bibnamefont {Benali}}, \bibinfo {author} {\bibfnamefont {S.}~\bibnamefont {Pezzagna}}, \bibinfo {author} {\bibfnamefont {J.}~\bibnamefont {Meijer}}, \bibinfo {author} {\bibfnamefont {A.~Yu}\ \bibnamefont {Kuznetsov}}, \bibinfo {author} {\bibfnamefont {H.~S.}\ \bibnamefont {Nguyen}}, \bibinfo {author} {\bibfnamefont {S.}~\bibnamefont {Cueff}}, \bibinfo {author} {\bibfnamefont {J.~M.}\ \bibnamefont {Gérard}}, \bibinfo {author} {\bibfnamefont {I.}~\bibnamefont {Robert-Philip}}, \bibinfo {author} {\bibfnamefont {B.}~\bibnamefont {Gil}}, \bibinfo {author} {\bibfnamefont {D.}~\bibnamefont {Caliste}}, \bibinfo {author} {\bibfnamefont {P.}~\bibnamefont {Pochet}}, \bibinfo {author} {\bibfnamefont {M.}~\bibnamefont {Abbarchi}}, \bibinfo {author} {\bibfnamefont
  {V.}~\bibnamefont {Jacques}}, \bibinfo {author} {\bibfnamefont {A.}~\bibnamefont {Dréau}}, \ and\ \bibinfo {author} {\bibfnamefont {G.}~\bibnamefont {Cassabois}},\ }\bibfield  {title} {\enquote {\bibinfo {title} {Single artificial atoms in silicon emitting at telecom wavelengths},}\ }\href@noop {} {\bibfield  {journal} {\bibinfo  {journal} {Nature Electronics}\ }\textbf {\bibinfo {volume} {3}},\ \bibinfo {pages} {738–743} (\bibinfo {year} {2020})}\BibitemShut {NoStop}%
\bibitem [{\citenamefont {Higginbottom}\ \emph {et~al.}(2022)\citenamefont {Higginbottom}, \citenamefont {Kurkjian}, \citenamefont {Chartrand}, \citenamefont {Kazemi}, \citenamefont {Brunelle}, \citenamefont {MacQuarrie}, \citenamefont {Klein}, \citenamefont {Lee-Hone}, \citenamefont {Stacho}, \citenamefont {Ruether}, \citenamefont {Bowness}, \citenamefont {Bergeron}, \citenamefont {DeAbreu}, \citenamefont {Harrigan}, \citenamefont {Kanaganayagam}, \citenamefont {Marsden}, \citenamefont {Richards}, \citenamefont {Stott}, \citenamefont {Roorda}, \citenamefont {Morse}, \citenamefont {Thewalt},\ and\ \citenamefont {Simmons}}]{Higginbottom2022}%
  \BibitemOpen
  \bibfield  {author} {\bibinfo {author} {\bibfnamefont {D.~B.}\ \bibnamefont {Higginbottom}}, \bibinfo {author} {\bibfnamefont {A.~T.K.}\ \bibnamefont {Kurkjian}}, \bibinfo {author} {\bibfnamefont {C.}~\bibnamefont {Chartrand}}, \bibinfo {author} {\bibfnamefont {M.}~\bibnamefont {Kazemi}}, \bibinfo {author} {\bibfnamefont {N.~A.}\ \bibnamefont {Brunelle}}, \bibinfo {author} {\bibfnamefont {E.~R.}\ \bibnamefont {MacQuarrie}}, \bibinfo {author} {\bibfnamefont {J.~R.}\ \bibnamefont {Klein}}, \bibinfo {author} {\bibfnamefont {N.~R.}\ \bibnamefont {Lee-Hone}}, \bibinfo {author} {\bibfnamefont {J.}~\bibnamefont {Stacho}}, \bibinfo {author} {\bibfnamefont {M.}~\bibnamefont {Ruether}}, \bibinfo {author} {\bibfnamefont {C.}~\bibnamefont {Bowness}}, \bibinfo {author} {\bibfnamefont {L.}~\bibnamefont {Bergeron}}, \bibinfo {author} {\bibfnamefont {A.}~\bibnamefont {DeAbreu}}, \bibinfo {author} {\bibfnamefont {S.~R.}\ \bibnamefont {Harrigan}}, \bibinfo {author} {\bibfnamefont {J.}~\bibnamefont {Kanaganayagam}}, \bibinfo
  {author} {\bibfnamefont {D.~W.}\ \bibnamefont {Marsden}}, \bibinfo {author} {\bibfnamefont {T.~S.}\ \bibnamefont {Richards}}, \bibinfo {author} {\bibfnamefont {L.~A.}\ \bibnamefont {Stott}}, \bibinfo {author} {\bibfnamefont {S.}~\bibnamefont {Roorda}}, \bibinfo {author} {\bibfnamefont {K.~J.}\ \bibnamefont {Morse}}, \bibinfo {author} {\bibfnamefont {M.~L.~W.}\ \bibnamefont {Thewalt}}, \ and\ \bibinfo {author} {\bibfnamefont {S.}~\bibnamefont {Simmons}},\ }\bibfield  {title} {\enquote {\bibinfo {title} {Optical observation of single spins in silicon},}\ }\href@noop {} {\bibfield  {journal} {\bibinfo  {journal} {Nature}\ }\textbf {\bibinfo {volume} {607}},\ \bibinfo {pages} {266–270} (\bibinfo {year} {2022})}\BibitemShut {NoStop}%
\bibitem [{\citenamefont {Simmons}(2024)}]{Simmons2024}%
  \BibitemOpen
  \bibfield  {author} {\bibinfo {author} {\bibfnamefont {S.}~\bibnamefont {Simmons}},\ }\bibfield  {title} {\enquote {\bibinfo {title} {Scalable fault-tolerant quantum technologies with silicon color centers},}\ }\href@noop {} {\bibfield  {journal} {\bibinfo  {journal} {PRX Quantum}\ }\textbf {\bibinfo {volume} {5}},\ \bibinfo {pages} {010102} (\bibinfo {year} {2024})}\BibitemShut {NoStop}%
\bibitem [{\citenamefont {Durand}\ \emph {et~al.}(2024)\citenamefont {Durand}, \citenamefont {Baron}, \citenamefont {Cache}, \citenamefont {Herzig}, \citenamefont {Khoury}, \citenamefont {Pezzagna}, \citenamefont {Meijer}, \citenamefont {Hartmann}, \citenamefont {Reboh}, \citenamefont {Abbarchi}, \citenamefont {Robert-Philip}, \citenamefont {G\'{e}rard}, \citenamefont {Jacques}, \citenamefont {Cassabois},\ and\ \citenamefont {Dr\'{e}au}}]{Durand2024_silicon}%
  \BibitemOpen
  \bibfield  {author} {\bibinfo {author} {\bibfnamefont {A.}~\bibnamefont {Durand}}, \bibinfo {author} {\bibfnamefont {Y.}~\bibnamefont {Baron}}, \bibinfo {author} {\bibfnamefont {F.}~\bibnamefont {Cache}}, \bibinfo {author} {\bibfnamefont {T.}~\bibnamefont {Herzig}}, \bibinfo {author} {\bibfnamefont {M.}~\bibnamefont {Khoury}}, \bibinfo {author} {\bibfnamefont {S.}~\bibnamefont {Pezzagna}}, \bibinfo {author} {\bibfnamefont {J.}~\bibnamefont {Meijer}}, \bibinfo {author} {\bibfnamefont {J-M.}\ \bibnamefont {Hartmann}}, \bibinfo {author} {\bibfnamefont {S.}~\bibnamefont {Reboh}}, \bibinfo {author} {\bibfnamefont {M.}~\bibnamefont {Abbarchi}}, \bibinfo {author} {\bibfnamefont {I.}~\bibnamefont {Robert-Philip}}, \bibinfo {author} {\bibfnamefont {J-M.}\ \bibnamefont {G\'{e}rard}}, \bibinfo {author} {\bibfnamefont {V.}~\bibnamefont {Jacques}}, \bibinfo {author} {\bibfnamefont {G.}~\bibnamefont {Cassabois}}, \ and\ \bibinfo {author} {\bibfnamefont {A.}~\bibnamefont {Dr\'{e}au}},\ }\bibfield  {title} {\enquote
  {\bibinfo {title} {Genuine and faux single g centers in carbon-implanted silicon},}\ }\href@noop {} {\bibfield  {journal} {\bibinfo  {journal} {Physical Review B}\ }\textbf {\bibinfo {volume} {110}},\ \bibinfo {pages} {L020102} (\bibinfo {year} {2024})}\BibitemShut {NoStop}%
\bibitem [{\citenamefont {Luo}\ \emph {et~al.}(2024)\citenamefont {Luo}, \citenamefont {Geng}, \citenamefont {Rana},\ and\ \citenamefont {Fuchs}}]{Luo2024}%
  \BibitemOpen
  \bibfield  {author} {\bibinfo {author} {\bibfnamefont {J.}~\bibnamefont {Luo}}, \bibinfo {author} {\bibfnamefont {Y.}~\bibnamefont {Geng}}, \bibinfo {author} {\bibfnamefont {F.}~\bibnamefont {Rana}}, \ and\ \bibinfo {author} {\bibfnamefont {G.~D.}\ \bibnamefont {Fuchs}},\ }\bibfield  {title} {\enquote {\bibinfo {title} {Room temperature optically detected magnetic resonance of single spins in gan},}\ }\href@noop {} {\bibfield  {journal} {\bibinfo  {journal} {Nature Materials}\ }\textbf {\bibinfo {volume} {23}},\ \bibinfo {pages} {512–518} (\bibinfo {year} {2024})}\BibitemShut {NoStop}%
\bibitem [{\citenamefont {Tran}\ \emph {et~al.}(2016{\natexlab{a}})\citenamefont {Tran}, \citenamefont {Bray}, \citenamefont {Ford}, \citenamefont {Toth},\ and\ \citenamefont {Aharonovich}}]{Tran2016a}%
  \BibitemOpen
  \bibfield  {author} {\bibinfo {author} {\bibfnamefont {T.~T.}\ \bibnamefont {Tran}}, \bibinfo {author} {\bibfnamefont {K.}~\bibnamefont {Bray}}, \bibinfo {author} {\bibfnamefont {M.~J.}\ \bibnamefont {Ford}}, \bibinfo {author} {\bibfnamefont {M.}~\bibnamefont {Toth}}, \ and\ \bibinfo {author} {\bibfnamefont {I.}~\bibnamefont {Aharonovich}},\ }\bibfield  {title} {\enquote {\bibinfo {title} {Quantum emission from hexagonal boron nitride monolayers},}\ }\href@noop {} {\bibfield  {journal} {\bibinfo  {journal} {Nature Nanotechnology}\ }\textbf {\bibinfo {volume} {11}},\ \bibinfo {pages} {37–41} (\bibinfo {year} {2016}{\natexlab{a}})}\BibitemShut {NoStop}%
\bibitem [{\citenamefont {Aharonovich}\ \emph {et~al.}(2016)\citenamefont {Aharonovich}, \citenamefont {Englund},\ and\ \citenamefont {Toth}}]{Aharonovich2016}%
  \BibitemOpen
  \bibfield  {author} {\bibinfo {author} {\bibfnamefont {I.}~\bibnamefont {Aharonovich}}, \bibinfo {author} {\bibfnamefont {D.}~\bibnamefont {Englund}}, \ and\ \bibinfo {author} {\bibfnamefont {M.ilos}\ \bibnamefont {Toth}},\ }\bibfield  {title} {\enquote {\bibinfo {title} {Solid-state single-photon emitters},}\ }\href@noop {} {\bibfield  {journal} {\bibinfo  {journal} {Nature Photonics}\ }\textbf {\bibinfo {volume} {10}},\ \bibinfo {pages} {631--641} (\bibinfo {year} {2016})}\BibitemShut {NoStop}%
\bibitem [{\citenamefont {Gottscholl}\ \emph {et~al.}(2020)\citenamefont {Gottscholl}, \citenamefont {Kianinia}, \citenamefont {Soltamov}, \citenamefont {Orlinskii}, \citenamefont {Mamin}, \citenamefont {Bradac}, \citenamefont {Kasper}, \citenamefont {Krambrock}, \citenamefont {Sperlich}, \citenamefont {Toth}, \citenamefont {Aharonovich},\ and\ \citenamefont {Dyakonov}}]{Gottscholl2020}%
  \BibitemOpen
  \bibfield  {author} {\bibinfo {author} {\bibfnamefont {A.}~\bibnamefont {Gottscholl}}, \bibinfo {author} {\bibfnamefont {M.}~\bibnamefont {Kianinia}}, \bibinfo {author} {\bibfnamefont {V.}~\bibnamefont {Soltamov}}, \bibinfo {author} {\bibfnamefont {S.}~\bibnamefont {Orlinskii}}, \bibinfo {author} {\bibfnamefont {G.}~\bibnamefont {Mamin}}, \bibinfo {author} {\bibfnamefont {C.}~\bibnamefont {Bradac}}, \bibinfo {author} {\bibfnamefont {C.}~\bibnamefont {Kasper}}, \bibinfo {author} {\bibfnamefont {K.}~\bibnamefont {Krambrock}}, \bibinfo {author} {\bibfnamefont {A.}~\bibnamefont {Sperlich}}, \bibinfo {author} {\bibfnamefont {M.}~\bibnamefont {Toth}}, \bibinfo {author} {\bibfnamefont {I.}~\bibnamefont {Aharonovich}}, \ and\ \bibinfo {author} {\bibfnamefont {V.}~\bibnamefont {Dyakonov}},\ }\bibfield  {title} {\enquote {\bibinfo {title} {Initialisation and read-out of intrinsic spin defects in a van der waals crystal at room temperature},}\ }\href@noop {} {\bibfield  {journal} {\bibinfo  {journal} {Nature
  Materials}\ }\textbf {\bibinfo {volume} {19}},\ \bibinfo {pages} {540–545} (\bibinfo {year} {2020})}\BibitemShut {NoStop}%
\bibitem [{\citenamefont {Chejanovsky}\ \emph {et~al.}(2021)\citenamefont {Chejanovsky}, \citenamefont {Mukherjee}, \citenamefont {Geng}, \citenamefont {Chen}, \citenamefont {Kim}, \citenamefont {Denisenko}, \citenamefont {Finkler}, \citenamefont {Taniguchi}, \citenamefont {Watanabe}, \citenamefont {Dasari}, \citenamefont {P.Auburger}, \citenamefont {Gali}, \citenamefont {Smet},\ and\ \citenamefont {Wrachtrup}}]{Chejanovsky2021}%
  \BibitemOpen
  \bibfield  {author} {\bibinfo {author} {\bibfnamefont {N.}~\bibnamefont {Chejanovsky}}, \bibinfo {author} {\bibfnamefont {A.}~\bibnamefont {Mukherjee}}, \bibinfo {author} {\bibfnamefont {J.}~\bibnamefont {Geng}}, \bibinfo {author} {\bibfnamefont {Y.~C.}\ \bibnamefont {Chen}}, \bibinfo {author} {\bibfnamefont {Y.}~\bibnamefont {Kim}}, \bibinfo {author} {\bibfnamefont {A.}~\bibnamefont {Denisenko}}, \bibinfo {author} {\bibfnamefont {A.}~\bibnamefont {Finkler}}, \bibinfo {author} {\bibfnamefont {T.}~\bibnamefont {Taniguchi}}, \bibinfo {author} {\bibfnamefont {K.}~\bibnamefont {Watanabe}}, \bibinfo {author} {\bibfnamefont {D.~Bhaktavatsala~Rao}\ \bibnamefont {Dasari}}, \bibinfo {author} {\bibnamefont {P.Auburger}}, \bibinfo {author} {\bibfnamefont {A.}~\bibnamefont {Gali}}, \bibinfo {author} {\bibfnamefont {J.~H.}\ \bibnamefont {Smet}}, \ and\ \bibinfo {author} {\bibfnamefont {J.}~\bibnamefont {Wrachtrup}},\ }\bibfield  {title} {\enquote {\bibinfo {title} {Single-spin resonance in a van der waals embedded
  paramagnetic defect},}\ }\href@noop {} {\bibfield  {journal} {\bibinfo  {journal} {Nature Materials}\ }\textbf {\bibinfo {volume} {20}},\ \bibinfo {pages} {1079–1084} (\bibinfo {year} {2021})}\BibitemShut {NoStop}%
\bibitem [{\citenamefont {Stern}\ \emph {et~al.}(2022)\citenamefont {Stern}, \citenamefont {Gu}, \citenamefont {Jarman}, \citenamefont {Barker}, \citenamefont {Mendelson}, \citenamefont {Chugh}, \citenamefont {Schott}, \citenamefont {Tan}, \citenamefont {Sirringhaus}, \citenamefont {Aharonovich},\ and\ \citenamefont {Atatüre}}]{Stern2022}%
  \BibitemOpen
  \bibfield  {author} {\bibinfo {author} {\bibfnamefont {H.~L.}\ \bibnamefont {Stern}}, \bibinfo {author} {\bibfnamefont {Q.}~\bibnamefont {Gu}}, \bibinfo {author} {\bibfnamefont {J.}~\bibnamefont {Jarman}}, \bibinfo {author} {\bibfnamefont {S.~Eizagirre}\ \bibnamefont {Barker}}, \bibinfo {author} {\bibfnamefont {N.}~\bibnamefont {Mendelson}}, \bibinfo {author} {\bibfnamefont {D.}~\bibnamefont {Chugh}}, \bibinfo {author} {\bibfnamefont {S.}~\bibnamefont {Schott}}, \bibinfo {author} {\bibfnamefont {H.~H.}\ \bibnamefont {Tan}}, \bibinfo {author} {\bibfnamefont {H.}~\bibnamefont {Sirringhaus}}, \bibinfo {author} {\bibfnamefont {I.}~\bibnamefont {Aharonovich}}, \ and\ \bibinfo {author} {\bibfnamefont {M.}~\bibnamefont {Atatüre}},\ }\bibfield  {title} {\enquote {\bibinfo {title} {Room-temperature optically detected magnetic resonance of single defects in hexagonal boron nitride},}\ }\href@noop {} {\bibfield  {journal} {\bibinfo  {journal} {Nature Communications}\ }\textbf {\bibinfo {volume} {13}},\ \bibinfo
  {pages} {618} (\bibinfo {year} {2022})}\BibitemShut {NoStop}%
\bibitem [{\citenamefont {Stern}\ \emph {et~al.}(2024)\citenamefont {Stern}, \citenamefont {Gilardoni}, \citenamefont {Gu}, \citenamefont {Barker}, \citenamefont {Powell}, \citenamefont {Deng}, \citenamefont {Fraser}, \citenamefont {Follet}, \citenamefont {Li}, \citenamefont {Ramsay}, \citenamefont {Tan}, \citenamefont {Aharonovich},\ and\ \citenamefont {Atat\"{u}re}}]{Stern2024}%
  \BibitemOpen
  \bibfield  {author} {\bibinfo {author} {\bibfnamefont {H.~L.}\ \bibnamefont {Stern}}, \bibinfo {author} {\bibfnamefont {C.~M.}\ \bibnamefont {Gilardoni}}, \bibinfo {author} {\bibfnamefont {Q.}~\bibnamefont {Gu}}, \bibinfo {author} {\bibfnamefont {S.~Eizagirre}\ \bibnamefont {Barker}}, \bibinfo {author} {\bibfnamefont {O.~F.~J.}\ \bibnamefont {Powell}}, \bibinfo {author} {\bibfnamefont {X.}~\bibnamefont {Deng}}, \bibinfo {author} {\bibfnamefont {S.~A.}\ \bibnamefont {Fraser}}, \bibinfo {author} {\bibfnamefont {L.}~\bibnamefont {Follet}}, \bibinfo {author} {\bibfnamefont {C.}~\bibnamefont {Li}}, \bibinfo {author} {\bibfnamefont {A.~J.}\ \bibnamefont {Ramsay}}, \bibinfo {author} {\bibfnamefont {H.~H.}\ \bibnamefont {Tan}}, \bibinfo {author} {\bibfnamefont {I.}~\bibnamefont {Aharonovich}}, \ and\ \bibinfo {author} {\bibfnamefont {M.}~\bibnamefont {Atat\"{u}re}},\ }\bibfield  {title} {\enquote {\bibinfo {title} {A quantum coherent spin in hexagonal boron nitride at ambient conditions},}\ }\href@noop {}
  {\bibfield  {journal} {\bibinfo  {journal} {Nature Materials}\ }\textbf {\bibinfo {volume} {23}},\ \bibinfo {pages} {1379–1385} (\bibinfo {year} {2024})}\BibitemShut {NoStop}%
\bibitem [{\citenamefont {Kim}\ \emph {et~al.}(2015)\citenamefont {Kim}, \citenamefont {Hsu}, \citenamefont {Park}, \citenamefont {Chae}, \citenamefont {Yun}, \citenamefont {Lee}, \citenamefont {Cho}, \citenamefont {Fang}, \citenamefont {Lee}, \citenamefont {Palacios}, \citenamefont {Dresselhaus}, \citenamefont {Kim}, \citenamefont {Lee},\ and\ \citenamefont {Kong}}]{Kim2015}%
  \BibitemOpen
  \bibfield  {author} {\bibinfo {author} {\bibfnamefont {S.~M.}\ \bibnamefont {Kim}}, \bibinfo {author} {\bibfnamefont {A.}~\bibnamefont {Hsu}}, \bibinfo {author} {\bibfnamefont {M.~Ho}\ \bibnamefont {Park}}, \bibinfo {author} {\bibfnamefont {S.~H.}\ \bibnamefont {Chae}}, \bibinfo {author} {\bibfnamefont {S.~J.}\ \bibnamefont {Yun}}, \bibinfo {author} {\bibfnamefont {J.~S.}\ \bibnamefont {Lee}}, \bibinfo {author} {\bibfnamefont {D.~H.}\ \bibnamefont {Cho}}, \bibinfo {author} {\bibfnamefont {W.}~\bibnamefont {Fang}}, \bibinfo {author} {\bibfnamefont {C.}~\bibnamefont {Lee}}, \bibinfo {author} {\bibfnamefont {T.}~\bibnamefont {Palacios}}, \bibinfo {author} {\bibfnamefont {M.}~\bibnamefont {Dresselhaus}}, \bibinfo {author} {\bibfnamefont {K.~K.}\ \bibnamefont {Kim}}, \bibinfo {author} {\bibfnamefont {Y.~H.}\ \bibnamefont {Lee}}, \ and\ \bibinfo {author} {\bibfnamefont {J.}~\bibnamefont {Kong}},\ }\bibfield  {title} {\enquote {\bibinfo {title} {Synthesis of large-area multilayer hexagonal boron nitride for high
  material performance},}\ }\href@noop {} {\bibfield  {journal} {\bibinfo  {journal} {Nature Communications}\ }\textbf {\bibinfo {volume} {6}},\ \bibinfo {pages} {8662} (\bibinfo {year} {2015})}\BibitemShut {NoStop}%
\bibitem [{\citenamefont {Chugh}\ \emph {et~al.}(2018)\citenamefont {Chugh}, \citenamefont {Wong-Leung}, \citenamefont {Li}, \citenamefont {Lysevych}, \citenamefont {Tan},\ and\ \citenamefont {Jagadish}}]{Chugh2018}%
  \BibitemOpen
  \bibfield  {author} {\bibinfo {author} {\bibfnamefont {D.}~\bibnamefont {Chugh}}, \bibinfo {author} {\bibfnamefont {J.}~\bibnamefont {Wong-Leung}}, \bibinfo {author} {\bibfnamefont {L.}~\bibnamefont {Li}}, \bibinfo {author} {\bibfnamefont {M.}~\bibnamefont {Lysevych}}, \bibinfo {author} {\bibfnamefont {H.~H.}\ \bibnamefont {Tan}}, \ and\ \bibinfo {author} {\bibfnamefont {C.}~\bibnamefont {Jagadish}},\ }\bibfield  {title} {\enquote {\bibinfo {title} {Flow modulation epitaxy of hexagonal boron nitride},}\ }\href@noop {} {\bibfield  {journal} {\bibinfo  {journal} {2D Materials}\ }\textbf {\bibinfo {volume} {5}},\ \bibinfo {pages} {045018} (\bibinfo {year} {2018})}\BibitemShut {NoStop}%
\bibitem [{\citenamefont {Lee}\ \emph {et~al.}(2018)\citenamefont {Lee}, \citenamefont {Choi}, \citenamefont {Yun}, \citenamefont {Kim}, \citenamefont {Boandoh}, \citenamefont {Park}, \citenamefont {Shin}, \citenamefont {Ko}, \citenamefont {Lee}, \citenamefont {Kim}, \citenamefont {Lee}, \citenamefont {Kim},\ and\ \citenamefont {Kim}}]{Lee2018}%
  \BibitemOpen
  \bibfield  {author} {\bibinfo {author} {\bibfnamefont {J.~Song}\ \bibnamefont {Lee}}, \bibinfo {author} {\bibfnamefont {S.~Ho}\ \bibnamefont {Choi}}, \bibinfo {author} {\bibfnamefont {S.~Joon}\ \bibnamefont {Yun}}, \bibinfo {author} {\bibfnamefont {Y.~In}\ \bibnamefont {Kim}}, \bibinfo {author} {\bibfnamefont {S.}~\bibnamefont {Boandoh}}, \bibinfo {author} {\bibfnamefont {J.~Hoon}\ \bibnamefont {Park}}, \bibinfo {author} {\bibfnamefont {B.~Gyu}\ \bibnamefont {Shin}}, \bibinfo {author} {\bibfnamefont {H.}~\bibnamefont {Ko}}, \bibinfo {author} {\bibfnamefont {S.~Hee}\ \bibnamefont {Lee}}, \bibinfo {author} {\bibfnamefont {Y.~Min}\ \bibnamefont {Kim}}, \bibinfo {author} {\bibfnamefont {Y.~Hee}\ \bibnamefont {Lee}}, \bibinfo {author} {\bibfnamefont {K.~Kang}\ \bibnamefont {Kim}}, \ and\ \bibinfo {author} {\bibfnamefont {S.~Min}\ \bibnamefont {Kim}},\ }\bibfield  {title} {\enquote {\bibinfo {title} {Wafer-scale single-crystal hexagonal boron nitride film via self-collimated grain formation},}\ }\href@noop {}
  {\bibfield  {journal} {\bibinfo  {journal} {Science}\ }\textbf {\bibinfo {volume} {362}},\ \bibinfo {pages} {817--821} (\bibinfo {year} {2018})}\BibitemShut {NoStop}%
\bibitem [{\citenamefont {Wang}\ \emph {et~al.}(2019)\citenamefont {Wang}, \citenamefont {Xu}, \citenamefont {Zhang}, \citenamefont {Qiao}, \citenamefont {Wu}, \citenamefont {Wang}, \citenamefont {Zhang}, \citenamefont {Liang}, \citenamefont {Zhang}, \citenamefont {Zhang}, \citenamefont {Chen}, \citenamefont {Xie}, \citenamefont {Zong}, \citenamefont {Shan}, \citenamefont {Guo}, \citenamefont {Willinger}, \citenamefont {Wu}, \citenamefont {Li}, \citenamefont {Wang}, \citenamefont {Gao}, \citenamefont {Wu}, \citenamefont {Zhang}, \citenamefont {Jiang}, \citenamefont {Yu}, \citenamefont {Wang}, \citenamefont {Bai}, \citenamefont {Wang}, \citenamefont {Ding},\ and\ \citenamefont {Liu}}]{Wang2019}%
  \BibitemOpen
  \bibfield  {author} {\bibinfo {author} {\bibfnamefont {L.}~\bibnamefont {Wang}}, \bibinfo {author} {\bibfnamefont {X.}~\bibnamefont {Xu}}, \bibinfo {author} {\bibfnamefont {L.}~\bibnamefont {Zhang}}, \bibinfo {author} {\bibfnamefont {R.}~\bibnamefont {Qiao}}, \bibinfo {author} {\bibfnamefont {M.}~\bibnamefont {Wu}}, \bibinfo {author} {\bibfnamefont {Z.}~\bibnamefont {Wang}}, \bibinfo {author} {\bibfnamefont {S.}~\bibnamefont {Zhang}}, \bibinfo {author} {\bibfnamefont {J.}~\bibnamefont {Liang}}, \bibinfo {author} {\bibfnamefont {Z.}~\bibnamefont {Zhang}}, \bibinfo {author} {\bibfnamefont {Z.}~\bibnamefont {Zhang}}, \bibinfo {author} {\bibfnamefont {W.}~\bibnamefont {Chen}}, \bibinfo {author} {\bibfnamefont {X.}~\bibnamefont {Xie}}, \bibinfo {author} {\bibfnamefont {J.}~\bibnamefont {Zong}}, \bibinfo {author} {\bibfnamefont {Y.}~\bibnamefont {Shan}}, \bibinfo {author} {\bibfnamefont {Y.}~\bibnamefont {Guo}}, \bibinfo {author} {\bibfnamefont {M.}~\bibnamefont {Willinger}}, \bibinfo {author} {\bibfnamefont
  {H.}~\bibnamefont {Wu}}, \bibinfo {author} {\bibfnamefont {Q.}~\bibnamefont {Li}}, \bibinfo {author} {\bibfnamefont {W.}~\bibnamefont {Wang}}, \bibinfo {author} {\bibfnamefont {P.}~\bibnamefont {Gao}}, \bibinfo {author} {\bibfnamefont {S.}~\bibnamefont {Wu}}, \bibinfo {author} {\bibfnamefont {Y.}~\bibnamefont {Zhang}}, \bibinfo {author} {\bibfnamefont {Y.}~\bibnamefont {Jiang}}, \bibinfo {author} {\bibfnamefont {D.}~\bibnamefont {Yu}}, \bibinfo {author} {\bibfnamefont {E.}~\bibnamefont {Wang}}, \bibinfo {author} {\bibfnamefont {X.}~\bibnamefont {Bai}}, \bibinfo {author} {\bibfnamefont {Z.~J.}\ \bibnamefont {Wang}}, \bibinfo {author} {\bibfnamefont {F.}~\bibnamefont {Ding}}, \ and\ \bibinfo {author} {\bibfnamefont {K.}~\bibnamefont {Liu}},\ }\bibfield  {title} {\enquote {\bibinfo {title} {Epitaxial growth of a 100-square-centimetre single-crystal hexagonal boron nitride monolayer on copper},}\ }\href@noop {} {\bibfield  {journal} {\bibinfo  {journal} {Nature}\ }\textbf {\bibinfo {volume} {570}},\ \bibinfo
  {pages} {91–95} (\bibinfo {year} {2019})}\BibitemShut {NoStop}%
\bibitem [{\citenamefont {Wang}\ \emph {et~al.}(2024)\citenamefont {Wang}, \citenamefont {Zhao}, \citenamefont {Gao}, \citenamefont {Zheng}, \citenamefont {Qian}, \citenamefont {Gao}, \citenamefont {Li}, \citenamefont {Tang}, \citenamefont {Tan}, \citenamefont {Wang}, \citenamefont {Zhu}, \citenamefont {Guo}, \citenamefont {Liu}, \citenamefont {Ding},\ and\ \citenamefont {Peng}}]{Wang2024}%
  \BibitemOpen
  \bibfield  {author} {\bibinfo {author} {\bibfnamefont {Y.}~\bibnamefont {Wang}}, \bibinfo {author} {\bibfnamefont {C.}~\bibnamefont {Zhao}}, \bibinfo {author} {\bibfnamefont {X.}~\bibnamefont {Gao}}, \bibinfo {author} {\bibfnamefont {L.}~\bibnamefont {Zheng}}, \bibinfo {author} {\bibfnamefont {J.}~\bibnamefont {Qian}}, \bibinfo {author} {\bibfnamefont {X.}~\bibnamefont {Gao}}, \bibinfo {author} {\bibfnamefont {J.}~\bibnamefont {Li}}, \bibinfo {author} {\bibfnamefont {J.}~\bibnamefont {Tang}}, \bibinfo {author} {\bibfnamefont {C.}~\bibnamefont {Tan}}, \bibinfo {author} {\bibfnamefont {J.}~\bibnamefont {Wang}}, \bibinfo {author} {\bibfnamefont {X.}~\bibnamefont {Zhu}}, \bibinfo {author} {\bibfnamefont {J.}~\bibnamefont {Guo}}, \bibinfo {author} {\bibfnamefont {Z.}~\bibnamefont {Liu}}, \bibinfo {author} {\bibfnamefont {F.}~\bibnamefont {Ding}}, \ and\ \bibinfo {author} {\bibfnamefont {H.}~\bibnamefont {Peng}},\ }\bibfield  {title} {\enquote {\bibinfo {title} {Ultraflat single-crystal hexagonal boron nitride
  for wafer-scale integration of a 2d-compatible high-$\kappa$ metal gate},}\ }\href@noop {} {\bibfield  {journal} {\bibinfo  {journal} {Nature Materials}\ }\textbf {\bibinfo {volume} {23}},\ \bibinfo {pages} {1495} (\bibinfo {year} {2024})}\BibitemShut {NoStop}%
\bibitem [{\citenamefont {Novoselov}\ \emph {et~al.}(2016)\citenamefont {Novoselov}, \citenamefont {Mishchenko}, \citenamefont {Carvalho},\ and\ \citenamefont {Neto}}]{Novoselov2016}%
  \BibitemOpen
  \bibfield  {author} {\bibinfo {author} {\bibfnamefont {K.~S.}\ \bibnamefont {Novoselov}}, \bibinfo {author} {\bibfnamefont {A.}~\bibnamefont {Mishchenko}}, \bibinfo {author} {\bibfnamefont {A.}~\bibnamefont {Carvalho}}, \ and\ \bibinfo {author} {\bibfnamefont {A.~H.~Castro}\ \bibnamefont {Neto}},\ }\bibfield  {title} {\enquote {\bibinfo {title} {2d materials and van der waals heterostructures},}\ }\href@noop {} {\bibfield  {journal} {\bibinfo  {journal} {Science}\ }\textbf {\bibinfo {volume} {353}},\ \bibinfo {pages} {9439} (\bibinfo {year} {2016})}\BibitemShut {NoStop}%
\bibitem [{\citenamefont {Fukamachi}\ \emph {et~al.}(2023)\citenamefont {Fukamachi}, \citenamefont {Solís-Fernández}, \citenamefont {Kawahara}, \citenamefont {Tanaka}, \citenamefont {Otake}, \citenamefont {Lin}, \citenamefont {Suenaga},\ and\ \citenamefont {Ago}}]{Fukamachi2023}%
  \BibitemOpen
  \bibfield  {author} {\bibinfo {author} {\bibfnamefont {S.}~\bibnamefont {Fukamachi}}, \bibinfo {author} {\bibfnamefont {P.}~\bibnamefont {Solís-Fernández}}, \bibinfo {author} {\bibfnamefont {K.}~\bibnamefont {Kawahara}}, \bibinfo {author} {\bibfnamefont {D.}~\bibnamefont {Tanaka}}, \bibinfo {author} {\bibfnamefont {T.}~\bibnamefont {Otake}}, \bibinfo {author} {\bibfnamefont {Y-Ch.}\ \bibnamefont {Lin}}, \bibinfo {author} {\bibfnamefont {K.}~\bibnamefont {Suenaga}}, \ and\ \bibinfo {author} {\bibfnamefont {H.}~\bibnamefont {Ago}},\ }\bibfield  {title} {\enquote {\bibinfo {title} {Large-area synthesis and transfer of multilayer hexagonal boron nitride for enhanced graphene device arrays},}\ }\href@noop {} {\bibfield  {journal} {\bibinfo  {journal} {Nature Electronics}\ }\textbf {\bibinfo {volume} {6}},\ \bibinfo {pages} {126--136} (\bibinfo {year} {2023})}\BibitemShut {NoStop}%
\bibitem [{\citenamefont {Li}\ \emph {et~al.}(2021)\citenamefont {Li}, \citenamefont {Fröch}, \citenamefont {Nonahal}, \citenamefont {Tran}, \citenamefont {Toth}, \citenamefont {Kim},\ and\ \citenamefont {Aharonovich}}]{Li2021}%
  \BibitemOpen
  \bibfield  {author} {\bibinfo {author} {\bibfnamefont {C.}~\bibnamefont {Li}}, \bibinfo {author} {\bibfnamefont {J.~E.}\ \bibnamefont {Fröch}}, \bibinfo {author} {\bibfnamefont {M.}~\bibnamefont {Nonahal}}, \bibinfo {author} {\bibfnamefont {T.~N.}\ \bibnamefont {Tran}}, \bibinfo {author} {\bibfnamefont {M.}~\bibnamefont {Toth}}, \bibinfo {author} {\bibfnamefont {S.}~\bibnamefont {Kim}}, \ and\ \bibinfo {author} {\bibfnamefont {I.}~\bibnamefont {Aharonovich}},\ }\bibfield  {title} {\enquote {\bibinfo {title} {Integration of hbn quantum emitters in monolithically fabricated waveguides},}\ }\href@noop {} {\bibfield  {journal} {\bibinfo  {journal} {ACS Photonics}\ }\textbf {\bibinfo {volume} {8}},\ \bibinfo {pages} {2966–2972} (\bibinfo {year} {2021})}\BibitemShut {NoStop}%
\bibitem [{\citenamefont {Kim}\ \emph {et~al.}(2018)\citenamefont {Kim}, \citenamefont {Fröch}, \citenamefont {Christian}, \citenamefont {Straw}, \citenamefont {Bishop}, \citenamefont {Totonjian}, \citenamefont {Watanabe}, \citenamefont {Taniguchi}, \citenamefont {Toth},\ and\ \citenamefont {Aharonovich}}]{Kim2018}%
  \BibitemOpen
  \bibfield  {author} {\bibinfo {author} {\bibfnamefont {S.}~\bibnamefont {Kim}}, \bibinfo {author} {\bibfnamefont {J.~E.}\ \bibnamefont {Fröch}}, \bibinfo {author} {\bibfnamefont {J.}~\bibnamefont {Christian}}, \bibinfo {author} {\bibfnamefont {M.}~\bibnamefont {Straw}}, \bibinfo {author} {\bibfnamefont {J.}~\bibnamefont {Bishop}}, \bibinfo {author} {\bibfnamefont {D.}~\bibnamefont {Totonjian}}, \bibinfo {author} {\bibfnamefont {K.}~\bibnamefont {Watanabe}}, \bibinfo {author} {\bibfnamefont {T.}~\bibnamefont {Taniguchi}}, \bibinfo {author} {\bibfnamefont {M.}~\bibnamefont {Toth}}, \ and\ \bibinfo {author} {\bibfnamefont {I.}~\bibnamefont {Aharonovich}},\ }\bibfield  {title} {\enquote {\bibinfo {title} {Photonic crystal cavities from hexagonal boron nitride},}\ }\href@noop {} {\bibfield  {journal} {\bibinfo  {journal} {Nature Communications}\ }\textbf {\bibinfo {volume} {9}},\ \bibinfo {pages} {2623} (\bibinfo {year} {2018})}\BibitemShut {NoStop}%
\bibitem [{\citenamefont {Vogl}\ \emph {et~al.}(2019{\natexlab{a}})\citenamefont {Vogl}, \citenamefont {Lecamwasam}, \citenamefont {Buchler}, \citenamefont {Lu},\ and\ \citenamefont {Lam}}]{Vogl2019}%
  \BibitemOpen
  \bibfield  {author} {\bibinfo {author} {\bibfnamefont {Tobias}\ \bibnamefont {Vogl}}, \bibinfo {author} {\bibfnamefont {Ruvi}\ \bibnamefont {Lecamwasam}}, \bibinfo {author} {\bibfnamefont {Ben~C.}\ \bibnamefont {Buchler}}, \bibinfo {author} {\bibfnamefont {Yuerui}\ \bibnamefont {Lu}}, \ and\ \bibinfo {author} {\bibfnamefont {Ping~Koy}\ \bibnamefont {Lam}},\ }\bibfield  {title} {\enquote {\bibinfo {title} {Compact cavity-enhanced single-photon generation with hexagonal boron nitride},}\ }\href@noop {} {\bibfield  {journal} {\bibinfo  {journal} {ACS Photonics}\ }\textbf {\bibinfo {volume} {6}},\ \bibinfo {pages} {1955--1962} (\bibinfo {year} {2019}{\natexlab{a}})}\BibitemShut {NoStop}%
\bibitem [{\citenamefont {Fröch}\ \emph {et~al.}(2021)\citenamefont {Fröch}, \citenamefont {Spencer}, \citenamefont {Kianinia}, \citenamefont {Totonjian}, \citenamefont {Nguyen}, \citenamefont {Gottscholl}, \citenamefont {Dyakonov}, \citenamefont {Toth}, \citenamefont {Kim},\ and\ \citenamefont {Aharonovich}}]{Frch2021}%
  \BibitemOpen
  \bibfield  {author} {\bibinfo {author} {\bibfnamefont {J.~E.}\ \bibnamefont {Fröch}}, \bibinfo {author} {\bibfnamefont {L.~P.}\ \bibnamefont {Spencer}}, \bibinfo {author} {\bibfnamefont {M.}~\bibnamefont {Kianinia}}, \bibinfo {author} {\bibfnamefont {D.~D.}\ \bibnamefont {Totonjian}}, \bibinfo {author} {\bibfnamefont {M.}~\bibnamefont {Nguyen}}, \bibinfo {author} {\bibfnamefont {A.}~\bibnamefont {Gottscholl}}, \bibinfo {author} {\bibfnamefont {V.}~\bibnamefont {Dyakonov}}, \bibinfo {author} {\bibfnamefont {M.}~\bibnamefont {Toth}}, \bibinfo {author} {\bibfnamefont {S.}~\bibnamefont {Kim}}, \ and\ \bibinfo {author} {\bibfnamefont {I.}~\bibnamefont {Aharonovich}},\ }\bibfield  {title} {\enquote {\bibinfo {title} {Coupling spin defects in hexagonal boron nitride to monolithic bullseye cavities},}\ }\href@noop {} {\bibfield  {journal} {\bibinfo  {journal} {Nano Letters}\ }\textbf {\bibinfo {volume} {21}},\ \bibinfo {pages} {6549–6555} (\bibinfo {year} {2021})}\BibitemShut {NoStop}%
\bibitem [{\citenamefont {Blascetta}\ \emph {et~al.}(2020)\citenamefont {Blascetta}, \citenamefont {Liebel}, \citenamefont {Lu}, \citenamefont {Taniguchi}, \citenamefont {Watanabe}, \citenamefont {Efetov},\ and\ \citenamefont {Hulst}}]{PalomboBlascetta2020}%
  \BibitemOpen
  \bibfield  {author} {\bibinfo {author} {\bibfnamefont {N.~Palombo}\ \bibnamefont {Blascetta}}, \bibinfo {author} {\bibfnamefont {M.}~\bibnamefont {Liebel}}, \bibinfo {author} {\bibfnamefont {X.}~\bibnamefont {Lu}}, \bibinfo {author} {\bibfnamefont {T.}~\bibnamefont {Taniguchi}}, \bibinfo {author} {\bibfnamefont {K.}~\bibnamefont {Watanabe}}, \bibinfo {author} {\bibfnamefont {D.~K.}\ \bibnamefont {Efetov}}, \ and\ \bibinfo {author} {\bibfnamefont {N.~F.~Van}\ \bibnamefont {Hulst}},\ }\bibfield  {title} {\enquote {\bibinfo {title} {Nanoscale imaging and control of hexagonal boron nitride single photon emitters by a resonant nanoantenna},}\ }\href@noop {} {\bibfield  {journal} {\bibinfo  {journal} {Nano Letters}\ }\textbf {\bibinfo {volume} {20}},\ \bibinfo {pages} {1992–1999} (\bibinfo {year} {2020})}\BibitemShut {NoStop}%
\bibitem [{\citenamefont {Noh}\ \emph {et~al.}(2018)\citenamefont {Noh}, \citenamefont {Choi}, \citenamefont {Kim}, \citenamefont {Im}, \citenamefont {Kim}, \citenamefont {Seo},\ and\ \citenamefont {Lee}}]{Noh2018}%
  \BibitemOpen
  \bibfield  {author} {\bibinfo {author} {\bibfnamefont {G.}~\bibnamefont {Noh}}, \bibinfo {author} {\bibfnamefont {D.}~\bibnamefont {Choi}}, \bibinfo {author} {\bibfnamefont {J-H.}\ \bibnamefont {Kim}}, \bibinfo {author} {\bibfnamefont {D-G.}\ \bibnamefont {Im}}, \bibinfo {author} {\bibfnamefont {Y-H.}\ \bibnamefont {Kim}}, \bibinfo {author} {\bibfnamefont {H.}~\bibnamefont {Seo}}, \ and\ \bibinfo {author} {\bibfnamefont {J.}~\bibnamefont {Lee}},\ }\bibfield  {title} {\enquote {\bibinfo {title} {Stark tuning of single-photon emitters in hexagonal boron nitride},}\ }\href@noop {} {\bibfield  {journal} {\bibinfo  {journal} {Nano Letters}\ }\textbf {\bibinfo {volume} {18}},\ \bibinfo {pages} {4710--4715} (\bibinfo {year} {2018})}\BibitemShut {NoStop}%
\bibitem [{\citenamefont {Akbari}\ \emph {et~al.}(2022)\citenamefont {Akbari}, \citenamefont {Biswas}, \citenamefont {Jha}, \citenamefont {Wong}, \citenamefont {Vest},\ and\ \citenamefont {Atwater}}]{Akbari2022}%
  \BibitemOpen
  \bibfield  {author} {\bibinfo {author} {\bibfnamefont {H.}~\bibnamefont {Akbari}}, \bibinfo {author} {\bibfnamefont {S.}~\bibnamefont {Biswas}}, \bibinfo {author} {\bibfnamefont {P.~Kumar}\ \bibnamefont {Jha}}, \bibinfo {author} {\bibfnamefont {J.}~\bibnamefont {Wong}}, \bibinfo {author} {\bibfnamefont {B.}~\bibnamefont {Vest}}, \ and\ \bibinfo {author} {\bibfnamefont {H.~A.}\ \bibnamefont {Atwater}},\ }\bibfield  {title} {\enquote {\bibinfo {title} {Lifetime-limited and tunable quantum light emission in h-bn via electric field modulation},}\ }\href@noop {} {\bibfield  {journal} {\bibinfo  {journal} {Nano Letters}\ }\textbf {\bibinfo {volume} {22}},\ \bibinfo {pages} {7798–7803} (\bibinfo {year} {2022})}\BibitemShut {NoStop}%
\bibitem [{\citenamefont {Zhigulin}\ \emph {et~al.}(2025)\citenamefont {Zhigulin}, \citenamefont {Park}, \citenamefont {Yamamura}, \citenamefont {Watanabe}, \citenamefont {Taniguchi}, \citenamefont {Toth}, \citenamefont {Kim},\ and\ \citenamefont {Aharonovich}}]{Zhigulin2025}%
  \BibitemOpen
  \bibfield  {author} {\bibinfo {author} {\bibfnamefont {I.}~\bibnamefont {Zhigulin}}, \bibinfo {author} {\bibfnamefont {G.}~\bibnamefont {Park}}, \bibinfo {author} {\bibfnamefont {K.}~\bibnamefont {Yamamura}}, \bibinfo {author} {\bibfnamefont {K.}~\bibnamefont {Watanabe}}, \bibinfo {author} {\bibfnamefont {T.}~\bibnamefont {Taniguchi}}, \bibinfo {author} {\bibfnamefont {M.}~\bibnamefont {Toth}}, \bibinfo {author} {\bibfnamefont {J.}~\bibnamefont {Kim}}, \ and\ \bibinfo {author} {\bibfnamefont {I.}~\bibnamefont {Aharonovich}},\ }\bibfield  {title} {\enquote {\bibinfo {title} {Electrical generation of color centers in hexagonal boron nitride},}\ }\href@noop {} {\bibfield  {journal} {\bibinfo  {journal} {ACS Applied Materials and Interfaces}\ }\textbf {\bibinfo {volume} {17}},\ \bibinfo {pages} {24129–24136} (\bibinfo {year} {2025})}\BibitemShut {NoStop}%
\bibitem [{\citenamefont {R.}\ \emph {et~al.}(2017)\citenamefont {R.}, \citenamefont {Jungwirth},\ and\ \citenamefont {Fuchs}}]{Jungwirth2017}%
  \BibitemOpen
  \bibfield  {author} {\bibinfo {author} {\bibfnamefont {N.}~\bibnamefont {R.}}, \bibinfo {author} {\bibnamefont {Jungwirth}}, \ and\ \bibinfo {author} {\bibfnamefont {G.~D.}\ \bibnamefont {Fuchs}},\ }\bibfield  {title} {\enquote {\bibinfo {title} {Optical absorption and emission mechanisms of single defects in hexagonal boron nitride},}\ }\href@noop {} {\bibfield  {journal} {\bibinfo  {journal} {Physical Review Letters}\ }\textbf {\bibinfo {volume} {119}},\ \bibinfo {pages} {057401} (\bibinfo {year} {2017})}\BibitemShut {NoStop}%
\bibitem [{\citenamefont {Arı}\ \emph {et~al.}(2025)\citenamefont {Arı}, \citenamefont {Polat}, \citenamefont {Fırat}, \citenamefont {{\c{C}}akır},\ and\ \citenamefont {Ateş}}]{Ari2025}%
  \BibitemOpen
  \bibfield  {author} {\bibinfo {author} {\bibfnamefont {O.}~\bibnamefont {Arı}}, \bibinfo {author} {\bibfnamefont {N.}~\bibnamefont {Polat}}, \bibinfo {author} {\bibfnamefont {V.}~\bibnamefont {Fırat}}, \bibinfo {author} {\bibfnamefont {{\"O}.}~\bibnamefont {{\c{C}}akır}}, \ and\ \bibinfo {author} {\bibfnamefont {S.}~\bibnamefont {Ateş}},\ }\bibfield  {title} {\enquote {\bibinfo {title} {Temperature-dependent spectral properties of hexagonal boron nitride color centers},}\ }\href@noop {} {\bibfield  {journal} {\bibinfo  {journal} {ACS Photonics}\ }\textbf {\bibinfo {volume} {12}},\ \bibinfo {pages} {1676--1682} (\bibinfo {year} {2025})}\BibitemShut {NoStop}%
\bibitem [{\citenamefont {Grosso}\ \emph {et~al.}(2017)\citenamefont {Grosso}, \citenamefont {Moon}, \citenamefont {Lienhard}, \citenamefont {Ali}, \citenamefont {Efetov}, \citenamefont {Furchi}, \citenamefont {Jarillo-Herrero}, \citenamefont {Ford}, \citenamefont {Aharonovich},\ and\ \citenamefont {Englund}}]{Grosso2017}%
  \BibitemOpen
  \bibfield  {author} {\bibinfo {author} {\bibfnamefont {G.}~\bibnamefont {Grosso}}, \bibinfo {author} {\bibfnamefont {H.}~\bibnamefont {Moon}}, \bibinfo {author} {\bibfnamefont {B.}~\bibnamefont {Lienhard}}, \bibinfo {author} {\bibfnamefont {S.}~\bibnamefont {Ali}}, \bibinfo {author} {\bibfnamefont {D.~K.}\ \bibnamefont {Efetov}}, \bibinfo {author} {\bibfnamefont {M.~M.}\ \bibnamefont {Furchi}}, \bibinfo {author} {\bibfnamefont {P.}~\bibnamefont {Jarillo-Herrero}}, \bibinfo {author} {\bibfnamefont {M.~J.}\ \bibnamefont {Ford}}, \bibinfo {author} {\bibfnamefont {I.}~\bibnamefont {Aharonovich}}, \ and\ \bibinfo {author} {\bibfnamefont {D.}~\bibnamefont {Englund}},\ }\bibfield  {title} {\enquote {\bibinfo {title} {Tunable and high-purity room temperature single-photon emission from atomic defects in hexagonal boron nitride},}\ }\href@noop {} {\bibfield  {journal} {\bibinfo  {journal} {Nature Communications}\ }\textbf {\bibinfo {volume} {8}},\ \bibinfo {pages} {705} (\bibinfo {year} {2017})}\BibitemShut
  {NoStop}%
\bibitem [{\citenamefont {Liu}\ \emph {et~al.}(2020)\citenamefont {Liu}, \citenamefont {Wang}, \citenamefont {Li}, \citenamefont {Yu}, \citenamefont {Ke}, \citenamefont {Meng}, \citenamefont {Tang}, \citenamefont {Li},\ and\ \citenamefont {Guo}}]{Liu2020}%
  \BibitemOpen
  \bibfield  {author} {\bibinfo {author} {\bibfnamefont {W.}~\bibnamefont {Liu}}, \bibinfo {author} {\bibfnamefont {Y.~T.}\ \bibnamefont {Wang}}, \bibinfo {author} {\bibfnamefont {Z.~P.}\ \bibnamefont {Li}}, \bibinfo {author} {\bibfnamefont {S.}~\bibnamefont {Yu}}, \bibinfo {author} {\bibfnamefont {Z.~J.}\ \bibnamefont {Ke}}, \bibinfo {author} {\bibfnamefont {Y.}~\bibnamefont {Meng}}, \bibinfo {author} {\bibfnamefont {J.~S.}\ \bibnamefont {Tang}}, \bibinfo {author} {\bibfnamefont {C.~F.}\ \bibnamefont {Li}}, \ and\ \bibinfo {author} {\bibfnamefont {G.~C.}\ \bibnamefont {Guo}},\ }\bibfield  {title} {\enquote {\bibinfo {title} {An ultrastable and robust single-photon emitter in hexagonal boron nitride},}\ }\href@noop {} {\bibfield  {journal} {\bibinfo  {journal} {Physica E: Low-Dimensional Systems and Nanostructures}\ }\textbf {\bibinfo {volume} {124}},\ \bibinfo {pages} {114251} (\bibinfo {year} {2020})}\BibitemShut {NoStop}%
\bibitem [{\citenamefont {Guo}\ \emph {et~al.}(2023)\citenamefont {Guo}, \citenamefont {Li}, \citenamefont {Liu}, \citenamefont {Yang}, \citenamefont {Zeng}, \citenamefont {Yu}, \citenamefont {Meng}, \citenamefont {Li}, \citenamefont {Wang}, \citenamefont {Xie}, \citenamefont {Ge}, \citenamefont {Wang}, \citenamefont {Li}, \citenamefont {Xu}, \citenamefont {Wang}, \citenamefont {Tang}, \citenamefont {Gali}, \citenamefont {Li},\ and\ \citenamefont {Guo}}]{Guo2023}%
  \BibitemOpen
  \bibfield  {author} {\bibinfo {author} {\bibfnamefont {N.~J.}\ \bibnamefont {Guo}}, \bibinfo {author} {\bibfnamefont {S.}~\bibnamefont {Li}}, \bibinfo {author} {\bibfnamefont {W.}~\bibnamefont {Liu}}, \bibinfo {author} {\bibfnamefont {Y.~Ze}\ \bibnamefont {Yang}}, \bibinfo {author} {\bibfnamefont {X.~Dong}\ \bibnamefont {Zeng}}, \bibinfo {author} {\bibfnamefont {S.}~\bibnamefont {Yu}}, \bibinfo {author} {\bibfnamefont {Y.}~\bibnamefont {Meng}}, \bibinfo {author} {\bibfnamefont {Z.~Peng}\ \bibnamefont {Li}}, \bibinfo {author} {\bibfnamefont {Z.~An}\ \bibnamefont {Wang}}, \bibinfo {author} {\bibfnamefont {L.~Ke}\ \bibnamefont {Xie}}, \bibinfo {author} {\bibfnamefont {R.~Chun}\ \bibnamefont {Ge}}, \bibinfo {author} {\bibfnamefont {J.~Feng}\ \bibnamefont {Wang}}, \bibinfo {author} {\bibfnamefont {Q.}~\bibnamefont {Li}}, \bibinfo {author} {\bibfnamefont {J.~Shi}\ \bibnamefont {Xu}}, \bibinfo {author} {\bibfnamefont {Y.~Tao}\ \bibnamefont {Wang}}, \bibinfo {author} {\bibfnamefont {J.~Shun}\ \bibnamefont {Tang}},
  \bibinfo {author} {\bibfnamefont {A.}~\bibnamefont {Gali}}, \bibinfo {author} {\bibfnamefont {C.~Feng}\ \bibnamefont {Li}}, \ and\ \bibinfo {author} {\bibfnamefont {G.~Can}\ \bibnamefont {Guo}},\ }\bibfield  {title} {\enquote {\bibinfo {title} {Coherent control of an ultrabright single spin in hexagonal boron nitride at room temperature},}\ }\href@noop {} {\bibfield  {journal} {\bibinfo  {journal} {Nature Communications}\ }\textbf {\bibinfo {volume} {14}},\ \bibinfo {pages} {2893} (\bibinfo {year} {2023})}\BibitemShut {NoStop}%
\bibitem [{\citenamefont {Kumar}\ \emph {et~al.}(2024)\citenamefont {Kumar}, \citenamefont {Samaner}, \citenamefont {Cholsuk}, \citenamefont {Matthes}, \citenamefont {Paçal}, \citenamefont {Oyun}, \citenamefont {Zand}, \citenamefont {Chapman}, \citenamefont {Saerens}, \citenamefont {Grange}, \citenamefont {Suwanna}, \citenamefont {Ateş},\ and\ \citenamefont {Vogl}}]{Kumar2024}%
  \BibitemOpen
  \bibfield  {author} {\bibinfo {author} {\bibfnamefont {A.}~\bibnamefont {Kumar}}, \bibinfo {author} {\bibfnamefont {Ç.}\ \bibnamefont {Samaner}}, \bibinfo {author} {\bibfnamefont {C.}~\bibnamefont {Cholsuk}}, \bibinfo {author} {\bibfnamefont {T.}~\bibnamefont {Matthes}}, \bibinfo {author} {\bibfnamefont {S.}~\bibnamefont {Paçal}}, \bibinfo {author} {\bibfnamefont {Y.}~\bibnamefont {Oyun}}, \bibinfo {author} {\bibfnamefont {A.}~\bibnamefont {Zand}}, \bibinfo {author} {\bibfnamefont {R.~J.}\ \bibnamefont {Chapman}}, \bibinfo {author} {\bibfnamefont {G.}~\bibnamefont {Saerens}}, \bibinfo {author} {\bibfnamefont {R.}~\bibnamefont {Grange}}, \bibinfo {author} {\bibfnamefont {S.}~\bibnamefont {Suwanna}}, \bibinfo {author} {\bibfnamefont {S.}~\bibnamefont {Ateş}}, \ and\ \bibinfo {author} {\bibfnamefont {T.}~\bibnamefont {Vogl}},\ }\bibfield  {title} {\enquote {\bibinfo {title} {Polarization dynamics of solid-state quantum emitters},}\ }\href@noop {} {\bibfield  {journal} {\bibinfo  {journal} {ACS Nano}\
  }\textbf {\bibinfo {volume} {18}},\ \bibinfo {pages} {5270--5281} (\bibinfo {year} {2024})}\BibitemShut {NoStop}%
\bibitem [{\citenamefont {Gottscholl}\ \emph {et~al.}(2021{\natexlab{a}})\citenamefont {Gottscholl}, \citenamefont {Diez}, \citenamefont {Soltamov}, \citenamefont {Kasper}, \citenamefont {Sperlich}, \citenamefont {Kianinia}, \citenamefont {Bradac}, \citenamefont {Aharonovich},\ and\ \citenamefont {Dyakonov}}]{Gottscholl2021a}%
  \BibitemOpen
  \bibfield  {author} {\bibinfo {author} {\bibfnamefont {A.}~\bibnamefont {Gottscholl}}, \bibinfo {author} {\bibfnamefont {M.}~\bibnamefont {Diez}}, \bibinfo {author} {\bibfnamefont {V.}~\bibnamefont {Soltamov}}, \bibinfo {author} {\bibfnamefont {C.}~\bibnamefont {Kasper}}, \bibinfo {author} {\bibfnamefont {A.}~\bibnamefont {Sperlich}}, \bibinfo {author} {\bibfnamefont {M.}~\bibnamefont {Kianinia}}, \bibinfo {author} {\bibfnamefont {C.}~\bibnamefont {Bradac}}, \bibinfo {author} {\bibfnamefont {I.}~\bibnamefont {Aharonovich}}, \ and\ \bibinfo {author} {\bibfnamefont {V.}~\bibnamefont {Dyakonov}},\ }\bibfield  {title} {\enquote {\bibinfo {title} {Room temperature coherent control of spin defects in hexagonal boron nitride},}\ }\href@noop {} {\bibfield  {journal} {\bibinfo  {journal} {Science Advances}\ }\textbf {\bibinfo {volume} {7}},\ \bibinfo {pages} {3630} (\bibinfo {year} {2021}{\natexlab{a}})}\BibitemShut {NoStop}%
\bibitem [{\citenamefont {Exarhos}\ \emph {et~al.}(2019)\citenamefont {Exarhos}, \citenamefont {Hopper}, \citenamefont {Patel}, \citenamefont {Doherty},\ and\ \citenamefont {Bassett}}]{Exarhos2019}%
  \BibitemOpen
  \bibfield  {author} {\bibinfo {author} {\bibfnamefont {A.~L.}\ \bibnamefont {Exarhos}}, \bibinfo {author} {\bibfnamefont {D.~A.}\ \bibnamefont {Hopper}}, \bibinfo {author} {\bibfnamefont {R.~N.}\ \bibnamefont {Patel}}, \bibinfo {author} {\bibfnamefont {M.~W.}\ \bibnamefont {Doherty}}, \ and\ \bibinfo {author} {\bibfnamefont {L.~C.}\ \bibnamefont {Bassett}},\ }\bibfield  {title} {\enquote {\bibinfo {title} {Magnetic-field-dependent quantum emission in hexagonal boron nitride at room temperature},}\ }\href@noop {} {\bibfield  {journal} {\bibinfo  {journal} {Nature Communications}\ }\textbf {\bibinfo {volume} {10}},\ \bibinfo {pages} {222} (\bibinfo {year} {2019})}\BibitemShut {NoStop}%
\bibitem [{\citenamefont {Ivády}\ \emph {et~al.}(2020)\citenamefont {Ivády}, \citenamefont {Barcza}, \citenamefont {Thiering}, \citenamefont {Li}, \citenamefont {Hamdi}, \citenamefont {Chou}, \citenamefont {Legeza},\ and\ \citenamefont {Gali}}]{Ivady2020}%
  \BibitemOpen
  \bibfield  {author} {\bibinfo {author} {\bibfnamefont {V.}~\bibnamefont {Ivády}}, \bibinfo {author} {\bibfnamefont {G.}~\bibnamefont {Barcza}}, \bibinfo {author} {\bibfnamefont {G.}~\bibnamefont {Thiering}}, \bibinfo {author} {\bibfnamefont {S.}~\bibnamefont {Li}}, \bibinfo {author} {\bibfnamefont {H.}~\bibnamefont {Hamdi}}, \bibinfo {author} {\bibfnamefont {J.~P.}\ \bibnamefont {Chou}}, \bibinfo {author} {\bibfnamefont {Ö.}\ \bibnamefont {Legeza}}, \ and\ \bibinfo {author} {\bibfnamefont {A.}~\bibnamefont {Gali}},\ }\bibfield  {title} {\enquote {\bibinfo {title} {Ab initio theory of the negatively charged boron vacancy qubit in hexagonal boron nitride},}\ }\href@noop {} {\bibfield  {journal} {\bibinfo  {journal} {npj Computational Materials}\ }\textbf {\bibinfo {volume} {6}} (\bibinfo {year} {2020})}\BibitemShut {NoStop}%
\bibitem [{\citenamefont {Gong}\ \emph {et~al.}(2024)\citenamefont {Gong}, \citenamefont {Du}, \citenamefont {Janzen}, \citenamefont {Liu}, \citenamefont {Liu}, \citenamefont {He}, \citenamefont {Ye}, \citenamefont {Li}, \citenamefont {Yao}, \citenamefont {Edgar}, \citenamefont {Henriksen},\ and\ \citenamefont {Zu}}]{Gong2024}%
  \BibitemOpen
  \bibfield  {author} {\bibinfo {author} {\bibfnamefont {R.}~\bibnamefont {Gong}}, \bibinfo {author} {\bibfnamefont {X.}~\bibnamefont {Du}}, \bibinfo {author} {\bibfnamefont {E.}~\bibnamefont {Janzen}}, \bibinfo {author} {\bibfnamefont {V.}~\bibnamefont {Liu}}, \bibinfo {author} {\bibfnamefont {Z.}~\bibnamefont {Liu}}, \bibinfo {author} {\bibfnamefont {G.}~\bibnamefont {He}}, \bibinfo {author} {\bibfnamefont {B.}~\bibnamefont {Ye}}, \bibinfo {author} {\bibfnamefont {T.}~\bibnamefont {Li}}, \bibinfo {author} {\bibfnamefont {N.~Y.}\ \bibnamefont {Yao}}, \bibinfo {author} {\bibfnamefont {J.~H.}\ \bibnamefont {Edgar}}, \bibinfo {author} {\bibfnamefont {E.~A.}\ \bibnamefont {Henriksen}}, \ and\ \bibinfo {author} {\bibfnamefont {C.}~\bibnamefont {Zu}},\ }\bibfield  {title} {\enquote {\bibinfo {title} {Isotope engineering for spin defects in van der waals materials},}\ }\href@noop {} {\bibfield  {journal} {\bibinfo  {journal} {Nature Communications}\ }\textbf {\bibinfo {volume} {15}},\ \bibinfo {pages} {104}
  (\bibinfo {year} {2024})}\BibitemShut {NoStop}%
\bibitem [{\citenamefont {Scholten}\ \emph {et~al.}(2024)\citenamefont {Scholten}, \citenamefont {Singh}, \citenamefont {Healey}, \citenamefont {Robertson}, \citenamefont {Haim}, \citenamefont {Tan}, \citenamefont {Broadway}, \citenamefont {Wang}, \citenamefont {Abe}, \citenamefont {Ohshima}, \citenamefont {Kianinia}, \citenamefont {Reineck}, \citenamefont {Aharonovich},\ and\ \citenamefont {Tetienne}}]{Scholten2024}%
  \BibitemOpen
  \bibfield  {author} {\bibinfo {author} {\bibfnamefont {S.~C.}\ \bibnamefont {Scholten}}, \bibinfo {author} {\bibfnamefont {P.}~\bibnamefont {Singh}}, \bibinfo {author} {\bibfnamefont {A.~J.}\ \bibnamefont {Healey}}, \bibinfo {author} {\bibfnamefont {I.~O.}\ \bibnamefont {Robertson}}, \bibinfo {author} {\bibfnamefont {G.}~\bibnamefont {Haim}}, \bibinfo {author} {\bibfnamefont {C.}~\bibnamefont {Tan}}, \bibinfo {author} {\bibfnamefont {D.~A}\ \bibnamefont {Broadway}}, \bibinfo {author} {\bibfnamefont {L.}~\bibnamefont {Wang}}, \bibinfo {author} {\bibfnamefont {H.}~\bibnamefont {Abe}}, \bibinfo {author} {\bibfnamefont {T.}~\bibnamefont {Ohshima}}, \bibinfo {author} {\bibfnamefont {M.}~\bibnamefont {Kianinia}}, \bibinfo {author} {\bibfnamefont {P.}~\bibnamefont {Reineck}}, \bibinfo {author} {\bibfnamefont {I.}~\bibnamefont {Aharonovich}}, \ and\ \bibinfo {author} {\bibfnamefont {J-P.}\ \bibnamefont {Tetienne}},\ }\bibfield  {title} {\enquote {\bibinfo {title} {Multi-species optically addressable spin defects in
  a van der waals material},}\ }\href@noop {} {\bibfield  {journal} {\bibinfo  {journal} {Nature Communications}\ }\textbf {\bibinfo {volume} {15}},\ \bibinfo {pages} {6727} (\bibinfo {year} {2024})}\BibitemShut {NoStop}%
\bibitem [{\citenamefont {Gao}\ \emph {et~al.}(2025{\natexlab{a}})\citenamefont {Gao}, , \citenamefont {Sumukh}, \citenamefont {Li}, \citenamefont {Ge}, \citenamefont {Dikshit}, \citenamefont {Zhang}, \citenamefont {amd K.~Shen}, \citenamefont {Jin}, \citenamefont {Ping},\ and\ \citenamefont {Li}}]{Gao2025}%
  \BibitemOpen
  \bibfield  {author} {\bibinfo {author} {\bibfnamefont {X.}~\bibnamefont {Gao}}, , \bibinfo {author} {\bibfnamefont {S.}~\bibnamefont {Sumukh}}, \bibinfo {author} {\bibfnamefont {K.}~\bibnamefont {Li}}, \bibinfo {author} {\bibfnamefont {Z.}~\bibnamefont {Ge}}, \bibinfo {author} {\bibfnamefont {S.}~\bibnamefont {Dikshit}}, \bibinfo {author} {\bibfnamefont {S.}~\bibnamefont {Zhang}}, \bibinfo {author} {\bibfnamefont {P.~Ju}\ \bibnamefont {amd K.~Shen}}, \bibinfo {author} {\bibfnamefont {Y.}~\bibnamefont {Jin}}, \bibinfo {author} {\bibfnamefont {Y.}~\bibnamefont {Ping}}, \ and\ \bibinfo {author} {\bibfnamefont {T.}~\bibnamefont {Li}},\ }\bibfield  {title} {\enquote {\bibinfo {title} {Single nuclear spin detection and control in a van der waals material},}\ }\href@noop {} {\bibfield  {journal} {\bibinfo  {journal} {Nature}\ }\textbf {\bibinfo {volume} {643}},\ \bibinfo {pages} {943–949} (\bibinfo {year} {2025}{\natexlab{a}})}\BibitemShut {NoStop}%
\bibitem [{\citenamefont {Gisin}\ \emph {et~al.}(2002)\citenamefont {Gisin}, \citenamefont {Ribordy}, \citenamefont {w.~Tittel},\ and\ \citenamefont {Zbinden}}]{Gisin2002}%
  \BibitemOpen
  \bibfield  {author} {\bibinfo {author} {\bibfnamefont {N.}~\bibnamefont {Gisin}}, \bibinfo {author} {\bibfnamefont {G.}~\bibnamefont {Ribordy}}, \bibinfo {author} {\bibnamefont {w.~Tittel}}, \ and\ \bibinfo {author} {\bibfnamefont {H.}~\bibnamefont {Zbinden}},\ }\bibfield  {title} {\enquote {\bibinfo {title} {Quantum cryptography},}\ }\href@noop {} {\bibfield  {journal} {\bibinfo  {journal} {Reviews Modern Physics}\ }\textbf {\bibinfo {volume} {74}},\ \bibinfo {pages} {145--195} (\bibinfo {year} {2002})}\BibitemShut {NoStop}%
\bibitem [{\citenamefont {Bourrellier}\ \emph {et~al.}(2016)\citenamefont {Bourrellier}, \citenamefont {Meuret}, \citenamefont {Tararan}, \citenamefont {Stéphan}, \citenamefont {Kociak}, \citenamefont {Tizei},\ and\ \citenamefont {Zobelli}}]{Bourrellier2016}%
  \BibitemOpen
  \bibfield  {author} {\bibinfo {author} {\bibfnamefont {R.}~\bibnamefont {Bourrellier}}, \bibinfo {author} {\bibfnamefont {S.}~\bibnamefont {Meuret}}, \bibinfo {author} {\bibfnamefont {A.}~\bibnamefont {Tararan}}, \bibinfo {author} {\bibfnamefont {O.}~\bibnamefont {Stéphan}}, \bibinfo {author} {\bibfnamefont {M.}~\bibnamefont {Kociak}}, \bibinfo {author} {\bibfnamefont {L.~H.~G.}\ \bibnamefont {Tizei}}, \ and\ \bibinfo {author} {\bibfnamefont {A.}~\bibnamefont {Zobelli}},\ }\bibfield  {title} {\enquote {\bibinfo {title} {Bright uv single photon emission at point defects in h-bn},}\ }\href@noop {} {\bibfield  {journal} {\bibinfo  {journal} {Nano Letters}\ }\textbf {\bibinfo {volume} {16}},\ \bibinfo {pages} {4317--4321} (\bibinfo {year} {2016})}\BibitemShut {NoStop}%
\bibitem [{\citenamefont {Tran}\ \emph {et~al.}(2016{\natexlab{b}})\citenamefont {Tran}, \citenamefont {Elbadawi}, \citenamefont {Totonjian}, \citenamefont {Lobo}, \citenamefont {Grosso}, \citenamefont {Moon}, \citenamefont {Englund}, \citenamefont {Ford}, \citenamefont {Aharonovich},\ and\ \citenamefont {Toth}}]{Tran2016b}%
  \BibitemOpen
  \bibfield  {author} {\bibinfo {author} {\bibfnamefont {T.~T.}\ \bibnamefont {Tran}}, \bibinfo {author} {\bibfnamefont {C.}~\bibnamefont {Elbadawi}}, \bibinfo {author} {\bibfnamefont {D.}~\bibnamefont {Totonjian}}, \bibinfo {author} {\bibfnamefont {C.~J.}\ \bibnamefont {Lobo}}, \bibinfo {author} {\bibfnamefont {G.}~\bibnamefont {Grosso}}, \bibinfo {author} {\bibfnamefont {H.}~\bibnamefont {Moon}}, \bibinfo {author} {\bibfnamefont {D.~R.}\ \bibnamefont {Englund}}, \bibinfo {author} {\bibfnamefont {M.~J.}\ \bibnamefont {Ford}}, \bibinfo {author} {\bibfnamefont {I.}~\bibnamefont {Aharonovich}}, \ and\ \bibinfo {author} {\bibfnamefont {M.}~\bibnamefont {Toth}},\ }\bibfield  {title} {\enquote {\bibinfo {title} {Robust multicolor single photon emission from point defects in hexagonal boron nitride},}\ }\href@noop {} {\bibfield  {journal} {\bibinfo  {journal} {ACS Nano}\ }\textbf {\bibinfo {volume} {10}},\ \bibinfo {pages} {7331--7338} (\bibinfo {year} {2016}{\natexlab{b}})}\BibitemShut {NoStop}%
\bibitem [{\citenamefont {Camphausen}\ \emph {et~al.}(2020)\citenamefont {Camphausen}, \citenamefont {Marini}, \citenamefont {Tawfik}, \citenamefont {Tran}, \citenamefont {Ford},\ and\ \citenamefont {Palomba}}]{Camphausen2020}%
  \BibitemOpen
  \bibfield  {author} {\bibinfo {author} {\bibfnamefont {R.}~\bibnamefont {Camphausen}}, \bibinfo {author} {\bibfnamefont {L.}~\bibnamefont {Marini}}, \bibinfo {author} {\bibfnamefont {S.~A.}\ \bibnamefont {Tawfik}}, \bibinfo {author} {\bibfnamefont {T.~T.}\ \bibnamefont {Tran}}, \bibinfo {author} {\bibfnamefont {M.~J.}\ \bibnamefont {Ford}}, \ and\ \bibinfo {author} {\bibfnamefont {S.}~\bibnamefont {Palomba}},\ }\bibfield  {title} {\enquote {\bibinfo {title} {Observation of near-infrared sub-poissonian photon emission in hexagonal boron nitride at room temperature},}\ }\href@noop {} {\bibfield  {journal} {\bibinfo  {journal} {APL Photonics}\ }\textbf {\bibinfo {volume} {5}},\ \bibinfo {pages} {076103} (\bibinfo {year} {2020})}\BibitemShut {NoStop}%
\bibitem [{\citenamefont {Gale}\ \emph {et~al.}(2022)\citenamefont {Gale}, \citenamefont {Li}, \citenamefont {Chen}, \citenamefont {Watanabe}, \citenamefont {Taniguchi}, \citenamefont {Aharonovich},\ and\ \citenamefont {Toth}}]{Gale2022}%
  \BibitemOpen
  \bibfield  {author} {\bibinfo {author} {\bibfnamefont {A.}~\bibnamefont {Gale}}, \bibinfo {author} {\bibfnamefont {C.}~\bibnamefont {Li}}, \bibinfo {author} {\bibfnamefont {Y.}~\bibnamefont {Chen}}, \bibinfo {author} {\bibfnamefont {K.}~\bibnamefont {Watanabe}}, \bibinfo {author} {\bibfnamefont {T.}~\bibnamefont {Taniguchi}}, \bibinfo {author} {\bibfnamefont {I.}~\bibnamefont {Aharonovich}}, \ and\ \bibinfo {author} {\bibfnamefont {M.}~\bibnamefont {Toth}},\ }\bibfield  {title} {\enquote {\bibinfo {title} {Site-specific fabrication of blue quantum emitters in hexagonal boron nitride},}\ }\href@noop {} {\bibfield  {journal} {\bibinfo  {journal} {ACS Photonics}\ }\textbf {\bibinfo {volume} {9}} (\bibinfo {year} {2022})}\BibitemShut {NoStop}%
\bibitem [{\citenamefont {Kumar}\ \emph {et~al.}(2023)\citenamefont {Kumar}, \citenamefont {Cholsuk}, \citenamefont {Zand}, \citenamefont {Mishuk}, \citenamefont {Matthes}, \citenamefont {Eilenberger}, \citenamefont {Suwanna},\ and\ \citenamefont {Vogl}}]{Kumar2023}%
  \BibitemOpen
  \bibfield  {author} {\bibinfo {author} {\bibfnamefont {A.}~\bibnamefont {Kumar}}, \bibinfo {author} {\bibfnamefont {C.}~\bibnamefont {Cholsuk}}, \bibinfo {author} {\bibfnamefont {A.}~\bibnamefont {Zand}}, \bibinfo {author} {\bibfnamefont {M.~N.}\ \bibnamefont {Mishuk}}, \bibinfo {author} {\bibfnamefont {T.}~\bibnamefont {Matthes}}, \bibinfo {author} {\bibfnamefont {F.}~\bibnamefont {Eilenberger}}, \bibinfo {author} {\bibfnamefont {S.}~\bibnamefont {Suwanna}}, \ and\ \bibinfo {author} {\bibfnamefont {T.}~\bibnamefont {Vogl}},\ }\bibfield  {title} {\enquote {\bibinfo {title} {Localized creation of yellow single photon emitting carbon complexes in hexagonal boron nitride},}\ }\href@noop {} {\bibfield  {journal} {\bibinfo  {journal} {APL Materials}\ }\textbf {\bibinfo {volume} {11}},\ \bibinfo {pages} {2170–2177} (\bibinfo {year} {2023})}\BibitemShut {NoStop}%
\bibitem [{\citenamefont {Vuong}\ \emph {et~al.}(2016)\citenamefont {Vuong}, \citenamefont {Cassabois}, \citenamefont {Valvin}, \citenamefont {Ouerghi}, \citenamefont {Chassagneux}, \citenamefont {Voisin},\ and\ \citenamefont {Gil}}]{Vuong2016}%
  \BibitemOpen
  \bibfield  {author} {\bibinfo {author} {\bibfnamefont {T.~Q.P.}\ \bibnamefont {Vuong}}, \bibinfo {author} {\bibfnamefont {G.}~\bibnamefont {Cassabois}}, \bibinfo {author} {\bibfnamefont {P.}~\bibnamefont {Valvin}}, \bibinfo {author} {\bibfnamefont {A.}~\bibnamefont {Ouerghi}}, \bibinfo {author} {\bibfnamefont {Y.}~\bibnamefont {Chassagneux}}, \bibinfo {author} {\bibfnamefont {C.}~\bibnamefont {Voisin}}, \ and\ \bibinfo {author} {\bibfnamefont {B.}~\bibnamefont {Gil}},\ }\bibfield  {title} {\enquote {\bibinfo {title} {Phonon-photon mapping in a color center in hexagonal boron nitride},}\ }\href@noop {} {\bibfield  {journal} {\bibinfo  {journal} {Physical Review Letters}\ }\textbf {\bibinfo {volume} {117}},\ \bibinfo {pages} {071108} (\bibinfo {year} {2016})}\BibitemShut {NoStop}%
\bibitem [{\citenamefont {Li}\ \emph {et~al.}(2022{\natexlab{a}})\citenamefont {Li}, \citenamefont {Pershin}, \citenamefont {Thiering}, \citenamefont {Udvarhelyi},\ and\ \citenamefont {Gali}}]{Li2022}%
  \BibitemOpen
  \bibfield  {author} {\bibinfo {author} {\bibfnamefont {S.}~\bibnamefont {Li}}, \bibinfo {author} {\bibfnamefont {A.}~\bibnamefont {Pershin}}, \bibinfo {author} {\bibfnamefont {G.}~\bibnamefont {Thiering}}, \bibinfo {author} {\bibfnamefont {P.}~\bibnamefont {Udvarhelyi}}, \ and\ \bibinfo {author} {\bibfnamefont {A.}~\bibnamefont {Gali}},\ }\bibfield  {title} {\enquote {\bibinfo {title} {Ultraviolet quantum emitters in hexagonal boron nitride from carbon clusters},}\ }\href@noop {} {\bibfield  {journal} {\bibinfo  {journal} {Journal of Physical Chemistry Letters}\ }\textbf {\bibinfo {volume} {13}},\ \bibinfo {pages} {3150–3157} (\bibinfo {year} {2022}{\natexlab{a}})}\BibitemShut {NoStop}%
\bibitem [{\citenamefont {Zhigulin}\ \emph {et~al.}(2023{\natexlab{a}})\citenamefont {Zhigulin}, \citenamefont {Yamamura}, \citenamefont {Iv\'{a}dy}, \citenamefont {Gale}, \citenamefont {Horder}, \citenamefont {Lobo}, \citenamefont {Kianinia}, \citenamefont {Toth},\ and\ \citenamefont {Aharonovich}}]{zhigulin_photophysics_2023}%
  \BibitemOpen
  \bibfield  {author} {\bibinfo {author} {\bibfnamefont {I.}~\bibnamefont {Zhigulin}}, \bibinfo {author} {\bibfnamefont {K.}~\bibnamefont {Yamamura}}, \bibinfo {author} {\bibfnamefont {V.}~\bibnamefont {Iv\'{a}dy}}, \bibinfo {author} {\bibfnamefont {A.}~\bibnamefont {Gale}}, \bibinfo {author} {\bibfnamefont {J.}~\bibnamefont {Horder}}, \bibinfo {author} {\bibfnamefont {C.~J.}\ \bibnamefont {Lobo}}, \bibinfo {author} {\bibfnamefont {M.}~\bibnamefont {Kianinia}}, \bibinfo {author} {\bibfnamefont {M.}~\bibnamefont {Toth}}, \ and\ \bibinfo {author} {\bibfnamefont {I.}~\bibnamefont {Aharonovich}},\ }\bibfield  {title} {\enquote {\bibinfo {title} {Photophysics of blue quantum emitters in hexagonal boron nitride},}\ }\href@noop {} {\bibfield  {journal} {\bibinfo  {journal} {Materials for Quantum Technology}\ }\textbf {\bibinfo {volume} {3}},\ \bibinfo {pages} {015002} (\bibinfo {year} {2023}{\natexlab{a}})}\BibitemShut {NoStop}%
\bibitem [{\citenamefont {Yamamura}\ \emph {et~al.}(2025)\citenamefont {Yamamura}, \citenamefont {Coste}, \citenamefont {Zeng}, \citenamefont {Toth}, \citenamefont {Kianinia},\ and\ \citenamefont {Aharonovich}}]{Yamamura2024}%
  \BibitemOpen
  \bibfield  {author} {\bibinfo {author} {\bibfnamefont {K.}~\bibnamefont {Yamamura}}, \bibinfo {author} {\bibfnamefont {N.}~\bibnamefont {Coste}}, \bibinfo {author} {\bibfnamefont {H.~Zhi~Jie}\ \bibnamefont {Zeng}}, \bibinfo {author} {\bibfnamefont {M.}~\bibnamefont {Toth}}, \bibinfo {author} {\bibfnamefont {M.}~\bibnamefont {Kianinia}}, \ and\ \bibinfo {author} {\bibfnamefont {I.}~\bibnamefont {Aharonovich}},\ }\bibfield  {title} {\enquote {\bibinfo {title} {Quantum efficiency of the b-centre in hexagonal boron nitride},}\ }\href@noop {} {\bibfield  {journal} {\bibinfo  {journal} {Nanophotonics}\ }\textbf {\bibinfo {volume} {14}},\ \bibinfo {pages} {1715–1720} (\bibinfo {year} {2025})}\BibitemShut {NoStop}%
\bibitem [{\citenamefont {Mendelson}\ \emph {et~al.}(2021)\citenamefont {Mendelson}, \citenamefont {Chugh}, \citenamefont {Reimers}, \citenamefont {Cheng}, \citenamefont {Gottscholl}, \citenamefont {Long}, \citenamefont {Mellor}, \citenamefont {Zettl}, \citenamefont {Dyakonov}, \citenamefont {Beton}, \citenamefont {Novikov}, \citenamefont {Jagadish}, \citenamefont {Tan}, \citenamefont {Ford}, \citenamefont {Toth}, \citenamefont {Bradac},\ and\ \citenamefont {Aharonovich}}]{Mendelson2021}%
  \BibitemOpen
  \bibfield  {author} {\bibinfo {author} {\bibfnamefont {N.}~\bibnamefont {Mendelson}}, \bibinfo {author} {\bibfnamefont {D.}~\bibnamefont {Chugh}}, \bibinfo {author} {\bibfnamefont {J.~R.}\ \bibnamefont {Reimers}}, \bibinfo {author} {\bibfnamefont {T.~S.}\ \bibnamefont {Cheng}}, \bibinfo {author} {\bibfnamefont {A.}~\bibnamefont {Gottscholl}}, \bibinfo {author} {\bibfnamefont {H.}~\bibnamefont {Long}}, \bibinfo {author} {\bibfnamefont {C.~J.}\ \bibnamefont {Mellor}}, \bibinfo {author} {\bibfnamefont {A.}~\bibnamefont {Zettl}}, \bibinfo {author} {\bibfnamefont {V.}~\bibnamefont {Dyakonov}}, \bibinfo {author} {\bibfnamefont {P.~H.}\ \bibnamefont {Beton}}, \bibinfo {author} {\bibfnamefont {S.~V.}\ \bibnamefont {Novikov}}, \bibinfo {author} {\bibfnamefont {C.}~\bibnamefont {Jagadish}}, \bibinfo {author} {\bibfnamefont {H.~H.}\ \bibnamefont {Tan}}, \bibinfo {author} {\bibfnamefont {M.~J.}\ \bibnamefont {Ford}}, \bibinfo {author} {\bibfnamefont {M.}~\bibnamefont {Toth}}, \bibinfo {author} {\bibfnamefont
  {C.}~\bibnamefont {Bradac}}, \ and\ \bibinfo {author} {\bibfnamefont {I.}~\bibnamefont {Aharonovich}},\ }\bibfield  {title} {\enquote {\bibinfo {title} {Identifying carbon as the source of visible single-photon emission from hexagonal boron nitride},}\ }\href@noop {} {\bibfield  {journal} {\bibinfo  {journal} {Nature Materials}\ }\textbf {\bibinfo {volume} {20}},\ \bibinfo {pages} {321–328} (\bibinfo {year} {2021})}\BibitemShut {NoStop}%
\bibitem [{\citenamefont {Plo}\ \emph {et~al.}(2025)\citenamefont {Plo}, \citenamefont {Pershin}, \citenamefont {Li}, \citenamefont {Poirier}, \citenamefont {Janzen}, \citenamefont {Schutte}, \citenamefont {Tian}, \citenamefont {Wynn}, \citenamefont {Bernard}, \citenamefont {Rousseau}, \citenamefont {Ibanez}, \citenamefont {Valvin}, \citenamefont {Desrat}, \citenamefont {Michel}, \citenamefont {Jacques}, \citenamefont {Gil}, \citenamefont {Kaminska}, \citenamefont {Wan}, \citenamefont {Edgar}, \citenamefont {Gali},\ and\ \citenamefont {Cassabois}}]{Plo2025}%
  \BibitemOpen
  \bibfield  {author} {\bibinfo {author} {\bibfnamefont {J.}~\bibnamefont {Plo}}, \bibinfo {author} {\bibfnamefont {A.}~\bibnamefont {Pershin}}, \bibinfo {author} {\bibfnamefont {S.}~\bibnamefont {Li}}, \bibinfo {author} {\bibfnamefont {T.}~\bibnamefont {Poirier}}, \bibinfo {author} {\bibfnamefont {E.}~\bibnamefont {Janzen}}, \bibinfo {author} {\bibfnamefont {H.}~\bibnamefont {Schutte}}, \bibinfo {author} {\bibfnamefont {M.}~\bibnamefont {Tian}}, \bibinfo {author} {\bibfnamefont {M.}~\bibnamefont {Wynn}}, \bibinfo {author} {\bibfnamefont {S.}~\bibnamefont {Bernard}}, \bibinfo {author} {\bibfnamefont {A.}~\bibnamefont {Rousseau}}, \bibinfo {author} {\bibfnamefont {A.}~\bibnamefont {Ibanez}}, \bibinfo {author} {\bibfnamefont {P.}~\bibnamefont {Valvin}}, \bibinfo {author} {\bibfnamefont {W.}~\bibnamefont {Desrat}}, \bibinfo {author} {\bibfnamefont {T.}~\bibnamefont {Michel}}, \bibinfo {author} {\bibfnamefont {V.}~\bibnamefont {Jacques}}, \bibinfo {author} {\bibfnamefont {B.}~\bibnamefont {Gil}}, \bibinfo
  {author} {\bibfnamefont {A.}~\bibnamefont {Kaminska}}, \bibinfo {author} {\bibfnamefont {N.}~\bibnamefont {Wan}}, \bibinfo {author} {\bibfnamefont {J.~H.}\ \bibnamefont {Edgar}}, \bibinfo {author} {\bibfnamefont {A.}~\bibnamefont {Gali}}, \ and\ \bibinfo {author} {\bibfnamefont {G.}~\bibnamefont {Cassabois}},\ }\bibfield  {title} {\enquote {\bibinfo {title} {Isotope substitution and polytype control for point defects identification: The case of the ultraviolet color center in hexagonal boron nitride},}\ }\href@noop {} {\bibfield  {journal} {\bibinfo  {journal} {Physical Review X}\ }\textbf {\bibinfo {volume} {15}},\ \bibinfo {pages} {021045} (\bibinfo {year} {2025})}\BibitemShut {NoStop}%
\bibitem [{\citenamefont {Maciaszek}\ and\ \citenamefont {Razinkovas}(2024)}]{Maciaszek2024}%
  \BibitemOpen
  \bibfield  {author} {\bibinfo {author} {\bibfnamefont {M.}~\bibnamefont {Maciaszek}}\ and\ \bibinfo {author} {\bibfnamefont {L}~\bibnamefont {Razinkovas}},\ }\bibfield  {title} {\enquote {\bibinfo {title} {Blue quantum emitter in hexagonal boron nitride and a carbon chain tetramer: a first-principles study},}\ }\href@noop {} {\bibfield  {journal} {\bibinfo  {journal} {ACS Applied Nano Materials}\ }\textbf {\bibinfo {volume} {7}},\ \bibinfo {pages} {18979--18985} (\bibinfo {year} {2024})}\BibitemShut {NoStop}%
\bibitem [{\citenamefont {Nikolay}\ \emph {et~al.}(2019)\citenamefont {Nikolay}, \citenamefont {Mendelson}, \citenamefont {\"{O}zelci}, \citenamefont {Sontheimer}, \citenamefont {B\"{o}hm}, \citenamefont {Kewes}, \citenamefont {Toth}, \citenamefont {Aharonovich},\ and\ \citenamefont {Benson}}]{Nikolay2019}%
  \BibitemOpen
  \bibfield  {author} {\bibinfo {author} {\bibfnamefont {N.}~\bibnamefont {Nikolay}}, \bibinfo {author} {\bibfnamefont {N.}~\bibnamefont {Mendelson}}, \bibinfo {author} {\bibfnamefont {E.}~\bibnamefont {\"{O}zelci}}, \bibinfo {author} {\bibfnamefont {B.}~\bibnamefont {Sontheimer}}, \bibinfo {author} {\bibfnamefont {F.}~\bibnamefont {B\"{o}hm}}, \bibinfo {author} {\bibfnamefont {G.}~\bibnamefont {Kewes}}, \bibinfo {author} {\bibfnamefont {M.}~\bibnamefont {Toth}}, \bibinfo {author} {\bibfnamefont {I.}~\bibnamefont {Aharonovich}}, \ and\ \bibinfo {author} {\bibfnamefont {O.}~\bibnamefont {Benson}},\ }\bibfield  {title} {\enquote {\bibinfo {title} {Direct measurement of quantum efficiency of single-photon emitters in hexagonal boron nitride},}\ }\href@noop {} {\bibfield  {journal} {\bibinfo  {journal} {Optica}\ }\textbf {\bibinfo {volume} {6}},\ \bibinfo {pages} {1084--1088} (\bibinfo {year} {2019})}\BibitemShut {NoStop}%
\bibitem [{\citenamefont {Bommer}\ and\ \citenamefont {Becher}(2019)}]{Bommer2019}%
  \BibitemOpen
  \bibfield  {author} {\bibinfo {author} {\bibfnamefont {A.}~\bibnamefont {Bommer}}\ and\ \bibinfo {author} {\bibfnamefont {C.}~\bibnamefont {Becher}},\ }\bibfield  {title} {\enquote {\bibinfo {title} {New insights into nonclassical light emission from defects in multi-layer hexagonal boron nitride},}\ }\href@noop {} {\bibfield  {journal} {\bibinfo  {journal} {Nanophotonics}\ }\textbf {\bibinfo {volume} {8}},\ \bibinfo {pages} {2041--2048} (\bibinfo {year} {2019})}\BibitemShut {NoStop}%
\bibitem [{\citenamefont {Chatterjee}\ \emph {et~al.}(2025)\citenamefont {Chatterjee}, \citenamefont {Biswas}, \citenamefont {Fuhr}, \citenamefont {Terlier}, \citenamefont {Sumpter}, \citenamefont {Ajayan}, \citenamefont {Aharonovich},\ and\ \citenamefont {Huang}}]{Chatterjee2025}%
  \BibitemOpen
  \bibfield  {author} {\bibinfo {author} {\bibfnamefont {A.}~\bibnamefont {Chatterjee}}, \bibinfo {author} {\bibfnamefont {A.}~\bibnamefont {Biswas}}, \bibinfo {author} {\bibfnamefont {A.~S.}\ \bibnamefont {Fuhr}}, \bibinfo {author} {\bibfnamefont {T.}~\bibnamefont {Terlier}}, \bibinfo {author} {\bibfnamefont {B.~G.}\ \bibnamefont {Sumpter}}, \bibinfo {author} {\bibfnamefont {P.~M.}\ \bibnamefont {Ajayan}}, \bibinfo {author} {\bibfnamefont {I.}~\bibnamefont {Aharonovich}}, \ and\ \bibinfo {author} {\bibfnamefont {S.}~\bibnamefont {Huang}},\ }\bibfield  {title} {\enquote {\bibinfo {title} {Room-temperature high-purity single-photon emission from carbon-doped boron nitride thin films},}\ }\href@noop {} {\bibfield  {journal} {\bibinfo  {journal} {Science Advances}\ }\textbf {\bibinfo {volume} {11}},\ \bibinfo {pages} {899} (\bibinfo {year} {2025})}\BibitemShut {NoStop}%
\bibitem [{\citenamefont {Dong}\ \emph {et~al.}(2023)\citenamefont {Dong}, \citenamefont {Mi}, \citenamefont {Hou}, \citenamefont {Huang}, \citenamefont {Wang}, \citenamefont {Yu}, \citenamefont {Wei}, \citenamefont {Zhang},\ and\ \citenamefont {Fang}}]{Dong2023}%
  \BibitemOpen
  \bibfield  {author} {\bibinfo {author} {\bibfnamefont {S.}~\bibnamefont {Dong}}, \bibinfo {author} {\bibfnamefont {S.}~\bibnamefont {Mi}}, \bibinfo {author} {\bibfnamefont {Q.}~\bibnamefont {Hou}}, \bibinfo {author} {\bibfnamefont {Y.}~\bibnamefont {Huang}}, \bibinfo {author} {\bibfnamefont {J.}~\bibnamefont {Wang}}, \bibinfo {author} {\bibfnamefont {Y.}~\bibnamefont {Yu}}, \bibinfo {author} {\bibfnamefont {Z.}~\bibnamefont {Wei}}, \bibinfo {author} {\bibfnamefont {Z.}~\bibnamefont {Zhang}}, \ and\ \bibinfo {author} {\bibfnamefont {J.}~\bibnamefont {Fang}},\ }\bibfield  {title} {\enquote {\bibinfo {title} {Decoy state semi-quantum key distribution},}\ }\href@noop {} {\bibfield  {journal} {\bibinfo  {journal} {EPJ Quantum Technology}\ }\textbf {\bibinfo {volume} {10}},\ \bibinfo {pages} {230504} (\bibinfo {year} {2023})}\BibitemShut {NoStop}%
\bibitem [{\citenamefont {Li}\ \emph {et~al.}(2022{\natexlab{b}})\citenamefont {Li}, \citenamefont {Zhang}, \citenamefont {Jiang},\ and\ \citenamefont {Cai}}]{Li2022b}%
  \BibitemOpen
  \bibfield  {author} {\bibinfo {author} {\bibfnamefont {H-W.}\ \bibnamefont {Li}}, \bibinfo {author} {\bibfnamefont {C-M.}\ \bibnamefont {Zhang}}, \bibinfo {author} {\bibfnamefont {M-S.}\ \bibnamefont {Jiang}}, \ and\ \bibinfo {author} {\bibfnamefont {Q-Y.}\ \bibnamefont {Cai}},\ }\bibfield  {title} {\enquote {\bibinfo {title} {Improving the performance of practical decoy-state quantum key distribution with advantage distillation technology},}\ }\href@noop {} {\bibfield  {journal} {\bibinfo  {journal} {Commununications Physics}\ }\textbf {\bibinfo {volume} {5}},\ \bibinfo {pages} {53} (\bibinfo {year} {2022}{\natexlab{b}})}\BibitemShut {NoStop}%
\bibitem [{\citenamefont {Samaner}\ \emph {et~al.}(2022)\citenamefont {Samaner}, \citenamefont {Paçal}, \citenamefont {Mutlu}, \citenamefont {Uyanık},\ and\ \citenamefont {Ateş}}]{Samaner2022}%
  \BibitemOpen
  \bibfield  {author} {\bibinfo {author} {\bibfnamefont {Ç.}\ \bibnamefont {Samaner}}, \bibinfo {author} {\bibfnamefont {S.}~\bibnamefont {Paçal}}, \bibinfo {author} {\bibfnamefont {G.}~\bibnamefont {Mutlu}}, \bibinfo {author} {\bibfnamefont {K.}~\bibnamefont {Uyanık}}, \ and\ \bibinfo {author} {\bibfnamefont {S.}~\bibnamefont {Ateş}},\ }\bibfield  {title} {\enquote {\bibinfo {title} {Free-space quantum key distribution with single photons from defects in hexagonal boron nitride},}\ }\href@noop {} {\bibfield  {journal} {\bibinfo  {journal} {Advanced Quantum Technologies}\ }\textbf {\bibinfo {volume} {5}},\ \bibinfo {pages} {2200059} (\bibinfo {year} {2022})}\BibitemShut {NoStop}%
\bibitem [{\citenamefont {Al-Juboori}\ \emph {et~al.}(2023)\citenamefont {Al-Juboori}, \citenamefont {Zeng}, \citenamefont {Nguyen}, \citenamefont {Ai}, \citenamefont {Laucht}, \citenamefont {Solntsev}, \citenamefont {Toth}, \citenamefont {Malaney},\ and\ \citenamefont {Aharonovich}}]{Al-Juboori2023}%
  \BibitemOpen
  \bibfield  {author} {\bibinfo {author} {\bibfnamefont {A.}~\bibnamefont {Al-Juboori}}, \bibinfo {author} {\bibfnamefont {H.~Z.~J.}\ \bibnamefont {Zeng}}, \bibinfo {author} {\bibfnamefont {M.~A.~P}\ \bibnamefont {Nguyen}}, \bibinfo {author} {\bibfnamefont {X.}~\bibnamefont {Ai}}, \bibinfo {author} {\bibfnamefont {A.}~\bibnamefont {Laucht}}, \bibinfo {author} {\bibfnamefont {A.}~\bibnamefont {Solntsev}}, \bibinfo {author} {\bibfnamefont {M.}~\bibnamefont {Toth}}, \bibinfo {author} {\bibfnamefont {R.}~\bibnamefont {Malaney}}, \ and\ \bibinfo {author} {\bibfnamefont {I.}~\bibnamefont {Aharonovich}},\ }\bibfield  {title} {\enquote {\bibinfo {title} {Quantum key distribution using a quantum emitter in hexagonal boron nitride},}\ }\href@noop {} {\bibfield  {journal} {\bibinfo  {journal} {Advanced Quantum Technologies}\ }\textbf {\bibinfo {volume} {6}},\ \bibinfo {pages} {2300038} (\bibinfo {year} {2023})}\BibitemShut {NoStop}%
\bibitem [{\citenamefont {Häußler}\ \emph {et~al.}(2021)\citenamefont {Häußler}, \citenamefont {Bayer}, \citenamefont {Waltrich}, \citenamefont {Mendelson}, \citenamefont {Li}, \citenamefont {Hunger}, \citenamefont {Aharonovich},\ and\ \citenamefont {Kubanek}}]{Häußler2021}%
  \BibitemOpen
  \bibfield  {author} {\bibinfo {author} {\bibfnamefont {S.}~\bibnamefont {Häußler}}, \bibinfo {author} {\bibfnamefont {G.}~\bibnamefont {Bayer}}, \bibinfo {author} {\bibfnamefont {R.}~\bibnamefont {Waltrich}}, \bibinfo {author} {\bibfnamefont {N.}~\bibnamefont {Mendelson}}, \bibinfo {author} {\bibfnamefont {C.}~\bibnamefont {Li}}, \bibinfo {author} {\bibfnamefont {D.}~\bibnamefont {Hunger}}, \bibinfo {author} {\bibfnamefont {I.}~\bibnamefont {Aharonovich}}, \ and\ \bibinfo {author} {\bibfnamefont {A.}~\bibnamefont {Kubanek}},\ }\bibfield  {title} {\enquote {\bibinfo {title} {Tunable fiber-cavity enhanced photon emission from defect centers in hbn},}\ }\href@noop {} {\bibfield  {journal} {\bibinfo  {journal} {Advanced Optical Materials}\ }\textbf {\bibinfo {volume} {9}},\ \bibinfo {pages} {2002218} (\bibinfo {year} {2021})}\BibitemShut {NoStop}%
\bibitem [{\citenamefont {Vogl}\ \emph {et~al.}(2021)\citenamefont {Vogl}, \citenamefont {Knopf}, \citenamefont {Weissflog}, \citenamefont {Lam},\ and\ \citenamefont {Eilenberger}}]{Vogl2021}%
  \BibitemOpen
  \bibfield  {author} {\bibinfo {author} {\bibfnamefont {T.}~\bibnamefont {Vogl}}, \bibinfo {author} {\bibfnamefont {H.}~\bibnamefont {Knopf}}, \bibinfo {author} {\bibfnamefont {M.}~\bibnamefont {Weissflog}}, \bibinfo {author} {\bibfnamefont {P.~K.}\ \bibnamefont {Lam}}, \ and\ \bibinfo {author} {\bibfnamefont {F.}~\bibnamefont {Eilenberger}},\ }\bibfield  {title} {\enquote {\bibinfo {title} {Sensitive single-photon test of extended quantum theory with two-dimensional hexagonal boron nitride},}\ }\href@noop {} {\bibfield  {journal} {\bibinfo  {journal} {Physical Review Research}\ }\textbf {\bibinfo {volume} {3}},\ \bibinfo {pages} {013296} (\bibinfo {year} {2021})}\BibitemShut {NoStop}%
\bibitem [{\citenamefont {Abasifard}\ \emph {et~al.}(2024)\citenamefont {Abasifard}, \citenamefont {Cholsuk}, \citenamefont {Pousa}, \citenamefont {Kumar}, \citenamefont {Zand}, \citenamefont {Riel}, \citenamefont {Oi},\ and\ \citenamefont {Vogl}}]{Abasifard2024}%
  \BibitemOpen
  \bibfield  {author} {\bibinfo {author} {\bibfnamefont {M.}~\bibnamefont {Abasifard}}, \bibinfo {author} {\bibfnamefont {C.}~\bibnamefont {Cholsuk}}, \bibinfo {author} {\bibfnamefont {R.~G.}\ \bibnamefont {Pousa}}, \bibinfo {author} {\bibfnamefont {A.}~\bibnamefont {Kumar}}, \bibinfo {author} {\bibfnamefont {A.}~\bibnamefont {Zand}}, \bibinfo {author} {\bibfnamefont {T.}~\bibnamefont {Riel}}, \bibinfo {author} {\bibfnamefont {D.~K.~L.}\ \bibnamefont {Oi}}, \ and\ \bibinfo {author} {\bibfnamefont {T.}~\bibnamefont {Vogl}},\ }\bibfield  {title} {\enquote {\bibinfo {title} {The ideal wavelength for daylight free-space quantum key distribution},}\ }\href@noop {} {\bibfield  {journal} {\bibinfo  {journal} {APL Quantum}\ }\textbf {\bibinfo {volume} {1}},\ \bibinfo {pages} {016113} (\bibinfo {year} {2024})}\BibitemShut {NoStop}%
\bibitem [{\citenamefont {Cholsuk}\ \emph {et~al.}(2022)\citenamefont {Cholsuk}, \citenamefont {Suwanna},\ and\ \citenamefont {Vogl}}]{Cholsuk2022}%
  \BibitemOpen
  \bibfield  {author} {\bibinfo {author} {\bibfnamefont {C.}~\bibnamefont {Cholsuk}}, \bibinfo {author} {\bibfnamefont {S.}~\bibnamefont {Suwanna}}, \ and\ \bibinfo {author} {\bibfnamefont {T.}~\bibnamefont {Vogl}},\ }\bibfield  {title} {\enquote {\bibinfo {title} {Tailoring the emission wavelength of color centers in hexagonal boron nitride for quantum applications},}\ }\href@noop {} {\bibfield  {journal} {\bibinfo  {journal} {Nanomaterials}\ }\textbf {\bibinfo {volume} {12}},\ \bibinfo {pages} {2427} (\bibinfo {year} {2022})}\BibitemShut {NoStop}%
\bibitem [{\citenamefont {A.~B. D.~Shaik}\ and\ \citenamefont {Jenkins}(2024)}]{Dhu-al2024}%
  \BibitemOpen
  \bibfield  {author} {\bibinfo {author} {\bibfnamefont {P.~Palla}\ \bibnamefont {A.~B. D.~Shaik}}\ and\ \bibinfo {author} {\bibfnamefont {D.}~\bibnamefont {Jenkins}},\ }\bibfield  {title} {\enquote {\bibinfo {title} {Electrical tuning of quantum light emitters in hbn for free space and telecom optical bands},}\ }\href@noop {} {\bibfield  {journal} {\bibinfo  {journal} {Scientific Reports}\ }\textbf {\bibinfo {volume} {14}},\ \bibinfo {pages} {811} (\bibinfo {year} {2024})}\BibitemShut {NoStop}%
\bibitem [{\citenamefont {Wang}\ \emph {et~al.}(2020)\citenamefont {Wang}, \citenamefont {Yan}, \citenamefont {Li}, \citenamefont {Liu}, \citenamefont {Liu}, \citenamefont {Guo}, \citenamefont {Guo}, \citenamefont {Zhou}, \citenamefont {Cui}, \citenamefont {Wang}, \citenamefont {Zhou}, \citenamefont {Xu}, \citenamefont {Xu}, \citenamefont {Li},\ and\ \citenamefont {Guo}}]{Wang2020}%
  \BibitemOpen
  \bibfield  {author} {\bibinfo {author} {\bibfnamefont {J.-F.}\ \bibnamefont {Wang}}, \bibinfo {author} {\bibfnamefont {F.-F.}\ \bibnamefont {Yan}}, \bibinfo {author} {\bibfnamefont {Q.}~\bibnamefont {Li}}, \bibinfo {author} {\bibfnamefont {Z.-H.}\ \bibnamefont {Liu}}, \bibinfo {author} {\bibfnamefont {H.}~\bibnamefont {Liu}}, \bibinfo {author} {\bibfnamefont {G.-P.}\ \bibnamefont {Guo}}, \bibinfo {author} {\bibfnamefont {L.-P.}\ \bibnamefont {Guo}}, \bibinfo {author} {\bibfnamefont {X.}~\bibnamefont {Zhou}}, \bibinfo {author} {\bibfnamefont {J.-M.}\ \bibnamefont {Cui}}, \bibinfo {author} {\bibfnamefont {J.}~\bibnamefont {Wang}}, \bibinfo {author} {\bibfnamefont {Z.-Q.}\ \bibnamefont {Zhou}}, \bibinfo {author} {\bibfnamefont {X.-Y.}\ \bibnamefont {Xu}}, \bibinfo {author} {\bibfnamefont {J.-S.}\ \bibnamefont {Xu}}, \bibinfo {author} {\bibfnamefont {C.-F.}\ \bibnamefont {Li}}, \ and\ \bibinfo {author} {\bibfnamefont {G.-C.}\ \bibnamefont {Guo}},\ }\bibfield  {title} {\enquote {\bibinfo {title} {Coherent
  control of nitrogen-vacancy center spins in silicon carbide at room temperature},}\ }\href@noop {} {\bibfield  {journal} {\bibinfo  {journal} {Physical Review Letters}\ }\textbf {\bibinfo {volume} {124}},\ \bibinfo {pages} {223601} (\bibinfo {year} {2020})}\BibitemShut {NoStop}%
\bibitem [{\citenamefont {Higginbottom}\ \emph {et~al.}(2023)\citenamefont {Higginbottom}, \citenamefont {Asadi}, \citenamefont {Chartrand}, \citenamefont {Ji}, \citenamefont {Bergeron}, \citenamefont {Thewalt}, \citenamefont {Simon},\ and\ \citenamefont {Simmons}}]{Higginbottom2023}%
  \BibitemOpen
  \bibfield  {author} {\bibinfo {author} {\bibfnamefont {D.~B.}\ \bibnamefont {Higginbottom}}, \bibinfo {author} {\bibfnamefont {F.~K.}\ \bibnamefont {Asadi}}, \bibinfo {author} {\bibfnamefont {C.}~\bibnamefont {Chartrand}}, \bibinfo {author} {\bibfnamefont {J.~W.}\ \bibnamefont {Ji}}, \bibinfo {author} {\bibfnamefont {L.}~\bibnamefont {Bergeron}}, \bibinfo {author} {\bibfnamefont {M.~L.W.}\ \bibnamefont {Thewalt}}, \bibinfo {author} {\bibfnamefont {C.}~\bibnamefont {Simon}}, \ and\ \bibinfo {author} {\bibfnamefont {S.}~\bibnamefont {Simmons}},\ }\bibfield  {title} {\enquote {\bibinfo {title} {Memory and transduction prospects for silicon t center devices},}\ }\href@noop {} {\bibfield  {journal} {\bibinfo  {journal} {PRX Quantum}\ }\textbf {\bibinfo {volume} {4}},\ \bibinfo {pages} {020308} (\bibinfo {year} {2023})}\BibitemShut {NoStop}%
\bibitem [{\citenamefont {Zhao}\ \emph {et~al.}(2021)\citenamefont {Zhao}, \citenamefont {Pettes}, \citenamefont {Zheng},\ and\ \citenamefont {Htoon}}]{Zhao2021}%
  \BibitemOpen
  \bibfield  {author} {\bibinfo {author} {\bibfnamefont {H.}~\bibnamefont {Zhao}}, \bibinfo {author} {\bibfnamefont {M.~T.}\ \bibnamefont {Pettes}}, \bibinfo {author} {\bibfnamefont {Y.}~\bibnamefont {Zheng}}, \ and\ \bibinfo {author} {\bibfnamefont {H.}~\bibnamefont {Htoon}},\ }\href@noop {} {\enquote {\bibinfo {title} {Site-controlled telecom-wavelength single-photon emitters in atomically-thin mote2},}\ } (\bibinfo {year} {2021})\BibitemShut {NoStop}%
\bibitem [{\citenamefont {Vogl}\ \emph {et~al.}(2017)\citenamefont {Vogl}, \citenamefont {Lu},\ and\ \citenamefont {Lam}}]{Vogl_2017}%
  \BibitemOpen
  \bibfield  {author} {\bibinfo {author} {\bibfnamefont {T.}~\bibnamefont {Vogl}}, \bibinfo {author} {\bibfnamefont {Y.}~\bibnamefont {Lu}}, \ and\ \bibinfo {author} {\bibfnamefont {P.~Koy}\ \bibnamefont {Lam}},\ }\bibfield  {title} {\enquote {\bibinfo {title} {Room temperature single photon source using fiber-integrated hexagonal boron nitride},}\ }\href@noop {} {\bibfield  {journal} {\bibinfo  {journal} {Journal of Physics D: Applied Physics}\ }\textbf {\bibinfo {volume} {50}},\ \bibinfo {pages} {295101} (\bibinfo {year} {2017})}\BibitemShut {NoStop}%
\bibitem [{\citenamefont {Schmitt-Manderbach}\ \emph {et~al.}(2007)\citenamefont {Schmitt-Manderbach}, \citenamefont {Weier}, \citenamefont {F\"urst}, \citenamefont {Ursin}, \citenamefont {Tiefenbacher}, \citenamefont {Scheidl}, \citenamefont {Perdigues}, \citenamefont {Sodnik}, \citenamefont {Kurtsiefer}, \citenamefont {Rarity}, \citenamefont {Zeilinger},\ and\ \citenamefont {Weinfurter}}]{Schmitt-Manderbach2007}%
  \BibitemOpen
  \bibfield  {author} {\bibinfo {author} {\bibfnamefont {T.}~\bibnamefont {Schmitt-Manderbach}}, \bibinfo {author} {\bibfnamefont {H.}~\bibnamefont {Weier}}, \bibinfo {author} {\bibfnamefont {M.}~\bibnamefont {F\"urst}}, \bibinfo {author} {\bibfnamefont {R.}~\bibnamefont {Ursin}}, \bibinfo {author} {\bibfnamefont {F.}~\bibnamefont {Tiefenbacher}}, \bibinfo {author} {\bibfnamefont {T.}~\bibnamefont {Scheidl}}, \bibinfo {author} {\bibfnamefont {J.}~\bibnamefont {Perdigues}}, \bibinfo {author} {\bibfnamefont {Z.}~\bibnamefont {Sodnik}}, \bibinfo {author} {\bibfnamefont {C.}~\bibnamefont {Kurtsiefer}}, \bibinfo {author} {\bibfnamefont {J.~G.}\ \bibnamefont {Rarity}}, \bibinfo {author} {\bibfnamefont {A.}~\bibnamefont {Zeilinger}}, \ and\ \bibinfo {author} {\bibfnamefont {H.}~\bibnamefont {Weinfurter}},\ }\bibfield  {title} {\enquote {\bibinfo {title} {Experimental demonstration of free-space decoy-state quantum key distribution over 144 km},}\ }\href@noop {} {\bibfield  {journal} {\bibinfo  {journal} {Physical
  Review Letters}\ }\textbf {\bibinfo {volume} {98}},\ \bibinfo {pages} {010504} (\bibinfo {year} {2007})}\BibitemShut {NoStop}%
\bibitem [{\citenamefont {Liao}\ \emph {et~al.}(2017)\citenamefont {Liao}, \citenamefont {Cai}, \citenamefont {Liu}, \citenamefont {Zhang}, \citenamefont {Li}, \citenamefont {Ren}, \citenamefont {Yin}, \citenamefont {Shen}, \citenamefont {Cao}, \citenamefont {Li}, \citenamefont {Li}, \citenamefont {Chen}, \citenamefont {Sun}, \citenamefont {Jia}, \citenamefont {Wu}, \citenamefont {Jiang}, \citenamefont {Wang}, \citenamefont {Huang}, \citenamefont {Wang}, \citenamefont {Zhou}, \citenamefont {Deng}, \citenamefont {Xi}, \citenamefont {Ma}, \citenamefont {Hu}, \citenamefont {Zhang}, \citenamefont {Chen}, \citenamefont {Liu}, \citenamefont {Wang}, \citenamefont {Zhu}, \citenamefont {Lu}, \citenamefont {Shu}, \citenamefont {Peng}, \citenamefont {Wang},\ and\ \citenamefont {Pan}}]{Liao2017}%
  \BibitemOpen
  \bibfield  {author} {\bibinfo {author} {\bibfnamefont {S-K.}\ \bibnamefont {Liao}}, \bibinfo {author} {\bibfnamefont {W-Q.}\ \bibnamefont {Cai}}, \bibinfo {author} {\bibfnamefont {W-Y.}\ \bibnamefont {Liu}}, \bibinfo {author} {\bibfnamefont {L.}~\bibnamefont {Zhang}}, \bibinfo {author} {\bibfnamefont {Y.}~\bibnamefont {Li}}, \bibinfo {author} {\bibfnamefont {J-G.}\ \bibnamefont {Ren}}, \bibinfo {author} {\bibfnamefont {J.}~\bibnamefont {Yin}}, \bibinfo {author} {\bibfnamefont {Q.}~\bibnamefont {Shen}}, \bibinfo {author} {\bibfnamefont {Y.}~\bibnamefont {Cao}}, \bibinfo {author} {\bibfnamefont {Z-P.}\ \bibnamefont {Li}}, \bibinfo {author} {\bibfnamefont {F-Z.}\ \bibnamefont {Li}}, \bibinfo {author} {\bibfnamefont {X-W.}\ \bibnamefont {Chen}}, \bibinfo {author} {\bibfnamefont {L-H.}\ \bibnamefont {Sun}}, \bibinfo {author} {\bibfnamefont {J-J.}\ \bibnamefont {Jia}}, \bibinfo {author} {\bibfnamefont {J-C.}\ \bibnamefont {Wu}}, \bibinfo {author} {\bibfnamefont {X-J.}\ \bibnamefont {Jiang}}, \bibinfo {author}
  {\bibfnamefont {J-F.}\ \bibnamefont {Wang}}, \bibinfo {author} {\bibfnamefont {Y-M.}\ \bibnamefont {Huang}}, \bibinfo {author} {\bibfnamefont {Q.}~\bibnamefont {Wang}}, \bibinfo {author} {\bibfnamefont {Y-L.}\ \bibnamefont {Zhou}}, \bibinfo {author} {\bibfnamefont {L.}~\bibnamefont {Deng}}, \bibinfo {author} {\bibfnamefont {T.}~\bibnamefont {Xi}}, \bibinfo {author} {\bibfnamefont {L.}~\bibnamefont {Ma}}, \bibinfo {author} {\bibfnamefont {T.}~\bibnamefont {Hu}}, \bibinfo {author} {\bibfnamefont {Q.}~\bibnamefont {Zhang}}, \bibinfo {author} {\bibfnamefont {Y-A.}\ \bibnamefont {Chen}}, \bibinfo {author} {\bibfnamefont {N-L.}\ \bibnamefont {Liu}}, \bibinfo {author} {\bibfnamefont {X-B.}\ \bibnamefont {Wang}}, \bibinfo {author} {\bibfnamefont {Z-C.}\ \bibnamefont {Zhu}}, \bibinfo {author} {\bibfnamefont {C-Y.}\ \bibnamefont {Lu}}, \bibinfo {author} {\bibfnamefont {R.}~\bibnamefont {Shu}}, \bibinfo {author} {\bibfnamefont {C-Z.}\ \bibnamefont {Peng}}, \bibinfo {author} {\bibfnamefont {J-Y.}\ \bibnamefont {Wang}},
  \ and\ \bibinfo {author} {\bibfnamefont {J-W.}\ \bibnamefont {Pan}},\ }\bibfield  {title} {\enquote {\bibinfo {title} {Satellite-to-ground quantum key distribution},}\ }\href@noop {} {\bibfield  {journal} {\bibinfo  {journal} {Nature}\ }\textbf {\bibinfo {volume} {549}},\ \bibinfo {pages} {43--47} (\bibinfo {year} {2017})}\BibitemShut {NoStop}%
\bibitem [{\citenamefont {Vogl}\ \emph {et~al.}(2018)\citenamefont {Vogl}, \citenamefont {Campbell}, \citenamefont {Buchler}, \citenamefont {Lu},\ and\ \citenamefont {Lam}}]{Vogl2018}%
  \BibitemOpen
  \bibfield  {author} {\bibinfo {author} {\bibfnamefont {T.}~\bibnamefont {Vogl}}, \bibinfo {author} {\bibfnamefont {G.}~\bibnamefont {Campbell}}, \bibinfo {author} {\bibfnamefont {B.~C.}\ \bibnamefont {Buchler}}, \bibinfo {author} {\bibfnamefont {Y.}~\bibnamefont {Lu}}, \ and\ \bibinfo {author} {\bibfnamefont {P.~K.}\ \bibnamefont {Lam}},\ }\bibfield  {title} {\enquote {\bibinfo {title} {Fabrication and deterministic transfer of high-quality quantum emitters in hexagonal boron nitride},}\ }\href@noop {} {\bibfield  {journal} {\bibinfo  {journal} {ACS Photonics}\ }\textbf {\bibinfo {volume} {5}},\ \bibinfo {pages} {2305--2312} (\bibinfo {year} {2018})}\BibitemShut {NoStop}%
\bibitem [{\citenamefont {Vest}\ \emph {et~al.}(2022)\citenamefont {Vest}, \citenamefont {Freiwang}, \citenamefont {Luhn}, \citenamefont {Vogl}, \citenamefont {Rau}, \citenamefont {Knips}, \citenamefont {Rosenfeld},\ and\ \citenamefont {Weinfurter}}]{Vest2022}%
  \BibitemOpen
  \bibfield  {author} {\bibinfo {author} {\bibfnamefont {G.}~\bibnamefont {Vest}}, \bibinfo {author} {\bibfnamefont {P.}~\bibnamefont {Freiwang}}, \bibinfo {author} {\bibfnamefont {J.}~\bibnamefont {Luhn}}, \bibinfo {author} {\bibfnamefont {T.}~\bibnamefont {Vogl}}, \bibinfo {author} {\bibfnamefont {M.}~\bibnamefont {Rau}}, \bibinfo {author} {\bibfnamefont {L.}~\bibnamefont {Knips}}, \bibinfo {author} {\bibfnamefont {W.}~\bibnamefont {Rosenfeld}}, \ and\ \bibinfo {author} {\bibfnamefont {H.}~\bibnamefont {Weinfurter}},\ }\bibfield  {title} {\enquote {\bibinfo {title} {Quantum key distribution with a hand-held sender unit},}\ }\href@noop {} {\bibfield  {journal} {\bibinfo  {journal} {Physics Review Applied}\ }\textbf {\bibinfo {volume} {18}},\ \bibinfo {pages} {024067} (\bibinfo {year} {2022})}\BibitemShut {NoStop}%
\bibitem [{\citenamefont {Vogl}\ \emph {et~al.}(2019{\natexlab{b}})\citenamefont {Vogl}, \citenamefont {Sripathy}, \citenamefont {Sharma}, \citenamefont {Reddy}, \citenamefont {Sullivan}, \citenamefont {Machacek}, \citenamefont {Zhang}, \citenamefont {Karouta}, \citenamefont {Buchler}, \citenamefont {Doherty}, \citenamefont {Lu},\ and\ \citenamefont {Lam}}]{Vogl2019_radiation}%
  \BibitemOpen
  \bibfield  {author} {\bibinfo {author} {\bibfnamefont {T.}~\bibnamefont {Vogl}}, \bibinfo {author} {\bibfnamefont {K.}~\bibnamefont {Sripathy}}, \bibinfo {author} {\bibfnamefont {A.}~\bibnamefont {Sharma}}, \bibinfo {author} {\bibfnamefont {P.}~\bibnamefont {Reddy}}, \bibinfo {author} {\bibfnamefont {J.}~\bibnamefont {Sullivan}}, \bibinfo {author} {\bibfnamefont {J.~R.}\ \bibnamefont {Machacek}}, \bibinfo {author} {\bibfnamefont {L.}~\bibnamefont {Zhang}}, \bibinfo {author} {\bibfnamefont {F.}~\bibnamefont {Karouta}}, \bibinfo {author} {\bibfnamefont {B.~C.}\ \bibnamefont {Buchler}}, \bibinfo {author} {\bibfnamefont {M.~W.}\ \bibnamefont {Doherty}}, \bibinfo {author} {\bibfnamefont {Y.}~\bibnamefont {Lu}}, \ and\ \bibinfo {author} {\bibfnamefont {P.~K.}\ \bibnamefont {Lam}},\ }\bibfield  {title} {\enquote {\bibinfo {title} {Radiation tolerance of two-dimensional material-based devices for space applications},}\ }\href@noop {} {\bibfield  {journal} {\bibinfo  {journal} {Nature Communications}\ }\textbf
  {\bibinfo {volume} {10}},\ \bibinfo {pages} {1202} (\bibinfo {year} {2019}{\natexlab{b}})}\BibitemShut {NoStop}%
\bibitem [{\citenamefont {Ahmadi}\ \emph {et~al.}(2024)\citenamefont {Ahmadi}, \citenamefont {Schwertfeger}, \citenamefont {Werner}, \citenamefont {Wiese}, \citenamefont {Lester}, \citenamefont {Ros}, \citenamefont {Krause}, \citenamefont {Ritter}, \citenamefont {Abasifard}, \citenamefont {Cholsuk}, \citenamefont {Krämer}, \citenamefont {Atzeni}, \citenamefont {G\"{u}ndo\u{g}an}, \citenamefont {Sachidananda}, \citenamefont {Pardo}, \citenamefont {Nolte}, \citenamefont {Lohrmann}, \citenamefont {Ling}, \citenamefont {Bartholom\u{a}us}, \citenamefont {Corrielli}, \citenamefont {Krutzik},\ and\ \citenamefont {Vogl}}]{Ahmadi2024}%
  \BibitemOpen
  \bibfield  {author} {\bibinfo {author} {\bibfnamefont {N.}~\bibnamefont {Ahmadi}}, \bibinfo {author} {\bibfnamefont {S.}~\bibnamefont {Schwertfeger}}, \bibinfo {author} {\bibfnamefont {P.}~\bibnamefont {Werner}}, \bibinfo {author} {\bibfnamefont {L.}~\bibnamefont {Wiese}}, \bibinfo {author} {\bibfnamefont {J.}~\bibnamefont {Lester}}, \bibinfo {author} {\bibfnamefont {E.~Da}\ \bibnamefont {Ros}}, \bibinfo {author} {\bibfnamefont {J.}~\bibnamefont {Krause}}, \bibinfo {author} {\bibfnamefont {S.}~\bibnamefont {Ritter}}, \bibinfo {author} {\bibfnamefont {M.}~\bibnamefont {Abasifard}}, \bibinfo {author} {\bibfnamefont {C.}~\bibnamefont {Cholsuk}}, \bibinfo {author} {\bibfnamefont {R.~G.}\ \bibnamefont {Krämer}}, \bibinfo {author} {\bibfnamefont {S.}~\bibnamefont {Atzeni}}, \bibinfo {author} {\bibfnamefont {M.}~\bibnamefont {G\"{u}ndo\u{g}an}}, \bibinfo {author} {\bibfnamefont {S.}~\bibnamefont {Sachidananda}}, \bibinfo {author} {\bibfnamefont {D.}~\bibnamefont {Pardo}}, \bibinfo {author} {\bibfnamefont
  {S.}~\bibnamefont {Nolte}}, \bibinfo {author} {\bibfnamefont {A.}~\bibnamefont {Lohrmann}}, \bibinfo {author} {\bibfnamefont {A.}~\bibnamefont {Ling}}, \bibinfo {author} {\bibfnamefont {J.}~\bibnamefont {Bartholom\u{a}us}}, \bibinfo {author} {\bibfnamefont {G.}~\bibnamefont {Corrielli}}, \bibinfo {author} {\bibfnamefont {M.}~\bibnamefont {Krutzik}}, \ and\ \bibinfo {author} {\bibfnamefont {T.}~\bibnamefont {Vogl}},\ }\bibfield  {title} {\enquote {\bibinfo {title} {Quick$^3$ - design of a satellite-based quantum light source for quantum communication and extended physical theory tests in space},}\ }\href@noop {} {\bibfield  {journal} {\bibinfo  {journal} {Advanced Quantum Technologies}\ }\textbf {\bibinfo {volume} {7}},\ \bibinfo {pages} {2300343} (\bibinfo {year} {2024})}\BibitemShut {NoStop}%
\bibitem [{\citenamefont {Zhigulin}\ \emph {et~al.}(2023{\natexlab{b}})\citenamefont {Zhigulin}, \citenamefont {Horder}, \citenamefont {Iv\'ady}, \citenamefont {Iteite}, \citenamefont {Gale}, \citenamefont {Li}, \citenamefont {Lobo}, \citenamefont {Toth}, \citenamefont {Aharonovich},\ and\ \citenamefont {Kianinia}}]{Zhigulin2023}%
  \BibitemOpen
  \bibfield  {author} {\bibinfo {author} {\bibfnamefont {I.}~\bibnamefont {Zhigulin}}, \bibinfo {author} {\bibfnamefont {J.}~\bibnamefont {Horder}}, \bibinfo {author} {\bibfnamefont {V.}~\bibnamefont {Iv\'ady}}, \bibinfo {author} {\bibfnamefont {S.~J.~U.}\ \bibnamefont {Iteite}}, \bibinfo {author} {\bibfnamefont {A.}~\bibnamefont {Gale}}, \bibinfo {author} {\bibfnamefont {C.}~\bibnamefont {Li}}, \bibinfo {author} {\bibfnamefont {C.~J.}\ \bibnamefont {Lobo}}, \bibinfo {author} {\bibfnamefont {M.}~\bibnamefont {Toth}}, \bibinfo {author} {\bibfnamefont {I.}~\bibnamefont {Aharonovich}}, \ and\ \bibinfo {author} {\bibfnamefont {M.}~\bibnamefont {Kianinia}},\ }\bibfield  {title} {\enquote {\bibinfo {title} {Stark effect of blue quantum emitters in hexagonal boron nitride},}\ }\href@noop {} {\bibfield  {journal} {\bibinfo  {journal} {Physical Review Applied}\ }\textbf {\bibinfo {volume} {19}},\ \bibinfo {pages} {044011} (\bibinfo {year} {2023}{\natexlab{b}})}\BibitemShut {NoStop}%
\bibitem [{\citenamefont {Grzeszczyk}\ \emph {et~al.}(2023)\citenamefont {Grzeszczyk}, \citenamefont {Vaklinova}, \citenamefont {Watanabe}, \citenamefont {Taniguchi}, \citenamefont {Novoselov},\ and\ \citenamefont {Koperski}}]{Grzeszczyk2023}%
  \BibitemOpen
  \bibfield  {author} {\bibinfo {author} {\bibfnamefont {M.}~\bibnamefont {Grzeszczyk}}, \bibinfo {author} {\bibfnamefont {K.}~\bibnamefont {Vaklinova}}, \bibinfo {author} {\bibfnamefont {K.}~\bibnamefont {Watanabe}}, \bibinfo {author} {\bibfnamefont {T.}~\bibnamefont {Taniguchi}}, \bibinfo {author} {\bibfnamefont {K.~S.}\ \bibnamefont {Novoselov}}, \ and\ \bibinfo {author} {\bibfnamefont {M.}~\bibnamefont {Koperski}},\ }\bibfield  {title} {\enquote {\bibinfo {title} {Electrical excitation of carbon centers in hexagonal boron nitride with tuneable quantum efficiency},}\ }\href@noop {} {\bibfield  {journal} {\bibinfo  {journal} {arXiv:2305.13679}\ } (\bibinfo {year} {2023})}\BibitemShut {NoStop}%
\bibitem [{\citenamefont {Yu}\ \emph {et~al.}(2024)\citenamefont {Yu}, \citenamefont {Lee}, \citenamefont {Watanabe}, \citenamefont {Taniguchi},\ and\ \citenamefont {Lee}}]{Yu2024}%
  \BibitemOpen
  \bibfield  {author} {\bibinfo {author} {\bibfnamefont {M.}~\bibnamefont {Yu}}, \bibinfo {author} {\bibfnamefont {J.}~\bibnamefont {Lee}}, \bibinfo {author} {\bibfnamefont {K.}~\bibnamefont {Watanabe}}, \bibinfo {author} {\bibfnamefont {T.}~\bibnamefont {Taniguchi}}, \ and\ \bibinfo {author} {\bibfnamefont {J.}~\bibnamefont {Lee}},\ }\bibfield  {title} {\enquote {\bibinfo {title} {Electrical pumping of h-bn single-photon sources in van der waals heterostructures},}\ }\href@noop {} {\bibfield  {journal} {\bibinfo  {journal} {arXiv:2407.14070}\ } (\bibinfo {year} {2024})}\BibitemShut {NoStop}%
\bibitem [{\citenamefont {Raussendorf}\ and\ \citenamefont {Briegel}(2001)}]{Raussendorf2001}%
  \BibitemOpen
  \bibfield  {author} {\bibinfo {author} {\bibfnamefont {R.}~\bibnamefont {Raussendorf}}\ and\ \bibinfo {author} {\bibfnamefont {H.~J.}\ \bibnamefont {Briegel}},\ }\bibfield  {title} {\enquote {\bibinfo {title} {A one-way quantum computer},}\ }\href@noop {} {\bibfield  {journal} {\bibinfo  {journal} {Physical Review Letters}\ }\textbf {\bibinfo {volume} {86}},\ \bibinfo {pages} {5188--5191} (\bibinfo {year} {2001})}\BibitemShut {NoStop}%
\bibitem [{\citenamefont {Fournier}\ \emph {et~al.}(2023)\citenamefont {Fournier}, \citenamefont {Roux}, \citenamefont {Watanabe}, \citenamefont {Taniguchi}, \citenamefont {Buil}, \citenamefont {Barjon}, \citenamefont {Hermier},\ and\ \citenamefont {Delteil}}]{Fournier2023}%
  \BibitemOpen
  \bibfield  {author} {\bibinfo {author} {\bibfnamefont {C.}~\bibnamefont {Fournier}}, \bibinfo {author} {\bibfnamefont {S.}~\bibnamefont {Roux}}, \bibinfo {author} {\bibfnamefont {K.}~\bibnamefont {Watanabe}}, \bibinfo {author} {\bibfnamefont {T.}~\bibnamefont {Taniguchi}}, \bibinfo {author} {\bibfnamefont {S.}~\bibnamefont {Buil}}, \bibinfo {author} {\bibfnamefont {J.}~\bibnamefont {Barjon}}, \bibinfo {author} {\bibfnamefont {J-P.}\ \bibnamefont {Hermier}}, \ and\ \bibinfo {author} {\bibfnamefont {A.}~\bibnamefont {Delteil}},\ }\bibfield  {title} {\enquote {\bibinfo {title} {Two-photon interference from a quantum emitter in hexagonal boron nitride},}\ }\href@noop {} {\bibfield  {journal} {\bibinfo  {journal} {Physical Review Applied}\ }\textbf {\bibinfo {volume} {19}},\ \bibinfo {pages} {L041003} (\bibinfo {year} {2023})}\BibitemShut {NoStop}%
\bibitem [{\citenamefont {White}\ \emph {et~al.}(2021)\citenamefont {White}, \citenamefont {Stewart}, \citenamefont {Solntsev}, \citenamefont {Li}, \citenamefont {Toth}, \citenamefont {Kianinia},\ and\ \citenamefont {Aharonovich}}]{White2021}%
  \BibitemOpen
  \bibfield  {author} {\bibinfo {author} {\bibfnamefont {S.}~\bibnamefont {White}}, \bibinfo {author} {\bibfnamefont {C.}~\bibnamefont {Stewart}}, \bibinfo {author} {\bibfnamefont {A.~S.}\ \bibnamefont {Solntsev}}, \bibinfo {author} {\bibfnamefont {C.}~\bibnamefont {Li}}, \bibinfo {author} {\bibfnamefont {M.}~\bibnamefont {Toth}}, \bibinfo {author} {\bibfnamefont {M.}~\bibnamefont {Kianinia}}, \ and\ \bibinfo {author} {\bibfnamefont {I.}~\bibnamefont {Aharonovich}},\ }\bibfield  {title} {\enquote {\bibinfo {title} {Phonon dephasing and spectral diffusion of quantum emitters in hexagonal boron nitride},}\ }\href@noop {} {\bibfield  {journal} {\bibinfo  {journal} {Optica}\ ,\ \bibinfo {pages} {1153--1158}} (\bibinfo {year} {2021})}\BibitemShut {NoStop}%
\bibitem [{\citenamefont {Koch}\ \emph {et~al.}(2024)\citenamefont {Koch}, \citenamefont {Bharadwaj},\ and\ \citenamefont {Kubanek}}]{Koch2024}%
  \BibitemOpen
  \bibfield  {author} {\bibinfo {author} {\bibfnamefont {M.~K.}\ \bibnamefont {Koch}}, \bibinfo {author} {\bibfnamefont {V.}~\bibnamefont {Bharadwaj}}, \ and\ \bibinfo {author} {\bibfnamefont {A.}~\bibnamefont {Kubanek}},\ }\bibfield  {title} {\enquote {\bibinfo {title} {Probing the limits for coherent optical control of a mechanically decoupled defect center in hexagonal boron nitride},}\ }\href@noop {} {\bibfield  {journal} {\bibinfo  {journal} {Communications Materials}\ }\textbf {\bibinfo {volume} {5}},\ \bibinfo {pages} {240} (\bibinfo {year} {2024})}\BibitemShut {NoStop}%
\bibitem [{\citenamefont {Hoese}\ \emph {et~al.}(2020)\citenamefont {Hoese}, \citenamefont {Reddy}, \citenamefont {Dietrich}, \citenamefont {Koch}, \citenamefont {Fehler}, \citenamefont {Doherty},\ and\ \citenamefont {Kubanek}}]{Hoese2020}%
  \BibitemOpen
  \bibfield  {author} {\bibinfo {author} {\bibfnamefont {M.}~\bibnamefont {Hoese}}, \bibinfo {author} {\bibfnamefont {P.}~\bibnamefont {Reddy}}, \bibinfo {author} {\bibfnamefont {A.}~\bibnamefont {Dietrich}}, \bibinfo {author} {\bibfnamefont {M.~K.}\ \bibnamefont {Koch}}, \bibinfo {author} {\bibfnamefont {K.~G.}\ \bibnamefont {Fehler}}, \bibinfo {author} {\bibfnamefont {M.~W.}\ \bibnamefont {Doherty}}, \ and\ \bibinfo {author} {\bibfnamefont {A.}~\bibnamefont {Kubanek}},\ }\bibfield  {title} {\enquote {\bibinfo {title} {Mechanical decoupling of quantum emitters in hexagonal boron nitride from low-energy phonon modes},}\ }\href@noop {} {\bibfield  {journal} {\bibinfo  {journal} {Science Advances}\ }\textbf {\bibinfo {volume} {6}},\ \bibinfo {pages} {6038} (\bibinfo {year} {2020})}\BibitemShut {NoStop}%
\bibitem [{\citenamefont {Dietrich}\ \emph {et~al.}(2020)\citenamefont {Dietrich}, \citenamefont {Doherty}, \citenamefont {Aharonovich},\ and\ \citenamefont {Kubanek}}]{Dietrich2020}%
  \BibitemOpen
  \bibfield  {author} {\bibinfo {author} {\bibfnamefont {A.}~\bibnamefont {Dietrich}}, \bibinfo {author} {\bibfnamefont {M.~W.}\ \bibnamefont {Doherty}}, \bibinfo {author} {\bibfnamefont {I.}~\bibnamefont {Aharonovich}}, \ and\ \bibinfo {author} {\bibfnamefont {A.}~\bibnamefont {Kubanek}},\ }\bibfield  {title} {\enquote {\bibinfo {title} {Solid-state single photon source with fourier transform limited lines at room temperature},}\ }\href@noop {} {\bibfield  {journal} {\bibinfo  {journal} {Physical Reviews B}\ }\textbf {\bibinfo {volume} {101}},\ \bibinfo {pages} {081401} (\bibinfo {year} {2020})}\BibitemShut {NoStop}%
\bibitem [{\citenamefont {Grange}\ \emph {et~al.}(2015)\citenamefont {Grange}, \citenamefont {Hornecker}, \citenamefont {Hunger}, \citenamefont {Poizat}, \citenamefont {G\'erard}, \citenamefont {Senellart},\ and\ \citenamefont {Auff\`eves}}]{Grange2015}%
  \BibitemOpen
  \bibfield  {author} {\bibinfo {author} {\bibfnamefont {T.}~\bibnamefont {Grange}}, \bibinfo {author} {\bibfnamefont {G.}~\bibnamefont {Hornecker}}, \bibinfo {author} {\bibfnamefont {D.}~\bibnamefont {Hunger}}, \bibinfo {author} {\bibfnamefont {J-P.}\ \bibnamefont {Poizat}}, \bibinfo {author} {\bibfnamefont {J-M.}\ \bibnamefont {G\'erard}}, \bibinfo {author} {\bibfnamefont {P.}~\bibnamefont {Senellart}}, \ and\ \bibinfo {author} {\bibfnamefont {A.}~\bibnamefont {Auff\`eves}},\ }\bibfield  {title} {\enquote {\bibinfo {title} {Cavity-funneled generation of indistinguishable single photons from strongly dissipative quantum emitters},}\ }\href@noop {} {\bibfield  {journal} {\bibinfo  {journal} {Physical Review Letters}\ }\textbf {\bibinfo {volume} {114}},\ \bibinfo {pages} {193601} (\bibinfo {year} {2015})}\BibitemShut {NoStop}%
\bibitem [{\citenamefont {Conlon}\ \emph {et~al.}(2023)\citenamefont {Conlon}, \citenamefont {Vogl}, \citenamefont {Marciniak}, \citenamefont {Pogorelov}, \citenamefont {Yung}, \citenamefont {Eilenberger}, \citenamefont {Berry}, \citenamefont {Santana}, \citenamefont {Blatt}, \citenamefont {Monz}, \citenamefont {Lam},\ and\ \citenamefont {Assad}}]{Conlon2023}%
  \BibitemOpen
  \bibfield  {author} {\bibinfo {author} {\bibfnamefont {L.~O.}\ \bibnamefont {Conlon}}, \bibinfo {author} {\bibfnamefont {T.}~\bibnamefont {Vogl}}, \bibinfo {author} {\bibfnamefont {C.~D.}\ \bibnamefont {Marciniak}}, \bibinfo {author} {\bibfnamefont {I.}~\bibnamefont {Pogorelov}}, \bibinfo {author} {\bibfnamefont {S.~K.}\ \bibnamefont {Yung}}, \bibinfo {author} {\bibfnamefont {F.}~\bibnamefont {Eilenberger}}, \bibinfo {author} {\bibfnamefont {D.~W.}\ \bibnamefont {Berry}}, \bibinfo {author} {\bibfnamefont {F.~S.}\ \bibnamefont {Santana}}, \bibinfo {author} {\bibfnamefont {R.}~\bibnamefont {Blatt}}, \bibinfo {author} {\bibfnamefont {T.}~\bibnamefont {Monz}}, \bibinfo {author} {\bibfnamefont {P.~K.}\ \bibnamefont {Lam}}, \ and\ \bibinfo {author} {\bibfnamefont {S.~M.}\ \bibnamefont {Assad}},\ }\bibfield  {title} {\enquote {\bibinfo {title} {Approaching optimal entangling collective measurements on quantum computing platforms},}\ }\href@noop {} {\bibfield  {journal} {\bibinfo  {journal} {Nature Physics}\
  }\textbf {\bibinfo {volume} {19}},\ \bibinfo {pages} {351--357} (\bibinfo {year} {2023})}\BibitemShut {NoStop}%
\bibitem [{\citenamefont {Gao}\ \emph {et~al.}(2021)\citenamefont {Gao}, \citenamefont {Pandey}, \citenamefont {Kianinia}, \citenamefont {Ahn}, \citenamefont {Ju}, \citenamefont {Aharonovich}, \citenamefont {Shivaram},\ and\ \citenamefont {Li}}]{Gao2020_LaserWritingVB}%
  \BibitemOpen
  \bibfield  {author} {\bibinfo {author} {\bibfnamefont {X.}~\bibnamefont {Gao}}, \bibinfo {author} {\bibfnamefont {S.}~\bibnamefont {Pandey}}, \bibinfo {author} {\bibfnamefont {M.}~\bibnamefont {Kianinia}}, \bibinfo {author} {\bibfnamefont {J.}~\bibnamefont {Ahn}}, \bibinfo {author} {\bibfnamefont {P.}~\bibnamefont {Ju}}, \bibinfo {author} {\bibfnamefont {I.}~\bibnamefont {Aharonovich}}, \bibinfo {author} {\bibfnamefont {N.}~\bibnamefont {Shivaram}}, \ and\ \bibinfo {author} {\bibfnamefont {T.}~\bibnamefont {Li}},\ }\bibfield  {title} {\enquote {\bibinfo {title} {Femtosecond laser writing of spin defects in hexagonal boron nitride},}\ }\href@noop {} {\bibfield  {journal} {\bibinfo  {journal} {ACS Photonics}\ }\textbf {\bibinfo {volume} {8}},\ \bibinfo {pages} {994--1000} (\bibinfo {year} {2021})}\BibitemShut {NoStop}%
\bibitem [{\citenamefont {Kianinia}\ \emph {et~al.}(2020)\citenamefont {Kianinia}, \citenamefont {White}, \citenamefont {Fröch}, \citenamefont {Bradac},\ and\ \citenamefont {Aharonovich}}]{Kianinia2020}%
  \BibitemOpen
  \bibfield  {author} {\bibinfo {author} {\bibfnamefont {M.}~\bibnamefont {Kianinia}}, \bibinfo {author} {\bibfnamefont {S.}~\bibnamefont {White}}, \bibinfo {author} {\bibfnamefont {J.~E.}\ \bibnamefont {Fröch}}, \bibinfo {author} {\bibfnamefont {C.}~\bibnamefont {Bradac}}, \ and\ \bibinfo {author} {\bibfnamefont {I.}~\bibnamefont {Aharonovich}},\ }\bibfield  {title} {\enquote {\bibinfo {title} {Generation of spin defects in hexagonal boron nitride},}\ }\href@noop {} {\bibfield  {journal} {\bibinfo  {journal} {ACS Photonics}\ }\textbf {\bibinfo {volume} {7}},\ \bibinfo {pages} {2147–2152} (\bibinfo {year} {2020})}\BibitemShut {NoStop}%
\bibitem [{\citenamefont {Guo}\ \emph {et~al.}(2022)\citenamefont {Guo}, \citenamefont {L.}, \citenamefont {Li}, \citenamefont {Yang}, \citenamefont {Yu}, \citenamefont {Meng}, \citenamefont {Wang}, \citenamefont {Zeng}, \citenamefont {Yan}, \citenamefont {Li}, \citenamefont {Wang}, \citenamefont {Xu}, \citenamefont {Wang}, \citenamefont {Tang}, \citenamefont {Li},\ and\ \citenamefont {Guo}}]{Guo2022}%
  \BibitemOpen
  \bibfield  {author} {\bibinfo {author} {\bibfnamefont {N.~J.}\ \bibnamefont {Guo}}, \bibinfo {author} {\bibfnamefont {W.}~\bibnamefont {L.}}, \bibinfo {author} {\bibfnamefont {Z.~P.}\ \bibnamefont {Li}}, \bibinfo {author} {\bibfnamefont {Y.~Z.}\ \bibnamefont {Yang}}, \bibinfo {author} {\bibfnamefont {S.}~\bibnamefont {Yu}}, \bibinfo {author} {\bibfnamefont {Y.}~\bibnamefont {Meng}}, \bibinfo {author} {\bibfnamefont {Z.~A.}\ \bibnamefont {Wang}}, \bibinfo {author} {\bibfnamefont {X.~Do.}\ \bibnamefont {Zeng}}, \bibinfo {author} {\bibfnamefont {F.~F.}\ \bibnamefont {Yan}}, \bibinfo {author} {\bibfnamefont {Q.}~\bibnamefont {Li}}, \bibinfo {author} {\bibfnamefont {J.~F.}\ \bibnamefont {Wang}}, \bibinfo {author} {\bibfnamefont {J.~S.}\ \bibnamefont {Xu}}, \bibinfo {author} {\bibfnamefont {Y.~T.}\ \bibnamefont {Wang}}, \bibinfo {author} {\bibfnamefont {J.~S.}\ \bibnamefont {Tang}}, \bibinfo {author} {\bibfnamefont {C.~F.}\ \bibnamefont {Li}}, \ and\ \bibinfo {author} {\bibfnamefont {G.~C.}\ \bibnamefont {Guo}},\
  }\bibfield  {title} {\enquote {\bibinfo {title} {Generation of spin defects by ion implantation in hexagonal boron nitride},}\ }\href@noop {} {\bibfield  {journal} {\bibinfo  {journal} {ACS Omega}\ }\textbf {\bibinfo {volume} {7}},\ \bibinfo {pages} {1733–1739} (\bibinfo {year} {2022})}\BibitemShut {NoStop}%
\bibitem [{\citenamefont {Baber}\ \emph {et~al.}(2022)\citenamefont {Baber}, \citenamefont {Malein}, \citenamefont {Khatri}, \citenamefont {Keatley}, \citenamefont {Guo}, \citenamefont {Withers}, \citenamefont {Ramsay},\ and\ \citenamefont {Luxmoore}}]{baber2021}%
  \BibitemOpen
  \bibfield  {author} {\bibinfo {author} {\bibfnamefont {S.}~\bibnamefont {Baber}}, \bibinfo {author} {\bibfnamefont {R.~N.~E.}\ \bibnamefont {Malein}}, \bibinfo {author} {\bibfnamefont {P.}~\bibnamefont {Khatri}}, \bibinfo {author} {\bibfnamefont {P.~S.}\ \bibnamefont {Keatley}}, \bibinfo {author} {\bibfnamefont {S.}~\bibnamefont {Guo}}, \bibinfo {author} {\bibfnamefont {F.}~\bibnamefont {Withers}}, \bibinfo {author} {\bibfnamefont {A.~J.}\ \bibnamefont {Ramsay}}, \ and\ \bibinfo {author} {\bibfnamefont {I.~J.}\ \bibnamefont {Luxmoore}},\ }\bibfield  {title} {\enquote {\bibinfo {title} {Excited state spectroscopy of boron vacancy defects in hexagonal boron nitride using time-resolved optically detected magnetic resonance},}\ }\href@noop {} {\bibfield  {journal} {\bibinfo  {journal} {Nano Letters}\ }\textbf {\bibinfo {volume} {22}},\ \bibinfo {pages} {461--467} (\bibinfo {year} {2022})}\BibitemShut {NoStop}%
\bibitem [{\citenamefont {Haykal}\ \emph {et~al.}(2022)\citenamefont {Haykal}, \citenamefont {Tanos}, \citenamefont {Minotto}, \citenamefont {Durand}, \citenamefont {Fabre}, \citenamefont {Li}, \citenamefont {Edgar}, \citenamefont {Ivády}, \citenamefont {Gali}, \citenamefont {Michel}, \citenamefont {Dréau}, \citenamefont {Gil}, \citenamefont {Cassabois},\ and\ \citenamefont {Jacques}}]{Haykal2022}%
  \BibitemOpen
  \bibfield  {author} {\bibinfo {author} {\bibfnamefont {A.}~\bibnamefont {Haykal}}, \bibinfo {author} {\bibfnamefont {R.}~\bibnamefont {Tanos}}, \bibinfo {author} {\bibfnamefont {N.}~\bibnamefont {Minotto}}, \bibinfo {author} {\bibfnamefont {A.}~\bibnamefont {Durand}}, \bibinfo {author} {\bibfnamefont {F.}~\bibnamefont {Fabre}}, \bibinfo {author} {\bibfnamefont {J.}~\bibnamefont {Li}}, \bibinfo {author} {\bibfnamefont {J.~H.}\ \bibnamefont {Edgar}}, \bibinfo {author} {\bibfnamefont {V.}~\bibnamefont {Ivády}}, \bibinfo {author} {\bibfnamefont {A.}~\bibnamefont {Gali}}, \bibinfo {author} {\bibfnamefont {T.}~\bibnamefont {Michel}}, \bibinfo {author} {\bibfnamefont {A.}~\bibnamefont {Dréau}}, \bibinfo {author} {\bibfnamefont {B.}~\bibnamefont {Gil}}, \bibinfo {author} {\bibfnamefont {G.}~\bibnamefont {Cassabois}}, \ and\ \bibinfo {author} {\bibfnamefont {V.}~\bibnamefont {Jacques}},\ }\bibfield  {title} {\enquote {\bibinfo {title} {Decoherence of vb- in defects in monoisotopic hexagonal boron nitride},}\
  }\href@noop {} {\bibfield  {journal} {\bibinfo  {journal} {Nature Communications}\ }\textbf {\bibinfo {volume} {13}},\ \bibinfo {pages} {4347} (\bibinfo {year} {2022})}\BibitemShut {NoStop}%
\bibitem [{\citenamefont {Ramsay}\ \emph {et~al.}(2023)\citenamefont {Ramsay}, \citenamefont {Hekmati}, \citenamefont {Patrickson}, \citenamefont {Baber}, \citenamefont {Arvidsson-Shukur}, \citenamefont {Bennett},\ and\ \citenamefont {Luxmoore}}]{Ramsay2023}%
  \BibitemOpen
  \bibfield  {author} {\bibinfo {author} {\bibfnamefont {A.~J.}\ \bibnamefont {Ramsay}}, \bibinfo {author} {\bibfnamefont {R.}~\bibnamefont {Hekmati}}, \bibinfo {author} {\bibfnamefont {C.~J.}\ \bibnamefont {Patrickson}}, \bibinfo {author} {\bibfnamefont {S.}~\bibnamefont {Baber}}, \bibinfo {author} {\bibfnamefont {D.~R.M.}\ \bibnamefont {Arvidsson-Shukur}}, \bibinfo {author} {\bibfnamefont {A.~J.}\ \bibnamefont {Bennett}}, \ and\ \bibinfo {author} {\bibfnamefont {I.~J.}\ \bibnamefont {Luxmoore}},\ }\bibfield  {title} {\enquote {\bibinfo {title} {Coherence protection of spin qubits in hexagonal boron nitride},}\ }\href@noop {} {\bibfield  {journal} {\bibinfo  {journal} {Nature Communications}\ }\textbf {\bibinfo {volume} {14}},\ \bibinfo {pages} {461} (\bibinfo {year} {2023})}\BibitemShut {NoStop}%
\bibitem [{\citenamefont {Reimers}\ \emph {et~al.}(2020)\citenamefont {Reimers}, \citenamefont {Shen}, \citenamefont {Kianinia}, \citenamefont {Bradac}, \citenamefont {Aharonovich}, \citenamefont {Ford},\ and\ \citenamefont {Piecuch}}]{Reimers2020}%
  \BibitemOpen
  \bibfield  {author} {\bibinfo {author} {\bibfnamefont {J.~R.}\ \bibnamefont {Reimers}}, \bibinfo {author} {\bibfnamefont {J.}~\bibnamefont {Shen}}, \bibinfo {author} {\bibfnamefont {M.}~\bibnamefont {Kianinia}}, \bibinfo {author} {\bibfnamefont {C.}~\bibnamefont {Bradac}}, \bibinfo {author} {\bibfnamefont {I.}~\bibnamefont {Aharonovich}}, \bibinfo {author} {\bibfnamefont {M.~J.}\ \bibnamefont {Ford}}, \ and\ \bibinfo {author} {\bibfnamefont {P.}~\bibnamefont {Piecuch}},\ }\bibfield  {title} {\enquote {\bibinfo {title} {Photoluminescence, photophysics, and photochemistry of the v b- defect in hexagonal boron nitride},}\ }\href@noop {} {\bibfield  {journal} {\bibinfo  {journal} {Physical Review B}\ }\textbf {\bibinfo {volume} {102}},\ \bibinfo {pages} {144105} (\bibinfo {year} {2020})}\BibitemShut {NoStop}%
\bibitem [{\citenamefont {Mendelson}\ \emph {et~al.}(2022)\citenamefont {Mendelson}, \citenamefont {Ritika}, \citenamefont {Kianinia}, \citenamefont {Scott}, \citenamefont {Kim}, \citenamefont {Fröch}, \citenamefont {Gazzana}, \citenamefont {Westerhausen}, \citenamefont {Xiao}, \citenamefont {Mohajerani}, \citenamefont {Strauf}, \citenamefont {Toth}, \citenamefont {Aharonovich},\ and\ \citenamefont {Xu}}]{Mendelson2022_Vbcavity}%
  \BibitemOpen
  \bibfield  {author} {\bibinfo {author} {\bibfnamefont {N.}~\bibnamefont {Mendelson}}, \bibinfo {author} {\bibfnamefont {R.}~\bibnamefont {Ritika}}, \bibinfo {author} {\bibfnamefont {M.}~\bibnamefont {Kianinia}}, \bibinfo {author} {\bibfnamefont {J.}~\bibnamefont {Scott}}, \bibinfo {author} {\bibfnamefont {S.}~\bibnamefont {Kim}}, \bibinfo {author} {\bibfnamefont {J.~E.}\ \bibnamefont {Fröch}}, \bibinfo {author} {\bibfnamefont {C.}~\bibnamefont {Gazzana}}, \bibinfo {author} {\bibfnamefont {M.}~\bibnamefont {Westerhausen}}, \bibinfo {author} {\bibfnamefont {L.}~\bibnamefont {Xiao}}, \bibinfo {author} {\bibfnamefont {S.~S.}\ \bibnamefont {Mohajerani}}, \bibinfo {author} {\bibfnamefont {S.}~\bibnamefont {Strauf}}, \bibinfo {author} {\bibfnamefont {M.}~\bibnamefont {Toth}}, \bibinfo {author} {\bibfnamefont {I.}~\bibnamefont {Aharonovich}}, \ and\ \bibinfo {author} {\bibfnamefont {Z-Q.}\ \bibnamefont {Xu}},\ }\bibfield  {title} {\enquote {\bibinfo {title} {Coupling spin defects in a layered material to nanoscale
  plasmonic cavities},}\ }\href@noop {} {\bibfield  {journal} {\bibinfo  {journal} {Advanced Materials}\ }\textbf {\bibinfo {volume} {34}},\ \bibinfo {pages} {2106046} (\bibinfo {year} {2022})}\BibitemShut {NoStop}%
\bibitem [{\citenamefont {Xu}\ \emph {et~al.}(2023)\citenamefont {Xu}, \citenamefont {Solanki}, \citenamefont {Sychev}, \citenamefont {Gao}, \citenamefont {Peana}, \citenamefont {Baburin}, \citenamefont {Pagadala}, \citenamefont {Martin}, \citenamefont {Chowdhury}, \citenamefont {Chen}, \citenamefont {Taniguchi}, \citenamefont {Watanabe}, \citenamefont {Rodionov}, \citenamefont {Kildishev}, \citenamefont {Li}, \citenamefont {Upadhyaya}, \citenamefont {Boltasseva},\ and\ \citenamefont {Shalaev}}]{Xu2023_VBcavity}%
  \BibitemOpen
  \bibfield  {author} {\bibinfo {author} {\bibfnamefont {X.}~\bibnamefont {Xu}}, \bibinfo {author} {\bibfnamefont {A.~B.}\ \bibnamefont {Solanki}}, \bibinfo {author} {\bibfnamefont {D.}~\bibnamefont {Sychev}}, \bibinfo {author} {\bibfnamefont {X.}~\bibnamefont {Gao}}, \bibinfo {author} {\bibfnamefont {S.}~\bibnamefont {Peana}}, \bibinfo {author} {\bibfnamefont {A.~S.}\ \bibnamefont {Baburin}}, \bibinfo {author} {\bibfnamefont {K.}~\bibnamefont {Pagadala}}, \bibinfo {author} {\bibfnamefont {Z.~O.}\ \bibnamefont {Martin}}, \bibinfo {author} {\bibfnamefont {S.~N.}\ \bibnamefont {Chowdhury}}, \bibinfo {author} {\bibfnamefont {Y.~P.}\ \bibnamefont {Chen}}, \bibinfo {author} {\bibfnamefont {T.}~\bibnamefont {Taniguchi}}, \bibinfo {author} {\bibfnamefont {K.}~\bibnamefont {Watanabe}}, \bibinfo {author} {\bibfnamefont {I.~A.}\ \bibnamefont {Rodionov}}, \bibinfo {author} {\bibfnamefont {A.~V.}\ \bibnamefont {Kildishev}}, \bibinfo {author} {\bibfnamefont {T.}~\bibnamefont {Li}}, \bibinfo {author} {\bibfnamefont
  {P.}~\bibnamefont {Upadhyaya}}, \bibinfo {author} {\bibfnamefont {A.}~\bibnamefont {Boltasseva}}, \ and\ \bibinfo {author} {\bibfnamefont {V.~M.}\ \bibnamefont {Shalaev}},\ }\bibfield  {title} {\enquote {\bibinfo {title} {Greatly enhanced emission from spin defects in hexagonal boron nitride enabled by a low-loss plasmonic nanocavity},}\ }\href@noop {} {\bibfield  {journal} {\bibinfo  {journal} {Nano Letters}\ }\textbf {\bibinfo {volume} {23}},\ \bibinfo {pages} {25--33} (\bibinfo {year} {2023})}\BibitemShut {NoStop}%
\bibitem [{\citenamefont {Clua-Provost}\ \emph {et~al.}(2024)\citenamefont {Clua-Provost}, \citenamefont {Mu}, \citenamefont {Durand}, \citenamefont {Schrader}, \citenamefont {Happacher}, \citenamefont {Bocquel}, \citenamefont {Maletinsky}, \citenamefont {Fraunié}, \citenamefont {Marie}, \citenamefont {Robert}, \citenamefont {Seine}, \citenamefont {Janzen}, \citenamefont {Edgar}, \citenamefont {Gil}, \citenamefont {Cassabois},\ and\ \citenamefont {Jacques}}]{Clua-provost2024}%
  \BibitemOpen
  \bibfield  {author} {\bibinfo {author} {\bibfnamefont {T.}~\bibnamefont {Clua-Provost}}, \bibinfo {author} {\bibfnamefont {Z.}~\bibnamefont {Mu}}, \bibinfo {author} {\bibfnamefont {A.}~\bibnamefont {Durand}}, \bibinfo {author} {\bibfnamefont {C.}~\bibnamefont {Schrader}}, \bibinfo {author} {\bibfnamefont {J.}~\bibnamefont {Happacher}}, \bibinfo {author} {\bibfnamefont {J.}~\bibnamefont {Bocquel}}, \bibinfo {author} {\bibfnamefont {P.}~\bibnamefont {Maletinsky}}, \bibinfo {author} {\bibfnamefont {J.}~\bibnamefont {Fraunié}}, \bibinfo {author} {\bibfnamefont {X.}~\bibnamefont {Marie}}, \bibinfo {author} {\bibfnamefont {C.}~\bibnamefont {Robert}}, \bibinfo {author} {\bibfnamefont {G.}~\bibnamefont {Seine}}, \bibinfo {author} {\bibfnamefont {E.}~\bibnamefont {Janzen}}, \bibinfo {author} {\bibfnamefont {J.~H.}\ \bibnamefont {Edgar}}, \bibinfo {author} {\bibfnamefont {B.}~\bibnamefont {Gil}}, \bibinfo {author} {\bibfnamefont {G.}~\bibnamefont {Cassabois}}, \ and\ \bibinfo {author} {\bibfnamefont
  {V.}~\bibnamefont {Jacques}},\ }\bibfield  {title} {\enquote {\bibinfo {title} {Spin-dependent photodynamics of boron-vacancy centers in hexagonal boron nitride},}\ }\href@noop {} {\bibfield  {journal} {\bibinfo  {journal} {Physical Review B}\ }\textbf {\bibinfo {volume} {110}},\ \bibinfo {pages} {014104} (\bibinfo {year} {2024})},\ \bibinfo {note} {publisher: American Physical Society}\BibitemShut {NoStop}%
\bibitem [{\citenamefont {Scholten}\ \emph {et~al.}(2025)\citenamefont {Scholten}, \citenamefont {Iwa\'{n}ski}, \citenamefont {Xing}, \citenamefont {Binder}, \citenamefont {D\k{a}browska}, \citenamefont {Tan}, \citenamefont {Cheng}, \citenamefont {Bradford}, \citenamefont {Mellor}, \citenamefont {Beton}, \citenamefont {Novikov}, \citenamefont {Mischke}, \citenamefont {Pasko}, \citenamefont {Yengel}, \citenamefont {Henning}, \citenamefont {Krotkus}, \citenamefont {Wysmolek},\ and\ \citenamefont {Tetienne}}]{Scholten2025}%
  \BibitemOpen
  \bibfield  {author} {\bibinfo {author} {\bibfnamefont {S.~C.}\ \bibnamefont {Scholten}}, \bibinfo {author} {\bibfnamefont {J.}~\bibnamefont {Iwa\'{n}ski}}, \bibinfo {author} {\bibfnamefont {K.}~\bibnamefont {Xing}}, \bibinfo {author} {\bibfnamefont {J.}~\bibnamefont {Binder}}, \bibinfo {author} {\bibfnamefont {A.~K.}\ \bibnamefont {D\k{a}browska}}, \bibinfo {author} {\bibfnamefont {H.~H.}\ \bibnamefont {Tan}}, \bibinfo {author} {\bibfnamefont {T.~S.}\ \bibnamefont {Cheng}}, \bibinfo {author} {\bibfnamefont {J.}~\bibnamefont {Bradford}}, \bibinfo {author} {\bibfnamefont {C.~J.}\ \bibnamefont {Mellor}}, \bibinfo {author} {\bibfnamefont {P.~H.}\ \bibnamefont {Beton}}, \bibinfo {author} {\bibfnamefont {S.~V.}\ \bibnamefont {Novikov}}, \bibinfo {author} {\bibfnamefont {J.}~\bibnamefont {Mischke}}, \bibinfo {author} {\bibfnamefont {S.}~\bibnamefont {Pasko}}, \bibinfo {author} {\bibfnamefont {E.}~\bibnamefont {Yengel}}, \bibinfo {author} {\bibfnamefont {A.}~\bibnamefont {Henning}}, \bibinfo {author} {\bibfnamefont
  {S.}~\bibnamefont {Krotkus}}, \bibinfo {author} {\bibfnamefont {A.}~\bibnamefont {Wysmolek}}, \ and\ \bibinfo {author} {\bibfnamefont {J-P}\ \bibnamefont {Tetienne}},\ }\bibfield  {title} {\enquote {\bibinfo {title} {Optically detected magnetic resonance of wafer-scale hexagonal boron nitride thin films},}\ }\href@noop {} {\bibfield  {journal} {\bibinfo  {journal} {arXiv:2505.21143}\ } (\bibinfo {year} {2025})}\BibitemShut {NoStop}%
\bibitem [{\citenamefont {Robertson}\ \emph {et~al.}(2024)\citenamefont {Robertson}, \citenamefont {Whitefield}, \citenamefont {Scholten}, \citenamefont {Singh}, \citenamefont {Healey}, \citenamefont {Reineck}, \citenamefont {Kianinia}, \citenamefont {Broadway}, \citenamefont {Aharonovich},\ and\ \citenamefont {Tetienne}}]{Robertson2024}%
  \BibitemOpen
  \bibfield  {author} {\bibinfo {author} {\bibfnamefont {I.~O.}\ \bibnamefont {Robertson}}, \bibinfo {author} {\bibfnamefont {B.}~\bibnamefont {Whitefield}}, \bibinfo {author} {\bibfnamefont {S.~C.}\ \bibnamefont {Scholten}}, \bibinfo {author} {\bibfnamefont {P.}~\bibnamefont {Singh}}, \bibinfo {author} {\bibfnamefont {A.~J.}\ \bibnamefont {Healey}}, \bibinfo {author} {\bibfnamefont {P.}~\bibnamefont {Reineck}}, \bibinfo {author} {\bibfnamefont {M.}~\bibnamefont {Kianinia}}, \bibinfo {author} {\bibfnamefont {D.~A.}\ \bibnamefont {Broadway}}, \bibinfo {author} {\bibfnamefont {I.}~\bibnamefont {Aharonovich}}, \ and\ \bibinfo {author} {\bibfnamefont {J-P.}\ \bibnamefont {Tetienne}},\ }\bibfield  {title} {\enquote {\bibinfo {title} {A universal mechanism for optically addressable solid-state spin pairs},}\ }\href@noop {} {\bibfield  {journal} {\bibinfo  {journal} {arXiv:2407.1314}\ } (\bibinfo {year} {2024})}\BibitemShut {NoStop}%
\bibitem [{\citenamefont {Woodward}(2002)}]{Woodward2002}%
  \BibitemOpen
  \bibfield  {author} {\bibinfo {author} {\bibfnamefont {J.~R.}\ \bibnamefont {Woodward}},\ }\bibfield  {title} {\enquote {\bibinfo {title} {Radical pairs in solution},}\ }\href@noop {} {\bibfield  {journal} {\bibinfo  {journal} {Progress in Reaction Kinetics and Mechanism}\ }\textbf {\bibinfo {volume} {27}},\ \bibinfo {pages} {165--207} (\bibinfo {year} {2002})}\BibitemShut {NoStop}%
\bibitem [{\citenamefont {Gilardoni}\ \emph {et~al.}(2025)\citenamefont {Gilardoni}, \citenamefont {Barker}, \citenamefont {Curtin}, \citenamefont {Fraser}, \citenamefont {Powell}, \citenamefont {Lewis}, \citenamefont {Deng}, \citenamefont {Ramsay}, \citenamefont {Li}, \citenamefont {Aharonovich}, \citenamefont {Tan}, \citenamefont {Atatüre},\ and\ \citenamefont {Stern}}]{Gilardoni2025}%
  \BibitemOpen
  \bibfield  {author} {\bibinfo {author} {\bibfnamefont {C.~M.}\ \bibnamefont {Gilardoni}}, \bibinfo {author} {\bibfnamefont {S.~Eizagiree.}\ \bibnamefont {Barker}}, \bibinfo {author} {\bibfnamefont {C.~L.}\ \bibnamefont {Curtin}}, \bibinfo {author} {\bibfnamefont {S.~A.}\ \bibnamefont {Fraser}}, \bibinfo {author} {\bibfnamefont {O.~F.~J.}\ \bibnamefont {Powell}}, \bibinfo {author} {\bibfnamefont {D.~K.}\ \bibnamefont {Lewis}}, \bibinfo {author} {\bibfnamefont {X.}~\bibnamefont {Deng}}, \bibinfo {author} {\bibfnamefont {A.~J.}\ \bibnamefont {Ramsay}}, \bibinfo {author} {\bibfnamefont {C.}~\bibnamefont {Li}}, \bibinfo {author} {\bibfnamefont {I.}~\bibnamefont {Aharonovich}}, \bibinfo {author} {\bibfnamefont {H.~H.}\ \bibnamefont {Tan}}, \bibinfo {author} {\bibfnamefont {M.}~\bibnamefont {Atatüre}}, \ and\ \bibinfo {author} {\bibfnamefont {H.~L.}\ \bibnamefont {Stern}},\ }\bibfield  {title} {\enquote {\bibinfo {title} {A single spin in hexagonal boron nitride for vectorial quantum magnetometry},}\ }\href@noop
  {} {\bibfield  {journal} {\bibinfo  {journal} {Nature Communications}\ }\textbf {\bibinfo {volume} {16}},\ \bibinfo {pages} {4947} (\bibinfo {year} {2025})}\BibitemShut {NoStop}%
\bibitem [{\citenamefont {Whitefield}\ \emph {et~al.}(2025)\citenamefont {Whitefield}, \citenamefont {Zeng}, \citenamefont {Liddle-Wesolowski}, \citenamefont {Robertson}, \citenamefont {Iv\'{a}dy}, \citenamefont {Watanabe}, \citenamefont {Taniguchi}, \citenamefont {Toth}, \citenamefont {Tetienne}, \citenamefont {Aharonovich},\ and\ \citenamefont {Kianinia}}]{Whitefield2025}%
  \BibitemOpen
  \bibfield  {author} {\bibinfo {author} {\bibfnamefont {B.}~\bibnamefont {Whitefield}}, \bibinfo {author} {\bibfnamefont {H.~Zhi~Jie}\ \bibnamefont {Zeng}}, \bibinfo {author} {\bibfnamefont {J.}~\bibnamefont {Liddle-Wesolowski}}, \bibinfo {author} {\bibfnamefont {I.~O.}\ \bibnamefont {Robertson}}, \bibinfo {author} {\bibfnamefont {V.}~\bibnamefont {Iv\'{a}dy}}, \bibinfo {author} {\bibfnamefont {K.}~\bibnamefont {Watanabe}}, \bibinfo {author} {\bibfnamefont {T.}~\bibnamefont {Taniguchi}}, \bibinfo {author} {\bibfnamefont {M.}~\bibnamefont {Toth}}, \bibinfo {author} {\bibfnamefont {J-P.}\ \bibnamefont {Tetienne}}, \bibinfo {author} {\bibfnamefont {I.}~\bibnamefont {Aharonovich}}, \ and\ \bibinfo {author} {\bibfnamefont {M.}~\bibnamefont {Kianinia}},\ }\bibfield  {title} {\enquote {\bibinfo {title} {Generation of narrowband quantum emitters in hbn with optically addressable spins},}\ }\href@noop {} {\bibfield  {journal} {\bibinfo  {journal} {arXiv:2501.15341}\ } (\bibinfo {year} {2025})}\BibitemShut {NoStop}%
\bibitem [{\citenamefont {Azuma}\ \emph {et~al.}(2023)\citenamefont {Azuma}, \citenamefont {Economou}, \citenamefont {Elkouss}, \citenamefont {Hilaire}, \citenamefont {Jiang}, \citenamefont {Lo},\ and\ \citenamefont {Tzitrin}}]{Azuma2023}%
  \BibitemOpen
  \bibfield  {author} {\bibinfo {author} {\bibfnamefont {K.}~\bibnamefont {Azuma}}, \bibinfo {author} {\bibfnamefont {S.~E.}\ \bibnamefont {Economou}}, \bibinfo {author} {\bibfnamefont {D.}~\bibnamefont {Elkouss}}, \bibinfo {author} {\bibfnamefont {P.}~\bibnamefont {Hilaire}}, \bibinfo {author} {\bibfnamefont {L.}~\bibnamefont {Jiang}}, \bibinfo {author} {\bibfnamefont {H.~Kwong}\ \bibnamefont {Lo}}, \ and\ \bibinfo {author} {\bibfnamefont {I.}~\bibnamefont {Tzitrin}},\ }\bibfield  {title} {\enquote {\bibinfo {title} {Quantum repeaters: From quantum networks to the quantum internet},}\ }\href@noop {} {\bibfield  {journal} {\bibinfo  {journal} {Reviews of Modern Physics}\ }\textbf {\bibinfo {volume} {95}},\ \bibinfo {pages} {045006} (\bibinfo {year} {2023})}\BibitemShut {NoStop}%
\bibitem [{\citenamefont {Childress}\ \emph {et~al.}(2005)\citenamefont {Childress}, \citenamefont {Taylor}, \citenamefont {Sørensen},\ and\ \citenamefont {Lukin}}]{Childress2005}%
  \BibitemOpen
  \bibfield  {author} {\bibinfo {author} {\bibfnamefont {L.}~\bibnamefont {Childress}}, \bibinfo {author} {\bibfnamefont {J.~M.}\ \bibnamefont {Taylor}}, \bibinfo {author} {\bibfnamefont {A.~S.}\ \bibnamefont {Sørensen}}, \ and\ \bibinfo {author} {\bibfnamefont {M.~D.}\ \bibnamefont {Lukin}},\ }\bibfield  {title} {\enquote {\bibinfo {title} {Fault-tolerant quantum repeaters with minimal physical resources and implementations based on single-photon emitters},}\ }\href@noop {} {\bibfield  {journal} {\bibinfo  {journal} {Physical Review A}\ }\textbf {\bibinfo {volume} {72}},\ \bibinfo {pages} {052330} (\bibinfo {year} {2005})}\BibitemShut {NoStop}%
\bibitem [{\citenamefont {Fang}\ \emph {et~al.}(2024)\citenamefont {Fang}, \citenamefont {Lai}, \citenamefont {Li}, \citenamefont {Su}, \citenamefont {Lu}, \citenamefont {Yang}, \citenamefont {Liu}, \citenamefont {Qiao}, \citenamefont {Li}, \citenamefont {He}, \citenamefont {Huang}, \citenamefont {Li}, \citenamefont {You}, \citenamefont {Huo}, \citenamefont {Bao},\ and\ \citenamefont {Pan}}]{Fang2024}%
  \BibitemOpen
  \bibfield  {author} {\bibinfo {author} {\bibfnamefont {R-Z.}\ \bibnamefont {Fang}}, \bibinfo {author} {\bibfnamefont {X-Y.}\ \bibnamefont {Lai}}, \bibinfo {author} {\bibfnamefont {T.}~\bibnamefont {Li}}, \bibinfo {author} {\bibfnamefont {R-Z.}\ \bibnamefont {Su}}, \bibinfo {author} {\bibfnamefont {B-W.}\ \bibnamefont {Lu}}, \bibinfo {author} {\bibfnamefont {C-W.}\ \bibnamefont {Yang}}, \bibinfo {author} {\bibfnamefont {R-Z.}\ \bibnamefont {Liu}}, \bibinfo {author} {\bibfnamefont {Y-K.}\ \bibnamefont {Qiao}}, \bibinfo {author} {\bibfnamefont {C.}~\bibnamefont {Li}}, \bibinfo {author} {\bibfnamefont {Z-G.}\ \bibnamefont {He}}, \bibinfo {author} {\bibfnamefont {J.}~\bibnamefont {Huang}}, \bibinfo {author} {\bibfnamefont {H.}~\bibnamefont {Li}}, \bibinfo {author} {\bibfnamefont {L-X.}\ \bibnamefont {You}}, \bibinfo {author} {\bibfnamefont {Y-H.}\ \bibnamefont {Huo}}, \bibinfo {author} {\bibfnamefont {X-H.}\ \bibnamefont {Bao}}, \ and\ \bibinfo {author} {\bibfnamefont {J-W.}\ \bibnamefont {Pan}},\ }\bibfield
  {title} {\enquote {\bibinfo {title} {Experimental generation of spin-photon entanglement in silicon carbide},}\ }\href@noop {} {\bibfield  {journal} {\bibinfo  {journal} {Physical Review Letters}\ }\textbf {\bibinfo {volume} {132}},\ \bibinfo {pages} {160801} (\bibinfo {year} {2024})}\BibitemShut {NoStop}%
\bibitem [{\citenamefont {Cholsuk}\ \emph {et~al.}(2024{\natexlab{a}})\citenamefont {Cholsuk}, \citenamefont {\k{C}akan}, \citenamefont {Suwanna},\ and\ \citenamefont {Vogl}}]{Cholsuk2024}%
  \BibitemOpen
  \bibfield  {author} {\bibinfo {author} {\bibfnamefont {C.}~\bibnamefont {Cholsuk}}, \bibinfo {author} {\bibfnamefont {A.}~\bibnamefont {\k{C}akan}}, \bibinfo {author} {\bibfnamefont {S.}~\bibnamefont {Suwanna}}, \ and\ \bibinfo {author} {\bibfnamefont {T.}~\bibnamefont {Vogl}},\ }\bibfield  {title} {\enquote {\bibinfo {title} {Identifying electronic transitions of defects in hexagonal boron nitride for quantum memories},}\ }\href@noop {} {\bibfield  {journal} {\bibinfo  {journal} {Advanced Optical Materials}\ }\textbf {\bibinfo {volume} {12}} (\bibinfo {year} {2024}{\natexlab{a}})}\BibitemShut {NoStop}%
\bibitem [{\citenamefont {Wehner}\ \emph {et~al.}(2018)\citenamefont {Wehner}, \citenamefont {Elkouss},\ and\ \citenamefont {Hanson}}]{Wehner2018}%
  \BibitemOpen
  \bibfield  {author} {\bibinfo {author} {\bibfnamefont {S.}~\bibnamefont {Wehner}}, \bibinfo {author} {\bibfnamefont {D.}~\bibnamefont {Elkouss}}, \ and\ \bibinfo {author} {\bibfnamefont {R.}~\bibnamefont {Hanson}},\ }\bibfield  {title} {\enquote {\bibinfo {title} {Quantum internet: A vision for the road ahead},}\ }\href@noop {} {\bibfield  {journal} {\bibinfo  {journal} {Science}\ }\textbf {\bibinfo {volume} {362}},\ \bibinfo {pages} {2302760} (\bibinfo {year} {2018})}\BibitemShut {NoStop}%
\bibitem [{\citenamefont {McMahon}\ and\ \citenamefont {Greve}(2015)}]{McMahon2015}%
  \BibitemOpen
  \bibfield  {author} {\bibinfo {author} {\bibfnamefont {P.~L.}\ \bibnamefont {McMahon}}\ and\ \bibinfo {author} {\bibfnamefont {K.~De}\ \bibnamefont {Greve}},\ }\enquote {\bibinfo {title} {Towards quantum repeaters with solid-state qubits: Spin-photon entanglement generation using self-assembled quantum dots},}\ in\ \href@noop {} {\emph {\bibinfo {booktitle} {Engineering the Atom-Photon Interaction: Controlling Fundamental Processes with Photons, Atoms and Solids}}}\ (\bibinfo  {publisher} {Springer International Publishing},\ \bibinfo {address} {Cham},\ \bibinfo {year} {2015})\ pp.\ \bibinfo {pages} {365--402}\BibitemShut {NoStop}%
\bibitem [{\citenamefont {Knaut}\ \emph {et~al.}(2024)\citenamefont {Knaut}, \citenamefont {Suleymanzade}, \citenamefont {Wei}, \citenamefont {Assumpcao}, \citenamefont {Stas}, \citenamefont {Huan}, \citenamefont {Machielse}, \citenamefont {Knall}, \citenamefont {Sutula}, \citenamefont {Baranes}, \citenamefont {Sinclair}, \citenamefont {De-Eknamkul}, \citenamefont {Levonian}, \citenamefont {Bhaskar}, \citenamefont {Park}, \citenamefont {Lončar},\ and\ \citenamefont {Lukin}}]{Knaut2024}%
  \BibitemOpen
  \bibfield  {author} {\bibinfo {author} {\bibfnamefont {C.~M.}\ \bibnamefont {Knaut}}, \bibinfo {author} {\bibfnamefont {A.}~\bibnamefont {Suleymanzade}}, \bibinfo {author} {\bibfnamefont {Y.~C.}\ \bibnamefont {Wei}}, \bibinfo {author} {\bibfnamefont {D.~R.}\ \bibnamefont {Assumpcao}}, \bibinfo {author} {\bibfnamefont {P.~J.}\ \bibnamefont {Stas}}, \bibinfo {author} {\bibfnamefont {Y.~Q.}\ \bibnamefont {Huan}}, \bibinfo {author} {\bibfnamefont {B.}~\bibnamefont {Machielse}}, \bibinfo {author} {\bibfnamefont {E.~N.}\ \bibnamefont {Knall}}, \bibinfo {author} {\bibfnamefont {M.}~\bibnamefont {Sutula}}, \bibinfo {author} {\bibfnamefont {G.}~\bibnamefont {Baranes}}, \bibinfo {author} {\bibfnamefont {N.}~\bibnamefont {Sinclair}}, \bibinfo {author} {\bibfnamefont {C.}~\bibnamefont {De-Eknamkul}}, \bibinfo {author} {\bibfnamefont {D.~S.}\ \bibnamefont {Levonian}}, \bibinfo {author} {\bibfnamefont {M.~K.}\ \bibnamefont {Bhaskar}}, \bibinfo {author} {\bibfnamefont {H.}~\bibnamefont {Park}}, \bibinfo {author}
  {\bibfnamefont {M.}~\bibnamefont {Lončar}}, \ and\ \bibinfo {author} {\bibfnamefont {M.~D.}\ \bibnamefont {Lukin}},\ }\bibfield  {title} {\enquote {\bibinfo {title} {Entanglement of nanophotonic quantum memory nodes in a telecom network},}\ }\href@noop {} {\bibfield  {journal} {\bibinfo  {journal} {Nature}\ }\textbf {\bibinfo {volume} {629}},\ \bibinfo {pages} {573--578} (\bibinfo {year} {2024})}\BibitemShut {NoStop}%
\bibitem [{\citenamefont {Ye}\ \emph {et~al.}(2019)\citenamefont {Ye}, \citenamefont {Seo},\ and\ \citenamefont {Galli}}]{Ye2019}%
  \BibitemOpen
  \bibfield  {author} {\bibinfo {author} {\bibfnamefont {M.}~\bibnamefont {Ye}}, \bibinfo {author} {\bibfnamefont {H.}~\bibnamefont {Seo}}, \ and\ \bibinfo {author} {\bibfnamefont {G.}~\bibnamefont {Galli}},\ }\bibfield  {title} {\enquote {\bibinfo {title} {Spin coherence in two-dimensional materials},}\ }\href@noop {} {\bibfield  {journal} {\bibinfo  {journal} {npj Computational Materials}\ }\textbf {\bibinfo {volume} {5}},\ \bibinfo {pages} {44} (\bibinfo {year} {2019})}\BibitemShut {NoStop}%
\bibitem [{\citenamefont {András~Tárkányi}(2025)}]{Tarkanyi2025}%
  \BibitemOpen
  \bibfield  {author} {\bibinfo {author} {\bibfnamefont {Viktor~Ivády}\ \bibnamefont {András~Tárkányi}},\ }\bibfield  {title} {\enquote {\bibinfo {title} {Understanding decoherence of the boron vacancy center in hexagonal boron nitride},}\ }\href@noop {} {\bibfield  {journal} {\bibinfo  {journal} {arXiv:2505.03292}\ } (\bibinfo {year} {2025})}\BibitemShut {NoStop}%
\bibitem [{\citenamefont {Itoh}\ and\ \citenamefont {Watanabe}(2014)}]{Itoh2014}%
  \BibitemOpen
  \bibfield  {author} {\bibinfo {author} {\bibfnamefont {K.~M.}\ \bibnamefont {Itoh}}\ and\ \bibinfo {author} {\bibfnamefont {H.}~\bibnamefont {Watanabe}},\ }\bibfield  {title} {\enquote {\bibinfo {title} {Isotope engineering of silicon and diamond for quantum computing and sensing applications},}\ }\href@noop {} {\bibfield  {journal} {\bibinfo  {journal} {MRS Communications}\ }\textbf {\bibinfo {volume} {4}},\ \bibinfo {pages} {143–157} (\bibinfo {year} {2014})}\BibitemShut {NoStop}%
\bibitem [{\citenamefont {Viola}\ and\ \citenamefont {Lloyd}(1998)}]{Viola1998}%
  \BibitemOpen
  \bibfield  {author} {\bibinfo {author} {\bibfnamefont {L.}~\bibnamefont {Viola}}\ and\ \bibinfo {author} {\bibfnamefont {S.}~\bibnamefont {Lloyd}},\ }\bibfield  {title} {\enquote {\bibinfo {title} {Dynamical suppression of decoherence in two-state quantum systems},}\ }\href@noop {} {\bibfield  {journal} {\bibinfo  {journal} {Physical Review A - Atomic, Molecular, and Optical Physics}\ }\textbf {\bibinfo {volume} {58}},\ \bibinfo {pages} {2733} (\bibinfo {year} {1998})}\BibitemShut {NoStop}%
\bibitem [{\citenamefont {Rizzato}\ \emph {et~al.}(2023{\natexlab{a}})\citenamefont {Rizzato}, \citenamefont {Schalk}, \citenamefont {Mohr}, \citenamefont {Hermann}, \citenamefont {Leibold}, \citenamefont {Bruckmaier}, \citenamefont {Salvitti}, \citenamefont {Qian}, \citenamefont {Ji}, \citenamefont {Astakhov}, \citenamefont {Kentsch}, \citenamefont {Helm}, \citenamefont {Stier}, \citenamefont {Finley},\ and\ \citenamefont {Bucher}}]{Rizzato2023_coherence}%
  \BibitemOpen
  \bibfield  {author} {\bibinfo {author} {\bibfnamefont {R.}~\bibnamefont {Rizzato}}, \bibinfo {author} {\bibfnamefont {M.}~\bibnamefont {Schalk}}, \bibinfo {author} {\bibfnamefont {S.}~\bibnamefont {Mohr}}, \bibinfo {author} {\bibfnamefont {J.~C.}\ \bibnamefont {Hermann}}, \bibinfo {author} {\bibfnamefont {J.~P.}\ \bibnamefont {Leibold}}, \bibinfo {author} {\bibfnamefont {F.}~\bibnamefont {Bruckmaier}}, \bibinfo {author} {\bibfnamefont {G.}~\bibnamefont {Salvitti}}, \bibinfo {author} {\bibfnamefont {C.}~\bibnamefont {Qian}}, \bibinfo {author} {\bibfnamefont {P.}~\bibnamefont {Ji}}, \bibinfo {author} {\bibfnamefont {G.~V.}\ \bibnamefont {Astakhov}}, \bibinfo {author} {\bibfnamefont {U.}~\bibnamefont {Kentsch}}, \bibinfo {author} {\bibfnamefont {M.}~\bibnamefont {Helm}}, \bibinfo {author} {\bibfnamefont {A.~V.}\ \bibnamefont {Stier}}, \bibinfo {author} {\bibfnamefont {J.~J.}\ \bibnamefont {Finley}}, \ and\ \bibinfo {author} {\bibfnamefont {D.~B.}\ \bibnamefont {Bucher}},\ }\bibfield  {title} {\enquote
  {\bibinfo {title} {Extending the coherence of spin defects in hbn enables advanced qubit control and quantum sensing},}\ }\href@noop {} {\bibfield  {journal} {\bibinfo  {journal} {Nature Communications}\ }\textbf {\bibinfo {volume} {14}},\ \bibinfo {pages} {5089} (\bibinfo {year} {2023}{\natexlab{a}})}\BibitemShut {NoStop}%
\bibitem [{\citenamefont {Gao}\ \emph {et~al.}(2025{\natexlab{b}})\citenamefont {Gao}, \citenamefont {Ge}, \citenamefont {Dikshit}, \citenamefont {Vaidya}, \citenamefont {Ju},\ and\ \citenamefont {Li}}]{Gao2025arxiv}%
  \BibitemOpen
  \bibfield  {author} {\bibinfo {author} {\bibfnamefont {X.}~\bibnamefont {Gao}}, \bibinfo {author} {\bibfnamefont {Z.}~\bibnamefont {Ge}}, \bibinfo {author} {\bibfnamefont {S.}~\bibnamefont {Dikshit}}, \bibinfo {author} {\bibfnamefont {S.}~\bibnamefont {Vaidya}}, \bibinfo {author} {\bibfnamefont {P.}~\bibnamefont {Ju}}, \ and\ \bibinfo {author} {\bibfnamefont {T.}~\bibnamefont {Li}},\ }\bibfield  {title} {\enquote {\bibinfo {title} {Room-temperature quantum entanglement in a van der waals material},}\ }\href@noop {} {\bibfield  {journal} {\bibinfo  {journal} {arXiv:2509.23170v1}\ } (\bibinfo {year} {2025}{\natexlab{b}})}\BibitemShut {NoStop}%
\bibitem [{\citenamefont {Bar-Gill}\ \emph {et~al.}(2013)\citenamefont {Bar-Gill}, \citenamefont {Pham}, \citenamefont {Jarmola}, \citenamefont {Budker},\ and\ \citenamefont {Walsworth}}]{Bar-Gill2013}%
  \BibitemOpen
  \bibfield  {author} {\bibinfo {author} {\bibfnamefont {N.}~\bibnamefont {Bar-Gill}}, \bibinfo {author} {\bibfnamefont {L.~M.}\ \bibnamefont {Pham}}, \bibinfo {author} {\bibfnamefont {A.}~\bibnamefont {Jarmola}}, \bibinfo {author} {\bibfnamefont {D.}~\bibnamefont {Budker}}, \ and\ \bibinfo {author} {\bibfnamefont {R.~L.}\ \bibnamefont {Walsworth}},\ }\bibfield  {title} {\enquote {\bibinfo {title} {Solid-state electronic spin coherence time approaching one second},}\ }\href@noop {} {\bibfield  {journal} {\bibinfo  {journal} {Nature Communications}\ }\textbf {\bibinfo {volume} {4}},\ \bibinfo {pages} {1743} (\bibinfo {year} {2013})}\BibitemShut {NoStop}%
\bibitem [{\citenamefont {Seo}\ \emph {et~al.}(2016)\citenamefont {Seo}, \citenamefont {Falk}, \citenamefont {Klimov}, \citenamefont {Miao}, \citenamefont {Galli},\ and\ \citenamefont {Awschalom}}]{Seo2016}%
  \BibitemOpen
  \bibfield  {author} {\bibinfo {author} {\bibfnamefont {H.}~\bibnamefont {Seo}}, \bibinfo {author} {\bibfnamefont {A.~L.}\ \bibnamefont {Falk}}, \bibinfo {author} {\bibfnamefont {P.~V.}\ \bibnamefont {Klimov}}, \bibinfo {author} {\bibfnamefont {K.~C.}\ \bibnamefont {Miao}}, \bibinfo {author} {\bibfnamefont {G.}~\bibnamefont {Galli}}, \ and\ \bibinfo {author} {\bibfnamefont {D.~D.}\ \bibnamefont {Awschalom}},\ }\bibfield  {title} {\enquote {\bibinfo {title} {Quantum decoherence dynamics of divacancy spins in silicon carbide},}\ }\href@noop {} {\bibfield  {journal} {\bibinfo  {journal} {Nature Communications}\ }\textbf {\bibinfo {volume} {7}},\ \bibinfo {pages} {12935} (\bibinfo {year} {2016})}\BibitemShut {NoStop}%
\bibitem [{\citenamefont {Murzakhanov}\ \emph {et~al.}(2022)\citenamefont {Murzakhanov}, \citenamefont {Mamin}, \citenamefont {Orlinskii}, \citenamefont {Gerstmann}, \citenamefont {Schmidt}, \citenamefont {Biktagirov}, \citenamefont {Aharonovich}, \citenamefont {Gottscholl}, \citenamefont {Sperlich}, \citenamefont {Dyakonov},\ and\ \citenamefont {Soltamov}}]{Murzakhanov2022}%
  \BibitemOpen
  \bibfield  {author} {\bibinfo {author} {\bibfnamefont {F.~F.}\ \bibnamefont {Murzakhanov}}, \bibinfo {author} {\bibfnamefont {G.~Vladimirovich}\ \bibnamefont {Mamin}}, \bibinfo {author} {\bibfnamefont {S.~Borisovich}\ \bibnamefont {Orlinskii}}, \bibinfo {author} {\bibfnamefont {U.}~\bibnamefont {Gerstmann}}, \bibinfo {author} {\bibfnamefont {W.~Gero}\ \bibnamefont {Schmidt}}, \bibinfo {author} {\bibfnamefont {T.}~\bibnamefont {Biktagirov}}, \bibinfo {author} {\bibfnamefont {I.}~\bibnamefont {Aharonovich}}, \bibinfo {author} {\bibfnamefont {A.}~\bibnamefont {Gottscholl}}, \bibinfo {author} {\bibfnamefont {A.}~\bibnamefont {Sperlich}}, \bibinfo {author} {\bibfnamefont {V.}~\bibnamefont {Dyakonov}}, \ and\ \bibinfo {author} {\bibfnamefont {V.~A.}\ \bibnamefont {Soltamov}},\ }\bibfield  {title} {\enquote {\bibinfo {title} {Electron-nuclear coherent coupling and nuclear spin readout through optically polarized v$_\text{B}$ spin states in hbn},}\ }\href@noop {} {\bibfield  {journal} {\bibinfo  {journal} {Nano
  Letters}\ }\textbf {\bibinfo {volume} {22}},\ \bibinfo {pages} {2718--2724} (\bibinfo {year} {2022})}\BibitemShut {NoStop}%
\bibitem [{\citenamefont {Lee}\ \emph {et~al.}(2022)\citenamefont {Lee}, \citenamefont {Park},\ and\ \citenamefont {Seo}}]{Lee2022}%
  \BibitemOpen
  \bibfield  {author} {\bibinfo {author} {\bibfnamefont {Jaewook}\ \bibnamefont {Lee}}, \bibinfo {author} {\bibfnamefont {Huijin}\ \bibnamefont {Park}}, \ and\ \bibinfo {author} {\bibfnamefont {Hosung}\ \bibnamefont {Seo}},\ }\bibfield  {title} {\enquote {\bibinfo {title} {First-principles theory of extending the spin qubit coherence time in hexagonal boron nitride},}\ }\href@noop {} {\bibfield  {journal} {\bibinfo  {journal} {npj 2D Materials and Applications}\ }\textbf {\bibinfo {volume} {6}},\ \bibinfo {pages} {60} (\bibinfo {year} {2022})}\BibitemShut {NoStop}%
\bibitem [{\citenamefont {Wolfowicz}\ \emph {et~al.}(2013)\citenamefont {Wolfowicz}, \citenamefont {Tyryshkin}, \citenamefont {George}, \citenamefont {Riemann}, \citenamefont {Abrosimov}, \citenamefont {Becker}, \citenamefont {Pohl}, \citenamefont {Thewalt}, \citenamefont {Lyon},\ and\ \citenamefont {Morton}}]{Wolfowicz2013}%
  \BibitemOpen
  \bibfield  {author} {\bibinfo {author} {\bibfnamefont {G.}~\bibnamefont {Wolfowicz}}, \bibinfo {author} {\bibfnamefont {A.~M.}\ \bibnamefont {Tyryshkin}}, \bibinfo {author} {\bibfnamefont {R.~E.}\ \bibnamefont {George}}, \bibinfo {author} {\bibfnamefont {H.}~\bibnamefont {Riemann}}, \bibinfo {author} {\bibfnamefont {N.~V.}\ \bibnamefont {Abrosimov}}, \bibinfo {author} {\bibfnamefont {P.}~\bibnamefont {Becker}}, \bibinfo {author} {\bibfnamefont {H.~J.}\ \bibnamefont {Pohl}}, \bibinfo {author} {\bibfnamefont {M.~L.W.}\ \bibnamefont {Thewalt}}, \bibinfo {author} {\bibfnamefont {S.~A.}\ \bibnamefont {Lyon}}, \ and\ \bibinfo {author} {\bibfnamefont {J.L.}\ \bibnamefont {Morton}},\ }\bibfield  {title} {\enquote {\bibinfo {title} {Atomic clock transitions in silicon-based spin qubits},}\ }\href@noop {} {\bibfield  {journal} {\bibinfo  {journal} {Nature Nanotechnology}\ }\textbf {\bibinfo {volume} {8}},\ \bibinfo {pages} {561–564} (\bibinfo {year} {2013})}\BibitemShut {NoStop}%
\bibitem [{\citenamefont {Neumann}\ \emph {et~al.}(2008)\citenamefont {Neumann}, \citenamefont {Mizuochi}, \citenamefont {Rempp}, \citenamefont {Hemmer}, \citenamefont {Watanabe}, \citenamefont {Yamasaki}, \citenamefont {Jacques}, \citenamefont {Gaebel}, \citenamefont {Jelezko},\ and\ \citenamefont {Wrachtrup}}]{Neumann2008}%
  \BibitemOpen
  \bibfield  {author} {\bibinfo {author} {\bibfnamefont {P.}~\bibnamefont {Neumann}}, \bibinfo {author} {\bibfnamefont {N.}~\bibnamefont {Mizuochi}}, \bibinfo {author} {\bibfnamefont {F.}~\bibnamefont {Rempp}}, \bibinfo {author} {\bibfnamefont {P.}~\bibnamefont {Hemmer}}, \bibinfo {author} {\bibfnamefont {H.}~\bibnamefont {Watanabe}}, \bibinfo {author} {\bibfnamefont {S.}~\bibnamefont {Yamasaki}}, \bibinfo {author} {\bibfnamefont {V.}~\bibnamefont {Jacques}}, \bibinfo {author} {\bibfnamefont {T.}~\bibnamefont {Gaebel}}, \bibinfo {author} {\bibfnamefont {F.}~\bibnamefont {Jelezko}}, \ and\ \bibinfo {author} {\bibfnamefont {J.}~\bibnamefont {Wrachtrup}},\ }\bibfield  {title} {\enquote {\bibinfo {title} {Multipartite entanglement among single spins in diamond},}\ }\href@noop {} {\bibfield  {journal} {\bibinfo  {journal} {Science}\ }\textbf {\bibinfo {volume} {320}},\ \bibinfo {pages} {1326--1329} (\bibinfo {year} {2008})}\BibitemShut {NoStop}%
\bibitem [{\citenamefont {Gao}\ \emph {et~al.}(2022)\citenamefont {Gao}, \citenamefont {Vaidya}, \citenamefont {Li}, \citenamefont {Ju}, \citenamefont {Jiang}, \citenamefont {Xu}, \citenamefont {Allcca}, \citenamefont {Shen}, \citenamefont {Taniguchi}, \citenamefont {Watanabe}, \citenamefont {Bhave}, \citenamefont {Chen}, \citenamefont {Ping},\ and\ \citenamefont {Li}}]{Gao2022}%
  \BibitemOpen
  \bibfield  {author} {\bibinfo {author} {\bibfnamefont {X.}~\bibnamefont {Gao}}, \bibinfo {author} {\bibfnamefont {S.}~\bibnamefont {Vaidya}}, \bibinfo {author} {\bibfnamefont {K.}~\bibnamefont {Li}}, \bibinfo {author} {\bibfnamefont {P.}~\bibnamefont {Ju}}, \bibinfo {author} {\bibfnamefont {B.}~\bibnamefont {Jiang}}, \bibinfo {author} {\bibfnamefont {Z.}~\bibnamefont {Xu}}, \bibinfo {author} {\bibfnamefont {A.~E.Llacsahuanga}\ \bibnamefont {Allcca}}, \bibinfo {author} {\bibfnamefont {K.}~\bibnamefont {Shen}}, \bibinfo {author} {\bibfnamefont {T.}~\bibnamefont {Taniguchi}}, \bibinfo {author} {\bibfnamefont {K.}~\bibnamefont {Watanabe}}, \bibinfo {author} {\bibfnamefont {S.~A.}\ \bibnamefont {Bhave}}, \bibinfo {author} {\bibfnamefont {Y.~P.}\ \bibnamefont {Chen}}, \bibinfo {author} {\bibfnamefont {Y.}~\bibnamefont {Ping}}, \ and\ \bibinfo {author} {\bibfnamefont {T.}~\bibnamefont {Li}},\ }\bibfield  {title} {\enquote {\bibinfo {title} {Nuclear spin polarization and control in hexagonal boron nitride},}\
  }\href@noop {} {\bibfield  {journal} {\bibinfo  {journal} {Nature Materials}\ }\textbf {\bibinfo {volume} {21}},\ \bibinfo {pages} {1024–1028} (\bibinfo {year} {2022})}\BibitemShut {NoStop}%
\bibitem [{\citenamefont {Wei}\ \emph {et~al.}(2022)\citenamefont {Wei}, \citenamefont {Jing}, \citenamefont {Zhang}, \citenamefont {Liao}, \citenamefont {Yuan}, \citenamefont {Fan}, \citenamefont {Lyu}, \citenamefont {Zhou}, \citenamefont {Wang}, \citenamefont {Deng}, \citenamefont {Song}, \citenamefont {Oblak}, \citenamefont {Guo},\ and\ \citenamefont {Zhou}}]{Wei2022}%
  \BibitemOpen
  \bibfield  {author} {\bibinfo {author} {\bibfnamefont {S.~H.}\ \bibnamefont {Wei}}, \bibinfo {author} {\bibfnamefont {B.}~\bibnamefont {Jing}}, \bibinfo {author} {\bibfnamefont {X.~Y.}\ \bibnamefont {Zhang}}, \bibinfo {author} {\bibfnamefont {J.~Y.}\ \bibnamefont {Liao}}, \bibinfo {author} {\bibfnamefont {C.~Z.}\ \bibnamefont {Yuan}}, \bibinfo {author} {\bibfnamefont {B.~Y.}\ \bibnamefont {Fan}}, \bibinfo {author} {\bibfnamefont {C.}~\bibnamefont {Lyu}}, \bibinfo {author} {\bibfnamefont {D.~L.}\ \bibnamefont {Zhou}}, \bibinfo {author} {\bibfnamefont {Y.}~\bibnamefont {Wang}}, \bibinfo {author} {\bibfnamefont {G.~W.}\ \bibnamefont {Deng}}, \bibinfo {author} {\bibfnamefont {H.~Z.}\ \bibnamefont {Song}}, \bibinfo {author} {\bibfnamefont {D.}~\bibnamefont {Oblak}}, \bibinfo {author} {\bibfnamefont {G.~C.}\ \bibnamefont {Guo}}, \ and\ \bibinfo {author} {\bibfnamefont {Q.}~\bibnamefont {Zhou}},\ }\bibfield  {title} {\enquote {\bibinfo {title} {Towards real-world quantum networks: A review},}\ }\href@noop {}
  {\bibfield  {journal} {\bibinfo  {journal} {Laser and Photonics Reviews}\ }\textbf {\bibinfo {volume} {16}},\ \bibinfo {pages} {2100219} (\bibinfo {year} {2022})}\BibitemShut {NoStop}%
\bibitem [{\citenamefont {Aslam}\ \emph {et~al.}(2023)\citenamefont {Aslam}, \citenamefont {Zhou}, \citenamefont {Urbach}, \citenamefont {Turner}, \citenamefont {Walsworth}, \citenamefont {Lukin},\ and\ \citenamefont {Park}}]{Aslam2023}%
  \BibitemOpen
  \bibfield  {author} {\bibinfo {author} {\bibfnamefont {N.}~\bibnamefont {Aslam}}, \bibinfo {author} {\bibfnamefont {H.}~\bibnamefont {Zhou}}, \bibinfo {author} {\bibfnamefont {E.~K.}\ \bibnamefont {Urbach}}, \bibinfo {author} {\bibfnamefont {M.~J.}\ \bibnamefont {Turner}}, \bibinfo {author} {\bibfnamefont {R.~L.}\ \bibnamefont {Walsworth}}, \bibinfo {author} {\bibfnamefont {M.D.}\ \bibnamefont {Lukin}}, \ and\ \bibinfo {author} {\bibfnamefont {Hongkun}\ \bibnamefont {Park}},\ }\bibfield  {title} {\enquote {\bibinfo {title} {Quantum sensors for biomedical applications},}\ }\href@noop {} {\bibfield  {journal} {\bibinfo  {journal} {Nature Reviews Physics}\ }\textbf {\bibinfo {volume} {5}},\ \bibinfo {pages} {157–169} (\bibinfo {year} {2023})}\BibitemShut {NoStop}%
\bibitem [{\citenamefont {Rizzato}\ \emph {et~al.}(2023{\natexlab{b}})\citenamefont {Rizzato}, \citenamefont {von Grafenstein},\ and\ \citenamefont {Bucher}}]{Rizzato2023_sensors}%
  \BibitemOpen
  \bibfield  {author} {\bibinfo {author} {\bibfnamefont {R.}~\bibnamefont {Rizzato}}, \bibinfo {author} {\bibfnamefont {N.~R.}\ \bibnamefont {von Grafenstein}}, \ and\ \bibinfo {author} {\bibfnamefont {D.~B.}\ \bibnamefont {Bucher}},\ }\bibfield  {title} {\enquote {\bibinfo {title} {Quantum sensors in diamonds for magnetic resonance spectroscopy: Current applications and future prospects},}\ }\href@noop {} {\bibfield  {journal} {\bibinfo  {journal} {Applied Physics Letters}\ }\textbf {\bibinfo {volume} {123}},\ \bibinfo {pages} {260502} (\bibinfo {year} {2023}{\natexlab{b}})}\BibitemShut {NoStop}%
\bibitem [{\citenamefont {Rovny}\ \emph {et~al.}(2024)\citenamefont {Rovny}, \citenamefont {Gopalakrishnan}, \citenamefont {Jayich}, \citenamefont {Maletinsky}, \citenamefont {Demler},\ and\ \citenamefont {de~Leon}}]{Rovny2024}%
  \BibitemOpen
  \bibfield  {author} {\bibinfo {author} {\bibfnamefont {J.}~\bibnamefont {Rovny}}, \bibinfo {author} {\bibfnamefont {S.}~\bibnamefont {Gopalakrishnan}}, \bibinfo {author} {\bibfnamefont {J.~A. C.~Bleszynski}\ \bibnamefont {Jayich}}, \bibinfo {author} {\bibfnamefont {P.}~\bibnamefont {Maletinsky}}, \bibinfo {author} {\bibfnamefont {E.}~\bibnamefont {Demler}}, \ and\ \bibinfo {author} {\bibfnamefont {N.~P.}\ \bibnamefont {de~Leon}},\ }\bibfield  {title} {\enquote {\bibinfo {title} {Nanoscale diamond quantum sensors for many-body physics},}\ }\href@noop {} {\bibfield  {journal} {\bibinfo  {journal} {Nature Reviews Physics}\ }\textbf {\bibinfo {volume} {6}},\ \bibinfo {pages} {753–768} (\bibinfo {year} {2024})}\BibitemShut {NoStop}%
\bibitem [{\citenamefont {Budker}\ and\ \citenamefont {Romalis}(2007)}]{Budker2007}%
  \BibitemOpen
  \bibfield  {author} {\bibinfo {author} {\bibfnamefont {D.}~\bibnamefont {Budker}}\ and\ \bibinfo {author} {\bibfnamefont {M.}~\bibnamefont {Romalis}},\ }\bibfield  {title} {\enquote {\bibinfo {title} {Optical magnetometry},}\ }\href@noop {} {\bibfield  {journal} {\bibinfo  {journal} {Nature Physics}\ }\textbf {\bibinfo {volume} {3}},\ \bibinfo {pages} {227} (\bibinfo {year} {2007})}\BibitemShut {NoStop}%
\bibitem [{\citenamefont {Balasubramanian}\ \emph {et~al.}(2008)\citenamefont {Balasubramanian}, \citenamefont {Chan}, \citenamefont {Kolesov}, \citenamefont {Al-Hmoud}, \citenamefont {Tisler}, \citenamefont {Shin}, \citenamefont {Kim}, \citenamefont {Wojcik}, \citenamefont {Hemmer}, \citenamefont {Krueger}, \citenamefont {Hanke}, \citenamefont {Leitenstorfer}, \citenamefont {Bratschitsch}, \citenamefont {Jelezko},\ and\ \citenamefont {Wrachtrup}}]{Balasubramanian2008}%
  \BibitemOpen
  \bibfield  {author} {\bibinfo {author} {\bibfnamefont {G.}~\bibnamefont {Balasubramanian}}, \bibinfo {author} {\bibfnamefont {I.~Y.}\ \bibnamefont {Chan}}, \bibinfo {author} {\bibfnamefont {R.}~\bibnamefont {Kolesov}}, \bibinfo {author} {\bibfnamefont {M.}~\bibnamefont {Al-Hmoud}}, \bibinfo {author} {\bibfnamefont {J.}~\bibnamefont {Tisler}}, \bibinfo {author} {\bibfnamefont {C.}~\bibnamefont {Shin}}, \bibinfo {author} {\bibfnamefont {C.}~\bibnamefont {Kim}}, \bibinfo {author} {\bibfnamefont {A.}~\bibnamefont {Wojcik}}, \bibinfo {author} {\bibfnamefont {P.R.}\ \bibnamefont {Hemmer}}, \bibinfo {author} {\bibfnamefont {A.}~\bibnamefont {Krueger}}, \bibinfo {author} {\bibfnamefont {T.}~\bibnamefont {Hanke}}, \bibinfo {author} {\bibfnamefont {A.}~\bibnamefont {Leitenstorfer}}, \bibinfo {author} {\bibfnamefont {R.}~\bibnamefont {Bratschitsch}}, \bibinfo {author} {\bibfnamefont {F.}~\bibnamefont {Jelezko}}, \ and\ \bibinfo {author} {\bibfnamefont {J.}~\bibnamefont {Wrachtrup}},\ }\bibfield  {title} {\enquote
  {\bibinfo {title} {Nanoscale imaging magnetometry with diamond spins under ambient conditions},}\ }\href@noop {} {\bibfield  {journal} {\bibinfo  {journal} {Nature}\ }\textbf {\bibinfo {volume} {455}},\ \bibinfo {pages} {648} (\bibinfo {year} {2008})}\BibitemShut {NoStop}%
\bibitem [{\citenamefont {Maze}\ \emph {et~al.}(2008)\citenamefont {Maze}, \citenamefont {Stanwix}, \citenamefont {Hodges}, \citenamefont {Hong}, \citenamefont {Taylor}, \citenamefont {Cappellaro}, \citenamefont {Jiang}, \citenamefont {Dutt}, \citenamefont {Togan}, \citenamefont {Zibrov}, \citenamefont {Yacoby}, \citenamefont {Walsworth},\ and\ \citenamefont {Lukin}}]{Maze2008}%
  \BibitemOpen
  \bibfield  {author} {\bibinfo {author} {\bibfnamefont {J.~R.}\ \bibnamefont {Maze}}, \bibinfo {author} {\bibfnamefont {P.~L.}\ \bibnamefont {Stanwix}}, \bibinfo {author} {\bibfnamefont {J.~S.}\ \bibnamefont {Hodges}}, \bibinfo {author} {\bibfnamefont {S.}~\bibnamefont {Hong}}, \bibinfo {author} {\bibfnamefont {J.~M.}\ \bibnamefont {Taylor}}, \bibinfo {author} {\bibfnamefont {P.}~\bibnamefont {Cappellaro}}, \bibinfo {author} {\bibfnamefont {L.}~\bibnamefont {Jiang}}, \bibinfo {author} {\bibfnamefont {M.~V.~G.}\ \bibnamefont {Dutt}}, \bibinfo {author} {\bibfnamefont {E.}~\bibnamefont {Togan}}, \bibinfo {author} {\bibfnamefont {A.~S.}\ \bibnamefont {Zibrov}}, \bibinfo {author} {\bibfnamefont {A.}~\bibnamefont {Yacoby}}, \bibinfo {author} {\bibfnamefont {R.~L.}\ \bibnamefont {Walsworth}}, \ and\ \bibinfo {author} {\bibfnamefont {M.~D.}\ \bibnamefont {Lukin}},\ }\bibfield  {title} {\enquote {\bibinfo {title} {Nanoscale magnetic sensing with an individual electronic spin in diamond},}\ }\href@noop {} {\bibfield
  {journal} {\bibinfo  {journal} {Nature}\ }\textbf {\bibinfo {volume} {455}},\ \bibinfo {pages} {644} (\bibinfo {year} {2008})}\BibitemShut {NoStop}%
\bibitem [{\citenamefont {Rondin}\ \emph {et~al.}(2012)\citenamefont {Rondin}, \citenamefont {Tetienne}, \citenamefont {Spinicelli}, \citenamefont {Savio}, \citenamefont {Karrai}, \citenamefont {Dantelle}, \citenamefont {Thiaville}, \citenamefont {Rohart}, \citenamefont {Roch.},\ and\ \citenamefont {Jacques}}]{Rondin2012}%
  \BibitemOpen
  \bibfield  {author} {\bibinfo {author} {\bibfnamefont {L.}~\bibnamefont {Rondin}}, \bibinfo {author} {\bibfnamefont {J-P.}\ \bibnamefont {Tetienne}}, \bibinfo {author} {\bibfnamefont {P.}~\bibnamefont {Spinicelli}}, \bibinfo {author} {\bibfnamefont {C.~Dal}\ \bibnamefont {Savio}}, \bibinfo {author} {\bibfnamefont {K.}~\bibnamefont {Karrai}}, \bibinfo {author} {\bibfnamefont {G.}~\bibnamefont {Dantelle}}, \bibinfo {author} {\bibfnamefont {A.}~\bibnamefont {Thiaville}}, \bibinfo {author} {\bibfnamefont {S.}~\bibnamefont {Rohart}}, \bibinfo {author} {\bibfnamefont {J.~F.}\ \bibnamefont {Roch.}}, \ and\ \bibinfo {author} {\bibfnamefont {V.}~\bibnamefont {Jacques}},\ }\bibfield  {title} {\enquote {\bibinfo {title} {Nanoscale magnetic field mapping with a single spin scanning probe magnetometer},}\ }\href@noop {} {\bibfield  {journal} {\bibinfo  {journal} {Appl. Phys. Lett.}\ }\textbf {\bibinfo {volume} {100}},\ \bibinfo {pages} {153118} (\bibinfo {year} {2012})}\BibitemShut {NoStop}%
\bibitem [{\citenamefont {Rondin}\ \emph {et~al.}(2014)\citenamefont {Rondin}, \citenamefont {Tetienne}, \citenamefont {Hingant}, \citenamefont {Roch}, \citenamefont {Maletinsky},\ and\ \citenamefont {Jacques}}]{Rondin2014}%
  \BibitemOpen
  \bibfield  {author} {\bibinfo {author} {\bibfnamefont {L.}~\bibnamefont {Rondin}}, \bibinfo {author} {\bibfnamefont {J-P.}\ \bibnamefont {Tetienne}}, \bibinfo {author} {\bibfnamefont {T.}~\bibnamefont {Hingant}}, \bibinfo {author} {\bibfnamefont {J.~F.}\ \bibnamefont {Roch}}, \bibinfo {author} {\bibfnamefont {P.}~\bibnamefont {Maletinsky}}, \ and\ \bibinfo {author} {\bibfnamefont {V.}~\bibnamefont {Jacques}},\ }\bibfield  {title} {\enquote {\bibinfo {title} {Magnetometry with nitrogen-vacancy defects in diamond},}\ }\href@noop {} {\bibfield  {journal} {\bibinfo  {journal} {Rep. Prog. Phys.}\ }\textbf {\bibinfo {volume} {77}},\ \bibinfo {pages} {056503} (\bibinfo {year} {2014})}\BibitemShut {NoStop}%
\bibitem [{\citenamefont {Barry}\ \emph {et~al.}(2020)\citenamefont {Barry}, \citenamefont {Schloss}, \citenamefont {Bauch}, \citenamefont {Turner}, \citenamefont {Hart}, \citenamefont {Pham},\ and\ \citenamefont {Walsworth}}]{Barry2020}%
  \BibitemOpen
  \bibfield  {author} {\bibinfo {author} {\bibfnamefont {J.~F.}\ \bibnamefont {Barry}}, \bibinfo {author} {\bibfnamefont {J.~M.}\ \bibnamefont {Schloss}}, \bibinfo {author} {\bibfnamefont {E.}~\bibnamefont {Bauch}}, \bibinfo {author} {\bibfnamefont {M.~J.}\ \bibnamefont {Turner}}, \bibinfo {author} {\bibfnamefont {C.~A.}\ \bibnamefont {Hart}}, \bibinfo {author} {\bibfnamefont {L.~M.}\ \bibnamefont {Pham}}, \ and\ \bibinfo {author} {\bibfnamefont {R.~L.}\ \bibnamefont {Walsworth}},\ }\bibfield  {title} {\enquote {\bibinfo {title} {Sensitivity optimization for nv-diamond magnetometry},}\ }\href@noop {} {\bibfield  {journal} {\bibinfo  {journal} {Reviews of Modern Physics}\ }\textbf {\bibinfo {volume} {92}},\ \bibinfo {pages} {015004} (\bibinfo {year} {2020})}\BibitemShut {NoStop}%
\bibitem [{\citenamefont {Kavčič}\ \emph {et~al.}(2024)\citenamefont {Kavčič}, \citenamefont {Podlipec}, \citenamefont {Krišelj}, \citenamefont {Jelen}, \citenamefont {Vella},\ and\ \citenamefont {Humar}}]{Kavi2024}%
  \BibitemOpen
  \bibfield  {author} {\bibinfo {author} {\bibfnamefont {A.}~\bibnamefont {Kavčič}}, \bibinfo {author} {\bibfnamefont {R.ok}\ \bibnamefont {Podlipec}}, \bibinfo {author} {\bibfnamefont {A.}~\bibnamefont {Krišelj}}, \bibinfo {author} {\bibfnamefont {A.}~\bibnamefont {Jelen}}, \bibinfo {author} {\bibfnamefont {D.}~\bibnamefont {Vella}}, \ and\ \bibinfo {author} {\bibfnamefont {M.}~\bibnamefont {Humar}},\ }\bibfield  {title} {\enquote {\bibinfo {title} {Intracellular biocompatible hexagonal boron nitride quantum emitters as single-photon sources and barcodes},}\ }\href@noop {} {\bibfield  {journal} {\bibinfo  {journal} {Nanoscale}\ }\textbf {\bibinfo {volume} {16}},\ \bibinfo {pages} {4691--4702} (\bibinfo {year} {2024})}\BibitemShut {NoStop}%
\bibitem [{\citenamefont {Gottscholl}\ \emph {et~al.}(2021{\natexlab{b}})\citenamefont {Gottscholl}, \citenamefont {Diez}, \citenamefont {Soltamov}, \citenamefont {Kasper}, \citenamefont {Krauße}, \citenamefont {Sperlich}, \citenamefont {Kianinia}, \citenamefont {Bradac}, \citenamefont {Aharonovich},\ and\ \citenamefont {Dyakonov}}]{Gottscholl2021b}%
  \BibitemOpen
  \bibfield  {author} {\bibinfo {author} {\bibfnamefont {A.}~\bibnamefont {Gottscholl}}, \bibinfo {author} {\bibfnamefont {M.}~\bibnamefont {Diez}}, \bibinfo {author} {\bibfnamefont {V.}~\bibnamefont {Soltamov}}, \bibinfo {author} {\bibfnamefont {C.}~\bibnamefont {Kasper}}, \bibinfo {author} {\bibfnamefont {D.}~\bibnamefont {Krauße}}, \bibinfo {author} {\bibfnamefont {A.}~\bibnamefont {Sperlich}}, \bibinfo {author} {\bibfnamefont {M.}~\bibnamefont {Kianinia}}, \bibinfo {author} {\bibfnamefont {C.}~\bibnamefont {Bradac}}, \bibinfo {author} {\bibfnamefont {I.}~\bibnamefont {Aharonovich}}, \ and\ \bibinfo {author} {\bibfnamefont {V.}~\bibnamefont {Dyakonov}},\ }\bibfield  {title} {\enquote {\bibinfo {title} {Spin defects in hbn as promising temperature, pressure and magnetic field quantum sensors},}\ }\href@noop {} {\bibfield  {journal} {\bibinfo  {journal} {Nature Communications}\ }\textbf {\bibinfo {volume} {12}},\ \bibinfo {pages} {4480} (\bibinfo {year} {2021}{\natexlab{b}})}\BibitemShut {NoStop}%
\bibitem [{\citenamefont {Liu}\ \emph {et~al.}(2025{\natexlab{a}})\citenamefont {Liu}, \citenamefont {Gong}, \citenamefont {Huang}, \citenamefont {Jin}, \citenamefont {Du}, \citenamefont {He}, \citenamefont {Janzen}, \citenamefont {Yang}, \citenamefont {Henriksen}, \citenamefont {Edgar}, \citenamefont {Galli},\ and\ \citenamefont {Zu}}]{liu2025}%
  \BibitemOpen
  \bibfield  {author} {\bibinfo {author} {\bibfnamefont {Z.}~\bibnamefont {Liu}}, \bibinfo {author} {\bibfnamefont {R.}~\bibnamefont {Gong}}, \bibinfo {author} {\bibfnamefont {B.}~\bibnamefont {Huang}}, \bibinfo {author} {\bibfnamefont {Y.}~\bibnamefont {Jin}}, \bibinfo {author} {\bibfnamefont {X.}~\bibnamefont {Du}}, \bibinfo {author} {\bibfnamefont {G.}~\bibnamefont {He}}, \bibinfo {author} {\bibfnamefont {E.}~\bibnamefont {Janzen}}, \bibinfo {author} {\bibfnamefont {L.}~\bibnamefont {Yang}}, \bibinfo {author} {\bibfnamefont {E.~A.}\ \bibnamefont {Henriksen}}, \bibinfo {author} {\bibfnamefont {J.~H.}\ \bibnamefont {Edgar}}, \bibinfo {author} {\bibfnamefont {G.}~\bibnamefont {Galli}}, \ and\ \bibinfo {author} {\bibfnamefont {C.}~\bibnamefont {Zu}},\ }\bibfield  {title} {\enquote {\bibinfo {title} {Temperature-dependent spin-phonon coupling of boron-vacancy centers in hexagonal boron nitride},}\ }\href@noop {} {\bibfield  {journal} {\bibinfo  {journal} {Physical Review B}\ }\textbf {\bibinfo {volume} {111}},\
  \bibinfo {pages} {024108} (\bibinfo {year} {2025}{\natexlab{a}})}\BibitemShut {NoStop}%
\bibitem [{\citenamefont {He}\ \emph {et~al.}(2025)\citenamefont {He}, \citenamefont {Gong}, \citenamefont {Wang}, \citenamefont {Liu}, \citenamefont {Hong}, \citenamefont {Zhang}, \citenamefont {Riofrio}, \citenamefont {Rehfuss}, \citenamefont {Chen}, \citenamefont {Yao}, \citenamefont {Poirier}, \citenamefont {Ye}, \citenamefont {Wang}, \citenamefont {Ran}, \citenamefont {Edgar}, \citenamefont {Zhang}, \citenamefont {Yao},\ and\ \citenamefont {Zu}}]{He2025}%
  \BibitemOpen
  \bibfield  {author} {\bibinfo {author} {\bibfnamefont {G.}~\bibnamefont {He}}, \bibinfo {author} {\bibfnamefont {R.}~\bibnamefont {Gong}}, \bibinfo {author} {\bibfnamefont {Z.}~\bibnamefont {Wang}}, \bibinfo {author} {\bibfnamefont {Z.}~\bibnamefont {Liu}}, \bibinfo {author} {\bibfnamefont {J.}~\bibnamefont {Hong}}, \bibinfo {author} {\bibfnamefont {T.}~\bibnamefont {Zhang}}, \bibinfo {author} {\bibfnamefont {A.~L.}\ \bibnamefont {Riofrio}}, \bibinfo {author} {\bibfnamefont {Z.}~\bibnamefont {Rehfuss}}, \bibinfo {author} {\bibfnamefont {M.}~\bibnamefont {Chen}}, \bibinfo {author} {\bibfnamefont {C.}~\bibnamefont {Yao}}, \bibinfo {author} {\bibfnamefont {T.}~\bibnamefont {Poirier}}, \bibinfo {author} {\bibfnamefont {B.}~\bibnamefont {Ye}}, \bibinfo {author} {\bibfnamefont {X.}~\bibnamefont {Wang}}, \bibinfo {author} {\bibfnamefont {S.}~\bibnamefont {Ran}}, \bibinfo {author} {\bibfnamefont {J.~H.}\ \bibnamefont {Edgar}}, \bibinfo {author} {\bibfnamefont {S.}~\bibnamefont {Zhang}}, \bibinfo {author}
  {\bibfnamefont {N.~Y.}\ \bibnamefont {Yao}}, \ and\ \bibinfo {author} {\bibfnamefont {C.}~\bibnamefont {Zu}},\ }\bibfield  {title} {\enquote {\bibinfo {title} {Probing stress and magnetism at high pressures with two-dimensional quantum sensors},}\ }\href@noop {} {\bibfield  {journal} {\bibinfo  {journal} {arXiv:2501.03319v1}\ } (\bibinfo {year} {2025})}\BibitemShut {NoStop}%
\bibitem [{\citenamefont {T.Yang}\ \emph {et~al.}(2022)\citenamefont {T.Yang}, \citenamefont {Mendelson}, \citenamefont {Li}, \citenamefont {Gottscholl}, \citenamefont {Scott}, \citenamefont {Kianinia}, \citenamefont {Dyakonov}, \citenamefont {Toth},\ and\ \citenamefont {Aharonovich}}]{Yang2022}%
  \BibitemOpen
  \bibfield  {author} {\bibinfo {author} {\bibnamefont {T.Yang}}, \bibinfo {author} {\bibfnamefont {N.}~\bibnamefont {Mendelson}}, \bibinfo {author} {\bibfnamefont {C.}~\bibnamefont {Li}}, \bibinfo {author} {\bibfnamefont {A.}~\bibnamefont {Gottscholl}}, \bibinfo {author} {\bibfnamefont {J.}~\bibnamefont {Scott}}, \bibinfo {author} {\bibfnamefont {M.}~\bibnamefont {Kianinia}}, \bibinfo {author} {\bibfnamefont {V.}~\bibnamefont {Dyakonov}}, \bibinfo {author} {\bibfnamefont {M.}~\bibnamefont {Toth}}, \ and\ \bibinfo {author} {\bibfnamefont {I.}~\bibnamefont {Aharonovich}},\ }\bibfield  {title} {\enquote {\bibinfo {title} {Spin defects in hexagonal boron nitride for strain sensing on nanopillar arrays},}\ }\href@noop {} {\bibfield  {journal} {\bibinfo  {journal} {Nanoscale}\ }\textbf {\bibinfo {volume} {14}},\ \bibinfo {pages} {5239--5244} (\bibinfo {year} {2022})}\BibitemShut {NoStop}%
\bibitem [{\citenamefont {Lyu}\ \emph {et~al.}(2022)\citenamefont {Lyu}, \citenamefont {Tan}, \citenamefont {Wu}, \citenamefont {Zhang}, \citenamefont {Zhang}, \citenamefont {Mu}, \citenamefont {Zúñiga-Pérez}, \citenamefont {Cai},\ and\ \citenamefont {Gao}}]{Lyu2022}%
  \BibitemOpen
  \bibfield  {author} {\bibinfo {author} {\bibfnamefont {X.}~\bibnamefont {Lyu}}, \bibinfo {author} {\bibfnamefont {Q.}~\bibnamefont {Tan}}, \bibinfo {author} {\bibfnamefont {L.}~\bibnamefont {Wu}}, \bibinfo {author} {\bibfnamefont {C.}~\bibnamefont {Zhang}}, \bibinfo {author} {\bibfnamefont {Z.}~\bibnamefont {Zhang}}, \bibinfo {author} {\bibfnamefont {Z.}~\bibnamefont {Mu}}, \bibinfo {author} {\bibfnamefont {J.}~\bibnamefont {Zúñiga-Pérez}}, \bibinfo {author} {\bibfnamefont {H.}~\bibnamefont {Cai}}, \ and\ \bibinfo {author} {\bibfnamefont {W.}~\bibnamefont {Gao}},\ }\bibfield  {title} {\enquote {\bibinfo {title} {Strain quantum sensing with spin defects in hexagonal boron nitride},}\ }\href@noop {} {\bibfield  {journal} {\bibinfo  {journal} {Nano Letters}\ }\textbf {\bibinfo {volume} {22}},\ \bibinfo {pages} {6553–6559} (\bibinfo {year} {2022})}\BibitemShut {NoStop}%
\bibitem [{\citenamefont {Kumar}\ \emph {et~al.}(2022)\citenamefont {Kumar}, \citenamefont {Fabre}, \citenamefont {Durand}, \citenamefont {Clua-Provost}, \citenamefont {Li}, \citenamefont {Edgar}, \citenamefont {Rougemaille}, \citenamefont {Coraux}, \citenamefont {Marie}, \citenamefont {Renucci}, \citenamefont {Robert}, \citenamefont {Robert-Philip}, \citenamefont {Gil}, \citenamefont {Cassabois}, \citenamefont {Finco},\ and\ \citenamefont {Jacques}}]{Kumar2022}%
  \BibitemOpen
  \bibfield  {author} {\bibinfo {author} {\bibfnamefont {P.}~\bibnamefont {Kumar}}, \bibinfo {author} {\bibfnamefont {F.}~\bibnamefont {Fabre}}, \bibinfo {author} {\bibfnamefont {A.}~\bibnamefont {Durand}}, \bibinfo {author} {\bibfnamefont {T.}~\bibnamefont {Clua-Provost}}, \bibinfo {author} {\bibfnamefont {J.}~\bibnamefont {Li}}, \bibinfo {author} {\bibfnamefont {J.~H.}\ \bibnamefont {Edgar}}, \bibinfo {author} {\bibfnamefont {N.}~\bibnamefont {Rougemaille}}, \bibinfo {author} {\bibfnamefont {J.}~\bibnamefont {Coraux}}, \bibinfo {author} {\bibfnamefont {X.}~\bibnamefont {Marie}}, \bibinfo {author} {\bibfnamefont {P.}~\bibnamefont {Renucci}}, \bibinfo {author} {\bibfnamefont {C.}~\bibnamefont {Robert}}, \bibinfo {author} {\bibfnamefont {I.}~\bibnamefont {Robert-Philip}}, \bibinfo {author} {\bibfnamefont {B.}~\bibnamefont {Gil}}, \bibinfo {author} {\bibfnamefont {G.}~\bibnamefont {Cassabois}}, \bibinfo {author} {\bibfnamefont {A.}~\bibnamefont {Finco}}, \ and\ \bibinfo {author} {\bibfnamefont {V.}~\bibnamefont
  {Jacques}},\ }\bibfield  {title} {\enquote {\bibinfo {title} {Magnetic imaging with spin defects in hexagonal boron nitride},}\ }\href@noop {} {\bibfield  {journal} {\bibinfo  {journal} {Physical Review Applied}\ }\textbf {\bibinfo {volume} {18}},\ \bibinfo {pages} {L061002} (\bibinfo {year} {2022})}\BibitemShut {NoStop}%
\bibitem [{\citenamefont {Healey}\ \emph {et~al.}(2023)\citenamefont {Healey}, \citenamefont {Scholten}, \citenamefont {Yang}, \citenamefont {Scott}, \citenamefont {Abrahams}, \citenamefont {Robertson}, \citenamefont {Hou}, \citenamefont {Guo}, \citenamefont {Rahman}, \citenamefont {Lu}, \citenamefont {Kianinia}, \citenamefont {Aharonovich},\ and\ \citenamefont {Tetienne}}]{Healey2023}%
  \BibitemOpen
  \bibfield  {author} {\bibinfo {author} {\bibfnamefont {A.~J.}\ \bibnamefont {Healey}}, \bibinfo {author} {\bibfnamefont {S.~C.}\ \bibnamefont {Scholten}}, \bibinfo {author} {\bibfnamefont {T.}~\bibnamefont {Yang}}, \bibinfo {author} {\bibfnamefont {J.~A.}\ \bibnamefont {Scott}}, \bibinfo {author} {\bibfnamefont {G.~J.}\ \bibnamefont {Abrahams}}, \bibinfo {author} {\bibfnamefont {I.~O.}\ \bibnamefont {Robertson}}, \bibinfo {author} {\bibfnamefont {X.~F.}\ \bibnamefont {Hou}}, \bibinfo {author} {\bibfnamefont {Y.~F.}\ \bibnamefont {Guo}}, \bibinfo {author} {\bibfnamefont {S.}~\bibnamefont {Rahman}}, \bibinfo {author} {\bibfnamefont {Y.}~\bibnamefont {Lu}}, \bibinfo {author} {\bibfnamefont {M.}~\bibnamefont {Kianinia}}, \bibinfo {author} {\bibfnamefont {I.}~\bibnamefont {Aharonovich}}, \ and\ \bibinfo {author} {\bibfnamefont {J-P.}\ \bibnamefont {Tetienne}},\ }\bibfield  {title} {\enquote {\bibinfo {title} {Quantum microscopy with van der waals heterostructures},}\ }\href@noop {} {\bibfield  {journal}
  {\bibinfo  {journal} {Nature Physics}\ }\textbf {\bibinfo {volume} {19}},\ \bibinfo {pages} {87–91} (\bibinfo {year} {2023})}\BibitemShut {NoStop}%
\bibitem [{\citenamefont {Zhou}\ \emph {et~al.}(2024)\citenamefont {Zhou}, \citenamefont {Lu}, \citenamefont {Chen}, \citenamefont {Huang}, \citenamefont {Yan}, \citenamefont {Al-matouq}, \citenamefont {Chang}, \citenamefont {Djugba}, \citenamefont {Jiang}, \citenamefont {Wang},\ and\ \citenamefont {Du}}]{Zhou2024}%
  \BibitemOpen
  \bibfield  {author} {\bibinfo {author} {\bibfnamefont {Jingcheng}\ \bibnamefont {Zhou}}, \bibinfo {author} {\bibfnamefont {Hanyi}\ \bibnamefont {Lu}}, \bibinfo {author} {\bibfnamefont {Di}~\bibnamefont {Chen}}, \bibinfo {author} {\bibfnamefont {Mengqi}\ \bibnamefont {Huang}}, \bibinfo {author} {\bibfnamefont {Gerald~Q.}\ \bibnamefont {Yan}}, \bibinfo {author} {\bibfnamefont {Faris}\ \bibnamefont {Al-matouq}}, \bibinfo {author} {\bibfnamefont {Jiu}\ \bibnamefont {Chang}}, \bibinfo {author} {\bibfnamefont {Dziga}\ \bibnamefont {Djugba}}, \bibinfo {author} {\bibfnamefont {Zhigang}\ \bibnamefont {Jiang}}, \bibinfo {author} {\bibfnamefont {Hailong}\ \bibnamefont {Wang}}, \ and\ \bibinfo {author} {\bibfnamefont {Chunhui~Rita}\ \bibnamefont {Du}},\ }\bibfield  {title} {\enquote {\bibinfo {title} {Sensing spin wave excitations by spin defects in few-layer-thick hexagonal boron nitride},}\ }\href@noop {} {\bibfield  {journal} {\bibinfo  {journal} {Science Advances}\ }\textbf {\bibinfo {volume} {10}},\ \bibinfo
  {pages} {8495} (\bibinfo {year} {2024})}\BibitemShut {NoStop}%
\bibitem [{\citenamefont {Huang}\ \emph {et~al.}(2022)\citenamefont {Huang}, \citenamefont {Zhou}, \citenamefont {Chen}, \citenamefont {Lu}, \citenamefont {McLaughlin}, \citenamefont {Li}, \citenamefont {Alghamdi}, \citenamefont {Djugba}, \citenamefont {Shi}, \citenamefont {Wang},\ and\ \citenamefont {Du}}]{Huang2022}%
  \BibitemOpen
  \bibfield  {author} {\bibinfo {author} {\bibfnamefont {M.}~\bibnamefont {Huang}}, \bibinfo {author} {\bibfnamefont {J.}~\bibnamefont {Zhou}}, \bibinfo {author} {\bibfnamefont {D.}~\bibnamefont {Chen}}, \bibinfo {author} {\bibfnamefont {H.}~\bibnamefont {Lu}}, \bibinfo {author} {\bibfnamefont {N.~J.}\ \bibnamefont {McLaughlin}}, \bibinfo {author} {\bibfnamefont {S.}~\bibnamefont {Li}}, \bibinfo {author} {\bibfnamefont {M.}~\bibnamefont {Alghamdi}}, \bibinfo {author} {\bibfnamefont {D.}~\bibnamefont {Djugba}}, \bibinfo {author} {\bibfnamefont {J.}~\bibnamefont {Shi}}, \bibinfo {author} {\bibfnamefont {H.}~\bibnamefont {Wang}}, \ and\ \bibinfo {author} {\bibfnamefont {C.~R.}\ \bibnamefont {Du}},\ }\bibfield  {title} {\enquote {\bibinfo {title} {Wide field imaging of van der waals ferromagnet fe3gete2 by spin defects in hexagonal boron nitride},}\ }\href@noop {} {\bibfield  {journal} {\bibinfo  {journal} {Nature Communications}\ }\textbf {\bibinfo {volume} {13}},\ \bibinfo {pages} {5369} (\bibinfo {year}
  {2022})}\BibitemShut {NoStop}%
\bibitem [{\citenamefont {Zang}\ \emph {et~al.}(2025)\citenamefont {Zang}, \citenamefont {Jiang}, \citenamefont {Guo}, \citenamefont {Liu}, \citenamefont {Ma}, \citenamefont {Liu}, \citenamefont {Shan}, \citenamefont {Dong}, \citenamefont {Zhang}, \citenamefont {Tangn}, \citenamefont {Chen}, \citenamefont {Guo},\ and\ \citenamefont {Sun}}]{Zang_AFM2024_VBmagnon}%
  \BibitemOpen
  \bibfield  {author} {\bibinfo {author} {\bibfnamefont {H-X.}\ \bibnamefont {Zang}}, \bibinfo {author} {\bibfnamefont {W.}~\bibnamefont {Jiang}}, \bibinfo {author} {\bibfnamefont {N-J}\ \bibnamefont {Guo}}, \bibinfo {author} {\bibfnamefont {Y.}~\bibnamefont {Liu}}, \bibinfo {author} {\bibfnamefont {M-Q.}\ \bibnamefont {Ma}}, \bibinfo {author} {\bibfnamefont {Z-W.}\ \bibnamefont {Liu}}, \bibinfo {author} {\bibfnamefont {L-K.}\ \bibnamefont {Shan}}, \bibinfo {author} {\bibfnamefont {Y.}~\bibnamefont {Dong}}, \bibinfo {author} {\bibfnamefont {S-C.}\ \bibnamefont {Zhang}}, \bibinfo {author} {\bibfnamefont {J-S.}\ \bibnamefont {Tangn}}, \bibinfo {author} {\bibfnamefont {X-D.}\ \bibnamefont {Chen}}, \bibinfo {author} {\bibfnamefont {G-C.}\ \bibnamefont {Guo}}, \ and\ \bibinfo {author} {\bibfnamefont {F-W.}\ \bibnamefont {Sun}},\ }\bibfield  {title} {\enquote {\bibinfo {title} {Detecting and imaging of magnons at nanoscale with van der waals quantum sensor},}\ }\href@noop {} {\bibfield  {journal} {\bibinfo
  {journal} {Advanced Functional Materials}\ }\textbf {\bibinfo {volume} {35}},\ \bibinfo {pages} {2412166} (\bibinfo {year} {2025})}\BibitemShut {NoStop}%
\bibitem [{\citenamefont {Mañas-Valero}\ \emph {et~al.}(2025)\citenamefont {Mañas-Valero}, \citenamefont {Doedes}, \citenamefont {Bondarenko}, \citenamefont {Borst}, \citenamefont {Kurdi}, \citenamefont {Poirier}, \citenamefont {Edgar}, \citenamefont {Jacques}, \citenamefont {Blanter},\ and\ \citenamefont {van~der Sar}}]{Manas-Valero2025}%
  \BibitemOpen
  \bibfield  {author} {\bibinfo {author} {\bibfnamefont {S.}~\bibnamefont {Mañas-Valero}}, \bibinfo {author} {\bibfnamefont {Y.~C.}\ \bibnamefont {Doedes}}, \bibinfo {author} {\bibfnamefont {A.}~\bibnamefont {Bondarenko}}, \bibinfo {author} {\bibfnamefont {M.}~\bibnamefont {Borst}}, \bibinfo {author} {\bibfnamefont {S.}~\bibnamefont {Kurdi}}, \bibinfo {author} {\bibfnamefont {T.}~\bibnamefont {Poirier}}, \bibinfo {author} {\bibfnamefont {J.~H.}\ \bibnamefont {Edgar}}, \bibinfo {author} {\bibfnamefont {V.}~\bibnamefont {Jacques}}, \bibinfo {author} {\bibfnamefont {Y.~M.}\ \bibnamefont {Blanter}}, \ and\ \bibinfo {author} {\bibfnamefont {T.}~\bibnamefont {van~der Sar}},\ }\bibfield  {title} {\enquote {\bibinfo {title} {Isofrequency spin-wave imaging using color center magnetometry for magnon spintronics},}\ }\href@noop {} {\bibfield  {journal} {\bibinfo  {journal} {arXiv:2508.18775}\ } (\bibinfo {year} {2025})}\BibitemShut {NoStop}%
\bibitem [{\citenamefont {Zhou}\ \emph {et~al.}(2023)\citenamefont {Zhou}, \citenamefont {Jiang}, \citenamefont {Liang}, \citenamefont {Ru}, \citenamefont {Bettiol},\ and\ \citenamefont {Gao}}]{Zhou_NL2023_VBsensitivity}%
  \BibitemOpen
  \bibfield  {author} {\bibinfo {author} {\bibfnamefont {F.}~\bibnamefont {Zhou}}, \bibinfo {author} {\bibfnamefont {Z.}~\bibnamefont {Jiang}}, \bibinfo {author} {\bibfnamefont {H.}~\bibnamefont {Liang}}, \bibinfo {author} {\bibfnamefont {S.}~\bibnamefont {Ru}}, \bibinfo {author} {\bibfnamefont {A.~A.}\ \bibnamefont {Bettiol}}, \ and\ \bibinfo {author} {\bibfnamefont {W.}~\bibnamefont {Gao}},\ }\bibfield  {title} {\enquote {\bibinfo {title} {Dc magnetic field sensitivity optimization of spin defects in hexagonal boron nitride},}\ }\href@noop {} {\bibfield  {journal} {\bibinfo  {journal} {Nano Letters}\ }\textbf {\bibinfo {volume} {23}},\ \bibinfo {pages} {6209--6215} (\bibinfo {year} {2023})}\BibitemShut {NoStop}%
\bibitem [{\citenamefont {Tang}\ \emph {et~al.}(2023)\citenamefont {Tang}, \citenamefont {Yin}, \citenamefont {Hart}, \citenamefont {Blanchard}, \citenamefont {Oon}, \citenamefont {Bhalerao}, \citenamefont {Schloss}, \citenamefont {Turner},\ and\ \citenamefont {Walsworth}}]{Tang2023}%
  \BibitemOpen
  \bibfield  {author} {\bibinfo {author} {\bibfnamefont {J.}~\bibnamefont {Tang}}, \bibinfo {author} {\bibfnamefont {Z.}~\bibnamefont {Yin}}, \bibinfo {author} {\bibfnamefont {C.~A.}\ \bibnamefont {Hart}}, \bibinfo {author} {\bibfnamefont {J.~W.}\ \bibnamefont {Blanchard}}, \bibinfo {author} {\bibfnamefont {J.~T.}\ \bibnamefont {Oon}}, \bibinfo {author} {\bibfnamefont {S.}~\bibnamefont {Bhalerao}}, \bibinfo {author} {\bibfnamefont {J.~M.}\ \bibnamefont {Schloss}}, \bibinfo {author} {\bibfnamefont {M.~J.}\ \bibnamefont {Turner}}, \ and\ \bibinfo {author} {\bibfnamefont {R.~L.}\ \bibnamefont {Walsworth}},\ }\bibfield  {title} {\enquote {\bibinfo {title} {Quantum diamond microscope for dynamic imaging of magnetic fields},}\ }\href@noop {} {\bibfield  {journal} {\bibinfo  {journal} {AVS Quantum Science}\ }\textbf {\bibinfo {volume} {5}},\ \bibinfo {pages} {044403} (\bibinfo {year} {2023})}\BibitemShut {NoStop}%
\bibitem [{\citenamefont {Gong}\ \emph {et~al.}(2023)\citenamefont {Gong}, \citenamefont {He}, \citenamefont {Gao}, \citenamefont {Ju}, \citenamefont {Liu}, \citenamefont {Ye}, \citenamefont {Henriksen}, \citenamefont {Li},\ and\ \citenamefont {Zu}}]{Gong2023}%
  \BibitemOpen
  \bibfield  {author} {\bibinfo {author} {\bibfnamefont {R.}~\bibnamefont {Gong}}, \bibinfo {author} {\bibfnamefont {G.}~\bibnamefont {He}}, \bibinfo {author} {\bibfnamefont {X.}~\bibnamefont {Gao}}, \bibinfo {author} {\bibfnamefont {P.}~\bibnamefont {Ju}}, \bibinfo {author} {\bibfnamefont {Z.}~\bibnamefont {Liu}}, \bibinfo {author} {\bibfnamefont {B.}~\bibnamefont {Ye}}, \bibinfo {author} {\bibfnamefont {E.~A.}\ \bibnamefont {Henriksen}}, \bibinfo {author} {\bibfnamefont {T.}~\bibnamefont {Li}}, \ and\ \bibinfo {author} {\bibfnamefont {C.}~\bibnamefont {Zu}},\ }\bibfield  {title} {\enquote {\bibinfo {title} {Coherent dynamics of strongly interacting electronic spin defects in hexagonal boron nitride},}\ }\href@noop {} {\bibfield  {journal} {\bibinfo  {journal} {Nature Communications}\ }\textbf {\bibinfo {volume} {14}},\ \bibinfo {pages} {3299} (\bibinfo {year} {2023})}\BibitemShut {NoStop}%
\bibitem [{\citenamefont {Patrickson}\ \emph {et~al.}(2024)\citenamefont {Patrickson}, \citenamefont {Baber}, \citenamefont {Ga\'{a}l}, \citenamefont {Ramsay},\ and\ \citenamefont {Luxmoore}}]{Patrickson2024}%
  \BibitemOpen
  \bibfield  {author} {\bibinfo {author} {\bibfnamefont {C.~J.}\ \bibnamefont {Patrickson}}, \bibinfo {author} {\bibfnamefont {S.}~\bibnamefont {Baber}}, \bibinfo {author} {\bibfnamefont {B.~B.}\ \bibnamefont {Ga\'{a}l}}, \bibinfo {author} {\bibfnamefont {A.~J.}\ \bibnamefont {Ramsay}}, \ and\ \bibinfo {author} {\bibfnamefont {I.~J.}\ \bibnamefont {Luxmoore}},\ }\bibfield  {title} {\enquote {\bibinfo {title} {High frequency magnetometry with an ensemble of spin qubits in hexagonal boron nitride},}\ }\href@noop {} {\bibfield  {journal} {\bibinfo  {journal} {npj Quantum Information}\ }\textbf {\bibinfo {volume} {10}},\ \bibinfo {pages} {5} (\bibinfo {year} {2024})}\BibitemShut {NoStop}%
\bibitem [{\citenamefont {Patrickson}\ \emph {et~al.}(2025)\citenamefont {Patrickson}, \citenamefont {Haemmerli}, \citenamefont {Guo}, \citenamefont {Ramsay},\ and\ \citenamefont {Luxmoore}}]{Patrickson2025}%
  \BibitemOpen
  \bibfield  {author} {\bibinfo {author} {\bibfnamefont {C.~J.}\ \bibnamefont {Patrickson}}, \bibinfo {author} {\bibfnamefont {V.}~\bibnamefont {Haemmerli}}, \bibinfo {author} {\bibfnamefont {S.}~\bibnamefont {Guo}}, \bibinfo {author} {\bibfnamefont {A.J.}\ \bibnamefont {Ramsay}}, \ and\ \bibinfo {author} {\bibfnamefont {I.J.}\ \bibnamefont {Luxmoore}},\ }\bibfield  {title} {\enquote {\bibinfo {title} {Microwave quantum heterodyne sensing using a continuous concatenated dynamical decoupling protocol},}\ }\href@noop {} {\bibfield  {journal} {\bibinfo  {journal} {Nature Communications}\ }\textbf {\bibinfo {volume} {16}},\ \bibinfo {pages} {4380} (\bibinfo {year} {2025})}\BibitemShut {NoStop}%
\bibitem [{\citenamefont {Udvarhelyi}\ \emph {et~al.}(2023)\citenamefont {Udvarhelyi}, \citenamefont {Clua-Provost}, \citenamefont {Durand}, \citenamefont {Li}, \citenamefont {Edgar}, \citenamefont {Gil}, \citenamefont {Cassabois}, \citenamefont {Jacques},\ and\ \citenamefont {Gali}}]{Udvarhelyi2023}%
  \BibitemOpen
  \bibfield  {author} {\bibinfo {author} {\bibfnamefont {P.}~\bibnamefont {Udvarhelyi}}, \bibinfo {author} {\bibfnamefont {T.}~\bibnamefont {Clua-Provost}}, \bibinfo {author} {\bibfnamefont {A.}~\bibnamefont {Durand}}, \bibinfo {author} {\bibfnamefont {J.}~\bibnamefont {Li}}, \bibinfo {author} {\bibfnamefont {J.H.}\ \bibnamefont {Edgar}}, \bibinfo {author} {\bibfnamefont {B.}~\bibnamefont {Gil}}, \bibinfo {author} {\bibfnamefont {G.}~\bibnamefont {Cassabois}}, \bibinfo {author} {\bibfnamefont {V.}~\bibnamefont {Jacques}}, \ and\ \bibinfo {author} {\bibfnamefont {Adam}\ \bibnamefont {Gali}},\ }\bibfield  {title} {\enquote {\bibinfo {title} {A planar defect spin sensor in a two-dimensional material susceptible to strain and electric fields},}\ }\href@noop {} {\bibfield  {journal} {\bibinfo  {journal} {npj Computational Materials}\ }\textbf {\bibinfo {volume} {9}},\ \bibinfo {pages} {150} (\bibinfo {year} {2023})}\BibitemShut {NoStop}%
\bibitem [{\citenamefont {Liu}\ \emph {et~al.}(2025{\natexlab{b}})\citenamefont {Liu}, \citenamefont {Gong}, \citenamefont {Huang}, \citenamefont {Jin}, \citenamefont {Du}, \citenamefont {He}, \citenamefont {Janzen}, \citenamefont {Yang}, \citenamefont {Henriksen}, \citenamefont {Edgar}, \citenamefont {Galli}, \citenamefont {Giulia},\ and\ \citenamefont {Zu}}]{Liu2025_sensing}%
  \BibitemOpen
  \bibfield  {author} {\bibinfo {author} {\bibfnamefont {Z.}~\bibnamefont {Liu}}, \bibinfo {author} {\bibfnamefont {R.}~\bibnamefont {Gong}}, \bibinfo {author} {\bibfnamefont {B.}~\bibnamefont {Huang}}, \bibinfo {author} {\bibfnamefont {Y.}~\bibnamefont {Jin}}, \bibinfo {author} {\bibfnamefont {X.}~\bibnamefont {Du}}, \bibinfo {author} {\bibfnamefont {G.}~\bibnamefont {He}}, \bibinfo {author} {\bibfnamefont {E.}~\bibnamefont {Janzen}}, \bibinfo {author} {\bibfnamefont {L.}~\bibnamefont {Yang}}, \bibinfo {author} {\bibfnamefont {E.~A.}\ \bibnamefont {Henriksen}}, \bibinfo {author} {\bibfnamefont {J.H.}\ \bibnamefont {Edgar}}, \bibinfo {author} {\bibfnamefont {G.}~\bibnamefont {Galli}}, \bibinfo {author} {\bibnamefont {Giulia}}, \ and\ \bibinfo {author} {\bibfnamefont {C.}~\bibnamefont {Zu}},\ }\bibfield  {title} {\enquote {\bibinfo {title} {Temperature-dependent spin-phonon coupling of boron-vacancy centers in hexagonal boron nitride},}\ }\href@noop {} {\bibfield  {journal} {\bibinfo  {journal} {Physical
  Review B}\ }\textbf {\bibinfo {volume} {111}},\ \bibinfo {pages} {024108} (\bibinfo {year} {2025}{\natexlab{b}})}\BibitemShut {NoStop}%
\bibitem [{\citenamefont {Mu}\ \emph {et~al.}(2025)\citenamefont {Mu}, \citenamefont {Frauni\'{e}}, \citenamefont {Durand}, \citenamefont {Cl\'{e}ment}, \citenamefont {Finco}, \citenamefont {Rouquette}, \citenamefont {Hadj-Azzem}, \citenamefont {Rougemaille}, \citenamefont {Coraux}, \citenamefont {Li}, \citenamefont {Poirier}, \citenamefont {Edgar}, \citenamefont {Gerber}, \citenamefont {Marie}, \citenamefont {Gil}, \citenamefont {Cassabois}, \citenamefont {Robert},\ and\ \citenamefont {Jacques}}]{Mu2025}%
  \BibitemOpen
  \bibfield  {author} {\bibinfo {author} {\bibfnamefont {Z.}~\bibnamefont {Mu}}, \bibinfo {author} {\bibfnamefont {J.}~\bibnamefont {Frauni\'{e}}}, \bibinfo {author} {\bibfnamefont {A.}~\bibnamefont {Durand}}, \bibinfo {author} {\bibfnamefont {S.}~\bibnamefont {Cl\'{e}ment}}, \bibinfo {author} {\bibfnamefont {A.}~\bibnamefont {Finco}}, \bibinfo {author} {\bibfnamefont {J.}~\bibnamefont {Rouquette}}, \bibinfo {author} {\bibfnamefont {A.}~\bibnamefont {Hadj-Azzem}}, \bibinfo {author} {\bibfnamefont {N.}~\bibnamefont {Rougemaille}}, \bibinfo {author} {\bibfnamefont {J.}~\bibnamefont {Coraux}}, \bibinfo {author} {\bibfnamefont {J.}~\bibnamefont {Li}}, \bibinfo {author} {\bibfnamefont {T.}~\bibnamefont {Poirier}}, \bibinfo {author} {\bibfnamefont {J.~H.}\ \bibnamefont {Edgar}}, \bibinfo {author} {\bibfnamefont {I.~C.}\ \bibnamefont {Gerber}}, \bibinfo {author} {\bibfnamefont {X.}~\bibnamefont {Marie}}, \bibinfo {author} {\bibfnamefont {B.}~\bibnamefont {Gil}}, \bibinfo {author} {\bibfnamefont {G.}~\bibnamefont
  {Cassabois}}, \bibinfo {author} {\bibfnamefont {C.}~\bibnamefont {Robert}}, \ and\ \bibinfo {author} {\bibfnamefont {V.}~\bibnamefont {Jacques}},\ }\bibfield  {title} {\enquote {\bibinfo {title} {Magnetic imaging under high pressure with a spin-based quantum sensor integrated in a van der waals heterostructure},}\ }\href@noop {} {\bibfield  {journal} {\bibinfo  {journal} {arXiv2501.03640}\ } (\bibinfo {year} {2025})}\BibitemShut {NoStop}%
\bibitem [{\citenamefont {Dolde}\ \emph {et~al.}(2011)\citenamefont {Dolde}, \citenamefont {Fedder}, \citenamefont {Doherty}, \citenamefont {N\"{o}bauer}, \citenamefont {Rempp}, \citenamefont {Balasubramanian}, \citenamefont {Wolf}, \citenamefont {Reinhard}, \citenamefont {Hollenberg}, \citenamefont {Jelezko},\ and\ \citenamefont {Wrachtrup}}]{Dolde2011}%
  \BibitemOpen
  \bibfield  {author} {\bibinfo {author} {\bibfnamefont {F.}~\bibnamefont {Dolde}}, \bibinfo {author} {\bibfnamefont {H.}~\bibnamefont {Fedder}}, \bibinfo {author} {\bibfnamefont {M.~W.}\ \bibnamefont {Doherty}}, \bibinfo {author} {\bibfnamefont {T.}~\bibnamefont {N\"{o}bauer}}, \bibinfo {author} {\bibfnamefont {F.}~\bibnamefont {Rempp}}, \bibinfo {author} {\bibfnamefont {G.}~\bibnamefont {Balasubramanian}}, \bibinfo {author} {\bibfnamefont {T.}~\bibnamefont {Wolf}}, \bibinfo {author} {\bibfnamefont {F.}~\bibnamefont {Reinhard}}, \bibinfo {author} {\bibfnamefont {L.~C.~L.}\ \bibnamefont {Hollenberg}}, \bibinfo {author} {\bibfnamefont {F.}~\bibnamefont {Jelezko}}, \ and\ \bibinfo {author} {\bibfnamefont {J.}~\bibnamefont {Wrachtrup}},\ }\bibfield  {title} {\enquote {\bibinfo {title} {Electric-field sensing using single diamond spins},}\ }\href@noop {} {\bibfield  {journal} {\bibinfo  {journal} {Nature Physics}\ }\textbf {\bibinfo {volume} {7}},\ \bibinfo {pages} {459–463} (\bibinfo {year}
  {2011})}\BibitemShut {NoStop}%
\bibitem [{\citenamefont {Durand}\ \emph {et~al.}(2023)\citenamefont {Durand}, \citenamefont {Clua-Provost}, \citenamefont {Fabre}, \citenamefont {Kumar}, \citenamefont {Li}, \citenamefont {Edgar}, \citenamefont {Udvarhelyi}, \citenamefont {Gali}, \citenamefont {Marie}, \citenamefont {Robert}, \citenamefont {G\'{e}rard}, \citenamefont {Gil}, \citenamefont {Cassabois},\ and\ \citenamefont {Jacques}}]{Durand2023}%
  \BibitemOpen
  \bibfield  {author} {\bibinfo {author} {\bibfnamefont {A.}~\bibnamefont {Durand}}, \bibinfo {author} {\bibfnamefont {T.}~\bibnamefont {Clua-Provost}}, \bibinfo {author} {\bibfnamefont {F.}~\bibnamefont {Fabre}}, \bibinfo {author} {\bibfnamefont {P.}~\bibnamefont {Kumar}}, \bibinfo {author} {\bibfnamefont {J.}~\bibnamefont {Li}}, \bibinfo {author} {\bibfnamefont {J.~H.}\ \bibnamefont {Edgar}}, \bibinfo {author} {\bibfnamefont {P.}~\bibnamefont {Udvarhelyi}}, \bibinfo {author} {\bibfnamefont {A.}~\bibnamefont {Gali}}, \bibinfo {author} {\bibfnamefont {X.}~\bibnamefont {Marie}}, \bibinfo {author} {\bibfnamefont {C.}~\bibnamefont {Robert}}, \bibinfo {author} {\bibfnamefont {J.~M.}\ \bibnamefont {G\'{e}rard}}, \bibinfo {author} {\bibfnamefont {B.}~\bibnamefont {Gil}}, \bibinfo {author} {\bibfnamefont {G.}~\bibnamefont {Cassabois}}, \ and\ \bibinfo {author} {\bibfnamefont {V.}~\bibnamefont {Jacques}},\ }\bibfield  {title} {\enquote {\bibinfo {title} {Optically active spin defects in few-layer thick hexagonal
  boron nitride},}\ }\href@noop {} {\bibfield  {journal} {\bibinfo  {journal} {Physical Review Letters}\ }\textbf {\bibinfo {volume} {131}},\ \bibinfo {pages} {116902} (\bibinfo {year} {2023})}\BibitemShut {NoStop}%
\bibitem [{\citenamefont {Sangtawesin}\ \emph {et~al.}(2019)\citenamefont {Sangtawesin}, \citenamefont {Dwyer}, \citenamefont {Srinivasan}, \citenamefont {Allred}, \citenamefont {Rodgers}, \citenamefont {Greve}, \citenamefont {Stacey}, \citenamefont {Dontschuk}, \citenamefont {O'Donnell}, \citenamefont {Hu}, \citenamefont {Evans}, \citenamefont {Jaye}, \citenamefont {Fischer}, \citenamefont {Markham}, \citenamefont {Twitchen}, \citenamefont {Park}, \citenamefont {Lukin},\ and\ \citenamefont {Leon}}]{Sangtawesin2019}%
  \BibitemOpen
  \bibfield  {author} {\bibinfo {author} {\bibfnamefont {S.}~\bibnamefont {Sangtawesin}}, \bibinfo {author} {\bibfnamefont {B.~L.}\ \bibnamefont {Dwyer}}, \bibinfo {author} {\bibfnamefont {S.}~\bibnamefont {Srinivasan}}, \bibinfo {author} {\bibfnamefont {J.~J.}\ \bibnamefont {Allred}}, \bibinfo {author} {\bibfnamefont {L.~V.H.}\ \bibnamefont {Rodgers}}, \bibinfo {author} {\bibfnamefont {K.~De}\ \bibnamefont {Greve}}, \bibinfo {author} {\bibfnamefont {A.}~\bibnamefont {Stacey}}, \bibinfo {author} {\bibfnamefont {N.}~\bibnamefont {Dontschuk}}, \bibinfo {author} {\bibfnamefont {K.~M.}\ \bibnamefont {O'Donnell}}, \bibinfo {author} {\bibfnamefont {D.}~\bibnamefont {Hu}}, \bibinfo {author} {\bibfnamefont {D.~A.}\ \bibnamefont {Evans}}, \bibinfo {author} {\bibfnamefont {C.}~\bibnamefont {Jaye}}, \bibinfo {author} {\bibfnamefont {D.~A.}\ \bibnamefont {Fischer}}, \bibinfo {author} {\bibfnamefont {M.~L.}\ \bibnamefont {Markham}}, \bibinfo {author} {\bibfnamefont {D.~J.}\ \bibnamefont {Twitchen}}, \bibinfo {author}
  {\bibfnamefont {H.}~\bibnamefont {Park}}, \bibinfo {author} {\bibfnamefont {M.~D.}\ \bibnamefont {Lukin}}, \ and\ \bibinfo {author} {\bibfnamefont {N.~P.~De}\ \bibnamefont {Leon}},\ }\bibfield  {title} {\enquote {\bibinfo {title} {Origins of diamond surface noise probed by correlating single-spin measurements with surface spectroscopy},}\ }\href@noop {} {\bibfield  {journal} {\bibinfo  {journal} {Physical Review X}\ }\textbf {\bibinfo {volume} {9}},\ \bibinfo {pages} {031052} (\bibinfo {year} {2019})}\BibitemShut {NoStop}%
\bibitem [{\citenamefont {Tetienne}\ \emph {et~al.}(2012)\citenamefont {Tetienne}, \citenamefont {Rondin}, \citenamefont {Spinicelli}, \citenamefont {Chipaux}, \citenamefont {Debuisschert}, \citenamefont {Roch},\ and\ \citenamefont {Jacques}}]{Tetienne2012}%
  \BibitemOpen
  \bibfield  {author} {\bibinfo {author} {\bibfnamefont {J-P.}\ \bibnamefont {Tetienne}}, \bibinfo {author} {\bibfnamefont {L.}~\bibnamefont {Rondin}}, \bibinfo {author} {\bibfnamefont {P.}~\bibnamefont {Spinicelli}}, \bibinfo {author} {\bibfnamefont {M.}~\bibnamefont {Chipaux}}, \bibinfo {author} {\bibfnamefont {T.}~\bibnamefont {Debuisschert}}, \bibinfo {author} {\bibfnamefont {J.~F.}\ \bibnamefont {Roch}}, \ and\ \bibinfo {author} {\bibfnamefont {V.}~\bibnamefont {Jacques}},\ }\bibfield  {title} {\enquote {\bibinfo {title} {Magnetic-field-dependent photodynamics of single nv defects in diamond: An application to qualitative all-optical magnetic imaging},}\ }\href@noop {} {\bibfield  {journal} {\bibinfo  {journal} {New Journal of Physics}\ }\textbf {\bibinfo {volume} {14}},\ \bibinfo {pages} {103033} (\bibinfo {year} {2012})}\BibitemShut {NoStop}%
\bibitem [{\citenamefont {Lee}\ \emph {et~al.}(2013)\citenamefont {Lee}, \citenamefont {Widmann}, \citenamefont {Rendler}, \citenamefont {Doherty}, \citenamefont {Babinec}, \citenamefont {Yang}, \citenamefont {Eyer}, \citenamefont {Siyushev}, \citenamefont {Hausmann}, \citenamefont {Loncar}, \citenamefont {Bodrog}, \citenamefont {Gali}, \citenamefont {Manson}, \citenamefont {Fedder},\ and\ \citenamefont {Wrachtrup}}]{Lee2013}%
  \BibitemOpen
  \bibfield  {author} {\bibinfo {author} {\bibfnamefont {S.~Y.}\ \bibnamefont {Lee}}, \bibinfo {author} {\bibfnamefont {M.}~\bibnamefont {Widmann}}, \bibinfo {author} {\bibfnamefont {T.}~\bibnamefont {Rendler}}, \bibinfo {author} {\bibfnamefont {M.~W.}\ \bibnamefont {Doherty}}, \bibinfo {author} {\bibfnamefont {T.~M.}\ \bibnamefont {Babinec}}, \bibinfo {author} {\bibfnamefont {S.}~\bibnamefont {Yang}}, \bibinfo {author} {\bibfnamefont {M.}~\bibnamefont {Eyer}}, \bibinfo {author} {\bibfnamefont {P.}~\bibnamefont {Siyushev}}, \bibinfo {author} {\bibfnamefont {B.~J.M.}\ \bibnamefont {Hausmann}}, \bibinfo {author} {\bibfnamefont {M.}~\bibnamefont {Loncar}}, \bibinfo {author} {\bibfnamefont {Z.}~\bibnamefont {Bodrog}}, \bibinfo {author} {\bibfnamefont {A.}~\bibnamefont {Gali}}, \bibinfo {author} {\bibfnamefont {N.~B.}\ \bibnamefont {Manson}}, \bibinfo {author} {\bibfnamefont {H.}~\bibnamefont {Fedder}}, \ and\ \bibinfo {author} {\bibfnamefont {J.}~\bibnamefont {Wrachtrup}},\ }\bibfield  {title} {\enquote {\bibinfo
  {title} {Readout and control of a single nuclear spin with a metastable electron spin ancilla},}\ }\href@noop {} {\bibfield  {journal} {\bibinfo  {journal} {Nature Nanotechnology}\ }\textbf {\bibinfo {volume} {8}},\ \bibinfo {pages} {487–492} (\bibinfo {year} {2013})}\BibitemShut {NoStop}%
\bibitem [{\citenamefont {Foglszinger}\ \emph {et~al.}(2022)\citenamefont {Foglszinger}, \citenamefont {Denisenko}, \citenamefont {Kornher}, \citenamefont {Schreck}, \citenamefont {Knolle}, \citenamefont {Yavkin}, \citenamefont {Kolesov},\ and\ \citenamefont {Wrachtrup}}]{Foglszinger2022}%
  \BibitemOpen
  \bibfield  {author} {\bibinfo {author} {\bibfnamefont {J.}~\bibnamefont {Foglszinger}}, \bibinfo {author} {\bibfnamefont {A.}~\bibnamefont {Denisenko}}, \bibinfo {author} {\bibfnamefont {T.}~\bibnamefont {Kornher}}, \bibinfo {author} {\bibfnamefont {M.}~\bibnamefont {Schreck}}, \bibinfo {author} {\bibfnamefont {W.}~\bibnamefont {Knolle}}, \bibinfo {author} {\bibfnamefont {B.}~\bibnamefont {Yavkin}}, \bibinfo {author} {\bibfnamefont {R.}~\bibnamefont {Kolesov}}, \ and\ \bibinfo {author} {\bibfnamefont {J.}~\bibnamefont {Wrachtrup}},\ }\bibfield  {title} {\enquote {\bibinfo {title} {Tr12 centers in diamond as a room temperature atomic scale vector magnetometer},}\ }\href@noop {} {\bibfield  {journal} {\bibinfo  {journal} {npj Quantum Information}\ }\textbf {\bibinfo {volume} {8}},\ \bibinfo {pages} {65} (\bibinfo {year} {2022})}\BibitemShut {NoStop}%
\bibitem [{\citenamefont {Gao}\ \emph {et~al.}(2024)\citenamefont {Gao}, \citenamefont {Vaidya}, \citenamefont {Dikshit}, \citenamefont {Ju}, \citenamefont {Shen}, \citenamefont {Jin}, \citenamefont {Zhang},\ and\ \citenamefont {Li}}]{Gao2024_sensing}%
  \BibitemOpen
  \bibfield  {author} {\bibinfo {author} {\bibfnamefont {Xingyu}\ \bibnamefont {Gao}}, \bibinfo {author} {\bibfnamefont {Sumukh}\ \bibnamefont {Vaidya}}, \bibinfo {author} {\bibfnamefont {Saakshi}\ \bibnamefont {Dikshit}}, \bibinfo {author} {\bibfnamefont {Peng}\ \bibnamefont {Ju}}, \bibinfo {author} {\bibfnamefont {Kunhong}\ \bibnamefont {Shen}}, \bibinfo {author} {\bibfnamefont {Yuanbin}\ \bibnamefont {Jin}}, \bibinfo {author} {\bibfnamefont {Shixiong}\ \bibnamefont {Zhang}}, \ and\ \bibinfo {author} {\bibfnamefont {Tongcang}\ \bibnamefont {Li}},\ }\bibfield  {title} {\enquote {\bibinfo {title} {Nanotube spin defects for omnidirectional magnetic field sensing},}\ }\href@noop {} {\bibfield  {journal} {\bibinfo  {journal} {Nature Communications}\ }\textbf {\bibinfo {volume} {15}},\ \bibinfo {pages} {7697} (\bibinfo {year} {2024})}\BibitemShut {NoStop}%
\bibitem [{\citenamefont {Robertson}\ \emph {et~al.}(2025)\citenamefont {Robertson}, \citenamefont {Johnson}, \citenamefont {Thalassinos}, \citenamefont {Scholten}, \citenamefont {Rietwyk}, \citenamefont {Gibson}, \citenamefont {Tetienne},\ and\ \citenamefont {Broadway}}]{Robertson2025_RF}%
  \BibitemOpen
  \bibfield  {author} {\bibinfo {author} {\bibfnamefont {I.~O.}\ \bibnamefont {Robertson}}, \bibinfo {author} {\bibfnamefont {B.~C.}\ \bibnamefont {Johnson}}, \bibinfo {author} {\bibfnamefont {G.}~\bibnamefont {Thalassinos}}, \bibinfo {author} {\bibfnamefont {S.~C.}\ \bibnamefont {Scholten}}, \bibinfo {author} {\bibfnamefont {K.~J.}\ \bibnamefont {Rietwyk}}, \bibinfo {author} {\bibfnamefont {B.~C.}\ \bibnamefont {Gibson}}, \bibinfo {author} {\bibfnamefont {J-P.}\ \bibnamefont {Tetienne}}, \ and\ \bibinfo {author} {\bibfnamefont {D.~A.}\ \bibnamefont {Broadway}},\ }\bibfield  {title} {\enquote {\bibinfo {title} {Radiofrequency receiver based on isotropic solid-state spins},}\ }\href@noop {} {\bibfield  {journal} {\bibinfo  {journal} {ACS Photonics}\ }\textbf {\bibinfo {volume} {12}},\ \bibinfo {pages} {581--587} (\bibinfo {year} {2025})}\BibitemShut {NoStop}%
\bibitem [{\citenamefont {Freysoldt}\ \emph {et~al.}(2014)\citenamefont {Freysoldt}, \citenamefont {Grabowski}, \citenamefont {Hickel}, \citenamefont {Neugebauer}, \citenamefont {Kresse}, \citenamefont {Janotti},\ and\ \citenamefont {de~Walle}}]{Freysoldt2014}%
  \BibitemOpen
  \bibfield  {author} {\bibinfo {author} {\bibfnamefont {C.}~\bibnamefont {Freysoldt}}, \bibinfo {author} {\bibfnamefont {B.}~\bibnamefont {Grabowski}}, \bibinfo {author} {\bibfnamefont {T.}~\bibnamefont {Hickel}}, \bibinfo {author} {\bibfnamefont {J.}~\bibnamefont {Neugebauer}}, \bibinfo {author} {\bibfnamefont {G.}~\bibnamefont {Kresse}}, \bibinfo {author} {\bibfnamefont {A.}~\bibnamefont {Janotti}}, \ and\ \bibinfo {author} {\bibfnamefont {C.~G.~Van}\ \bibnamefont {de~Walle}},\ }\bibfield  {title} {\enquote {\bibinfo {title} {First-principles calculations for point defects in solids},}\ }\href@noop {} {\bibfield  {journal} {\bibinfo  {journal} {Reviews of Modern Physics}\ }\textbf {\bibinfo {volume} {86}},\ \bibinfo {pages} {253--305} (\bibinfo {year} {2014})},\ \bibinfo {note} {publisher: American Physical Society}\BibitemShut {NoStop}%
\bibitem [{\citenamefont {Iv\'{a}dy}\ \emph {et~al.}(2018)\citenamefont {Iv\'{a}dy}, \citenamefont {Abrikosov},\ and\ \citenamefont {Gali}}]{Ivady2018}%
  \BibitemOpen
  \bibfield  {author} {\bibinfo {author} {\bibfnamefont {V.}~\bibnamefont {Iv\'{a}dy}}, \bibinfo {author} {\bibfnamefont {I.~A.}\ \bibnamefont {Abrikosov}}, \ and\ \bibinfo {author} {\bibfnamefont {A.}~\bibnamefont {Gali}},\ }\bibfield  {title} {\enquote {\bibinfo {title} {First principles calculation of spin-related quantities for point defect qubit research},}\ }\href@noop {} {\bibfield  {journal} {\bibinfo  {journal} {npj Computational Materials}\ }\textbf {\bibinfo {volume} {4}},\ \bibinfo {pages} {1--13} (\bibinfo {year} {2018})}\BibitemShut {NoStop}%
\bibitem [{\citenamefont {Gali}(2023)}]{gali_recent_2023}%
  \BibitemOpen
  \bibfield  {author} {\bibinfo {author} {\bibfnamefont {A.}~\bibnamefont {Gali}},\ }\bibfield  {title} {\enquote {\bibinfo {title} {Recent advances in the ab initio theory of solid-state defect qubits},}\ }\href@noop {} {\bibfield  {journal} {\bibinfo  {journal} {Nanophotonics}\ }\textbf {\bibinfo {volume} {12}},\ \bibinfo {pages} {359--397} (\bibinfo {year} {2023})}\BibitemShut {NoStop}%
\bibitem [{\citenamefont {Alkauskas}\ \emph {et~al.}(2014)\citenamefont {Alkauskas}, \citenamefont {Buckley}, \citenamefont {Awschalom},\ and\ \citenamefont {de~Walle}}]{alkauskas_first-principles_2014}%
  \BibitemOpen
  \bibfield  {author} {\bibinfo {author} {\bibfnamefont {A.}~\bibnamefont {Alkauskas}}, \bibinfo {author} {\bibfnamefont {B.~B.}\ \bibnamefont {Buckley}}, \bibinfo {author} {\bibfnamefont {D.~D.}\ \bibnamefont {Awschalom}}, \ and\ \bibinfo {author} {\bibfnamefont {C.~G.~Van}\ \bibnamefont {de~Walle}},\ }\bibfield  {title} {\enquote {\bibinfo {title} {First-principles theory of the luminescence lineshape for the triplet transition in diamond nv centres},}\ }\href@noop {} {\bibfield  {journal} {\bibinfo  {journal} {New Journal of Physics}\ }\textbf {\bibinfo {volume} {16}},\ \bibinfo {pages} {073026} (\bibinfo {year} {2014})}\BibitemShut {NoStop}%
\bibitem [{\citenamefont {Barcza}\ \emph {et~al.}(2021)\citenamefont {Barcza}, \citenamefont {Iv\'{a}dy}, \citenamefont {Szilv\'{a}si}, \citenamefont {V\"{o}r\"{o}s}, \citenamefont {Veis}, \citenamefont {Gali},\ and\ \citenamefont {Legeza}}]{barcza_dmrg_2021}%
  \BibitemOpen
  \bibfield  {author} {\bibinfo {author} {\bibfnamefont {G.}~\bibnamefont {Barcza}}, \bibinfo {author} {\bibfnamefont {V.}~\bibnamefont {Iv\'{a}dy}}, \bibinfo {author} {\bibfnamefont {T.}~\bibnamefont {Szilv\'{a}si}}, \bibinfo {author} {\bibfnamefont {M.}~\bibnamefont {V\"{o}r\"{o}s}}, \bibinfo {author} {\bibfnamefont {L.}~\bibnamefont {Veis}}, \bibinfo {author} {\bibfnamefont {A.}~\bibnamefont {Gali}}, \ and\ \bibinfo {author} {\bibfnamefont {\"{O}.}\ \bibnamefont {Legeza}},\ }\bibfield  {title} {\enquote {\bibinfo {title} {Dmrg on top of plane-wave kohn–sham orbitals: A case study of defected boron nitride},}\ }\href@noop {} {\bibfield  {journal} {\bibinfo  {journal} {Journal of Chemical Theory and Computation}\ }\textbf {\bibinfo {volume} {17}},\ \bibinfo {pages} {1143--1154} (\bibinfo {year} {2021})}\BibitemShut {NoStop}%
\bibitem [{\citenamefont {Cholsuk}\ \emph {et~al.}(2023)\citenamefont {Cholsuk}, \citenamefont {Suwanna},\ and\ \citenamefont {Vogl}}]{Cholsuk2023}%
  \BibitemOpen
  \bibfield  {author} {\bibinfo {author} {\bibfnamefont {C.}~\bibnamefont {Cholsuk}}, \bibinfo {author} {\bibfnamefont {S.}~\bibnamefont {Suwanna}}, \ and\ \bibinfo {author} {\bibfnamefont {T.}~\bibnamefont {Vogl}},\ }\bibfield  {title} {\enquote {\bibinfo {title} {Comprehensive scheme for identifying defects in solid-state quantum systems},}\ }\href@noop {} {\bibfield  {journal} {\bibinfo  {journal} {The Journal of Physical Chemistry Letters}\ }\textbf {\bibinfo {volume} {14}},\ \bibinfo {pages} {6564--6571} (\bibinfo {year} {2023})}\BibitemShut {NoStop}%
\bibitem [{\citenamefont {Davidsson}(2020)}]{davidsson_theoretical_2020}%
  \BibitemOpen
  \bibfield  {author} {\bibinfo {author} {\bibfnamefont {J.}~\bibnamefont {Davidsson}},\ }\bibfield  {title} {\enquote {\bibinfo {title} {Theoretical polarization of zero phonon lines in point defects},}\ }\href@noop {} {\bibfield  {journal} {\bibinfo  {journal} {Journal of Physics: Condensed Matter}\ }\textbf {\bibinfo {volume} {32}},\ \bibinfo {pages} {385502} (\bibinfo {year} {2020})},\ \bibinfo {note} {publisher: IOP Publishing}\BibitemShut {NoStop}%
\bibitem [{\citenamefont {Cholsuk}\ \emph {et~al.}(2024{\natexlab{b}})\citenamefont {Cholsuk}, \citenamefont {Zand}, \citenamefont {\k{C}akan},\ and\ \citenamefont {Vogl}}]{doi:10.1021/acs.jpcc.4c03404}%
  \BibitemOpen
  \bibfield  {author} {\bibinfo {author} {\bibfnamefont {C.}~\bibnamefont {Cholsuk}}, \bibinfo {author} {\bibfnamefont {A.}~\bibnamefont {Zand}}, \bibinfo {author} {\bibfnamefont {A.}~\bibnamefont {\k{C}akan}}, \ and\ \bibinfo {author} {\bibfnamefont {T.}~\bibnamefont {Vogl}},\ }\bibfield  {title} {\enquote {\bibinfo {title} {The hbn defects database: A theoretical compilation of color centers in hexagonal boron nitride},}\ }\href@noop {} {\bibfield  {journal} {\bibinfo  {journal} {The Journal of Physical Chemistry C}\ }\textbf {\bibinfo {volume} {128}},\ \bibinfo {pages} {12716--12725} (\bibinfo {year} {2024}{\natexlab{b}})}\BibitemShut {NoStop}%
\bibitem [{\citenamefont {Szász}\ \emph {et~al.}(2013)\citenamefont {Szász}, \citenamefont {Hornos}, \citenamefont {Marsman},\ and\ \citenamefont {Gali}}]{szasz_hyperfine_2013}%
  \BibitemOpen
  \bibfield  {author} {\bibinfo {author} {\bibfnamefont {K.}~\bibnamefont {Szász}}, \bibinfo {author} {\bibfnamefont {T.}~\bibnamefont {Hornos}}, \bibinfo {author} {\bibfnamefont {M.}~\bibnamefont {Marsman}}, \ and\ \bibinfo {author} {\bibfnamefont {A.}~\bibnamefont {Gali}},\ }\bibfield  {title} {\enquote {\bibinfo {title} {Hyperfine coupling of point defects in semiconductors by hybrid density functional calculations: The role of core spin polarization},}\ }\href@noop {} {\bibfield  {journal} {\bibinfo  {journal} {Physical Review B}\ }\textbf {\bibinfo {volume} {88}},\ \bibinfo {pages} {075202} (\bibinfo {year} {2013})},\ \bibinfo {note} {publisher: American Physical Society}\BibitemShut {NoStop}%
\bibitem [{\citenamefont {Takács}\ and\ \citenamefont {Ivády}(2024)}]{takacs_accurate_2024}%
  \BibitemOpen
  \bibfield  {author} {\bibinfo {author} {\bibfnamefont {István}\ \bibnamefont {Takács}}\ and\ \bibinfo {author} {\bibfnamefont {Viktor}\ \bibnamefont {Ivády}},\ }\bibfield  {title} {\enquote {\bibinfo {title} {Accurate hyperfine tensors for solid state quantum applications: case of the nv center in diamond},}\ }\href@noop {} {\bibfield  {journal} {\bibinfo  {journal} {Communications Physics}\ }\textbf {\bibinfo {volume} {7}},\ \bibinfo {pages} {1--6} (\bibinfo {year} {2024})}\BibitemShut {NoStop}%
\bibitem [{\citenamefont {Benedek}\ \emph {et~al.}(2023)\citenamefont {Benedek}, \citenamefont {Babar}, \citenamefont {Ganyecz}, \citenamefont {Szilv\'{a}si}, \citenamefont {Legeza}, \citenamefont {Barcza},\ and\ \citenamefont {Iv\'{a}dy}}]{Benedek2023}%
  \BibitemOpen
  \bibfield  {author} {\bibinfo {author} {\bibfnamefont {Z.}~\bibnamefont {Benedek}}, \bibinfo {author} {\bibfnamefont {R.}~\bibnamefont {Babar}}, \bibinfo {author} {\bibfnamefont {A.}~\bibnamefont {Ganyecz}}, \bibinfo {author} {\bibfnamefont {T.}~\bibnamefont {Szilv\'{a}si}}, \bibinfo {author} {\bibfnamefont {\"{O}.}\ \bibnamefont {Legeza}}, \bibinfo {author} {\bibfnamefont {G.}~\bibnamefont {Barcza}}, \ and\ \bibinfo {author} {\bibfnamefont {V.}~\bibnamefont {Iv\'{a}dy}},\ }\bibfield  {title} {\enquote {\bibinfo {title} {Symmetric carbon tetramers forming spin qubits in hexagonal boron nitride},}\ }\href@noop {} {\bibfield  {journal} {\bibinfo  {journal} {npj Computational Materials}\ }\textbf {\bibinfo {volume} {9}},\ \bibinfo {pages} {187} (\bibinfo {year} {2023})}\BibitemShut {NoStop}%
\bibitem [{\citenamefont {Hayee}\ \emph {et~al.}(2020)\citenamefont {Hayee}, \citenamefont {Yu}, \citenamefont {Zhang}, \citenamefont {Ciccarino}, \citenamefont {Nguyen}, \citenamefont {Marshall}, \citenamefont {Aharonovich}, \citenamefont {Vučković}, \citenamefont {Narang}, \citenamefont {Heinz},\ and\ \citenamefont {Dionne}}]{Hayee2020}%
  \BibitemOpen
  \bibfield  {author} {\bibinfo {author} {\bibfnamefont {F.}~\bibnamefont {Hayee}}, \bibinfo {author} {\bibfnamefont {L.}~\bibnamefont {Yu}}, \bibinfo {author} {\bibfnamefont {J.~Linda}\ \bibnamefont {Zhang}}, \bibinfo {author} {\bibfnamefont {C.~J.}\ \bibnamefont {Ciccarino}}, \bibinfo {author} {\bibfnamefont {M.}~\bibnamefont {Nguyen}}, \bibinfo {author} {\bibfnamefont {A.~F.}\ \bibnamefont {Marshall}}, \bibinfo {author} {\bibfnamefont {I.}~\bibnamefont {Aharonovich}}, \bibinfo {author} {\bibfnamefont {J.}~\bibnamefont {Vučković}}, \bibinfo {author} {\bibfnamefont {P.}~\bibnamefont {Narang}}, \bibinfo {author} {\bibfnamefont {T.~F.}\ \bibnamefont {Heinz}}, \ and\ \bibinfo {author} {\bibfnamefont {J.~A.}\ \bibnamefont {Dionne}},\ }\bibfield  {title} {\enquote {\bibinfo {title} {Revealing multiple classes of stable quantum emitters in hexagonal boron nitride with correlated optical and electron microscopy},}\ }\href@noop {} {\bibfield  {journal} {\bibinfo  {journal} {Nature Materials}\ }\textbf {\bibinfo
  {volume} {19}},\ \bibinfo {pages} {534–539} (\bibinfo {year} {2020})}\BibitemShut {NoStop}%
\bibitem [{\citenamefont {Singla}\ \emph {et~al.}(2024)\citenamefont {Singla}, \citenamefont {Joshi}, \citenamefont {López-Morales}, \citenamefont {Sarkar}, \citenamefont {Flick},\ and\ \citenamefont {Chakraborty}}]{Singla2024}%
  \BibitemOpen
  \bibfield  {author} {\bibinfo {author} {\bibfnamefont {S.}~\bibnamefont {Singla}}, \bibinfo {author} {\bibfnamefont {P.}~\bibnamefont {Joshi}}, \bibinfo {author} {\bibfnamefont {G.~I.}\ \bibnamefont {López-Morales}}, \bibinfo {author} {\bibfnamefont {S.}~\bibnamefont {Sarkar}}, \bibinfo {author} {\bibfnamefont {J.}~\bibnamefont {Flick}}, \ and\ \bibinfo {author} {\bibfnamefont {B.}~\bibnamefont {Chakraborty}},\ }\bibfield  {title} {\enquote {\bibinfo {title} {Probing correlation of optical emission and defect sites in hexagonal boron nitride by high-resolution stem-eels},}\ }\href@noop {} {\bibfield  {journal} {\bibinfo  {journal} {Nano Letters}\ }\textbf {\bibinfo {volume} {24}},\ \bibinfo {pages} {9212–9220} (\bibinfo {year} {2024})}\BibitemShut {NoStop}%
\bibitem [{\citenamefont {Pelliciari}\ \emph {et~al.}(2024)\citenamefont {Pelliciari}, \citenamefont {Mejia}, \citenamefont {Woods}, \citenamefont {Gu}, \citenamefont {Li}, \citenamefont {Chand}, \citenamefont {Fan}, \citenamefont {Watanabe}, \citenamefont {Taniguchi}, \citenamefont {Bisogni},\ and\ \citenamefont {Grosso}}]{Pelliciari2024}%
  \BibitemOpen
  \bibfield  {author} {\bibinfo {author} {\bibfnamefont {J.}~\bibnamefont {Pelliciari}}, \bibinfo {author} {\bibfnamefont {E.}~\bibnamefont {Mejia}}, \bibinfo {author} {\bibfnamefont {J.~M.}\ \bibnamefont {Woods}}, \bibinfo {author} {\bibfnamefont {Y.}~\bibnamefont {Gu}}, \bibinfo {author} {\bibfnamefont {J.}~\bibnamefont {Li}}, \bibinfo {author} {\bibfnamefont {S.B.}\ \bibnamefont {Chand}}, \bibinfo {author} {\bibfnamefont {S.}~\bibnamefont {Fan}}, \bibinfo {author} {\bibfnamefont {K.}~\bibnamefont {Watanabe}}, \bibinfo {author} {\bibfnamefont {T.}~\bibnamefont {Taniguchi}}, \bibinfo {author} {\bibfnamefont {V.}~\bibnamefont {Bisogni}}, \ and\ \bibinfo {author} {\bibfnamefont {G.}~\bibnamefont {Grosso}},\ }\bibfield  {title} {\enquote {\bibinfo {title} {Elementary excitations of single-photon emitters in hexagonal boron nitride},}\ }\href@noop {} {\bibfield  {journal} {\bibinfo  {journal} {Nature Materials}\ }\textbf {\bibinfo {volume} {23}},\ \bibinfo {pages} {1230} (\bibinfo {year} {2024})}\BibitemShut
  {NoStop}%
\bibitem [{\citenamefont {Ziegler}\ \emph {et~al.}(2019)\citenamefont {Ziegler}, \citenamefont {Klaiss}, \citenamefont {Blaikie}, \citenamefont {Miller}, \citenamefont {Horowitz},\ and\ \citenamefont {Alemán}}]{Ziegler2019}%
  \BibitemOpen
  \bibfield  {author} {\bibinfo {author} {\bibfnamefont {J.}~\bibnamefont {Ziegler}}, \bibinfo {author} {\bibfnamefont {R.}~\bibnamefont {Klaiss}}, \bibinfo {author} {\bibfnamefont {A.}~\bibnamefont {Blaikie}}, \bibinfo {author} {\bibfnamefont {D.}~\bibnamefont {Miller}}, \bibinfo {author} {\bibfnamefont {V.~R.}\ \bibnamefont {Horowitz}}, \ and\ \bibinfo {author} {\bibfnamefont {B.~J.}\ \bibnamefont {Alemán}},\ }\bibfield  {title} {\enquote {\bibinfo {title} {Deterministic quantum emitter formation in hexagonal boron nitride via controlled edge creation},}\ }\href@noop {} {\bibfield  {journal} {\bibinfo  {journal} {Nano Letters}\ }\textbf {\bibinfo {volume} {19}},\ \bibinfo {pages} {2121–2127} (\bibinfo {year} {2019})}\BibitemShut {NoStop}%
\bibitem [{\citenamefont {Singh}\ \emph {et~al.}(2025)\citenamefont {Singh}, \citenamefont {Robertson}, \citenamefont {Scholten}, \citenamefont {Healey}, \citenamefont {Abe}, \citenamefont {Ohshima}, \citenamefont {Tan}, \citenamefont {Kianinia}, \citenamefont {Aharonovich}, \citenamefont {Broadway.}, \citenamefont {Reineck},\ and\ \citenamefont {Tetienne}}]{Singh2025}%
  \BibitemOpen
  \bibfield  {author} {\bibinfo {author} {\bibfnamefont {P.}~\bibnamefont {Singh}}, \bibinfo {author} {\bibfnamefont {I.~O.}\ \bibnamefont {Robertson}}, \bibinfo {author} {\bibfnamefont {S.~C.}\ \bibnamefont {Scholten}}, \bibinfo {author} {\bibfnamefont {A.~J.}\ \bibnamefont {Healey}}, \bibinfo {author} {\bibfnamefont {H.}~\bibnamefont {Abe}}, \bibinfo {author} {\bibfnamefont {T.}~\bibnamefont {Ohshima}}, \bibinfo {author} {\bibfnamefont {H.~H.}\ \bibnamefont {Tan}}, \bibinfo {author} {\bibfnamefont {M.}~\bibnamefont {Kianinia}}, \bibinfo {author} {\bibfnamefont {I.}~\bibnamefont {Aharonovich}}, \bibinfo {author} {\bibfnamefont {D.~A.}\ \bibnamefont {Broadway.}}, \bibinfo {author} {\bibfnamefont {P.}~\bibnamefont {Reineck}}, \ and\ \bibinfo {author} {\bibfnamefont {J-P.}\ \bibnamefont {Tetienne}},\ }\bibfield  {title} {\enquote {\bibinfo {title} {Violet to near-infrared optical addressing of spin pairs in hexagonal boron nitride},}\ }\href@noop {} {\bibfield  {journal} {\bibinfo  {journal} {Advanced
  Materials}\ }\textbf {\bibinfo {volume} {37}},\ \bibinfo {pages} {2414846} (\bibinfo {year} {2025})}\BibitemShut {NoStop}%
\end{thebibliography}%
\end{document}